\documentclass[11pt,a4paper]{article}
\pdfoutput=1
\usepackage{jheppub_mod}
\usepackage[T1]{fontenc}
\usepackage{inputenc}
\usepackage{amsmath,amssymb,amsfonts}
\usepackage{bm}
\usepackage{pifont}
\usepackage[dvipsnames]{xcolor}
\usepackage{mathtools}
\usepackage{float}
\usepackage{subcaption}
\usepackage{mathrsfs}
\usepackage{graphicx}

\newcommand{\symfootnote}[1]{%
\let\oldthefootnote=\thefootnote%
\stepcounter{mpfootnote}%
\addtocounter{footnote}{-1}%
\renewcommand{\thefootnote}{\fnsymbol{mpfootnote}}%
\footnote{#1}%
\let\thefootnote=\oldthefootnote%
}

\newcommand{\beq}{\begin{equation}}
\newcommand{\eeq}{\end{equation}}
\newcommand{\bes}{\begin{subequations}}
\newcommand{\ees}{\end{subequations}}

\newcommand{\e}{\epsilon}
\renewcommand{\d}{\partial}
\newcommand{\be}[1]{\begin{equation}\label{#1}}
\newcommand{\ee}{\end{equation}}
\newcommand{\bea}[1]{\begin{eqnarray}\label{#1}}
\newcommand{\eea}{\end{eqnarray}}
\newcommand{\refb}[1]{(\ref{#1})}

\renewcommand{\O}{{\mathcal{O}}}
\renewcommand{\L}{{\mathcal{L}}}

\newcommand{\bL}{\bar{{\mathcal{L}}}}
\newcommand{\z}{{\bar z}}
\newcommand{\h}{{\bar h}}
\renewcommand{\>}{\rangle}
\newcommand{\<}{\langle}
\newcommand{\w}{\omega}

\newcommand{\eps}{\varepsilon}

\newcommand{\de}{\delta}

\newcommand{\rw}{\rightarrow}

\newcommand{\g}{\gamma}
\newcommand{\D}{\Delta}

\renewcommand{\a}{\alpha}
\newcommand{\ta}{\tilde{\alpha}}

\renewcommand{\b}{\beta}
\renewcommand{\t}{\tau}

\def\cP{\mathcal{P}}

\newcommand{\gb}{\hat{\gamma}}
\newcommand{\gc}{\tilde{\gamma}}

\newcommand{\ts}{\tilde{\Sigma}}

\def\cJ{\mathcal{J}}
\def\cL{\mathcal{L}}

\def\cP{\mathcal{P}}

\def\cH{\mathcal{H}}

\def\be{\begin{equation}}
\def\ee{\end{equation}}
\def\ben{\begin{equation*}}
\def\een{\end{equation*}}

\definecolor{green}{rgb}{0.1,0.8,0.2}

\DeclareUnicodeCharacter{221E}{\ensuremath{\infty}}

\allowdisplaybreaks

\title{The Carrollian Kaleidoscope}

\author[a]{\small{Arjun Bagchi,}}
\author[b,c]{\small{Aritra Banerjee,}} 
\author[d]{\small{Prateksh Dhivakar,}}
\author[a]{\small{Saikat Mondal,}} 
\author[e,f]{\small{and Ashish Shukla}} 

\affiliation[a]{Indian Institute of Technology Kanpur, Kanpur 208016, India.} 
\affiliation[b]{Birla Institute of Technology and Science, Pilani Campus, Pilani, Rajasthan 333031, India.}
\affiliation[c]{Asia Pacific Center for Theoretical Physics, Postech, Pohang 37673, Korea.}
\affiliation[d]{Department of Physics and Astronomy, University of Victoria, Victoria, BC V8W 2Y2, Canada.}
\affiliation[e]{Theory Division, Saha Institute of Nuclear Physics, 1/AF Bidhan Nagar, Kolkata 700064, India.}
\affiliation[f]{Homi Bhabha National Institute, Anushakti Nagar, Mumbai 400094, India.}

\emailAdd{abagchi@iitk.ac.in}
\emailAdd{aritra.banerjee@pilani.bits-pilani.ac.in}
\emailAdd{pratekshd@uvic.ca}
\emailAdd{saikatmd@iitk.ac.in}
\emailAdd{ashish.shukla@saha.ac.in}

\abstract{The Carroll group arises in the vanishing speed of light limit of the Poincaré group and was initially discarded as just a mathematical curiosity. However, recent developments have proved otherwise. Carroll and conformal Carroll symmetries are now ubiquitous, appearing in diverse physical phenomena starting from condensed matter physics to quantum gravity. This review aims to provide the reader a gateway into this fast-developing field. After an introduction and setting the stage with basics of the symmetry in question, we detail the construction of Carrollian and Carrollian Conformal field theories (CCFT). 

\medskip

We then focus on applications. By far the most popular of these applications is in the context of the construction of holography in asymptotically flat spacetimes (AFS) in terms of a co-dimension one dual CCFT. We review the early work on AFS$_3$/CCFT$_2$ before delving into an in-depth analysis for the construction of the dual to 4D AFS. 

\medskip

Two other important sets of applications are in hydrodynamics and in condensed matter physics, which we discuss in detail. Carroll hydrodynamics is introduced as the $c\to0$ limit of relativistic hydrodynamics first and then reconstructed from a symmetry based approach. Relations to ultrarelativistic flows and connections to the quark-gluon plasma are discussed with concrete examples of the Bjorken and Gubser flow models. In condensed matter applications, we cover connections to fractons, flat bands, and phase separation in Luttinger liquid models. 
To conclude, we give very brief outlines of other topics of interest including string theory and black hole horizons.\\

~~~~~~~~~~~~~~~~~~~~~~~~~~~~~\emph{\textit{Invited Review for European Physical Journal C.}}
}

\medskip

\begin{document}
\maketitle

\newpage

\section{Introduction}

The term kaleidoscope is derived from the ancient Greek words kalos, eidos and skopeo which mean respectively beautiful, shape and to observe. Hence kaleidoscope literally translates to ``the observation of beautiful forms.'' And indeed this review is about the observation of beautiful forms that are associated with a novel symmetry, which to begin with was thought to be exotic and unphysical, but now is being discovered in numerous very different contexts. 

\subsection{Symmetries are a theorist's best friend}

Symmetries are a theorist's best friend (with due apologies to Leo Robin and Marilyn Monroe!). The repeated emergence of relativistic conformal symmetry in various physical systems of interest, starting from the theory of phase transitions in statistical physics to the Cosmic Microwave Background Radiation in cosmology and ultimately to the worldsheet of string theory, made the ideas and methods of relativistic conformal field theory (CFT) one of the principle tools in a theoretical physicist's toolkit. A particularly powerful tool in this toolkit is the infinite dimensional Virasoro algebra, two copies of which make up the symmetry of two dimensional CFTs. The infinite dimensional symmetry sometimes leads to completely solvable integrable systems, and also helps in computing various quantities like correlation functions from symmetries alone, without reference to any underlying Lagrangian.

\medskip

Now there is a ``new'' player in town. The symmetry we are interested in is {\em Carrollian} symmetry. First discovered and christened by Jean-Marc L\'{e}vy-Leblond \cite{LBLL}, and independently by Sen Gupta \cite{SenGupta:1966qer}, in the 1960s (and hence the quotations on the new in the first sentence of the paragraph) the symmetry group historically arose from taking a vanishing speed of light limit on the relativistic Poincaré group. The symmetry is named after Lewis Carroll in a rather wonderful acknowledgment of the Wonderland that he created in his famous works ``Alice's Adventures in Wonderland'' and ``Through the Looking Glass.'' The vanishing speed of light, as we will see in the following sections, means that movement becomes impossible{\footnote{This argument turns out to be a little too quick. We will see some interesting counter-examples.}} in the Carrollian world. This is very similar to the land of the Red Queen, where Alice runs and runs and discovers she is at the exact place where she started from.

\medskip

\begin{figure}[t]
    \centering
    \includegraphics[scale=0.42]{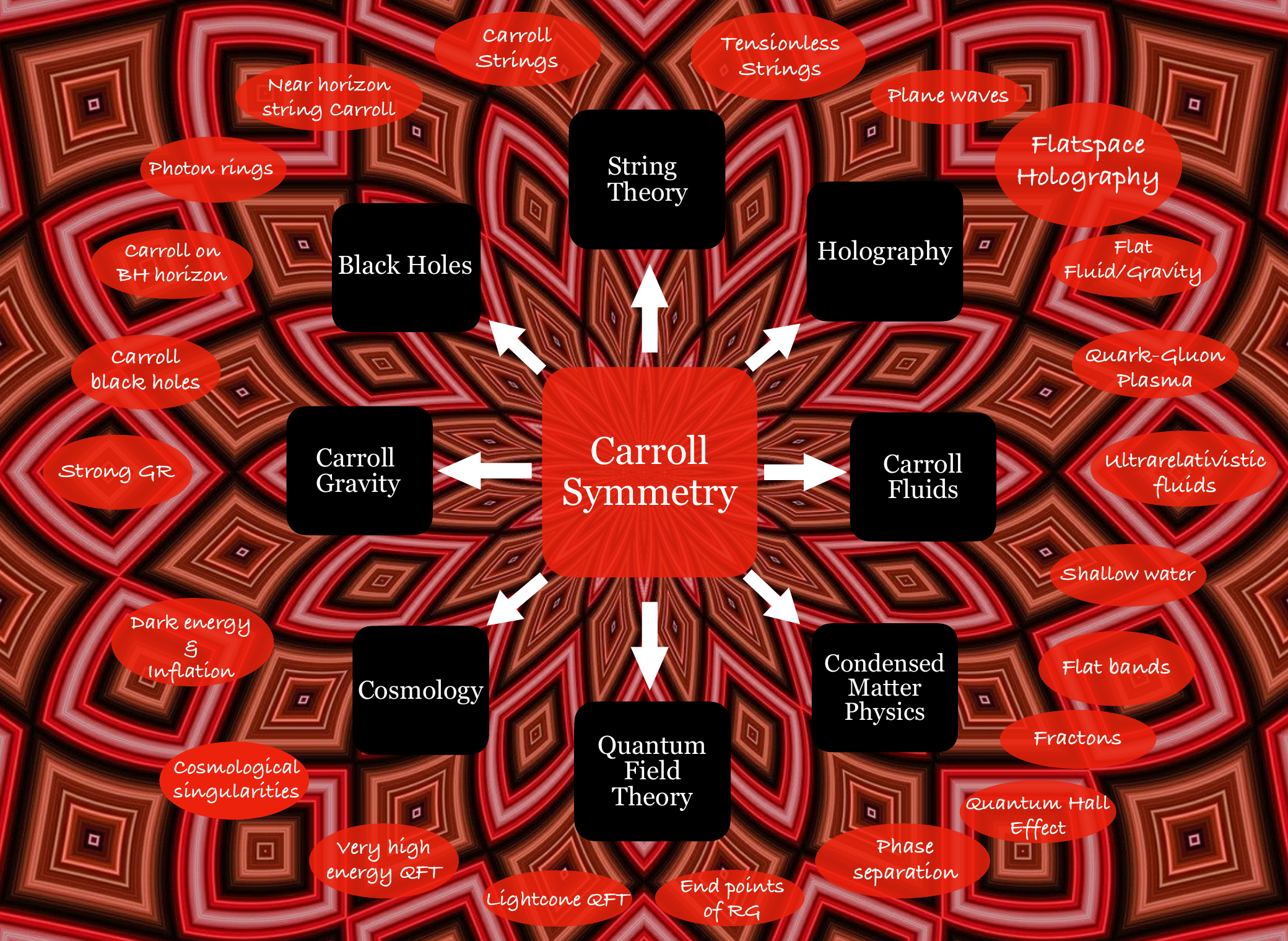}
    \caption{Through the Looking Glass: The Carrollian Kaleidoscope!}
    \label{fig:Avatar}
\end{figure}

Carrollian symmetries and their conformal cousins are showing up in various a priori unrelated corners of theoretical physics. After spending more than four decades in hibernation, more than twice the time Rip van Winkle spent sleeping, Carrollian symmetries have found a new lease on life, especially in the context of holography of asymptotically flat spacetimes (AFS) \cite{Bagchi:2010zz, Bagchi:2012cy, Duval:2014uva, Bagchi:2016bcd, Bagchi:2022emh, Donnay:2022aba}. More recently, numerous examples have come up where Carroll symmetries play a central role. Some of these arise in the study of ultra-relativistic hydrodynamics relevant for the quark-gluon plasma \cite{Bagchi:2023ysc, Bagchi:2023rwd}, and specific systems in condensed matter physics like fractons \cite{Bidussi:2021nmp} and systems with flat bands \cite{Bagchi:2022eui}. Our review will particularly focus on applications of Carrollian symmetries to flat holography and to condensed matter physics and hydrodynamics (we will be providing detailed references in the appropriate sections). Apart from these applications, Carroll symmetries have appeared in the context of black hole physics as symmetries on event horizons \cite{Penna:2018gfx, Donnay:2019jiz}, in cosmology in the context of inflation and dark energy \cite{deBoer:2021jej}, and in string theory on the tensionless limit of strings \cite{Bagchi:2013bga, Bagchi:2015nca} and strings near black holes \cite{Bagchi:2023cfp, Bagchi:2024rje}. We will briefly touch up on these and other emerging applications in the penultimate section of this review (and will provide references to the literature there). To provide the reader with a picture of the all-encompassing nature of Carrollian symmetries, in Fig.~\ref{fig:Avatar}, we present the Carrollian Kaleidoscope. It is red to reflect the land of the Red Queen, and perhaps also the eye of Sauron!  

\subsection{Carroll: Ultra-local, Ultra-high energy, and Ultra-relativistic}
\label{ssec:intro2}

Before we begin in earnest, let us offer some general remarks about why we may be interested in this peculiar limit at all. Although Nature is fundamentally relativistic, the world of our day-to-day observations is, to a large extent, non-relativistic. Building roads and bridges do not need special or general relativistic inputs. Newton's laws do just fine. Many real life condensed matter systems and hydrodynamic systems are inherently dictated by Galilean symmetries. The Galilean world can be reached from the relativistic one by sending the speed of light $c \to \infty$. Algebraically, this is an In{\"o}n{\"u}-Wigner contraction of the Poincaré algebra. We will discuss this limiting procedure in some detail at the very beginning of the main body of this review. 

\medskip

Galilean symmetries thus become relevant when things are moving very slowly with respect to the speed of light. This can be regarded as a very low energy limit of any relativistic system which forms a closed subsector of the relativistic theory governed by a different spacetime algebra. The sector can be defined by the Galilean symmetries without any reference to any superstructure or a parent relativistic theory. 

\medskip

What then is the meaning of a contraction of the Poincaré algebra where the speed of light $c \to 0$ instead? Since this seems like the diametrically opposite limit, this perhaps is the {\em very high energy} sector of a relativistic theory, which again is a closed subsector governed now by the Carrollian algebra. We will see that this is a point of view that we can make precise in the formulation of Carrollian theories and their various applications. The limit $c \to 0$ can be both {\em ultra-local} and {\em ultra-relativistic} depending on the context in which this is used. Since the lightcones close up in the $c \to 0$ limit, as we have remarked before, movement becomes impossible and hence the {\em ultra-local} nature manifests itself. We shall also see that zooming near to a null surface also results in a Carrollian algebra and hence this is associated with an ultra-relativistic $v\to c_{bulk}$ limit, where $c_{bulk}$ is the speed of light in a higher dimensional spacetime. This goes hand in hand with the observation that Carroll structures arise generically on null hypersurfaces. 

\medskip 

\ding{112} {\underline{\em{Carroll at Black Hole Event Horizons}}

\smallskip

To clarify our point above, let us take recourse to a simple example. Consider the metric of a Schwarzschild black hole in usual Schwarzschild coordinates:
\begin{align}
    ds^2 = - \left(1-\frac{2M}{r}\right) dt^2 + \left(1-\frac{2M}{r}\right)^{-1} dr^2 + r^2 d\Omega^2. 
\end{align}
Here $M$ is the mass of the black hole and $r=2M$ is the location of the event horizon. Consider the $r-t$ plane and the slope of the lightcones in there. We get
\begin{align}
    \frac{dr}{dt}= \pm \left(1-\frac{2M}{r}\right).
\end{align}
At $r\to \infty$, the lightcones are 45 degrees and the speed of light  in the co-dimension one hypersurface at $r\to \infty$, $c_{r\to\infty}$ is $1$. As one moves into the bulk of spacetime, at distances $r_1>r_2> \ldots >r_n$, the speeds of light on these lower dimensional surfaces decrease $1 > c_{r_1} > c_{r_2} > \ldots > c_{r_n}$. These are the speeds of light for an infalling observer as measured by the asymptotic observer at infinity. Of course, any finite speed of light can be rescaled back to one and physics remains unchanged. But as $r\to2M$, things begin to change. At the event horizon, the speed of light $c_{r=2M} = 0$. For the asymptotic observer, the infalling observer remains stuck to the event horizon forever. This is the Carroll structure emerging in the physics of the description of the infalling observer by the asymptotic observer. The point to remember is that there is always a bulk speed of light in this story which remains untouched. The co-dimension one speeds of light $c_r$ change and go to zero. This is related to the distance from the event horizon which plays the role of this effective lower dimensional speed of light. $r=2M$ is a null surface. Non-null hypersurfaces at some $r=r_n$ are boosted to a null hypersurface in the process of taking $c_r\to0$ (equivalently $r\to2M$) and hence the limit is ultra-relativistic. Hence $c_r \to 0$ is equivalent to $v/c_{bulk} \to 1$.

\begin{figure}[t]
\centering
\includegraphics[scale=1]{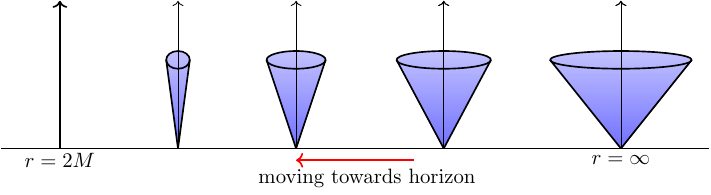}
\caption{Emergent Carroll physics as one approaches the Schwarzschild horizon.}
\end{figure}

\medskip

One may complain that this is a very observer and hence coordinate dependent picture. While that is true, we have done this to clarify the nomenclature ``ultra-relativistic'' which seems to be a source of much confusion even with researchers in the field. Also, the nature of the surface $r=2M$ is null irrespective of the coordinates used. So the emergence of Carroll structures on the event horizons of black holes is a coordinate independent statement. The approach to the event horizon would depend on the choice of coordinates, which in turn dictates the manner in which the codimension one lightcones close. But at the event horizon, the geometry of the null surface is dictated by Carroll physics. We will have a more mathematical way of understanding how Carroll can be viewed as ultra-relativistic in Sec. \ref{section2} (for the restless, see Sec. \ref{nullc}). 

\medskip

\ding{112} {\underline{\em{Effective speeds of light and null surfaces}}

\smallskip

In the above, the speeds of light $c_r$ in the co-dimension one constant $r$ surfaces are related to the distance from the horizon in the bulk. The bulk radial direction hence acts as an effective speed of light. This can be made more precise and we do so near the end of our review in Sec.~\ref{penult}. 

\medskip

In various applications of interest, there are various different quantities that act as the effective speed of light which goes to zero in the limit. For the black hole, as just discussed, it is the distance from the event horizon. For a null string, where Carroll structures arise on the worldsheet, this would be the string tension. In the case of flatspace holography, it is the inverse of the AdS radius which gets related to the speed of light in the dual CFT as the limit to flatspace is taken. In various condensed matter systems, the role of the speed of light can be played by characteristic velocities in the theory, like the Fermi velocity. 

\medskip

The ubiquitous nature of Carroll symmetries is intimately related to the fact that these symmetries show up on generic null surfaces. Hence anywhere one encounters a null surface, Carrollian symmetries will inevitably play an important role. This means that we will keep adding bubbles to the Kaleidoscope in Fig.~\ref{fig:Avatar}. The two most important null surfaces in physics are the event horizons of generic black holes and the asymptotic null infinity of Minkowski spacetimes where flatspace scattering in and out states are defined. It is for this reason Carroll symmetries are central to the understanding of black holes and holography in asymptotically flat spacetimes, as we go on to review in detail in this work. 

\subsection{A trilogy in four parts}

This is a rapidly developing field and the purpose of this review is to provide readers an entry point into the subject. We have strived to make this a self contained piece of work, i.e. the reader hopefully will not have to keep going to various other articles to understand the basic material covered in this review. But at the same time, we wanted to make this a place where one can readily look up references to older work and if required go back to them. We have, in places, also tried to give a more modern flavour to some of the older literature including some of the authors' own works. 

\medskip

A theoretical physicist's dream is to answer the Ultimate Question of Life, the Universe and Everything \cite{Adams:1979}. Of course, we cannot do better than 42 and we will not suggest that the answer is Carroll.\footnote{Hopefully that will take less than seven and a half million years of thought.} However,  our review would be a trilogy in four parts. 

\medskip

Part I will lay down the basics of Carroll symmetry. Here we will study contractions of the Poincaré algebra and differentiate between the more familiar Galilean algebra and the Carroll algebra. We will construct the geometry behind the symmetry and also study representations of the algebra. We will then look at conformal extensions. Other sections in this part would be devoted to constructing quantum field theories and conformal field theories with Carroll symmetry. 

\medskip

Part II will focus on applications of Carroll symmetry to holography in asymptotically flat spacetimes (AFS), which is, to date, the most popular field in Carrollian physics. After discussing generalities of holography in AFS and the role of Carrollian Conformal Field Theories (CCFT) as field theory duals of gravity in AFS, we will devote a section primarily to older literature on AFS$_3$/CCFT$_2$ that spans the last decade. The connection between 4D AFS and 3D CCFTs would be outlined in three other sections that follow. 

\medskip

Part III will focus on emerging connections to condensed matter physics and hydrodynamics. We will detail how, rather surprisingly, Carroll dynamics appears in seemingly unrelated branches of physics.

\medskip

We conclude in Part IV. Here, for completeness, we will devote a section for brief summaries of applications of Carrollian symmetries we don't deal with in detail in the main text. We end with some concluding remarks.  

\medskip

As with everything in life in general, this review is not unbiased. Try as we may to do otherwise, we will be partial to our way of understanding the subject and to the contributions we have made. There are some differing interpretations and points of view. We will try to list other contributions which we think are important after each part of the review. 

\newpage

\part{Carroll Basics}

\bigskip

\bigskip

\bigskip

\bigskip

\section*{Outline of Part I}

In the first part of our review, we focus on the essentials: aspects of symmetry and how to build theories with these symmetries. The first section, Sec.~\ref{section2}, outlines from very basics, what Carrollian and Carrollian Conformal symmetries are. Sec.~\ref{sec:carfieldtheories} then provides the first constructions of Carroll field theories where we address various spins $s =0, 1/2, 1$. 

\medskip

The focus shifts to Carrollian CFTs in Sec.~\ref{ssec:2dccfts}, which is entirely devoted to dimension $D=2$. This section builds on CCFTs in lines parallel to usual relativistic 2D CFTs and in various places uses the map from the relativistic to the Carrollian world. We also learn how to deform relativistic 2D CFTs with marginal operators to obtain CCFTs. Sec.~\ref{sec:carrfields4} is devoted to CCFTs in dimensions $D>2$. After initial discussions of the representation theory for arbitrary dimensions, we concentrate on $D=3$ and obtain correlation functions of CCFTs here, discovering two branches of Ward identities. 

\medskip

As will be a feature through out the review, each section would end with a subsection called ``Pointers to the literature'', where we attempt to provide the reader with other important developments in subject matter dealt with in the particular section. We will try to be detailed in our references here, but by no means claim to be exhaustive.  

\newpage

\section{Carrollian symmetries: a detailed look}
\label{section2}
This section will be devoted to a more detailed introduction to the Carroll group and its symmetry structure. We will begin with preliminaries, especially how one obtains these symmetries as an In{\"o}n{\"u}-Wigner contraction \cite{inonu1953contraction} of Poincaré symmetries. We will then discuss the group structure associated to Carroll symmetries and contrast this to the better known Galilean group.  

\subsection{Pre-preliminaries: Poincaré symmetries}\label{sec2}
We begin by setting up notation. As is of course very well known, Poincaré group is the isometry group of the Minkowski spacetime. The group consists of spacetime translations and Lorentz transformations, keeping the proper distance invariant. The action of these transformations in the spacetime coordinates is given by
\begin{equation}
    x^{\mu} \to x'^{\mu} = \Lambda^{\mu}_{\,\,\nu}\,x^{\nu}  + a^{\mu}
\end{equation}
where $a$ and $\Lambda$ are the parameters of the spacetime translations and Lorentz transformations respectively. If a general transformation is indicated by $(\Lambda, a)$, the group composition rule for the Poincaré group is given by
\begin{equation}
    (\Lambda_2,a_2) \cdot (\Lambda_1,a_1) = (\Lambda_2\Lambda_1 , a_2 + \Lambda_2a_1).
\end{equation}
The structure of the group has the form of a semi-direct product given by,
\begin{equation}
    ISO(1,d) = SO(1,d) \ltimes \mathbb{R}^{1,d}.
\end{equation}
The Lie algebra $\mathfrak{iso}(1,d)$ of the Poincaré group is spanned by translations $P_{\mu}$  and Lorentz generators $M_{\mu\nu}$ following the commutation relations
\begin{subequations}\label{Poinalg}
  \begin{align}
        [M_{\mu\nu}, M_{\rho\sigma}] &= \eta_{\mu\rho}M_{\nu\sigma} - \eta_{\mu\sigma}M_{\nu\rho} + \eta_{\nu\sigma}M_{\mu\rho} -\eta_{\nu\rho}M_{\mu\sigma},\\
        [M_{\mu\nu}, P_{\rho}] &= \eta_{\mu\rho}P_{\nu} - \eta_{\nu\rho}P_{\mu}, \quad [P_{\mu},P_{\nu}] = 0.
    \end{align}  
\end{subequations}    
Poincaré algebra has two Casimirs given by
\begin{eqnarray}
    \mathbb{C}_2 = P^{\mu}P_{\mu}\,,\qquad 
    \mathbb{C}_4 = W^{\mu}W_{\mu}\,.
\end{eqnarray}
Here $W_{\mu}$ is the Pauli-Lubansky (pseudo-)vector defined as $W = \star (P \wedge M)$. 
%$W_{\mu} = \frac{1}{2}\epsilon_{\mu\nu\rho\sigma}P^{\nu}M^{\rho\sigma}$. 
The Casimirs $\mathbb{C}_2$ and $\mathbb{C}_4$ are associated with the mass and spin of the states in the irreducible representations of the Poincaré group.

\medskip
 
Again, to set the stage for the later non-Lorentzian discussions, let us also remind the reader of some elementary geometric background. An infinitesimal co-ordinate transformation $x^{\mu} \to x'^{\mu} = x^{\mu} + \epsilon \xi^{\mu}$ is an isometry if it leaves the metric invariant i.e. $\pounds_{\xi}g_{\mu\nu} = 0$, i.e. the
Lie derivative of metric vanishes along the vector field $\xi$. Vector fields which satisfy this equation are called Killing vectors and the ensuing algebra of the Killing vectors is the Lie algebra of the isometry group of the metric. Considering the Minkowski spacetime with the standard metric $\eta_{\mu\nu} = \text{diag}(-1,1,...,1)$ and taking the co-ordinates $x^{\mu}$ as a basis for the Killing vectors, we get the generators as 
\begin{equation}
    P_{\mu} = \partial_{\mu}\,,\qquad M_{\mu\nu} = x_{\mu}\partial_{\nu} - x_{\nu}\partial_{\mu}.
\end{equation}

Generalization to the Conformal group is straightforward. By definition, a conformal transformation keeps the metric invariant up to a scale. The conformal group has the Poincaré group as a subgroup and in addition includes dilatations $(D)$ and special conformal transformations $(K_{\mu})$ which lead to additional non-zero commutation relations
\begin{subequations}
  \begin{align}
        [D,P_{\mu}] &= P_{\mu}, \qquad [K_{\mu},P_{\nu}] = 2\left(\eta_{\mu\nu}D - M_{\mu\nu}\right),\\
        [D,K_{\mu}] &= -K_{\mu}, \quad  
        [M_{\mu\nu}, K_{\rho}] = \left(\eta_{\mu\rho}K_{\nu} - \eta_{\nu\rho}K_{\mu}\right).
    \end{align}  
\end{subequations}
These generators can be found by solving the conformal Killing equation $\mathcal{L}_{\xi}g_{\mu\nu} = \Omega(x)g_{\mu\nu}$ for a flat background. The differential forms of the additional generators in the coordinate basis are
\begin{eqnarray}
    D = x^{\mu}\partial_{\mu}\,, \quad K_{\mu} = 2x_{\mu}x^{\nu}\partial_{\nu} - x^2\partial_{\mu}.
\end{eqnarray}

\subsection{Carroll from Poincaré}
\label{ssec:carpoincare}

Carroll group arises via an Inönü–Wigner contraction from the Poincaré group \cite{LBLL, SenGupta:1966qer}. This is performed by taking the singular limit, which is the speed of light $c \rightarrow 0$ limit of the Poincaré group. This is also equivalent to the co-ordinate scaling \cite{Bagchi:2012cy}: 
    \begin{equation}\label{carrscale}
		x^i\rw  x^i ,~~\quad t\rw \epsilon t, ~~\quad \epsilon\rw 0
    \end{equation}
where superscript $i$ denotes the number of spatial dimensions. Under this limit, the Carroll generators can be defined by rescaling the Poincaré generators in the following way 
\be{}
H \equiv \epsilon P_0^{\text{P}}, \,~~\quad P_i \equiv P_i^{\text{P}}, \,~~\quad  C_i \equiv \epsilon J_{0i}^{\text{P}}, \,~~\quad  J_{ij}\equiv J_{ij}^{\text{P}},
\ee
where the superscript ${\text{P}}$ refers to the Poincaré generators. In the position space basis, the Carroll generators take the form:
\begin{equation}\label{Car-gen}
H=\partial_{t} ,~\quad P_i=\partial_{i} ,~\quad C_i=x_i\partial_{t} ,~\quad J_{ij}=x_i\partial_{j}-x_j\partial_{i}.
\end{equation}
Here $H, P_i, C_i, J_{ij}$ are time translation, spatial translations, Carroll boosts and spatial rotations generators respectively. These generators act on space-time coordinate as
\be{}\label{cartrans}		
t{'}=t+a -\vec{v}.\vec{x}, ~~\quad \vec{x}{'}=\boldsymbol{R}\vec{x}+\vec{b}
\ee
generating Carroll transformations. Here the parameters of the group $(a, \vec{b}, \vec{v}, \boldsymbol{R})$ describes time and space translations, Carroll boosts and $SO(d)$ rotations respectively. We see that under Carroll boosts, the time coordinate changes but the spatial ones remain invariant. In other words, Carroll symmetry group can be thought of as symmetry group of absolute space, where time is relative. The group composition rule for the Carroll group is given by
\begin{equation}
    (a_2, \vec{b}_2, \vec{v}_2, \boldsymbol{R}_2) \cdot (a_1, \vec{b}_1, \vec{v}_1, \boldsymbol{R}_1) = (a_2+a_1-\vec{v}_2\cdot \vec{b}_1, \boldsymbol{R}_2\vec{b}_1+\vec{b}_2, \vec{v}_1+\boldsymbol{R}_1\vec{v}_2, \boldsymbol{R}_2\boldsymbol{R}_1)\,.
\end{equation}
The structure of the group is given by 
\begin{equation}
    \mathfrak{Carr}(1,d) =(SO(d) \ltimes R^d) \ltimes R^{d+1}.
\end{equation} 
The Lie algebra of the Carroll group is given by the following non-zero commutation relations
\begin{subequations}\label{carral}
\begin{align}
        [J_{ij},J_{kl}]&=so(d),~~\quad [C_i,P_j]=-\delta_{ij}H, \\
        [J_{ij},P_{k}]&=-\delta_{ik}P_{j} + \delta_{jk}P_{i},~~ \quad 
        [J_{ij},C_{k}]=-\delta_{ik}C_{j} + \delta_{jk}C_{i}.
    \end{align}     
\end{subequations}
     
What distinguishes the Carroll group from its parent relativistic group is the commutativity of the Carroll boosts, i.e. $[C_i, C_j]=0$, which reflects the non-Lorentzian nature of the Carroll algebra. It also has two Casimir invariants given by 
\begin{equation}\label{carrcasimir}
    \mathbb{C}_1 = H\,,\qquad \mathbb{C}_4 = \frac{1}{2}(HJ_{ij} + C_iP_j - C_jP_i)^2
\end{equation}

When we generalise to Conformal Carroll symmetry, we have additional generators, viz.~dilatation $(D)$, temporal $(K_0)$ and spatial $(K_i)$ special Conformal Carroll transformation. These can be found similarly by rescaling the relativistic Conformal generators.  
In the position space basis, they take the form
\begin{equation}\label{CCar-gen}
D=t\partial_{t} + x_i\partial_{i} ,\quad K_0 = x_i^2\partial_{t} ,\quad K_i = 2x_i(t\partial_{t} + x_j\partial_{j}) - x_j^2\partial_{i}.
\end{equation}
The additional non-vanishing commutation relations are 
\begin{eqnarray}
&&[D,P_i] = P_i, \,~~~ [D,H] = H, \,~~~ [D,K_i] = -K_i, \,~~~ [D,K_0] = -K_0, \nonumber\\
&&[K_0,P_i] = -2C_i, \,~~~ [K_i,H]=-2C_i, \,~~~ [K_i,P_j] = -2\delta_{ij}D-2J_{ij}. 
\end{eqnarray}
This is the global part of Conformal Carrollian Algebra (CCA). Just like the transformations and the generators, the algebra can also be obtained by the $c\to0$ limit of the relativistic Conformal algebra.  

\medskip

\ding{112} \underline{\em{A look at Carroll geometry}}

\smallskip

Our approach to Carrollian symmetry so far has been algebraic. Although this would be the main theme of our review, it is very important to understand the underlying geometric structures that come hand in hand with Carrollian symmetry. Below we discuss Carrollian manifolds which generalize Riemannian manifolds and the isometry structure of these manifolds would lead us back to the Carroll and Conformal Carroll algebras we have introduced above. 

\medskip

We start with the simplest setting, the $D$-dimensional \footnote{We are using the convention that $D$ denotes the full spacetime dimension while $d = D-1$ denotes the number of spatial dimensions.} 
flat space Minkowski metric with line element:
	\begin{equation}
		ds^2 =-c^2dt^2+(dx^i)^2.
	\end{equation}
	Then the covariant metric and its contravariant inverse are: 
	\begin{eqnarray}\label{Minkwmetric}
		\eta_{\mu\nu} = 
		\begin{pmatrix}
			-c^2  &0 \\
			0  &I_{D-1}
		\end{pmatrix}\, , ~~~
		\qquad	\eta^{\mu\nu} =
		\begin{pmatrix}
			-1/c^2 & 0 \\
			0 & I_{D-1} 
		\end{pmatrix} \, .
	\end{eqnarray}
	Now taking the Carroll limit $(c\rw0)$ we get a degenerate covariant spatial metric $h_{\mu\nu}$ with only one zero eigenvalue and a degenerate contravariant temporal metric $\Theta^{\mu\nu}$ with only one non-zero eigenvalue: 
\begin{eqnarray}\label{FlC}
\eta_{\mu\nu} \rightarrow h_{\mu\nu} = \begin{pmatrix}
			0 & 0 \\
			0 & \quad I_{D-1}
		\end{pmatrix}, 
		\qquad
		-c^2 \eta^{\mu\nu} \rightarrow \Theta^{\mu\nu} = 	\begin{pmatrix}
			1 & 0 \\
			0 & \quad 0_{D-1}
		\end{pmatrix}   \, .
\end{eqnarray}        
%These two are the invariant tensors for the transformations under the Carroll group. 
As $\Theta^{\mu\nu}$ is basically a 1$\times$1 matrix, we can define the vector $\theta^{\mu}$ such that 
\be\label{clockf}
\Theta^{\mu\nu}=\theta^{\mu}\theta^{\nu},
\ee 
and degeneracy simply implies $h_{\mu\nu}\theta^{\nu}=0$.
	
\medskip
	
We can formalize this structure from an intrinsic Carrollian perspective \cite{Henneaux:1979vn, Duval:2014uoa}. 

\smallskip

A {\em{strong Carroll structure}} is a geometric quadruplet $(\mathcal{C},h,\theta, \nabla)$, where $\mathcal{C}$ is a $D$ dimensional manifold, with an inherently defined coordinate chart $(t,x^{i})$. $h$ is a covariant, symmetric, positive tensor field of rank $D-1$ and of signature $(0,+1,\ldots,+1)$ which acts as a spatial metric. Further, $\theta$ is a nowhere vanishing vector field which generates the kernel of $h$. The symmetric affine connection $\nabla$ is responsible for parallel transporting both $h_{\mu\nu}$ and $\theta^{\nu}$. 

\smallskip

This is to be contrasted with a {\em{weak Carroll structure}} which makes no reference to a connection. 

\medskip

The $D-$dimensional Carrollian manifold can be described by a fibre bundle with the $(D-1)$ dimensional spatial directions forming the base space and the temporal direction forming a 1D fibre on top of this.

\medskip

As we will be mostly interested in flat Carrollian manifolds in our discussions, we take the following conventions for our geometric quantities: 
\begin{equation}\label{eq:flatman}
	\mathcal{C} = \underbrace{\mathbb{R}}_{fibre} \times \underbrace{\mathbb{R}^{D-1}}_{base} , \quad h = \delta_{ij}dx^i\otimes dx^j ,\quad \theta = \partial_{t}.
\end{equation} 
Carroll Lie algebra is identified with those vector fields $\xi = \xi^{\mu}\partial_{\mu}$ of $\mathcal{C}$, which satisfy the pertinent isometry conditions: 
	\begin{eqnarray}\label{CLaa}
		\pounds_{\xi}h_{\mu\nu}=0, ~~~\quad \pounds_{\xi}\theta^{\mu}=0.
	\end{eqnarray}
These conditions on the invariant fields can be easily solved and one thus gets the Carroll Killing vectors, 
	\begin{eqnarray}
		\xi^i = \omega^i_j x^j + b^i ,\qquad \xi^0 = a + f(x^k).
	\end{eqnarray}
From this one can conclude that isometries of a flat Carroll manifold are infinite-dimensional. If we further demand that the killing vector preserves the connection defined by $\nabla$, the function $f(x^k)$ will be restricted to be linear and it boils down to the finite-dimensional Carroll algebra given by \refb{carral} \cite{Duval:2014uoa, Penna:2018gfx}. To include conformal Carroll isometries in this setup, one has to modify the isometry  $(\ref{CLaa})$ to the condition for conformal isometries in the Carroll background: 
	\begin{eqnarray}\label{con-car}
		\pounds_{\xi}h_{\mu\nu}=\lambda_1 h_{\mu\nu},\quad \pounds_{\xi}\theta^{\mu}= {\lambda_2}\theta^{\mu},
	\end{eqnarray}
where $\lambda_i$ are constants. Here $\pounds_{\xi}$ represents the Lie derivative along the vector field $\xi$. The general solution to the differential equations implied by these isometries is given by,
\begin{align}
    \xi  = f(x)\d_t + \omega^i_{\,\,j}x^j\d_i + b^i\d_i &+ \Delta\left(x^i\d_i - \frac{2\lambda_2}{\lambda_1}t\d_t \right) \nonumber\\
    &+ k_i\left(2x^i\left(x^k\d_k - \frac{2\lambda_2}{\lambda_1}t\d_t \right) - x^kx_k\delta^{ij}\d_j\right).
\end{align}
where $\omega^i_{\,\,j}, b^i, \Delta, k_i$ are integration of constants. When the ratio $N \equiv {\lambda_1}/{\lambda_2} = -2$, the form gets further simplified
\be
\label{eq:carkilling}
\xi  = f(x)\d_t + \omega^i_{\,\,j}x^j\d_i + b^i\d_i + \Delta\left(x^i\d_i + t\d_t \right) + k_i\left(2x^i\left(x^k\d_k + t\d_t \right) - x^kx_k\delta^{ij}\d_j\right).
\ee

\medskip

In later sections, when discussing holography of asymptotically flat spacetimes (AFS), we will see that there is a celebrated isomorphism between the conformal Carroll algebras and the Bondi-Metzner-Sachs (BMS) algebras \cite{Bondi:1962px, Sachs:1962wk}, which form the asymptotic symmetry algebra in AFS in one higher dimension \cite{Bagchi:2010zz, Duval:2014uva}.  

\medskip

For this particular value of $N \equiv {\lambda_1}/{\lambda_2} = -2$, these conformal Carroll generators in $D$ dimensions close to form exactly the BMS$_{D+1}$ algebra \cite{Duval:2014uoa}. This specific value of the ratio is the one for which space and time scale in the same way under the Dilatation generator. For other values of $N$, one gets more generic Carroll conformal algebras. 

\subsection{Comparison with Galilean symmetry}
Here we will compare our analysis in the previous section to another non-Lorentzian symmetry, which perhaps is much more familiar to uninitiated reader, namely the Galilean symmetry. 
Galilean group also arises via an Inönü–Wigner contraction \cite{inonu1953contraction} from the Poincaré group. But instead of taking the speed of light to zero, here the contraction is performed by taking the speed of light $c\to \infty$ limit. This is equivalent to the co-ordinate scaling (compare with \eqref{carrscale}):
    \begin{equation}
		x^i\rw  \epsilon x^i ,\quad t\rw t, \quad \epsilon\rw 0
    \end{equation}
where superscript $i$ denotes the number of spatial dimensions. Under this limit, the Galilean generators are defined by rescaling the parent Poincaré generators in the following way 
\be{}
H \equiv P_0^{\text{P}}, \,~~\quad P_i \equiv \frac{1}{\epsilon} P_i^{\text{P}}, \,~~\quad  G_i \equiv \frac{1}{\epsilon} J_{0i}^{\text{P}}, \,~~\quad  J_{ij}\equiv J_{ij}^{\text{P}},
\ee
where the superscript ${\text{P}}$ refers to the Poincaré generators. The explicit forms of the Galilean generators in the position space basis are:
\begin{equation}
H=\partial_{t} ,~~\quad P_i=\partial_{i} ,~~\quad G_i=t\partial_{i} ,~~\quad J_{ij}=x_i\partial_{j}-x_j\partial_{i}.
\end{equation}
Here $H, P_i, G_i, J_{ij}$ are time translation, spatial translations, Galilean boosts and spatial rotations generators respectively. These generators generate the Galilean or non-relativistic transformation on space-time coordinate as
\be		
t{'}=t+a , ~~\quad \vec{x}{'}=\boldsymbol{R}\vec{x} -\vec{v}t + \vec{b},
\ee
where the parameters of the group $(a, \vec{b}, \vec{v}, \boldsymbol{R})$ describes time-translation, space-translation, Galilean boosts and $SO(d)$ rotation respectively. As can be seen from the transformation relations, under Galilean boosts, the time remains invariant, and space coordinates change, making it a symmetry group of absolute time. Notice the difference with the Carroll group, which was the symmetry group of absolute space. The group composition rule for the Galilean group is given by,
\begin{equation}
    (a_2, \vec{b}_2, \vec{v}_2, \boldsymbol{R}_2) \cdot (a_1, \vec{b}_1, \vec{v}_1, \boldsymbol{R}_1) = (a_2+a_1-\vec{v}_2\cdot \vec{b}_1, \boldsymbol{R}_2\vec{b}_1+\vec{b}_2, \vec{v}_1+\boldsymbol{R}_1\vec{v}_2, \boldsymbol{R}_2\boldsymbol{R}_1)
\end{equation}
The structure of the group is given by, 
\begin{equation}
    \mathfrak{Gal}(1,d) =(SO(d) \ltimes R^d) \ltimes R^{d+1}.
\end{equation} 
The Lie algebra of the Galilean group is given by following non-zero commutation relations
    \begin{align}
        [J_{ij},J_{kl}]&=-\delta_{ik}J_{jl} + \delta_{il}J_{jk} - \delta_{jl}J_{ik} +\delta_{jk}J_{il}, ~~\quad [G_i,H]=P_i,  \\
        [J_{ij},P_{k}]&=-\delta_{ik}P_{j} + \delta_{jk}P_{i},~~\quad [J_{ij},G_{k}]=-\delta_{ik}G_{j} + \delta_{jk}G_{i}.  
    \end{align}  
Galilean group is non-Lorentzian as can be seen, two Galilean boosts commute $[G_i, G_j]=0$. Now note that, this commutation of boost is true for Carrollian case as well. Despite that, the structure of Carroll and Galilei groups prominently differ in the algebraic level. To make this more apparent, here are the differences explicitly:
\begin{align}
    \textbf{Carroll:}&\quad [C_i,H] = 0,~\quad [C_i,P_j]= -\delta_{ij}H.\nonumber\\
    \textbf{Galilei:}&\quad [G_i,H] = P_i,~\quad [G_i,P_j]= 0.
\end{align}
In the Carroll algebra, Hamiltonian $H$ commutes with Carroll boosts $C_i$ along with other generators, making $H$ a central element for the Carroll algebra as mentioned earlier, whereas for the Galilean algebra, $H$ doesn't commute with Galilean boosts $B_i$ and hence is not central.  Also the boosts and spatial translations commute in the Galilean case while for Carroll the commutator between these two can be nonvanishing.

\medskip

One can generalise the Galilean symmetry to include conformal symmetry as well. Then we will have additional generators, Dilatation $(D)$, temporal $(K_0)$ and spatial $(K_i)$ special Conformal Galilean transformation. 
In the usual basis, they take the form
\begin{equation}
D=t\partial_{t} + x_i\partial_{i} ,~~\quad K_0 = 2tx_i\partial_i + t^2\partial_{t} ,~~\quad K_i = -t^2\partial_i.
\end{equation}
The additional commutation relations are 
\begin{eqnarray}
&&[D,P_i] = P_i, \,~~~ [D,H] = H, \,~~~ [D,K_i] = -K_i, \,~~~ [D,K_0] = -K_0, \nonumber\\
&&[K_0,P_i] = 2G_i, \,~~~ [K_i,H]=2G_i, \,~~~ [K_i,P_j] = 0. 
\end{eqnarray} 
This algebra is known as the Galilean conformal algebra (GCA) and has found applications in many places including as a dual for non-relativistic version of AdS/CFT \cite{Bagchi:2009my}.{\footnote{Interestingly the Galilean conformal algebra (GCA) admits an extension to infinite dimensional algebra in all dimensions \cite{Bagchi:2009my,Bagchi:2009ca, Bagchi:2009pe}. To see this, one can define the generators in the following way 
\begin{eqnarray}
    L^{(n)} = -(n+1)t^nx^i\partial_i\,,\quad M_i^{(n)} = t^{n+1}\partial_i.
\end{eqnarray}
They satisfy the following algebra
\begin{align}
[L^{(n)}, L^{(m)}] = (n-m)L^{(n+m)}, \quad [L^{(n)}, M_i^{(m)}] = (n-m)M_i^{(n+m)} , \quad [M_i^{(n)}, M_i^{(m)}]=0.
\end{align}
These generators together with $J_{ij}$ constitutes the infinite-dimensional GCA algebra. Here $n = 0,\pm1$ defines the global part of the algebra.}}

\medskip

From the geometric perspective, consider the $D$-dimensional Minkowski space with line element:
	\begin{equation}
		ds^2 =-c^2dt^2+(dx^i)^2.
	\end{equation}
Taking the Galilean limit on the metric \eqref{Minkwmetric}, one gets a degenerate contravariant spatial metric $h^{\mu\nu}$ and a degenerate covariant temporal metric $\tau_{\mu\nu}$: 
\begin{eqnarray}
-c^{-2} \eta_{\mu\nu} \rightarrow \tau_{\mu\nu} = \begin{pmatrix}
			1 & 0 \\
			0 & \quad 0_{D-1}
		\end{pmatrix}, ~~
		\qquad
		\eta^{\mu\nu} \rightarrow h^{\mu\nu} = 	\begin{pmatrix}
			0 & 0 \\
			0 & \quad I_{D-1}
		\end{pmatrix}.   
\end{eqnarray}        
These two are the invariant tensors for the Galilean group, which one can now juxtapose against their Carrollian cousins. Here one can write $\tau_{\mu\nu} = \tau_{\mu}\tau_{\nu}$ and because of degeneracy the condition $h^{\mu\nu}\tau_{\nu} = 0$ is always satisfied. 

\medskip

Following \cite{Duval:2014uoa}, we now briefly recount the structure of Newton-Cartan (NC) manifolds. The isometries of the flat version of these manifolds gives rise to the Galilean algebra. Our discussion would be very similar to the one about Carroll manifolds earlier. A NC structure is a quadruplet $(\mathcal{N},\tilde{h},\tau, \nabla)$, where $\mathcal{N}$ is a $D$ dimensional manifold, on which one can choose a coordinate chart $(t,x^{i})$. Also, $\tilde{h}$ is a contravariant, symmetric and positive tensor field of rank $D-1$ with a signature $(0,+1,\ldots,+1)$. $\tau$ is a non-vanishing one-form which generates the kernel of $\tilde{h}$.  $\nabla$ is a symmetric affine connection that parallel transports both $\tilde{h}^{\mu\nu}$ and $\tau_{\mu}$. This $D-$dimensional NC manifold can be described by a fibre bundle with the temporal direction forming the base space and the $(D-1)$ dimensional spatial directions forming the fibre on top of this. Remembering our earlier discussion, the flat Galilean manifolds are written as: 
\begin{equation}
	\mathcal{C} = \underbrace{\mathbb{R}}_{base} \times \underbrace{\mathbb{R}^{D-1}}_{fibre} , ~~\quad {h} = \delta^{ij}\partial_i\otimes \partial_j ,~~\quad \tau = dt.
\end{equation} 
Galilei Lie algebra is identified with those of the vector fields $\xi = \xi^{\mu}\partial_{\mu}$ of $\mathcal{C}$, which satisfy the isometry conditions: 
	\begin{eqnarray}\label{CLa}
		\pounds_{\xi}{h}^{\mu\nu}=0,~~~~\quad \pounds_{\xi}\tau_{\mu}=0.
	\end{eqnarray}
Solving these, one gets 
\begin{eqnarray}
    \xi^0 = a_t\, \text{(constant)},~~\qquad \xi^i = \omega^i_{\,\,j}x^j + f^i(t) + b^i.
\end{eqnarray}
If we further preserve the connection, the function $f^i(t)$ becomes linear in $t$, and we get finite dimensional Galilean algebra.
One can generalise the discussion to include conformal isometries the algebra of which then gives the Galilean Conformal Algebra for flat NC manifolds.

\subsection{Representations of Carroll algebra}\label{sec:repn}
Now that we have an understanding of the Carroll algebra, we turn to its representations. The discussion below follows \cite{deBoer:2021jej} closely and hence we only consider unitary irreducible representations. A natural choice for building the representation theory is to label representations by the eigenvalues of the Casimir operators, which for Carroll algebra is given by \eqref{carrcasimir}. As we have seen above, a defining feature of the Carroll algebra is that the Hamiltonian $H$ is a central element. So one can use this to separate Carroll representations into two distinct sectors: $H=0$ and $H\neq 0$. 

\medskip

Now remember that $[C_i, P_j] = -H \, \delta_{ij}$. For representations with $H=0$, the generators $P_i$ and $C_i$ can be simultaneously diagonalised and one can label the states by their eigenvalues. Spatial rotations act on these eigenvalues and by extracting an irreducible representation of $so(d)$ algebra, one can construct an irreducible representation of Carroll algebra. In the second case, i.e. $H\neq 0$, $P_i$ and $C_i$ cannot be diagonalised in the same basis. Either of these eigenfunctions can be combined with a representation of the rotation group. 

\medskip

For the moment, we focus our discussions on Carroll representations in $D=4$. The details of Carroll Conformal representations will be elucidated in a later section. For just the Carroll group, it is convenient to define $W_k$ and $S_k$ via
\begin{eqnarray}
    M_{ij} = \epsilon_{ijk}W_k,~~\quad J_{ij}=\epsilon_{ijk}S_k,
\end{eqnarray}
Here $M_{ij} = HJ_{ij} + C_iP_j - C_jP_i$ and so $W_i = HS_i - \eps_{ijk}P_jC_k$. 
Keeping helicity in mind, one can further define a similar operator $L=S_iP_i$. 
One can now use the little group method to examine induced representations. An important result to use here would be  the Carroll boost transformation in energy-momentum coordinates is given by,
\begin{eqnarray}
    E'=E,~~\quad \vec{p}' = \vec{p} - \vec{b}E.
\end{eqnarray}
Let's consider the two disjoint cases based on the Hamiltonian separately: 
\begin{itemize}
   \item[$\star$]  Consider the case when the eigenvalue of $H$, denoted as $E$, is non-zero. As long as this holds, one can always migrate to a frame where $\vec{p} = 0$, resulting to the little group being $SO(3)$. The eigenvalue of the Casimir $C_4$ is $E^2s(s+1)$, with $s=0,\frac{1}{2},1,...$ and so on. The eigenstates in this case are labelled as $|E\neq 0, \vec{p} = 0, s, m_s\rangle$, where $m_s$ are eigenvalues of $S_3$, the rotation operator along the z-axis. 

   \item[$\star$] Now when $E =0$, the momentum $\vec{p}$ is Carroll boost invariant. So without loss of generality, one can always choose $\vec{p} = p\hat{z}$. Then acting on such states, the $z$-component of $\vec{W}$ vanishes and $L = S_3P_3$. The little group will be then $ISO(2)$, generated by $L, W_1, W_2$. The relevant commutators for this group are given by  $$[W_i,W_j] = 0\,,\,\, [L,W_1] = pW_2\,,\,\,[L,W_1] = -pW_1$$ Now there can be two further cases, labelled by: $(1)\, W_iW_i = 0$ and $(2)\, W_iW_i > 0$. In the first case the helicity operator $S_3$ can be a good operator and its eigenvalue $\lambda$ can be used to label the state $|E = 0, \vec{p} = p\hat{z}, \lambda\rangle$. 
\end{itemize}

{\ding{112}} \underline{\emph{{An example: Carroll particles}}}
\medskip

To understand more about the two classes of representations discussed in the last section, we can focus on the dynamics of a single Carroll particle and see what these representations mean.
One of the way to study the dynamics of free Carroll particle can be obtained starting from the relativistic action of a massive point particle and taking the appropriate Carroll limit. The canonical action of such a relativistic particle in the Minkowski background is given by,
\begin{equation}
    \mathcal{S} = \int d\tau \left[p_{\mu}\dot{x}^{\mu} - \frac{e}{2}\left(p_{\mu}p^{\mu} + c^2m^2\right)\right].
\end{equation}
Here $e$ is the einbein, $\tau$ is some worldline affine parameter and we will be using the metric signature $(-,+,+,+)$. Writing $x^0 = ct$ and $p^0 = E/c$, the action takes the form, 
\begin{equation}
    \mathcal{S} = \int d\tau \left[-E\dot{t} + p_i\dot{x}^i - \frac{e}{2}\left(-\frac{E^2}{c^2} + p_ip^i + c^2m^2\right)\right].
\end{equation}
Now, in the first case, we rescale the einbein variable $e \to -c^2e$ and mass $m\to M/c^2$ and then take the Carroll limit $c\to 0$. The action becomes,
\begin{equation}\label{ppcarr}
    \mathcal{S}_{carr} = \int d\tau \left[-E\dot{t} + p_i\dot{x}^i - \frac{e}{2}\left(E^2 - M^2\right)\right].
\end{equation}
It can be easily verified the action is explicitly invariant under the Carroll transformation 
\begin{eqnarray}
    &&t \to t + b_i R^i_{\,j}x^j + a, ~~\quad x^i \to R^i_{\,j}x^j + a^i,\\
    &&E \to E,~~\quad p_i \to R_i^{\,j}p_j + b_iE.
\end{eqnarray}
The equations of motion one gets from the action \eqref{ppcarr} are
\begin{eqnarray}
    \dot{E} = 0, ~~\quad \dot{p}_i = 0,~~\quad \dot{x}^i = 0,~~\quad \dot{t} = -eE,
\end{eqnarray}
which implies this particular kind of single Carroll particle can not move and has constant energy. The mass-shell constraint of the Carroll particle is simply given by, $E^2 - M^2 = 0$, which tells us the energy of Carroll particles can be positive, negative or zero (for the massless case), which is also the consequence of the Hamiltonian being the central element of the Carroll algebra. Note that, if we were to now quantise the system and the physical states are subjected to the (Carrollian) mass-shell constraint, we get the wave equation 
\begin{equation}
    \left(\frac{d^2}{dt^2} + M^2\right)\psi(t,x) = 0.
\end{equation}
In the next chapter, we will see this very same equation arising in the context of electric Carroll scalar field theory. 
\medskip

Alternatively, we can scale the action differently. Before taking the Carroll limit, we can rescale the coordinates, einbein and mass as:
\begin{eqnarray}
    E\to cE\,,\quad t\to \frac{t}{c}\,,\quad p_i \to \frac{p_i}{c}\,,\quad x^i \to cx^i\,,\quad e \to c^2e\,,\quad m \to \frac{M}{c}.
\end{eqnarray}
Note that this scaling does not change the canonical commutation relations between the coordinates and momenta. 
Now if we take the Carroll limit, the action becomes, 
\begin{equation}\label{ppcarrmag}
    \mathcal{S} = \int d\tau \left[-E\dot{t} + p_i\dot{x}^i - \frac{e}{2}\left(p_ip^i + M^2\right) \right].
\end{equation}
This action is not by default Carroll invariant. In fact, the condition of invariance under Carroll boosts $\delta_c\mathcal{S} = 0$ implies the constraint $E=0$. This on the other hand is equivalent to adding a term $\lambda E$ in the action \eqref{ppcarrmag}, with $\lambda$ being a Lagrange multiplier\footnote{One can compare this with the covariant small $c$ expansion formalism in \cite{Bagchi:2024rje}, where one gets a similar constraint for the magnetic action.}. For a clear explanation of the underlying idea, refer to\cite{Bergshoeff:2022qkx}.  
In this case, the equations of motion one gets from the action \eqref{ppcarrmag} are
\begin{eqnarray}
    \quad \dot{p}_i = 0,~~\quad \dot{x}^i = ep^i.
\end{eqnarray}
The mass-shell constraint imposed on this action is $p_ip^i + M^2 =0$. It can be inferred that in the Carroll world, the particles having non-zero energy are always at rest, whereas the particles which can move must have zero energy.

\subsection{Null contractions}\label{nullc}

Our discussion so far has hinged on the appearance of Carrollian symmetries by a $c\to0$ limit of Poincar\'{e} symmetries and we have found that a $D$ dimensional Poincar\'{e} algebra upon contraction leads to a Carroll algebra in the same number of dimensions. In this subsection, we encounter a different route to obtain Carrollian algebras. 

\medskip

A point that has been stressed in the introduction and also repeated above is that the Carroll algebra is associated with null surfaces and the contraction that we have used to get to it has been called {\em ultra-relativistic}. In usual parlance, an ultra-relativistic limit is where the speed of the object in question becomes very close to the speed of light, i.e. $v/c \to 1$. At first sight, this is very different from $c\to0$. But as we discussed in the example of the black hole horizon in the introduction, in many physical contexts one can think of two distinct speeds of light, one for a bulk observer ($c_{bulk}$) and another for an observer on a co-dimension one surface ($c$). 
We can visualise a connection between the two: $c\to 0 \equiv v/c_{bulk} \to 1$. 

\medskip

It seems there should be different ways of getting at Carroll symmetries from Poincaré symmetries. As mentioned above, the contraction we have discussed so far gives us the Carroll algebra from the Poincar\'{e} algebra in the same number of dimensions. There should be another limit where one can zoom into a null surface and obtain a lower dimensional Carroll algebra. This is what we will discuss now following \cite{Bagchi:2024epw}. To study this, we will use the light-cone coordinate system and use contraction on these co-ordinates. Throughout this subsection, we will be working in 4D Minkowski spacetime. The generalisation to higher dimensions is immediate. The co-ordinates are labelled by $(t,x,y,z)$. We introduce the light-cone coordinates $x^{\pm} = \frac{1}{\sqrt{2}}(t\pm z)$, and re-define the Poincaré generators in the lightcone: 
\begin{subequations}
    \begin{eqnarray}
&& P_+ =\frac{1}{\sqrt{2}}( P_t + P_z) = \d_+, \quad P_- = \frac{1}{\sqrt{2}}(P_t - P_z )= \d_-, \quad P_i = \d_i\\
&& J^{1}_i =\frac{1}{\sqrt{2}}(J_{it} - J_{iz}) =  x^+\d_i + x_i\d_-, \quad J^{2}_i = \frac{1}{\sqrt{2}}(J_{it} + J_{iz})  =  x^-\d_i + x_i\d_+, \\
&& J_{ij} = x_i\d_j - x_j\d_i, \qquad B= J_{zt} = x^+\d_+ - x^-\d_-.
\end{eqnarray}
\end{subequations}
We want to perform contractions on the null direction here. So we can either focus on $x^+$ or $x^-$ direction and it can be shown in a straightforward manner that in both cases, the results are identical. Here let's focus on $x^-$. The contraction is 
\begin{eqnarray}\label{nullcontrac}
    x^+ \to x^+, ~~~~ x^- \to \e x^-, ~~~~ x^i \to x^i, ~~~\quad \e \to 0.
\end{eqnarray}
After contraction, the new generators are 
\begin{subequations}\label{2carr}
\bea{}
&& H_+=\lim_{\e\to 0} P_+= \d_+, ~~\quad C_{+i} = \lim_{\e\to 0} J^{2}_i = x_i\d_+, \\
&& H_-= \lim_{\e\to 0} \e P_-=\d_-, ~~\quad C_{-i} = \lim_{\e\to 0} \e J^{1}_i = x_i\d_-,  \\
&& P_i = \d_i, ~~\quad J_{ij} = x_i\d_j - x_j\d_i, ~~\quad B = x^+\d_+ - x^-\d_-.
\eea
\end{subequations}
Note that, we get two copies of overlapping Carroll algebra of codimension one spanned by these following set of generators:
\begin{equation}
\text{Carroll(+):}  \quad \{H_+, C_{+i}, P_i, J_{ij} \},~~\quad 
\text{Carroll(--):}  \quad \{H_-, C_{-i}, P_i, J_{ij}\}. 
\end{equation}
Each set follows \eqref{carral}.
The additional generator $B$ has the interesting property
\be
[\O_\pm, B] = \pm \O_\pm, ~~~\quad [\O_T, B] = 0, 
\ee
where $\O_\pm = (H_\pm, C_{i\pm})$ are the operators which resides on the lightcone and $\O_T= (P_i, J_{ij})$ are the transverse operators. Another thing is to note that the sets of operators $\O_+$ and $\O_-$ commute and these thus form two disjoint sectors of the algebra. 

\medskip

We can also perform the null contractions on relativistic Conformal algebra. The additional conformal generators in lightcone coordinates are given by
\begin{subequations}
\bea{}
&& D=x^+\partial_+ +x^-\partial_-+ x^i\partial_i, \\
&&K^1= - \frac{1}{\sqrt{2}}(K_t-K_z)=x^kx_k\partial_-+2(x^+)^2\partial_++2x^+x^k\partial_k ,\\
&&K^2=-\frac{1}{\sqrt{2}}(K_t+K_z)=x^kx_k\partial_++2(x^-)^2\partial_-+2x^-x^k\partial_k,\\
&&K_i=2x_i(x^+\partial_++x^-\partial_-+x^i\partial_i)+(2x^+x^--x^jx_j)\partial_i. 
\eea
\end{subequations}
Now if we perform the similar contraction \eqref{nullcontrac}, we will get
\bea{}
&&K_+=\lim_{\e\to 0} K^2= x^kx_k\partial_+,  ~~\quad  K_-=\lim_{\e\to 0}\e K^1= x^kx_k\partial_-,\\
&&\bar{K}_i=\lim_{\e\to 0}K_i =2x_i(x^+\partial_++x^-\partial_-+x^i\partial_i) -x^jx_j\partial_i,
\eea
while $D$ stays unchanged. We now get two copies of overlapping conformal Carroll algebra of codimension one spanned by the following set of generators
\begin{subequations}
\begin{align}
\text{Conf. Carroll($+$):} & \quad \{H_+, C_{+i}, P_i, J_{ij},D, K_+,\bar{K}_i \}, \\
\text{Conf. Carroll($-$):} & \quad \{H_-, C_{-i}, P_i, J_{ij},D,K_-,\bar{K}_i\}. 
\end{align}
\end{subequations}

Instead of the ``longitudinal'' contraction along one of the null directions, as performed above, we could also focus on the non-null directions in the theory. This other type of contraction is called the ``transverse'' contraction:
\begin{align}\label{trans-cont}
    x_i \to \e x_i, ~~\quad x^\pm \to x^\pm, ~~\quad \e\to0.
\end{align}
 This give rises to a similar set of generators to the ones encountered in the longitudinal case \eqref{2carr}, with the Carroll boosts replaced by Galilean ones:
 \begin{align}
     G_{i}^+ = x^+\d_i, ~~\quad  G_{i}^- = x^-\d_i.
 \end{align}
  This leads to two overlapping co-dimension one {\em Galilean} algebras spanned by these two sets of generators 
\begin{align}
    \text{Galilean($+$):} &\quad \{H_+, G_{i}^+, P_i, J_{ij} \}, \\  \text{Galilean($-$):} &\quad \{H_-, G_{i}^-, P_i, J_{ij}\}.
\end{align}    
Each of these two sets of Galilean algebras can be further extended to their Bargmann versions by including the Hamiltonian of the other set as the mass. 

\medskip

This construction has direct ties to the one proposed by Susskind \cite{Susskind:1967rg}, in the context of perturbative processes in QCD, where it was discovered that one can get a lower dimensional Galilean sub-algebra inside the Poincaré algebra when going to the lightcone coordinates. What \cite{Bagchi:2024epw} does is to identify not one, but two Galilean sub-algebras along with two Carroll subalgebras in the higher dimensional Poincaré algebra. From the point of view of the contractions above, the limit just disentangles the two subalgebras which previously had non-zero cross commutators in the parent Poincaré algebra. 

\medskip

In \cite{Susskind:1967rg}, Susskind related his discussion to an infinite momentum limit (see also \cite{Weinberg:1966jm}), where the Galilean invariance occurs in the direction transverse to the infinite momentum. Later important developments, including the construction of the Discrete Light Cone Quantisation (DLCQ) of various theories depended on the infinite momentum frame considerations. 
From \cite{Bagchi:2024epw}, it becomes clear that the infinite momentum limit is just the transverse contraction from the point of view of the algebra since \eqref{trans-cont} in terms of the momenta reads:
\begin{align}
    p_i \to \frac{1}{\e} p_i, ~~\quad p_\pm \to p_\pm, ~~ \quad \e\to 0,
\end{align}
where $p_a ~(a= i, \pm)$ are the various momenta associated with the respective directions. 
The discussion in \cite{Bagchi:2024epw} actually generalises Susskind's observation and provides a way to access both Galilean and Carrollian sub-algebras. It is expected that this may pave the way for an equivalent understanding of the Carroll algebra in various new avenues like the BFSS matrix model \cite{Banks:1996vh}.

\subsection{Pointers to literature}\label{pt:sec2}
Here we summarize some relevant literature, with the repeating caveat that we are not being exhaustive in our lists. 
\begin{itemize}
\item{\em{Representation theory}}
\begin{itemize}
    \item[$\star$] The very first well recognised papers to discuss the unitary irreducible representations (UIRs) of Conformal Carroll (BMS) group appeared in the 1970s written by McCarthy \cite{mccarthy1, mccarthy2, mccarthy3}. In these pioneering works, McCarthy studied the UIR of BMS$_4$. See further important work McCarthy and collaborators \cite{mccarthy4,mccarthy5,mccarthy6,mccarthy7,mccarthy8}. These discussions rely on a particular application of Wigner’s method \cite{wigner1939unitary} of induced representations, generalised by Mackey \cite{mackey1968induced}.
    \item[$\star$] Even before McCarthy, the representation theory of BMS was attempted at by Cantoni, who in a series of works \cite{cantoni1,cantoni2,cantoni3} discussed the subject at length. This also first included a discussion on faithful linear representation of the generalized Bondi-Metzner group induced from each faithful linear representation of the
inhomogeneous orthochronous Lorentz group. 
   \item[$\star$] For further references to representation theory, the reader is referred to Sec.~\ref{ssec:pointersotherholo}.  
\end{itemize}

\item{\em{Carroll particle dynamics}}
\begin{itemize}

    \item[$\star$] Carroll particle dynamics has had a long history, with various works having a go at it from representation theory motivations. One may first start looking at \cite{Bergshoeff:2014jla} where it was shown that single free (massless) Carroll particles may not have any non-trivial dynamics, but a set of interacting ones may. For other discussions along that direction, including conformal extension and `tachyonic' extension thereof, see \cite{Duval:2014uoa,deBoer:2021jej,Casalbuoni:2014ofa,Casalbuoni:2023bbh} and follow-ups. Carroll particles in Two-Time spacetimes have been recently discussed in \cite{Kamenshchik:2023kxi}.

\item[$\star$] In this review, we have discussed only about the single Carroll particle in Sec.~\ref{sec:repn}. As a further step, One can generalise this discussion to include two particle systems \cite{Bergshoeff:2014jla}. There it has been shown that in contrast to the single particle scenario, there are interacting Carroll particles, resulting in the non-trivial dynamics. The generalisation to many particle Carroll systems has been discussed in \cite{deBoer:2021jej}.

\end{itemize}

\item{\em{Aspects of Lightcone:}} Appearance of lower dimensional twin non-Lorentzian algebras have also been described in a recent set of works \cite{Barnich:2024aln, Majumdar:2024rxg}, focusing on the lightcone formulation of quantum field theories. 

\end{itemize}

\newpage

\section{Carroll field theories}
\label{sec:carfieldtheories}

In the previous section, we put forward the details of Carrollian and conformal Carrollian symmetries and touched upon some aspects of representation theory for Carrollian symmetries. Armed with this knowledge, we now explore systems which will realise these symmetries. To this end, in this section, we will discuss Carrollian field theories. We will consider scalar and vector fields first before moving to Carroll spinors. 

\medskip

In general, any Carroll field theory can be broadly categorized into two types, $(1)$ \textit{electric} Carroll, $(2)$ \textit{magnetic} Carroll\footnote{There is also a third one, which is the combination of electric and magnetic.}. The origin of this terminology goes to the Carrollian limits of electromagnetism \cite{Duval:2014uoa}; where depending on the limits, either the electric or the magnetic field survives. There are two approaches here: either one starts in a bottom-up way, starting from the basic ingredients, and uses symmetry principles to construct an invariant Carroll action; or one has to go in a top-down manner, starting from a relativistic field theory (written in Lagrangian or Hamiltonian formalism) and perform a consistent Carroll limit on it. In the second approach, there are also two ways to reach a Carroll invariant field theory. One can either carefully take suitable scaling limits on the fields or one can expand the fields in small powers of $c$, resulting in leading, sub-leading and higher order theories. Both formulations lead us to Electric and Magnetic types of Carroll field theory as we will see below.   
\medskip

Before we move on to more detailed discussion, let us pause for a moment and discuss a bit of the historical context of (conformal) Carroll invariant field theories\footnote{Even before Carroll symmetries were understood as something meaningful, old works by Klauder \cite{Klauder:1970cs} discussed ultra-local scalar field theories. In modern language, these theories are the electric version of Carroll theories. Similarly, ultra-local fermions were also discussed in \cite{Klauder:1973nc}.}. Such field theories were originally constructed from a Holographic point of view, to find putative duals to flat space, something we will be discussing in a later section. In \cite{Barnich:2012rz} such a candidate scalar theory was constructed by taking appropriate scaling limits of a two-dimensional Liouville field theory, which explicitly shows conformal Carroll invariance. Two different ways to take such limits were already spotted in that work. We will discuss this at length in Sec.~\ref{ssec:carlioville}. 

\medskip

More concrete ways to take contractions of relativistic field theories came into being with \cite{Duval:2014uoa}, using a directly opposite formalism from the original Galilean electrodynamics discussion in \cite{LeBellac:1973unm}. In the beginning of the current resurgence of Carrollian physics, limits of the equations of motion were considered \cite{Bagchi:2016bcd, Bagchi:2019xfx} and the theories were shown to possess infinite Carrollian (conformal) symmetries following similar considerations in Galilean theories \cite{Bagchi:2014ysa, Bagchi:2015qcw,Bagchi:2017yvj}, before the first actions were written down in \cite{Basu:2018dub, Bagchi:2019clu}. 

\medskip

We will discuss in this work the more modern ways to approach of constructing Carrollian theories below, both intrinsically and from limits.

\medskip

\subsection{Scalar field}\label{carrscalar}
In our first example, we will consider the construction of a Carroll scalar field. Let us consider a free relativistic real scalar field $\phi$, with Klein-Gordon Lagrangian
\begin{eqnarray}\label{KGaction}
    \mathcal{L}_{rel} = \frac{1}{2}\eta^{\mu\nu}\partial_{\mu}\phi\partial_{\nu}\phi - V(\phi) = \frac{1}{2c^2}(\partial_t\phi)^2 - \frac{1}{2}(\partial_i\phi)^2 - V(\phi). 
\end{eqnarray}
The simplest way to get a Carroll invariant theory from this is to take the explicit limit $c \to 0$. In this limit, after regularizing properly, the action becomes,
\begin{eqnarray}\label{CarrKG}
    \mathcal{L}_{carr} = \lim_{c\to 0}c^2\mathcal{L}_{rel} = \frac{1}{2}(\partial_t\phi)^2 - \tilde{V}(\phi).
\end{eqnarray}
Here the potential $\tilde{V}(\phi)$ contains properly regulated mass and interaction terms. In the covariant way, the only possible Carroll invariant kinetic term for scalar field one can write is $\theta^{\mu}\theta^{\nu}\d_{\mu}\phi\d_{\nu}\phi$, with $\theta^\mu$ being the Carrollian clock-form defined in \eqref{clockf}. This kinetic term also results to the action \eqref{CarrKG}, albeit without the potential term for the flat Carroll choice (i.e. $\mathcal{L}\sim \dot\phi^2$). The equation of motion in this case reads 
\be\label{eleceom}
\d_t^2\phi(x,t) + \frac{\partial \tilde{V}}{\partial \phi}= 0.
\ee
Now recall from the earlier section, we were getting a similar equation for an electric Carrollian particle. Keeping in tune with the convention, the above is the electric Carroll theory for a scalar field. In the next subsequent sections, we will see this action will also appear from different ways of taking the Carroll limit. However, as promised before, we will see, this is not the only Carroll invariant action one can write down for the scalar field. As can be seen from the action \eqref{CarrKG}, there are no spatial derivatives in this particular action, suggesting the theory to be ultra-local. 
\medskip

\subsubsection*{Getting the action: method of expansion}
One of the distinguishing characteristics of Carrollian structure is that the spacetime lightcone completely closes. The slope of lightcone in Lorentzian geometry is $1/c$, where $c$ is the speed of light. This implies, to get a Carrollian theory, one needs to carry out an expansion in small powers of $c$.  This small $c$ expansion permits a perturbative expansion around the $c=0$ point, which is similar to the opposite and very useful post-Newtonian expansion around large $c$ \cite{VandenBleeken:2017rij}. In what follows, all our expansions would be carried out in even powers of $c$ \footnote{One can expand keeping odd powers. The $c^2$-expansion remains a consistent sector, and it is expected that dropping odd powers does not change physical conclusions. See however \cite{Ergen:2020yop} for a similar discussion in the non-relativistic $1/c$ expansion.}. Assuming that up to an overall power of $c$, any field is analytic in $c$ i.e. they can be expanded around $c=0$ such that 
\begin{eqnarray}\label{phiexpand}
    \phi = c^\alpha\left(\phi_0 + c^2\phi_1 + c^4\phi_2 + ... \right)
\end{eqnarray}
for some arbitrary $\alpha$. We now focus on \eqref{KGaction}. The expansion in \eqref{phiexpand} now corresponds to scalar field \cite{deBoer:2021jej}.
 Plugging this expansion, one can find that the Lagrangian expands according to
\begin{eqnarray}
    \mathcal{L} = c^{\hat{\alpha}}\left( \mathcal{L}_0 +  c^2\mathcal{L}_1 +  c^4\mathcal{L}_2 + ... \right), \qquad \hat{\alpha} = 2\alpha -2.
\end{eqnarray}
The leading and sub-leading order terms are 
\begin{eqnarray}
    \mathcal{L}_0 = \frac{1}{2}(\partial_t\phi_0)^2\,,\qquad \mathcal{L}_1 = \partial_t\phi_0\partial_t\phi_1 - \frac{1}{2}(\partial_i\phi_0)^2. 
\end{eqnarray}
Under Lorentz boost transformations, the scalar field $\phi$ transforms as 
\begin{eqnarray}\label{relboost}
    \delta_{B}\phi = ct\beta^i\partial_i\phi + \frac{1}{c}\beta^ix_i\partial_t\phi\,,
\end{eqnarray}
here $\beta^i$ is the Lorentz boost parameter.
One can also substitute the field expansion in the Lorentz boost transformation \eqref{relboost}, and find the transformations for leading and subleading order fields as
\begin{eqnarray}
    \delta_{B}\phi_0 = b^ix_i\partial_t\phi_0\,, \qquad \delta_{B}\phi_1 = b^ix_i\partial_t\phi_1 + tb^i\partial_i\phi_0\,.
\end{eqnarray}
Here we define $\beta^i = c\,b^i$, where $b^i$ is the Carroll boost parameter.
It is straightforward to check that the Lagrangian $\mathcal{L}_0$ is Carroll boost invariant, while the Lagrangian $\mathcal{L}_1$ is not. However, one can make it Carroll boost invariant by adding a Lagrange multiplier to $\mathcal{L}_1$, which sets $\partial_t\phi_0$ to zero on-shell. The modified Lagrangian will be 
\begin{eqnarray}
    \mathcal{L}_1 = \chi\partial_t\phi_0 - \frac{1}{2}(\partial_i\phi_0)^2.
\end{eqnarray}
Here we have absorbed the $\partial_t\phi_1$ term into the Lagrange multiplier $\chi$. Now this Lagrangian will be Carroll boost invariant if we implement the transformation laws
\begin{eqnarray}
    \delta_{B}\phi_0 = b^ix_i\partial_t\phi_0\,, \qquad \delta_{B}\chi = b^ix_i\partial_t\chi + b^i\partial_i\phi_0\,.
\end{eqnarray}

In general, using Noether's prescription one can find the explicit expression for the energy-momentum tensor for any translationally invariant field theory:
\begin{align}\label{EM tensor}
    T^{\mu}_{\,\,\nu} = \frac{\partial \mathcal{L}}{\partial (\partial_{\mu} \phi)} \partial_{\nu} \phi - \delta^{\mu}_{\;\;\nu} \mathcal{L} \,.
\end{align}
The same can be applied to Carrollian field theories and we can obtain a Carrollian stress tensor in this way. We will have a lot more to say about stress tensors in the context of Carrollian CFTs later in the review. But let us point out a rather important feature for stress tensors for generic Carroll invariant field theories, not necessarily conformal. 

For generic Carroll field theories, the energy flux given by the stress tensor component $T^i_{\,\,t}$ vanishes. The reason can be found deeply rooted in the representation theory of Carroll group, where it can be proven that the vanishing of $T^i_{\,\,t}$ is actually the consequence of the Carroll boost Ward identity\footnote{Local Carroll boost invariance demands that $h^\mu_{\;\rho} T^\rho_{\;\nu} \theta^\nu = 0$, where $h$ and $\theta$ are defined in \eqref{FlC}. For details, please refer to \cite{Baiguera:2022lsw}.}. This implies a tell-tale sign of Carroll boost invariance in field theories on flat spacetime: simply $T^i_{\,\,t} = 0$. Throughout the review, we will be using this condition again and again as one of the crucial check for Carroll boost invariance. 

Getting back to the Carroll scalar, it can be checked that for the electric Lagrangian $\mathcal{L}_0$, the component $T^i_{\,\,t}$ is identically zero.  However, for magnetic $\mathcal{L}_1$, the component $T^i_{\,\,t} = -\partial_i\phi_0\partial_t\phi_0$, which vanishes only if we impose the leading order constraint given by $\partial_t\phi_0 = 0$. 

\subsubsection*{Getting the action: Hamiltonian formalism}
Our second method is based on Hamiltonian structures, more importantly zero-signature ones, which have been discussed in the literature quite early \cite{Henneaux:1979vn}.
Here we start again with the same free relativistic scalar field $\phi$ and write down the action in Hamiltonian formalism. Note that, the way to incorporate degenerate metric structures native to Carrollian world into Hamiltonian dynamics can be found in the seminal works \cite{Barnich:2012rz,Henneaux:2021yzg}, especially the latter one. We will mostly focus on examples here, and the reader is invited to look at the original papers for an intrinsic geometric structure. The canonical action for relativistic scalar is given by, 
\begin{eqnarray}\label{KGhamil}
    \mathcal{S}[\phi,\pi_{\phi}] = \int dt \left(\int d^dx\, \pi_{\phi}\dot{\phi} - H\right).
\end{eqnarray}
Here $H$ is Hamiltonian and $\mathcal{E}$ is the Hamiltonian density.
As usual, reinstating all the $c$ factors, we have: 
\begin{eqnarray}
    H = \int d^dx\, \mathcal{E} ,~~\qquad \mathcal{E} = \frac{1}{2}\left(c^2\pi_{\phi}^2 + (\partial_i\phi)^2\right).
\end{eqnarray}
With $\pi_\phi$ being the canonical momenta conjugate to $\phi$.
Now if we directly perform the contraction $c\to 0$ in the action, we get
\begin{eqnarray}
    &&\mathcal{S}_{mag}[\phi,\pi_{\phi}] = \lim_{c\to 0} \mathcal{S}[\phi,\pi_{\phi}] = \int dt \left(\int d^dx\, \pi_{\phi}\dot{\phi} - H_{mag}\right),\\
\text{where} \quad    && H_{mag} = \int d^dx\, \mathcal{E}_{mag} ,~\qquad \mathcal{E}_{mag} = \lim_{c\to 0}\mathcal{E} =  \frac{1}{2}(\partial_i\phi)^2.
\end{eqnarray}
One can also perform another contraction by first rescaling the field and its conjugate momentum such that the canonical structure is preserved and then taking the limit. After rescaling, we have $\phi = c\,\phi'$ and $\pi_{\phi} = \frac{1}{c}\pi'_{\phi}$, on which we take the limit to get,
\begin{eqnarray}
    &&\mathcal{S}_{el}[\phi',\pi'_{\phi}] = \int dt \left(\int d^dx\, \pi_{\phi}'\dot{\phi}' - H_{el}\right),\\
\text{where} \quad    && H_{el} = \int d^dx\, \mathcal{E}_{el} \,,\qquad \mathcal{E}_{el} = \lim_{c\to 0}\mathcal{E} =  \frac{1}{2}\pi_{\phi}'^2\,.
\end{eqnarray}
In this formalism \cite{Henneaux:2021yzg}, the electric scalar theory is the one where in the Hamiltonian density we keep only the momenta (time derivatives of the field) and drop all the spatial derivatives. The magnetic scalar theory is the opposite one, where we keep only the spatial gradients of the fields. The characteristics of these two classes of theories have been summarised in Table \eqref{carrscalartable}.

\begin{table}[H]
\centering
\renewcommand{\arraystretch}{2.7}
\begin{tabular}{|c|c|c|}
\hline
   \textbf{Electric scalar} & $\mathcal{S}_{el}[\phi,\pi_{\phi}] = \int dt d^dx \left(\pi_{\phi}\dot{\phi} - \frac{1}{2}\pi_{\phi}^2\right)$ & $\dot{\phi} = \pi_{\phi}\,,\dot{\pi}_{\phi} = 0$ \\[7pt]
\hline
  \textbf{Magnetic scalar} & $\mathcal{S}_{mag}[\phi,\pi_{\phi}] = \int dt d^dx \left(\pi_{\phi}\dot{\phi} - \frac{1}{2}(\partial_i{\phi})^2\right)$ & $\dot{\phi} = 0\,,\dot{\pi}_{\phi} = \partial_i\partial^i\phi$ \\[7pt]
\hline
\end{tabular}
\caption{Electric and Magnetic Carroll scalar theory in Hamiltonian formalism.}
\label{carrscalartable}
\end{table}

Now we should comment on the canonical structure of these two classes of Hamiltonians. 
Using the canonical commutation relation $\{\phi(x),\pi(x')\}_{\rm PB} =\delta(x-x'),$ one can calculate the Poisson bracket of Hamiltonian densities at different spatial points. For a relativistic field, this leads to 
\begin{eqnarray}\label{dsc}
    \left\{\mathcal{E}(x), \mathcal{E}\left(x^{\prime}\right)\right\}_{\mathrm{PB}}
=2 \mathcal{P}^k(x) \partial_{k}\delta\left(x-x^{\prime}\right)
+\left(\partial_k \mathcal{P}^k(x')\right) \delta\left(x-x^{\prime}\right),
\end{eqnarray}
where $\mathcal{P}^k$ is the momentum density given by $\mathcal{P}_k = c^2\pi \nabla_k\phi$. This is known as the Dirac-Schwinger condition \cite{ds1,ds2} in classical field theory, hailed as a sufficient condition for relativistic covariance. Now one of the trademark of the Carroll dynamics is the vanishing of the above Poisson bracket, i.e.
\be\label{hamizero}
\left\{\mathcal{E}(x), \mathcal{E}\left(x^{\prime}\right)\right\}_{\mathrm{PB}}
= 0.
\ee

%This is true since Carrollian momentum density is identically zero. 
In our case, both of the contractions (electric/magnetic) satisfy the vanishing Poisson bracket relation and hence are Carroll invariant.

\subsection{Vector field}

In this section, we will deal with the vector fields in 4D, in the similar spirit of scalars.  The Lagrangian for the vector fields in the absence of sources is given by $\mathcal{L} =- \frac{1}{4}F_{\mu\nu}F^{\mu\nu}$, where $F_{\mu\nu}$ is the field strength defined as $F_{\mu\nu} = \partial_{\mu}A_{\nu} - \partial_{\nu}A_{\mu}$. 
The vacuum Maxwell's equations can be derived from this Lagrangian and are given by
\begin{eqnarray}
    \nabla \cdot E = 0,~\quad\frac{\partial E}{\partial t} -c^2 \nabla \times B=0,~\quad \nabla \cdot B = 0,~\quad\frac{\partial B}{\partial t} + \nabla \times E = 0.
\end{eqnarray}
Of these, the first two come from varying the Lagrangian with respect to the fields $A_{\mu}$ and the other two is found from the Bianchi identity. Following the work of Le Bellac and Lévy-Leblond \cite{LeBellac:1973unm} in the context of Galilean electromagnetism, it was observed that Maxwell's electromagnetism also admits two different types of Carroll limits \cite{Duval:2014uoa}. If one takes the limit $c\to 0$ keeping $E$ and $B$ fixed, we are left with electric-type Carrollian electromagnetism, whose equations are
\begin{eqnarray}\label{elcarr}
    \nabla \cdot E = 0,~\quad\frac{\partial E}{\partial t} = 0,~\quad \nabla \cdot B = 0,~\quad\frac{\partial B}{\partial t} + \nabla \times E = 0  \,.
\end{eqnarray}
Taking this limit at the Lagrangian level leads to $\mathcal{L}_e = \frac{1}{2}F_{ti}F^{ti} = \frac{1}{2}E^2$. One can check that the variation of this gives rise to \eqref{elcarr}. To get the magnetic version of the Carrollian electromagnetism, we first perform a rescaling $E \to c^2E$ and $B \to B$ in the original equations and then take the Carroll limit. This gives us
\begin{eqnarray}\label{magacarr}
    \nabla \cdot E = 0,~\quad\frac{\partial E}{\partial t} - \nabla \times B=0,~\quad \nabla \cdot B = 0,~\quad\frac{\partial B}{\partial t} = 0.
\end{eqnarray}
Similar to how the relativistic Maxwell's equations are invariant under electromagnetic duality: $E \to c^2B\,,B\to -E$, the two types of Carrollian electromagnetism is also invariant under the following duality transformation: $E_\text{(el)} \to -B_\text{(mag)}\,,B_\text{(el)} \to E_\text{(mag)}$.

\medskip

\subsubsection*{Expansion method}\label{gaugeexp}
Just like in the case of scalars, we would like to discuss methods to get the Carroll electrodynamics actions. Here we will start with the Maxwell's action $\mathcal{L} =- \frac{1}{4}F_{\mu\nu}F^{\mu\nu}$. Here $A_{\mu}$ is the gauge field and we will expand this field in powers of $c$. However, here the components $A_0$ and $A_i$, will be expanded in different power of $c$. This is justified when we look at the one-form
\begin{eqnarray}
    A = A_{\mu}dx^{\mu} = A_0dx^0 + A_idx^i = A_0cdt + A_idx^i = A_tdt + A_idx^i.
\end{eqnarray}
We define $A_t = cA_0$ and the expansion around $c=0$ is given by,  
\begin{eqnarray}
    A_t = c^\alpha\sum_{n = 0}^{\infty}\left(A_t^{(n)}c^{2n}\right),~~\qquad A_i = c^\alpha\sum_{n = 0}^{\infty}\left(A_i^{(n)}c^{2n}\right),
\end{eqnarray}
for arbitrary values of $\alpha$. Substituting these in the Maxwell's action, one gets the leading and the subleading order Lagrangians as
\begin{eqnarray}
  &\mathcal{L}_0 = \frac{1}{2}\left(E_i^{(0)}\right)^2,\label{LOem} \\~~
  &\mathcal{L}_1 = E_i^{(0)}E_i^{(1)} - \frac{1}{4}\left(F_{ij}^{(0)}\right)^2.\label{NLOem}
\end{eqnarray}
Here, $E_i^{(n)} = cF_{0i}^{(n)} = \partial_tA_i^{(n)} - \partial_iA_t^{(n)}$ and $B_i^{(n)} = \frac{1}{2}\epsilon_{ijk}F_{jk}^{(n)}$. Notice that the $\mathcal{L}_0$ \eqref{LOem} is the same action as one gets directly taking the limit from Maxwell's action. Under Lorentz boost, the fields $A_{\mu}$ transform as
\begin{eqnarray}
    \delta_BA_{\mu} = ct\beta^i\partial_iA_{\mu} + \frac{1}{c}\beta^ix_i\partial_tA_{\mu} + \hat{\delta}A_{\mu}.
\end{eqnarray}
Here $\hat{\delta}A_{0} = \beta_iA_i$ and $\hat{\delta}A_{i} = \beta_iA_0$ and $\beta^i$ is the Lorentz boost parameter. Substituting the field expansions in the above equation, one gets the transformations for the $n$-th order gauge fields as
\begin{eqnarray}
    &&\delta A_t^{(n)} = b^ix_i\partial_tA_t^{(n)} + tb^i\partial_iA_t^{(n-1)} + b^iA_i^{(n-1)},\\
    &&\delta A_k^{(n)} = b^ix_i\partial_tA_k^{(n)} + tb^i\partial_iA_k^{(n-1)} + b^i\delta_{ik}A_t^{(n)},
\end{eqnarray}
where $A_{\mu}^{(-1)} = 0$. Varying \eqref{LOem} w.r.t $A_t$ and $A_i$ we will get the Carrollian version of Gauss's and Ampere's law in the leading and subleading order. The other two equations come from the Bianchi identity $\epsilon^{\mu\nu\rho\sigma}\partial_{\nu}F_{\rho\sigma} = 0$, which reads
\begin{eqnarray}
    \partial_iB_i^{(n)} = 0,~~\quad \partial_tB_i^{(n)} + (\nabla \times E^{(n)})_i = 0.
\end{eqnarray}

\subsubsection*{Hamiltonian formalism}
The Maxwell action in Hamiltonian formalism is given by the phase space one, 
\begin{eqnarray}
    \mathcal{S}[A_t, A_i, \pi^i ] = \int dt \left( \int d^dx\, \pi^i\dot{A}_i - H\right).
\end{eqnarray}
Here, Hamiltonian $H$ is defined as
\begin{eqnarray}
     H = \int d^dx\, \left(\mathcal{E} - A_t\partial_i\pi^i\right),~~\qquad \mathcal{E} = \frac{1}{2}\left(c^2\pi^i\pi_i + \frac{1}{2}F^{ij}F_{ij}\right).
\end{eqnarray}
Similar to the scalar field, magnetic Carroll theory is obtained by directly taking $c\to 0$ limit, 
\begin{eqnarray}
    &&\mathcal{S}_{mag}[A_t, A_i, \pi^i ] = \lim_{c\to 0} \mathcal{S}[A_t, A_i, \pi^i ] = \int dt \left(\int d^dx\, \pi^i\dot{A}_i - H_{mag}\right),\\
\text{where} \quad    && H_{mag} = \int d^dx\, \left(\mathcal{E}_{mag}- A_t\partial_i\pi^i\right) \,,\qquad \mathcal{E}_{mag} = \lim_{c\to 0}\mathcal{E} =  \frac{1}{4}F^{ij}F_{ij}\,.
\end{eqnarray}

The electric contraction is performed by first rescaling the field and its conjugate momentum $A_{(t,i)}' = cA_{(t,i)}\,, \pi'^i = \frac{1}{c}\pi^i$, preserving the canonical structure and then taking the limit, 
\begin{eqnarray}
    &&\mathcal{S}_{el}[A_t, A_i, \pi^i ] = \lim_{c\to 0} \mathcal{S}[A_t', A_i', \pi'^i ] = \int dt \left(\int d^dx\, \pi'^i\dot{A}_i' - H_{el}\right),\\
\text{where} \quad    && H_{el} = \int d^dx\, \left(\mathcal{E}_{el}- A_t'\partial_i\pi'^i\right) \,,\qquad \mathcal{E}_{el} = \lim_{c\to 0}\mathcal{E} =  \frac{1}{2}\pi'^i\pi_i'.
\end{eqnarray}
In both of these cases, it can be easily checked that the energy densities at two spatial points commute (see \eqref{hamizero}), i.e.
\begin{equation}
    [\mathcal{E}(x^k), \mathcal{E}(x'^k)] = 0.
\end{equation}
The equations of motion in four spacetime dimensions are given in the Table \eqref{carrgaugetable} below. These can be compared with \eqref{elcarr} and \eqref{magacarr}.
\begin{table}[H]
\centering
\renewcommand{\arraystretch}{2.7}
\begin{tabular}{|c|c|}
\hline
   \textbf{Electric sector} & $\nabla \cdot E = 0\,,\nabla \cdot B = 0\,,\frac{\partial B}{\partial t} - \nabla \times E = 0\,,\frac{\partial E}{\partial t} = 0  $ \\[7pt]
\hline
  \textbf{Magnetic sector} & $\nabla \cdot E = 0\,,\nabla \cdot B = 0\,,\frac{\partial E}{\partial t} + \nabla \times B = 0\,,\frac{\partial B}{\partial t} = 0  $ \\[7pt]
\hline
\end{tabular}
\caption{Field equations for Electric and Magnetic gauge fields}
\label{carrgaugetable}
\end{table}

\subsection{Spinors}\label{spinsec}
In this section, we investigate Carroll spinors in general dimension with the aim of building a coherent framework for their analysis.

\medskip

{\ding{112}} \underline{\em{Carrollian Clifford Algebra}}
\medskip

Let us first discuss the relevant representation theory for Carroll spinors, which will be our main driving power in this section.
Earlier we saw that for Carroll spacetimes, we have two invariant metrics given by $\tilde{h}_{\mu\nu}$ and $\Theta^{\mu\nu}$. In view of this, we propose that the Carrollian version of Clifford algebra is given by \cite{Bagchi:2022eui}:  
\begin{eqnarray}\label{CClf}
\big\{\gc_{\mu},\gc_{\nu}\big\}=2\tilde{h}_{\mu\nu},~~\quad \big\{\gb^{\mu},\gb^{\nu}\big\}=2\Theta^{\mu\nu}.
\end{eqnarray}
Note that, there exist two distinct sets of $\g$ matrices in contrast to the relativistic situation. 
In this review, we will only talk about $\gc$ matrices. For more details and classification of representations, readers can consult the original works \cite{Banerjee:2022ocj,Bagchi:2022eui}. Using these, we can construct a new set of matrices (generators analogue of the Lorentz ones) $\ts_{\mu\nu}$, defined in the following way
\be
\ts_{\mu\nu} \equiv \frac{1}{4}[\gc_{\mu}, \gc_{\nu}],
\ee
and compute the commutation relations between them,
\begin{subequations}\label{carr}
\begin{eqnarray}
&&[\ts_{0i},\ts_{0j}]=0, \\
&&[\ts_{0i},\ts_{jk}]=-\delta_{ij}\ts_{0k}+\delta_{ik}\ts_{0j},\\
&&[\ts_{ij},\ts_{kl}]=-\delta_{il}\ts_{jk}+\delta_{ik}\ts_{jl}-\delta_{jk}\ts_{il}+\delta_{jl}\ts_{ik}.
\end{eqnarray}
\end{subequations}
From the above, it is clear that the algebra indeed comprises the homogeneous component of the Carroll algebra. Here $\ts_{0i}$ are the Carroll boosts and $\ts_{ij}$ are spatial rotations.

\medskip

{\ding{112}} \underline{\em{Structure of Carroll spinors}}
\medskip

Our goal in this section is to construct spinor actions invariant under the Carroll group.
We first start with spinor $\Psi$ and wish to define its adjoint, to construct all possible bilinears that are Carroll invariant, all using the gamma matrices \eqref{CClf} we have constructed. Now, since Carroll Clifford algebra is manifestly degenerate, defining an adjoint with $\gc_0$ would be troublesome. So, we analogously define the adjoint by $\bar{\Psi}= \Psi^{\dagger}\Lambda$, 
where $\Lambda$ is a Hermitian matrix yet to be found by the symmetries. Under Carroll transformation, the spinor field $\Psi(x)$ transforms as 
	\begin{eqnarray}
		\Psi(x) \rw \mathcal{S}[\Sigma]\Psi(x),
	\end{eqnarray}
where $\mathcal{S}[\Sigma] = \exp\big(\frac{1}{2}\omega^{\mu\nu}\ts_{\mu\nu}\big).$
Here $\ts_{\mu\nu}$'s are the generator	of the Carrollian algebra and $\omega^{\mu\nu}$ are antisymmetric parameters. Demanding the Carroll invariance on the simplest bilinear $\bar{\Psi}\Psi$ then requires 
	\begin{equation}\label{sd}
		\ts_{\mu\nu}^\dagger\Lambda + \Lambda\ts_{\mu\nu} = 0.
	\end{equation}		
It is easy to check that $(\ref{sd})$ is satisfied if we can find $\Lambda$ such that 
\begin{subequations}
    \begin{eqnarray}\label{pm}
		&&\gc_{{\mu}}^\dagger = \pm \Lambda\gc_{\mu}\Lambda^{-1} \quad\text{(either + or -)},\\
		&&\ts_{\mu\nu}^\dagger = -\Lambda\ts_{\mu\nu}\Lambda^{-1}\label{pm1}.
	\end{eqnarray}
\end{subequations}

The last relation is merely $(\ref{sd})$ in disguise. This shows that $\bar{\Psi}\Psi$ transforms as a scalar under the Carroll symmetry. One can similarly show that the components of $\bar{\Psi}\gc_\mu\Psi$ change as a Carroll vector under boosts, with our choice of Dirac conjugate. This is a non-trivial statement for Carrollian fermions in the way that for Dirac fermions, the adjoint matrix was part of the Clifford algebra, whereas for the Carroll case it's not.\footnote{In fact in 2D, a faithful real representation of this matrix could be $\Lambda_{R}=-\sigma_2$ so the Carroll boost generators transform using $\ts_{01}^\dagger=-\Lambda_{R} \ts_{01} \Lambda_{R}^{-1}$}

\medskip

{\ding{112}} \underline{\em{Action for Carroll fermions and its symmetries}}\label{sec4}
\medskip

First, we will look at an action, built out from the $\gc_{\mu}$ matrices. We will look at the $D=4$ spacetime dimensions, although our analysis will be valid in other dimensions as well. 
Since the $\gc_{\mu}$ matrices provide a faithful representation, it is clear that the only possible Carroll invariant kinetic term we have is \footnote{One can also have another invariant kinetic term which doesn't involve gamma matrices given by $\Bar{\Psi}\theta^{\mu}\partial_{\mu}\Psi$.}
	\begin{equation}\label{CL}
		\mathcal{L}_{\text{kinetic}} =  \bar{\Psi}\,\theta^{\mu}\theta^{\nu}\,\tilde{\gamma}_{\mu}\partial_{\nu}\, \Psi = \bar{\Psi}\gc_{0}\partial_{t}\Psi . 
	\end{equation}
Now one can write two possible Carroll invariant mass terms, given by  
\begin{eqnarray}
   \mathcal{L}^{(1)}_{\text{mass}} = m\bar{\Psi}\Psi \,,\qquad
   \mathcal{L}^{(2)}_{\text{mass}} = m\bar{\Psi}\,\theta^{\mu}\tilde{\gamma}_{\mu}\,\Psi. 
\end{eqnarray}

\medskip

Due to the degenerate structure in the $\gc$ matrices, only two out of the four components of the Carroll-Dirac fermion $\Psi = \begin{pmatrix}
		\phi &~~~\chi
	\end{pmatrix}^T$ 
take part in the dynamics and the action, and we can add the second type of mass term from above with that. The full action for the lower fermions then read:
\begin{equation} \label{2complag}
S_{\text{lower}} = \int dt d^3x \left(i \phi^\dagger \partial_{t}\phi-m  \phi^\dagger \phi \right),
\end{equation}
where $\phi$ is a 2-component spinor, and as mentioned, there is no dependence on $\chi$.
Each space point in the massive fermion  case possesses a $SU(2)$ symmetry, which expands to $SO(4)$ in the massless case.  

\medskip

Because of the degeneracy in the representation of the Carrollian Clifford algebra, we see that we have a reduced theory of two-component spinors, which leads to the dynamics of a single degree of freedom. Additionally, the spinor components $\phi$ are decoupled in \eqref{2complag}, suggesting that the representation in question is reducible. Therefore half a degree of freedom is appropriate for the minimalistic Carrollian fermion theory\footnote{See, eg. chapter seven of \cite{Henneaux:1992ig}.}.

\medskip

{\em{Continuous symmetries:}} For massless case, the free theory \eqref{2complag} exhibits space-time translation, spatial rotation and the infinite number of supertranslation symmetries. The rotation generators act on the spinor components as:
	\begin{eqnarray} \label{rot_spin_half}
		&&\delta_{n} \phi = \epsilon^{ijk} n_i x_j \partial_k \phi - \frac{i}{2} n^l \sigma_l \phi,
	\end{eqnarray}
for a constant $3$-vector $n^i$.	
The supertranslations act on the 2-component spinor $\phi$ respectively as:
\begin{eqnarray}
	&& \delta_{f} \phi = f(\vec{x}) \partial_{t}\phi.
\end{eqnarray}
Here by $f$ we denote any tensor field on $\mathbb{R}^3$. The special cases $f =1, f = x^i $ and $x^i x_i$	respectively correspond to time translation, Carrollian boost and the temporal part of the special Carrollian conformal transformation.

\medskip

For the massless case, as expected, the dilatation and the spatial part of the special conformal transformation are symmetries as well:
\begin{subequations}
    \begin{eqnarray}
		&&\delta_{\Delta}\phi=(t\partial_{t}+x_{k}\partial_{k}+\Delta)\phi,   \\
		&& \delta_{k}\phi=k^j \left( 2\Delta x_{j}+2x_{j}t\partial_{t}+2x_{i}x_{j}\partial_{i}\phi-x_{i}x_{i}\partial_{j} -i \epsilon_{ijk} x^i \sigma_k \right)\phi.
	\end{eqnarray} 
\end{subequations}

The scaling invariance imposes that $\Delta=\frac{D-1}{2}$. For $D=4$, $\Delta=\frac{3}{2}$, as in the relativistic case, which matches with the previous exposition with Carrollian fermions \cite{Bagchi:2019xfx}. Interestingly, the infinite dimensional supertranslations retain a symmetry of \refb{CL}, despite the fact that the entire finite CCA does not turn out to be symmetries of the massive system because the mass factor clearly breaks scaling invariance.

The Hamiltonian for the Carroll fermions is given by 
\begin{eqnarray}\label{lowerH}
\mathcal{H}_{\text{lower}}=\Pi\dot{\Psi} - \mathcal{L} = \mathcal{L}_{\text{mass}},
\end{eqnarray}
where the momentum conjugate to the spinor is given by $\Pi = \bar{\Psi}\gc_{0}$. Note that this Hamiltonian vanishes for the massless case. The vanishing of the Hamiltonian in the massless case can also be seen after the computation of the energy-momentum tensor for the action. Starting from the Lagrangian \refb{CL}, one can use Noether's procedure to arrive at the energy-momentum tensor.  
Component-wise the above turns out to be: 
\begin{eqnarray}
T^t_{\,\,\,t} =  m  \phi^\dagger \phi, \quad T^i_{\,\,\,t} = 0, \quad T^t_{\,\,\,i} = i \phi^\dagger \d_{i}\phi, \quad T^i_{\,\,\,j}= -\delta^i_j \mathcal{L}.
\end{eqnarray}
A key indicator of a Carroll invariant system is the disappearance of $T^i_{\,\,\,t}$, which results from Carroll boost invariance as discussed earlier around \eqref{EM tensor}.  When $T^t_{\,\,\,t}$ further vanishes in the massless scenario, the system's Hamiltonian likewise vanishes.  In the massless situation, it is also evident that the stress tensor is traceless (on-shell), demonstrating the system's development of conformal Carroll symmetry.

\medskip

In summary, these Carroll fermions come together with the Carroll Clifford algebra which has only spatial legs. The fermions appear to be ``timelike'', however, in that they are not aware of the spatial slice. According to the action \refb{CL}, the spatial components just take a ride here.  Although they are anti-commuting, these fermions do not necessarily have half-integral spins. This only indicates that the fermions do not play a defining role in the spin portion, which is located in the $(D-1)$ dimensional spatial fibre of the $D$ dimensional Carrollian spacetime. Purely, they live along the null-timelike direction in the closed-up lightcone. Essentially, they exhibit the behaviour of a one-dimensional theory that ignores the spatial fibre and only exists on the null line.  For instance, lower fermions on the null boundary of 4D Minkowski spacetimes $\mathscr{I}^\pm$, which is determined by $\mathbb{R}_u \times \mathbb{S}^2$, would only see the null line parametrised by the retarded time $u$ and would not be able to see the celestial spheres connected to each point on the null line. 

\medskip

{\ding{112}} \underline{\em{The limiting approach}}
\medskip

We have so far discussed Carroll fermions in the bottom-up approach i.e. starting from the fundamental $\g$ matrices, building the representations and constructing the invariant action. There has been work on Carroll fermions constructed via the limiting approach \cite{Bergshoeff:2023vfd}, which we now review. Recall that under Lorentz transformation, a Dirac spinor transforms as 
\begin{eqnarray}
    \delta\Psi(X) = \Xi^A\partial_A\Psi(X) - \frac{1}{4}\Lambda_{AB}\Gamma^{AB}\Psi(X)
\end{eqnarray}
where $\Xi^A = \Lambda^{A}_{\,\,B}X^B$ and $\Gamma^{AB} = \frac{1}{2}[\Gamma^A, \Gamma^B] $ induces Lorentz algebra. Note that these generators can be mapped to the ones we used earlier via $\Gamma^{AB}= 2\Sigma^{AB}$, and we will use $\Gamma$ for simplicity in the rest of this subsection.  The indices $A,B$ denote co-ordinates in D-dim. Minkowski spacetime, running from $0$ to $D-1$. One can decompose the label $A = (0,a)$ and redefine the co-ordinates and parameters as 
\begin{eqnarray}
    X^0 = \frac{t}{\tilde{c}}\,,\quad X^a = x^a\,,\quad \Gamma^{AB}= \lambda^{ab}\,,\quad \Gamma^{0a} = \frac{1}{\tilde{c}}\lambda^{0a}
\end{eqnarray}
Now we can take the Carroll limit $\tilde{c} \to \infty$, note that here $\tilde{c}$ is defined as $\tilde{c} = 1/c$, where $c$ is the speed of light. In this way one can obtain the transformation rule for Carroll fermion $\psi = \Psi$ as 
\begin{eqnarray}\label{eltransB}
    \delta \psi = \xi^0\frac{\partial\psi}{\partial t} + \xi^a\frac{\partial\psi}{\partial x^a} - \frac{1}{4}\lambda^{ab}\Gamma_{ab}\psi
\end{eqnarray}
where the parameters $\xi^0, \xi^a$ parametrise Carroll boosts and spatial rotations in the spacetime coordinates. As can be seen, in the spin part of the transformation only spatial rotations appear, which the authors of \cite{Bergshoeff:2023vfd} claim to be a characteristic of electric Carroll fermion. However, there exists another way of taking the limit, after splitting the spinor into two-components using projection matrices:
\begin{eqnarray}
    \Psi_{\pm} = \boldsymbol{P}_{\pm}\Psi\,,\quad\text{where} \,\, \boldsymbol{P}_{\pm} = \frac{1}{2}\left(1 + i\Gamma^0\right)\,.
\end{eqnarray}
and redefining the new set of spinors as $\Psi_{\pm} = \tilde{c}^{\pm 1/2 + \epsilon}\boldsymbol{P}_{\pm}\psi_{\pm}$. Now if one takes the Carroll limit, the transformation rule reads,
\begin{subequations}
    \begin{eqnarray}
    &&\delta \psi_+ = \xi^0\frac{\partial\psi_+}{\partial t} + \xi^a\frac{\partial\psi_+}{\partial x^a} - \frac{1}{4}\lambda^{ab}\Gamma_{ab}\psi_+\label{magtransB1}\\
    &&\delta \psi_- = \xi^0\frac{\partial\psi_-}{\partial t} + \xi^a\frac{\partial\psi_-}{\partial x^a} - \frac{1}{4}\lambda^{ab}\Gamma_{ab}\psi_- - \frac{1}{2}\lambda^{0a}\Gamma_{0a}\psi_+\,\label{magtransB2}.
\end{eqnarray}
\end{subequations}

Note that with these variables, now Carroll boosts appear non-trivially in the spin part of the transformation. \cite{Bergshoeff:2023vfd} claims this as the transformation law for the magnetic Carroll fermion.
\medskip

Now let's look at the level of the Lagrangian. The starting point is the following off-diagonal Lagrangian consisting of two Dirac spinors $\Psi$ and $X$, 
\begin{equation}
    \mathcal{L}_{off-diag} = \bar{X}\Gamma^A\partial_A\Psi - \frac{M}{\tilde{c}}\bar{X}\Psi + h.c.
\end{equation}
Here $M$ is the complex mass parameter. So Carroll limit is obtained by taking limit $\tilde{c} \to \infty$. One introduces projected spinors defined as
\begin{eqnarray}
    \Psi_{\pm} = \boldsymbol{P}_{\pm}\Psi\,,\quad X_{\pm} = \boldsymbol{P}_{\pm}X, \quad \text{where} \,\, \boldsymbol{P}_{\pm} = \frac{1}{2}\left(\mathbb{I} + i\Gamma^0\right)\,.
\end{eqnarray}
The electric limit is then obtained by rescaling the mass parameter $M = \tilde{c}^2m$ together with appropriate rescaling of the fields: 
\begin{eqnarray}
    \Psi_+ = \sqrt{\tilde{c}}\tilde{c}^{-1}\psi_+\,,\quad \Psi_- = \frac{1}{\sqrt{\tilde{c}}}\tilde{c}^{-1}\psi_-\,,\quad X_+ = \sqrt{\tilde{c}}\tilde{c}^{-1}\chi_+\,,\quad X_- = \frac{1}{\sqrt{\tilde{c}}}\tilde{c}^{-1}\chi_-,
\end{eqnarray}
After taking the limit $\tilde{c} \to \infty$, one obtains
\begin{eqnarray}
    \mathcal{L}_{off-diag} = \bar{\chi}_+\Gamma^0\dot{\psi}_+ - m\bar{\chi}_+\psi_+ + h.c.
\end{eqnarray}
This is invariant when both $\psi_+$ and $\chi_+$ transform according to \eqref{eltransB} and thus suggests the truncation $\chi_+ = \psi_+$. Collecting all of this, one obtains the following electric Carroll Lagrangian
\begin{eqnarray}
    \mathcal{L}_{electric} = 2\bar{\psi}_+\Gamma^0\dot{\psi}_+ - 2\mathfrak{Re}(m)\bar{\psi}_+\psi_+, 
\end{eqnarray}
with $\mathfrak{Re}(m) = \frac{1}{2}(m + m^*)$. Similarly, the magnetic limit is performed by rescaling the mass parameter $M = \tilde{c}^2m$ together with the twisted rescaling of the fields
\begin{equation}
    \Psi_+ = \sqrt{\tilde{c}}\tilde{c}^{-1/2}\psi_+\,,\quad \Psi_- = \frac{1}{\sqrt{\tilde{c}}}\tilde{c}^{-1/2}\psi_-\,,\quad X_+ = \frac{1}{\sqrt{\tilde{c}}}\tilde{c}^{-1/2}\chi_+\,,\quad X_- = \sqrt{\tilde{c}}\tilde{c}^{-1/2}\chi_-,
\end{equation}
and taking the limit $\tilde{c} \to \infty$, one obtains
\begin{eqnarray}
    \mathcal{L}_{off-diag} = \bar{\chi}_+\Gamma^0\dot{\psi}_+ + \bar{\chi}_-\Gamma^0\dot{\psi}_- + \bar{\chi}_-\Gamma^a\partial_a\psi_+  - m(\bar{\chi}_+\psi_+ + \bar{\chi}_-\psi_-) + h.c.
\end{eqnarray}
with $\psi_+, \chi_-$ transforming following \eqref{magtransB1} and $\psi_-, \chi_+$ transforming following \eqref{magtransB2}. This suggests for even dimensions one can do the truncation $\chi_{\pm} = \Gamma_*\psi_{\pm}$,  where $$\Gamma_* = (-i)^{1+D/2}\Gamma^0\Gamma^1...\Gamma^{D-1}.$$ 
This leads to the following minimal Lagrangian
\begin{align}
    \mathcal{L}_{magnetic} = 2\bar{\psi}_-\Gamma^0\Gamma_*\dot{\psi}_+ + 2\bar{\psi}_+\Gamma^0\Gamma_*\dot{\psi}_- + 2\bar{\psi}_+\Gamma^a\Gamma_*\partial_a\psi_+  - 2i\mathfrak{Im}(m)(\bar{\psi}_+\Gamma_*\psi_+ + \bar{\psi}_-\Gamma_*\psi_-) .
\end{align}
with $\mathfrak{Im}(m) = \frac{1}{2i}(m-m^*)$. 

\medskip

Although the form of the actions in both these approaches (from intrinsic Carroll Clifford algebra and from limiting analysis) are similar, the main difference lies in the building blocks i.e. in the structure of $\g$ matrices. In the first case, we have started with the intrinsically Carrollian $\g$ matrices, whereas in the limiting analysis although the action and the transformation of the spinors follow Carrollian symmetry, the $\g$ matrices used are Lorentzian in nature.

\subsection{Pointers to literature}\label{pt-3}
There is more work on Carroll field theories available in the literature. We mention some of the more important ones below. Some of the discussions will be in the holography sections, viz. Sec.~\ref{pt-3dAFS} and Sec.~\ref{ssec:pointersotherholo} where they are more appropriate. 

\begin{itemize}
    
\item {\em{Carroll scalars}: } Various aspects of Carroll scalars have been discussed in \cite{Hao:2021urq, Bagchi:2022eav, Baiguera:2022lsw, Bekaert:2022oeh, Liu:2022mne, Banerjee:2023jpi, Ciambelli:2023xqk, deBoer:2023fnj, Afshar:2024llh, Chen:2024voz, Bekaert:2024itn}. We elaborate on some of this below. 
    \begin{itemize}
    
      \item[$\star$] Carroll scalar fields (in 2D) were discussed from a covariant point of view in \cite{Bagchi:2022eav}. Using the vielbein formulation on a flat Carroll manifold, electric/magnetic and a `mixed' Carroll boost invariant field theory action was written down here. 
      
        \item [$\star$] Other aspects of Carroll scalars, especially in 4D, have been discussed in separate works. In \cite{Rivera-Betancour:2022lkc} the authors discuss conformally coupled scalar fields in arbitrary dimensions. In \cite{Ciambelli:2023xqk} a new type of Carroll scalar was proposed which was between Electric and Magnetic representations. This has exciting consequences, like even a single particle with non-vanishing energy can move in this theory. Higher derivative Carroll scalar fields were introduced in \cite{Tadros:2024fgi} and associated Ostrogradsky instabilities were discussed.

        \item[$\star$] Quantisation of Carroll scalar fields, despite being a very much ongoing effort, has also been discussed in various papers. A clear computation occurs in \cite{Banerjee:2023jpi} where symmetries for interacting (electric) Carrollian scalar field theory and their quantum fate has been investigated. This work also discusses a Carrollian version of RG flow and associated fixed point analysis. In-depth discussion about quantising electric and magnetic scalar theories also appears in \cite{deBoer:2023fnj}. Both works also introduce a very special vacuum structure for the Carroll scalar, i.e. the exhibition of infinite degeneracies in the spectrum. Quantisation of Carrollian conformal scalars have been discussed in \cite{Chen:2024voz}.
    \end{itemize}
\item {\em{Carroll fermions}}
    \begin{itemize}
        \item [$\star$] In this review, we have discussed Carroll fermions via a bottom-up construction starting from the fundamental Carrollian $\g$ matrices. This work mainly focussed on the fermionic action constructed out of $\gc_{\mu}$ matrices, however one can also built a Carroll invariant action out of $\hat{\g}^{\mu}$ matrices. For details, the readers are referred to \cite{Bagchi:2022eui}. 
        
        \item[$\star$] In a companion paper \cite{Banerjee:2022ocj},  2D Carroll fermions were studied. Similar works include \cite{Hao:2022xhq} which also describes 2D fermions, including interactions and \cite{Yu:2022bcp} which discusses a BMS analogue of the free Ising model. 

        \item[$\star$] Carroll or ultra-local fermions have appeared in the lattice setting e.g. in a 1+1 D chain in \cite{Ara:2024fbr}. We will have more to say about this in our condensed matter applications. 

        \item[$\star$] Quantisation of Carroll fermions in interacting theory was addressed in \cite{Ekiz:2025hdn}. The fermions were also coupled to a Carrollian scalar field using  a Carrollian Yukawa theory. Furthermore, RG flow and analysis of fixed points have been presented. 
    \end{itemize}
\item {\em{Carroll gauge fields}}
    \begin{itemize}
        \item [$\star$] Carroll gauge fields have also received attention. Early discussions about symmetry structures of Carroll electrodynamics appeared in \cite{Duval:2014uoa, Bagchi:2016bcd} and further in \cite{Bagchi:2019xfx}. In \cite{Basu:2018dub} the canonical phase space and pre-symplectic analysis was performed on a Carroll electrodynamics theory on a flat manifold. \cite{Banerjee:2020qjj} investigated the construction of the most general (interacting) magnetic Carroll electrodynamics theory, focusing on the Helmholtz integrability criterion for differential equations arising out of consistency conditions imposed by CCFTs. 

        \item [$\star$] Beyond the Abelian paradigm, Carroll Yang-Mills theories have also received some attention, beginning with \cite{Bagchi:2016bcd}. In \cite{Islam:2023rnc} electric and magnetic sectors of $SU(N)$ Yang-Mills fields were considered. Depending on the colour index, each sector could have multiple sub-sectors in this case. 

        \item [$\star$] The seminal work \cite{Henneaux:2021yzg} has already discussed Carrollian version of $p$-form gauge theories and higher spin gauge fields. Similar constructions for non-linear conformal/non conformal Carroll electrodynamics has appeared in \cite{Mehra:2024zqv, Correa:2024qej,Chen:2024vho,Chen:2025ndc}.
    \end{itemize}
\item {\em{Carroll swiftons}}
    \begin{itemize}
        \item [$\star$] From representation theory, we know magnetic Carroll particles are the ones propagating strictly outside of the lightcone (`tachyonic'). In Carroll theories, such particles are of special interest since they can actually move \cite{deBoer:2021jej,Casalbuoni:2023bbh}. However, for field theories, finding propagating modes in both electric and magnetic sectors are a challenge. Such (scalar and vector) Carroll field theories with non-vanishing velocities have been recently constructed in \cite{Ecker:2024czx}. Remarkably, the Hamiltonians for such tachyonic field theories are found to be bounded from below, making them free from related instabilities in the Lorentzian cousin. Dubbed `swiftons'\footnote{With a due nod of appreciation to Taylor Swift, or, just to differentiate them from standard tachyons.}, these theories are of great interest, since they can actually carry information, like in the case for Carrollian analogue of Hawking quanta \cite{Aggarwal:2024yxy}.
    \end{itemize}

\item {\em{Generic construction of Carroll field theories}}
    \begin{itemize}

        \item[$\star$] Another way to construct Carroll invariant field theories turns out to be the null-reduction of the Bargmann invariant actions in one higher dimension. Such a construction has been introduced in \cite{Chen:2023pqf}.

        \item[$\star$] Interesting algorithms to build up Carroll invariant (scalar and gauge) field theories have also been discussed in \cite{Bergshoeff:2022qkx} (and follow-ups thereof). The main idea is to start with a so-called `seed' action, starting from which one could interpolate between Galilean and Carrollian theories. This requires one to add additional constraints iteratively via Lagrange multipliers to ensure the desired invariance. 

        \item [$\star$] In related discussions, algebraic expansion methods have been used to focus on theories that deviate one order in $c$ from the Carroll theories, but still far from effective Poincaré symmetries being restored. This results in representations where the commutator between the Hamiltonian and the Boost are non zero, but only in a minimalistic way. In a sense these are like relativistic corrections on Galilean physics, and relies on perturbative corrections to the contracted algebra itself. In this process the starting algebra gets new generators and gets expanded. Such Lie-algebraic expansions have been used to write down \textit{post-Carrollian} theories, both in worldline formalims and in gravity, in \cite{Gomis:2022spp, Ecker:2025ncp}.
    \end{itemize}

\end{itemize}

\newpage

\section{Carrollian Conformal Field Theories: Two dimensions}
\label{ssec:2dccfts}

After discussing the generalities of field theories with Carrollian symmetries (in generic dimensions), in this section we will focus on the conformal cousin of Carroll symmetries, which naturally imbibe a Conformal Carrollian Algebra (CCA). Relativistic conformal field theories are of utmost importance in all branches of physics, and a Carrollian version thereof would be even more interesting. We will start building up the discussion of CCA starting from two dimensions, and higher dimensional cases will follow in the next sections. 

\medskip

In the relativistic world, 2D CFTs occupy a special place with their underlying infinite dimensional Virasoro algebra. The power of infinite dimensional symmetries allows one to compute various quantities of interest often without referring to any Lagrangian description. One of the reasons why Carrollian CFTs hold the promise for being even more powerful is that the Conformal Carroll algebra is infinite dimensional in all dimensions. We saw a glimpse of this when discussing conformal isometries of flat Carrollian manifolds in Sec.\ref{section2}. 

\medskip

The limit from relativistic to Carrollian CFT would be particularly more interesting in $D=2$ as here we would be able to map infinite symmetry structures to each other. Our construction of 2D CCFTs would be inspired by analogous constructions in 2D relativistic CFTs. 

\medskip 

In this section, we first flesh out the limit from the infinite Virasoro algebra to the 2D conformal Carroll algebra and then elaborate on representations of the 2D CCA and the construction of lower point correlations from Ward identities. We then move to four point functions and Carrollian blocks. After this, we concentrate on the stress tensor and finally detail the construction of the partition function and Carroll modular transformations. 

\subsection{Tale of two contractions}\label{sec: Tale of two contraction}
In $D=2$, the Conformal Carrollian symmetries can be deduced from the conformal isometries of the Carroll manifold \eqref{con-car}. The 2D Conformal Carroll Algebra is given by 
\begin{subequations}\label{carr2}
\begin{align} 
& [L_n, L_m] = (n-m)L_{n+m} + \frac{c_L}{12}(n^3-n)\delta_{n+m,0}, \\
& [L_n, M_m] = (n-m)M_{n+m} + \frac{c_M}{12}(n^3-n)\delta_{n+m,0}, 
\quad [M_n, M_m]=0
\end{align}    
\end{subequations}
Here $c_L, c_M$ are two central terms allowed by Jacobi identities. 

\medskip

As well known, symmetries of a relativistic CFT in two dimensions are dictated by two copies of the Virasoro algebra, given by holomorphic and anti-holomorphic generators,  
\begin{subequations}\label{Vira}
    \begin{align}
    [\L_n, \L_m] &= (n-m)\L_{n+m} + \frac{c}{12}\delta_{n+m,0}(n^3-n). \\
    [\bL_n, \bL_m] &= (n-m)\bL_{n+m} + \frac{\bar{c}}{12}\delta_{n+m,0}(n^3-n), \quad [\L_n, \bL_m]=0.
\end{align}
\end{subequations}
Here $c,\bar{c}$ are the central charges associated to the two algebras.
Interestingly, there are two different possible contractions of the above which generates \refb{carr2}. These will be called the Ultra-Relativistic (UR) and Non-relativistic (NR) contractions respectively: 
\begin{subequations}\label{contr}
   \begin{align}
    &\text{UR contraction:} \quad L_n = \L_n - \bL_{-n}, \quad M_n = \e(\L_n + \bL_{-n}), \label{contr-ur} \\
    &\text{NR contraction:} \quad L_n = \L_n + \bL_{n}, \quad M_n = \e(\L_n - \bL_{n}). \label{contr-nr}
\end{align} 
\end{subequations}
Notice that the UR contractions mix creation and annihilation operators (positive and negative mode numbers) while the NR contraction does not. This will have profound consequences later. 
\medskip

First a bit about the nomenclature. We saw that in general dimensions, the non-relativistic contraction leads to the Galilean algebra from the Poincaré algebra in previous sections while the ultra-relativistic contraction generated the Carroll algebra. These were given by
\begin{align}
    &\text{UR contraction:} \quad c\to 0 \equiv \, t \to \e t, x^i \to x^i , \\
    &\text{NR contraction:} \quad  c\to \infty \equiv \, t \to t, x^i \to \e x^i.
\end{align}
If we choose to work on the cylinder for the 2D relativistic CFT, the Virasoro generators are given by
\begin{align}\label{virgen}
    \L_n = \exp(i n \omega)\partial_\omega, \quad \bL_n = \exp(i n \bar \omega)\partial_{\bar\omega}, \quad \text{with }   \omega, \bar \omega = \t \pm \sigma,
\end{align}
where $\t$ is the temporal direction and $\sigma$ is the angular direction on the cylinder. We now do the contractions by plugging in \eqref{virgen} in \eqref{contr}: 
\begin{subequations}
\begin{align}
    & \text{UR contraction: }(\t \to \e \t, \sigma \to \sigma) \quad 
     L_n = e^{in\sigma}\left( \partial_\sigma - in \tau \partial_\tau \right), \quad M_n = e^{in\sigma}\partial_\tau. \label{car-cyl}\\
    & \text{NR contraction: }(\t \to \t, \sigma \to \e \sigma) \quad 
     L_n = e^{in\t}\left( \partial_\t - in \sigma \partial_\sigma \right), \quad M_n = e^{in\t}\partial_\sigma. \label{gal-cyl}
\end{align}    
\end{subequations}
The generators go over to each other under a swap of spatial and temporal directions: $\sigma\leftrightarrow\t$ {\footnote{Although the above works like a charm for the vector field representation on the cylinder, for the more familiar representations on the plane, the NR contraction works, while the UR one blows up. One can remedy this by choosing a reality condition different from $\bar{z}=z^\star$. See \cite{Hao:2021urq} for details.}}. The magic of two dimensions means that the Carroll and Galilean contractions of the relativistic Conformal (and of course Poincaré) algebra give isomorphic algebras. This is because of the fact that there is one contracted and one non-contracted direction in the spacetime and the symmetry algebra is blind to the nature of these directions. 

\subsection{Representations of 2D CCA}
\label{ssec:2dreps}
We now construct representations of 2D  Carroll CFTs. The states are labelled with their weights under $L_0$ and since $[L_0, M_0]=0$, one also needs to label them under $M_0$. 
\begin{align}
    L_0 |\Delta, \xi \rangle = \Delta |\Delta, \xi \rangle, \quad M_0 |\Delta, \xi \rangle = \xi |\Delta, \xi \rangle. 
\end{align}
We will now discuss two very different looking representations. We begin with the more familiar highest weight representations. For this, notice
\begin{align}
   L_0 L_n |\Delta, \xi \rangle = (\D - n) L_n |\Delta, \xi \rangle, \quad L_0 M_n |\Delta, \xi \rangle = (\D - n) M_n |\Delta, \xi \rangle.
\end{align}
Hence $L_n, M_n$ with $n>0$ lower the $\D$ eigenvalue of the state. In parallel with usual 2D  relativistic CFTs, we will thus define Carrollian primaries $|\Delta, \xi \rangle_p$ as those for which the $\D$ values cannot be lowered further:
\begin{align}
    L_n |\Delta, \xi \rangle_p = 0, \quad M_n |\Delta, \xi \rangle_p = 0, \quad n>0. 
\end{align}
A generic state is created from a primary state by acting with raising operators. Although these representations seem most natural to consider from the point of view of a CFT, as we will discuss below, they are generically non-unitary \cite{Bagchi:2009pe}. 
\medskip

This brings us to another set of representations called the induced representations, where we define the so called ``rest frame'' state $|M, s \rangle$ \cite{Campoleoni:2016vsh}: 
\begin{align} \label{ind}
 M_0 |M, s \rangle = M |M, s \rangle, \quad L_0 |M, s \rangle = s |M, s \rangle, \quad  M_n |M, s \rangle = 0, \quad \forall n \neq 0. 
\end{align}
In this case, a generic state can be built out of the rest frame state by acting with $L_n$'s with arbitrary $n$ both positive and negative, 
\begin{align}
    |\psi\rangle = L_{n_1} L_{n_2} \ldots L_{n_m} |M, s \rangle. 
\end{align}
While at first sight the induced representation may come across as unfamiliar, especially from a field theoretic point of view, things become clearer once one considers the limit of the Virasoro representations. In a 2D  CFT, states are labeled under their $(\L_0, \bL_0)$ eigenvalues:
\begin{align}
    \L_0 |h, \h \> = h |h, \h \>, \quad \bL_0 |h, \h \> = \h |h, \h \>. 
\end{align}
Primary states are defined as 
\begin{align}
    \L_n |h, \h \>_p = 0, \quad \bL_n |h, \h \>_p = 0. 
\end{align}
Now consider the UR limit defined above \eqref{contr}. Following the mapping, the primary state condition becomes
\begin{subequations}\label{ind-lim}
    \begin{align}
    & \L_n |h, \h \>_p = 0 \Rightarrow \left(L_n + \frac{1}{\e} M_n\right) |h, \h \>_p = 0, \\
    & \bL_n |h, \h \>_p = 0 \Rightarrow \left(- L_{-n} + \frac{1}{\e} M_{-n}\right) |h, \h \>_p = 0.
\end{align}
\end{subequations}
Note that 
\begin{align}
    M = \e (h+\h), \quad s= h-\h. 
\end{align}
If the $\e \to 0$ limit is performed at fixed $M$ and $s$, then we see that \eqref{ind-lim} reduces to \eqref{ind}. So if some aspects of relativistic CFTs were to descend to CCFTs in $D=2$, it is likely that the induced representations would play an important role. 

\medskip

On the other hand, if we look at the NR limit in \eqref{contr}, the highest weights of the Virasoro algebra descend to highest weights of 2D CCFT. One expects both sets of representations to be important in the understanding of 2D CCFTs, perhaps in different physical contexts. 

Let us consider some curious aspects of the highest weight representations of 2D CCFTs now. Consider the action of $M_0$ on descendants of a primary $|\D, \xi\>$:
\begin{align}
    M_0 L_{-n} |\D, \xi\> = \xi \, L_{-n} |\D, \xi\> + M_{-n} |\D, \xi\>. 
\end{align}
The action of $M_0$ on highest weight modules of 2D CCFTs is thus generically non-diagonal. Hence $L_0$ and $M_0$ cannot be simultaneously diagonalised in a highest weight module, even if the primary states themselves are simultaneous eigenstates of the two operators. This opens up the possibility of more generic representations where $L_0$ and $M_0$ are not diagonal even on primaries and form a Jordan canonical structure. 
\medskip

The above feature is immediately reminiscent of similar non-diagonal structures found in logarithmic CFTs \footnote{See, for example, \cite{Creutzig:2013hma} for a self-contained introduction.}. It is known that log CFTs are generically non-unitary. Non-unitarity is also a feature of the highest weight representations of 2D CCFTs. We can see this if we consider the Kac determinant for the representations. This is negative for $\xi\neq0$. 

\medskip

In these generic non-trivial multiplet representations, it is possible to choose a basis where the action of $L_0$ is diagonal while $M_0$ has a Jordan block structure. When one considers spacetime representations, it becomes natural to consider this choice. To see this, let us look at the planar representation of the 2D CCA:
\begin{align}\label{CCA-pl}
    L_n = x^{n+1}\partial_x + (n+1) x^n t \partial_t, \quad M_n = -x^{n+1}\partial_t.
\end{align}
We note that one can get to these planar representations of the CCA from the cylinder representations \eqref{car-cyl} by the plane-to-cylinder map:
\begin{align}\label{p2c}
    x = e^{i \sigma}, \quad t = i\t e^{i\sigma}.
\end{align}

For the global subalgebra generated by $L_{0, \pm 1}, M_{0, \pm1}$, these plane generators are exactly the ones we obtained earlier in generic dimensions \eqref{Car-gen}, \eqref{CCar-gen}. As noted above, the centre of the algebra is generated by $(L_0, M_0)$. $L_0 = x\partial_x + t\partial_t$ gives rise to dilatation on the plane while $M_0 = - x\partial_t$ generates Carroll boosts. Carroll boosts make the following spacetime transformation:
\begin{align}
    x' = x, \quad t' = t + \beta x \quad \Rightarrow
    \begin{pmatrix} t' \\ x'   \end{pmatrix} =  \mathscr{C} \begin{pmatrix} t \\ x \end{pmatrix}, \quad \text{where} \quad \mathscr{C} = \begin{pmatrix}
        1 & \beta \\ 0 & 1 \end{pmatrix}. 
\end{align}
This is like a 2D Galilean boost with space and time interchanged, so that time is relative and space is absolute as we saw before \eqref{cartrans}. The action of this boost on spacetime vectors or tensors is thus not diagonal. Thus if we consider tensorial Carroll primaries $(\Phi)$, these would naturally transform in multiplets:
\begin{align}
    \Phi^{a_1 a_2 \ldots a_n } (t', x') = \mathscr{C}^{a_1}_{\, \, b_1} \mathscr{C}^{a_2}_{\, \, b_2} \ldots \mathscr{C}^{a_n}_{\, \, b_n} \, \, \Phi^{b_1 b_2 \ldots b_n } (t, x). 
\end{align}
Below, when reviewing correlation functions, we will focus only on the non-multiplet or singlet representations. More details of the multiplet representations can be found in \cite{Chen:2020vvn}. 

\medskip

We have thus discussed highest weight and induced representations. As mentioned above, the highest weight representations of a 2D CCFT are generically non-unitary. On the other hand, the induced representations are constructed from unitarity considerations and are manifestly unitary. An interesting point is that although these two representations seem very different, the characters of the two representations are identical (up to a modulus sign) \cite{Bagchi:2019unf}. We will return to this when we discuss modular transformations in the context of Carroll CFTs below. 

\medskip

It is important to stress that both highest weight and induced representations exist for 2D CCFTs. Just because one (highest weight) descends from the non-relativistic limit and the other (induced) from the ultra-relativistic limit of the Virasoro highest weight representations, it does not mean that in a Carrollian theory (which follows that UR limit) cannot have highest weight representations. It is just that these highest weight representations in a 2D CCFT are intrinsically defined and don't follow the UR limit. We will now focus primarily on highest weight representations. This is what was done historically and primarily because of close analogy with methods of 2D  CFT which could be imported to the 2D Carroll theory. 

\subsection{Correlation functions: 2 and 3 points}
In relativistic CFTs, conformal invariance is strong enough to determine two and three point correlation functions of the theory. We will see that this is a feature that is also true in CCFTs in arbitrary dimensions. For now, we focus on $D=2$. We will return to similar considerations in higher dimensions in Sec.~\ref{sec:carrfields4}. 

\medskip

We will work on the Carroll plane $\mathbb{R}_t \times \mathbb{R}_x$ remembering that $t$ is now the {\em null} time. We will use the plane representations of the 2D CCA \eqref{CCA-pl}. We begin by postulating a state-operator correspondence for the 2D CCFT {\footnote{Let us stress that this indeed is a postulate and there may well be intricacies in defining this on Carroll manifolds because of its non-compact nature. Most of what we discuss would be applicable without this state-operator correspondence. We adopt this for ease of computations.}}: 
\begin{align}
    \mathcal{O}_{\D, \xi}(0, 0) |0\> \equiv |\D, \xi\> 
\end{align}
where $\mathcal{O}_{\D, \xi}$ is a Carroll primary operator which is labelled at the origin by $(\Delta,\xi)$, the eigenvalues of $(L_0,M_0)$ 
\begin{eqnarray}
    [L_0, \mathcal{O}] = \Delta \mathcal{O}\,,\quad[M_0,\mathcal{O}] = \xi \mathcal{O}\,,
\end{eqnarray} 
where we have dropped the subscripts on the operator. We will continue to do so going forward. To find the operators at other spacetime positions, one can act on $\mathcal{O}$ by the translation operator $U$: 
\begin{equation}
    \mathcal{O}(t,x) = U\mathcal{O}(0,0)U^{-1}, \quad U = e^{tM_{-1} + xL_{-1}}.
\end{equation}
Using the Baker-Campbell-Hausdorff formula, it can be shown that 
\begin{subequations}
 \begin{align}
    &[L_n,\mathcal{O}(t,x)] = \left(x^{n+1}\d_x + (n+1)tx^n\d_t + (n+1)(x^n\Delta - ntx^{n-1}\xi)\right)\mathcal{O}(t,x) ,\\
    &[M_n,\mathcal{O}(t,x)] = \left(-x^{n+1}\d_t + (n+1)x^n\xi \right)\mathcal{O}(t,x)\,,\quad n\geq -1.
\end{align}   
\end{subequations}
We now want to look at the two-point functions between two primary operators $\mathcal{O}_1(t_1,x_1)$ and $\mathcal{O}_2(t_2,x_2)$, defined as 
\begin{eqnarray}
    G^{(2)}(t_1,x_1;t_2,x_2) = \langle 0|\mathcal{O}_1(t_1,x_1)\mathcal{O}_2(t_2,x_2)|0\rangle.
\end{eqnarray}
Requiring the vacuum state to be invariant under global Conformal Carroll symmetries i.e. $L_{0,\pm 1}$ and $M_{0,\pm 1}$, the form of the two-function is fully constrained by Ward identities. Using $L_{-1}$ and $M_{-1}$, $G^{(2)}$ is translation invariant in two coordinates and hence one can write 
\begin{equation}
    G^{(2)}(t_1,x_1;t_2,x_2) = G^{(2)}(t_{12},x_{12}),
\end{equation}
where $t_{12}=t_1-t_2$ and $x_{12}=x_1-x_2$. Under the Carrollian boost $M_{-1}$
\begin{eqnarray}
    &&\left(-x_1\d_{t_1} - x_2\d_{t_2} + \xi_1 + \xi_2\right)G^{(2)} = 0 \implies \left(-x_{12}\d_{t_{12}} + \xi_1 + \xi_2\right)G^{(2)}(t_{12},x_{12}) = 0\nonumber\\
    \implies &&G^{(2)}(t_{12},x_{12}) = f(x_{12})\exp\left((\xi^1+\xi^2)\frac{t_{12}}{x_{12}}\right).
\end{eqnarray}
The functional form of $f(x_{12})$ can be fixed by looking at the behavior under dilatation $L_0$, which similarly gives
\begin{eqnarray}
    &&\left(t_{12}\d_{t_{12}} + x_{12}\d_{x_{12}} + \Delta_1 + \Delta_2\right)G^{(2)}(t_{12},x_{12}) = 0\nonumber\\
    \implies && G^{(2)}(t_{12},x_{12}) = Cx_{12}^{-(\Delta_1 +\Delta_2)}\exp\left((\xi_1+\xi_2)\frac{t_{12}}{x_{12}}\right)\,,
\end{eqnarray}
$C$ being an arbitrary constant. Finally, under the action of $L_1$ and $M_1$, we get $\Delta_1 = \Delta_2$ and $\xi_1 = \xi_2$. So the two-point function reads as
\begin{equation}\label{bms32pt}
    G^{(2)}(t_{12},x_{12}) = C_{12}x_{12}^{-2\Delta_{1}}\exp\left(2\xi_1\frac{t_{12}}{x_{12}}\right) \delta_{\D_1, \D_2} \delta_{\xi_1, \xi_2} \,.
\end{equation}
Here $C_{12}$ is a normalization constant which can be chosen to be $\delta_{12}$. 

\medskip

In the same spirit, one can construct the three-point function of three Carroll primary operators. The explicit form is given by 

\begin{align}\label{bms33pt}
    &G^{(3)}(t_1,x_1;t_2,x_2;t_3,x_3) = \langle 0|\mathcal{O}_1(t_1,x_1)\mathcal{O}_2(t_2,x_2)\mathcal{O}_3(t_3,x_3)|0\rangle\nonumber\\
    &= C_{123}|x_{12}|^{\Delta_{123}}|x_{23}|^{\Delta_{231}}|x_{31}|^{\Delta_{312}}\exp\left(-\xi_{123}\frac{t_{12}}{x_{12}}\right)\exp\left(-\xi_{231}\frac{t_{23}}{x_{23}}\right)\exp\left(-\xi_{312}\frac{t_{31}}{x_{31}}\right)\,.
\end{align}
Here $C_{123}$ is called the structure constant and we have defined 
\begin{equation}
    x_{ij} = x_i-x_j\,,\quad t_{ij} = t_i-t_j\,,\quad \Delta_{ijk} = -\Delta_i - \Delta_j + \Delta_k\,,\quad \xi_{ijk} = -\xi_i-\xi_j+\xi_k\,.
\end{equation}
So we see that in a 2D CCFT, the two and three point functions are fixed by symmetry to constants. The structure constants of the three-point function will be important in the Carroll analogue of the relativistic bootstrap programme. 

\subsection{Four points: Crossing, Bootstrap and Carroll blocks}
So far, we have only considered two and three-point functions. One can also look at higher point correlation functions and check the extent to which symmetry alone fixes their form. 

\newpage
{\ding{112}} \underline{\em{Operator algebra}}
\medskip

Now, since all the features of correlation functions are included in the operator product algebra, it suffices to examine constraints on the OPE to understand how symmetries constrain the correlation functions. We consider the following ansatz for the OPE of two Carrollian primary field operators $\mathcal{O}_{1,2}$ with weights $(\D_1,\xi_1)$ and $(\D_2,\xi_2)$
\begin{align}\label{OPE}
    &\mathcal{O}_{1}(x_1,t_1)\mathcal{O}_{2}(x_2,t_2)=\sum_{p,\{\vec{k},\vec{q}\}} x_{12}^{\D_{12p}}\,e^{-\xi_{12p}\frac{t_{12}}{x_{12}}}\left(\sum_{\a=0}^{K+Q} C_{12}^{p\{\vec{k},\vec{q}\},\a}x_{12}^{K+Q-\a}t_{12}^{\a}\right)\,\mathcal{O}_{p}^{\{\vec{k},\vec{q}\}}(x_2,t_2).
\end{align}
where $\vec{k}=(k_1,k_2,...k_r)$ and $\vec{q}=(q_1,q_2,...q_s)$, and $K=\sum_{i}k_{i},\,Q=\sum_{i}q_{i}$ and the descendant fields $\mathcal{O}_{p}^{\{\vec{k},\vec{q}\}}(x_2,t_2)$ are denoted as, 
\begin{align}
    \mathcal{O}_{p}^{\{\vec{k},\vec{q}\}}(x,t)&=\left((L_{-1})^{k_1}...(L_{-l})^{k_l}(M_{-1})^{q_1}...(M_{-j})^{q_j}\mathcal{O}_p\right)(x,t)\equiv \left(L_{\vec{k}}M_{\vec{q}}\mathcal{O}_p\right)(x,t).
\end{align}
In \eqref{OPE}, the term in front of the parentheses is fixed by the need for the OPE to provide the correct two-point function and the term in parentheses is to guarantee when $L_0$ acts on the equation, both sides of the OPE change in the same manner. Furthermore, one can use this OPE inside the three-point function and compare it with \eqref{bms33pt}. It gives $C_{12}^{p\{0,0\},0}\equiv C_{12}^{p}=C_{p12}.$ So, one can rewrite 
\begin{eqnarray}
    C_{12}^{p\{\vec{k},\vec{q}\},\a}=C_{12}^{p}\,\b_{12}^{p\{\vec{k},\vec{q}\},\a},
\end{eqnarray}
where $\b_{12}^{p\{0,0\},0}=1$, by convention. To compute the coefficients $\b_{12}^{p\{\vec{k},\vec{q}\},\a}$, we demand that both sides of the OPE transform in the same manner when acted on by $L_n, M_n$. Without loss of generality, we consider the simplest case $\D_{1}=\D_{2}=\D,\,\xi_{1}=\xi_{2}=\xi$ and act both sides of \eqref{OPE} on the vacuum state. This gives us 
\begin{eqnarray}\label{opev}
    \mathcal{O}(x,t)|\D,\xi\>=\sum_{p}x^{-2\D+\D_{p}}\,e^{(2\xi-\xi_{p})\frac{t}{x}}\sum_{N\geq \a}C_{12}^{p}x^{N-\a}t^{\a}|N,\a\>_p\,.
\end{eqnarray}
Here the state $|N,\a\>_p$ is a descendant state at level N, given by, 
\begin{eqnarray}
    |N,\a\>_p=\sum_{\stackrel{ \{\vec{k},\vec{q}\},}{K+Q=N,\,\a\leq N}}\b_{12}^{p\{\vec{k},\vec{q}\},\a}L_{\vec{k}}M_{\vec{q}}|\D_p,\xi_p\rangle,
\end{eqnarray}
such that 
\be
L_{0}|N,\a\>_p=(\D_{p}+N)|N,\a\>_p.
\ee
Acting $L_n, M_0, M_n$ on both sides of \eqref{opev}, and after a little bit of calculation,  we get three recursion relations
\begin{align}
    &L_{n}|N+n,\a\>_p  =  \left(N+n \a -\D +n \D +\D _p\right)|N,\a\>_p 
  +\left(n \xi -n^2 \xi -n \xi _p\right)|N,\a -1\>_p\,,\nonumber\\
  &M_{0}|N,\a\>_p = \xi_{p} |N,\a,\>_p-(\a +1)|N,\a +1\>_p\,,\\
  &M_{n}|N+n,\a\>_p  =  \left((n-1) \xi +\xi _p\right)|N,\a\>_p-(\a+1)|N,\a+1\>_p. \nonumber
\end{align}
The coefficients $\b_{12}^{p\{\vec{k},\vec{q}\},\a}$ can be computed using the above recursion relations. For details of the calculation, please refer to \cite{Bagchi:2016geg, Bagchi:2017cpu}. 

\medskip
{\ding{112}} \underline{\em{Crossing symmetry and Bootstrap}}
\medskip

Next, we want to look at the four-point function. Just like usual relativistic CFT, a four-point function of Carroll primaries is not completely fixed by the global conformal Carroll group. They depend on the Carroll analogue of cross-ratios given by 
\begin{eqnarray}
    u = \frac{x_{12}x_{34}}{x_{13}x_{24}}\,,\quad \frac{v}{u} = \frac{t_{12}}{x_{12}} + \frac{t_{34}}{x_{34}}  -\frac{t_{13}}{x_{13}}  -\frac{t_{24}}{x_{24}}\,, 
\end{eqnarray}
which are invariant under the global subgroup. The functional form of a Carroll conformal four-point function looks like  
\begin{eqnarray}
    \langle\prod_{i=1}^4\mathcal{O}_i(x_i,t_i)\rangle = P(\{\D_i,\xi_i,t_{ij},x_{ij}\})\mathcal{G}_{Car}(u,v),
\end{eqnarray}
where $P(\{\D_i,\xi_i,x_{ij},t_{ij}\}) = \prod_{1\leq i<j\leq4}x_{ij}^{\sum_{k=1}^{4}\D_{ijk}/3}e^{-\frac{t_{ij}}{x_{ij}}\sum_{k=1}^{4}\xi_{ijk}/3}$. Since these new cross-ratios are invariant under global transformations, one can perform a transformation in order to set 
\be   
\{(x_i,t_i) \} \rightarrow \{(\infty,0), (1,0), (u,v), (0,0)\},
\ee 
and then the four-point function can be related to a matrix element between in and out states, defined as 
\begin{eqnarray}
    G_{34}^{21}(u,v) &&\equiv \lim_{x_{1}\rightarrow\infty,t_{1}\rightarrow0}x_1^{2\D_{1}}\exp\left(-\frac{2\xi_{1}t_{1}}{x_{1}}\right)\langle\mathcal{O}_{1}(x_{1},t_{1})\mathcal{O}_{2}(1,0)\mathcal{O}_{3}(u,v)\mathcal{O}_{4}(0,0)\rangle \\
    && = \langle\D_{1},\xi_{1}|\mathcal{O}_{2}(1,0)\mathcal{O}_{3}(u,v)|\D_{4},\xi_{4}\rangle.
\end{eqnarray}
Note that the sequence in which the indices appear is crucial. 
After a little bit of calculation, it can be easily checked that the four-point function can be expressed as 
\begin{eqnarray}
    \<\prod_{i=1}^{4}\mathcal{O}_{i}(x_{i},t_{i})\> = P(\{\D_i,\xi_i,x_{ij},t_{ij}\}) f(u,v)^{-1}G_{34}^{21}(u,v),
\end{eqnarray}
where 
\begin{eqnarray}
    f(u,v)&=&(1-u)^{\frac{1}{3}(\D_{231}+\D_{234})}u^{\frac{1}{3}(\D_{341}+\D_{342})} e^{\frac{v}{3(1-u)}(\xi_{231}+\xi_{234})}e^{-\frac{v}{3u}(\xi_{341}+\xi_{342})}\,.
\end{eqnarray}
Now to define the above function $G$, we have taken a particular order of the operators within the correlator. However, except a sign change for fermions, this ordering does not matter. So, one can otherwise define another function 
\begin{eqnarray}
    G_{32}^{41}(u,v)=\langle\D_{1},\xi_{1}|\mathcal{O}_{4}(1,0)\mathcal{O}_{3}(u,v)|\D_{2},\xi_{2}\rangle,
\end{eqnarray}
and it can be obtained by sending $\{(x_i,t_i) \} \rightarrow \{(\infty,0), (0,0), (1-u,-v), (1,0)\}$. Then we obtain the identity 
\begin{eqnarray}
    G_{34}^{21}(u,v)=G_{32}^{41}(1-u,-v).
\end{eqnarray}
This condition is manifestation of crossing symmetry. Using the operator algebra between $\mathcal{O_2}$ and $\mathcal{O}_3$, the function $G_{34}^{21}(u,v)$ can be written as 
\begin{eqnarray}
    G_{34}^{21}(u,v)=\sum_{p}C_{34}^{p}C_{12}^{p}A_{34}^{21}(p|u,v),
\end{eqnarray}
where we have introduced the four-point conformal block 
\begin{align}
    A_{34}^{21}(p|u,v)	&=	(C_{12}^{p})^{-1}u^{-\D_{3}-\D_{4}+\D_{p}}\,e^{(\xi_{3}+\xi_{4}-\xi_{p})\frac{v}{u}}\sum_{N\geq\a}u^{N-\a}v^{\a}\<\D_{1},\xi_{1}|\mathcal{O}_{2}(1,0)|N,\a\>_p \cr
	&=	u^{-\D_{3}-\D_{4}+\D_{p}}\,e^{(\xi_{3}+\xi_{4}-\xi_{p})\frac{v}{u}} \cr
	&	\times\sum_{\{\vec{k},\vec{q}\}}\left(\sum_{\a=0}^{K+Q}\b_{34}^{p\{\vec{k},\vec{q}\},\a}u^{K+Q-\a}v^{\a}\right)\frac{\<\D_{1},\xi_{1}|\mathcal{O}_{2}(1,0)L_{\vec{k}}M_{\vec{q}}|\D_{p},\xi_{p}\>}{\<\D_{1},\xi_{1}|\mathcal{O}_{2}(1,0)|\D_{p},\xi_{p}\>} \cr
\end{align}
Similar way using OPE of $\mathcal{O}_4$ and $\mathcal{O}_3$, one can write down 
\begin{eqnarray}
    G_{32}^{41}(u,v)=\sum_p C^p_{23}C^{p}_{14} A_{32}^{41}(p|u,v).
\end{eqnarray}
Using crossing symmetry, one can get 
\begin{eqnarray}
    \sum_{p}C_{34}^{p}C_{12}^{p}A_{34}^{21}(p|u,v)=\sum_{q}C_{32}^{q}C_{41}^{q}A_{32}^{41}(q|1-u,-v),
\end{eqnarray}
which is the Carroll bootstrap equation (known in the previous literature \cite{Bagchi:2017cpu} as the BMS bootstrap equation). We can solve it to identify every consistent Carroll invariant conformal theory provided we know the closed form expressions of the blocks. Nevertheless, even though symmetry fixes the Carroll blocks entirely, we can only solve them within a simplifying limit, which we now address. 

\medskip
{\ding{112}} \underline{\em{Large central charge limit and Global Carroll block}}
\medskip

Following \cite{Dolan:2000ut, Dolan:2003hv} for 2D CFTs, one can get a closed form expression for the global conformal blocks by taking the large central charge limit on the Virasoro conformal blocks. A similar technique can be employed to obtain the global Conformal Carroll blocks. We do this by taking the asymptotic limit $c_L,c_M\to \infty$ in \eqref{OPE}. Then the leading order term will contain those descendant fields generated by $L_{-1}, M_{-1}$: 
\begin{align}
    \mathcal{O}_{3}(x_3,t_3)\mathcal{O}_{4}(x_4,t_4) = &\sum_{p,\{k,q\}}x_{34}^{\D_{34p}}\,e^{-\xi_{34p}\frac{t_{34}}{x_{34}}}C_{34}^{p}\, \nonumber \\ & \sum_{\a=0}^{N=k+q}\b_{34}^{p\{k,q\},\a}u^{k+q-\a}v^{\a} \, (L_{-1})^{k}(M_{-1})^{q}\mathcal{O}_p(x_4,t_4) \nonumber\\
&+ \mathscr{O}\left(\frac{1}{c_L},\frac{1}{c_M}\right)+\ldots
\end{align}
This implies the function $G_{34}^{21}(u,v)$ exhibits an expansion of the form 
\begin{align}
    G_{34}^{21}(u,v)= \sum_{p}C_{12}^{p}C_{34}^{p}\,g_{34}^{21}(p|u,v) + \mathscr{O}\left(\frac{1}{c_L},\frac{1}{c_M}\right)+...\,.
\end{align}
Here $g_{ij}^{kl}(p|u,v)$ are the global Carroll blocks, which can be obtained by taking the large central charge limit of the blocks $A_{ij}^{kl}(p|u,v)$,
\begin{align}
    g_{ij}^{kl}(p|u,v)&=\lim_{c_L,c_M\to \infty} A_{ij}^{kl}(p|u,v)\\
    = u^{\D_{34p}}\,e^{-\xi_{34p}\frac{v}{u}} 
	& \sum_{\{k,q\}}\left(\sum_{\a=0}^{N=k+q}\b_{34}^{p\{k,q\},\a}u^{N-\a}v^{\a}\right)\frac{\langle\D_{1},\xi_{1}|\mathcal{O}_{2}(1,0)(L_{-1})^{k}(M_{-1})^{q}|\D_{p},\xi_{p}\rangle}{\langle\D_{1},\xi_{1}|\mathcal{O}_{2}(1,0)|\D_{p},\xi_{p}\rangle}.\nonumber
\end{align}
By requiring that both sides of the OPE transform in the same manner under the action of the quadratic Casimirs belonging to the global algebra formed by $\{L_{-1},L_0,L_1,M_{-1},M_0,M_1\}$, the expression of the blocks $g_{34}^{21}(p|u,v)$ can be computed. These Casimirs are 
\begin{eqnarray}
    \mathcal{C}_1 = M_0^2-M_{-1}M_1\,,\quad  
\mathcal{C}_2 = 2L_0M_0-\frac{1}{2}(L_{-1}M_1+L_1M_{-1}+M_1L_{-1}+M_{-1}L_1).
\end{eqnarray}
For simplicity, we consider the case where $\D_{i=1,2,3,4}=\D,\,\xi_{i=1,2,3,4}=\xi$. One can now act these Casimirs on the OPE and find differential equations corresponding to the blocks. However, these equations get much simpler when we use the following function 
\begin{eqnarray}
    h(p|u,v)=u^{2\D}e^{-\frac{2\xi v}{u}}g_{\D,\xi}(p|u,v).
\end{eqnarray}
This leads us to 
\begin{subequations}
    \begin{align}
    &\left[\d_v^2+\frac{\xi_p^2}{u^2(u-1)}\right]h(p|u,v)=0.\\
    &\left[ u^2 \d_v - (1-\frac{3}{2} u) u v \d_v^2 + (u-1) u^2 \d_u\d_v \right] h(p|u,v) =(\D_p - 1) \xi _p \ h(p|u,v). 
\end{align}
\end{subequations}
Solving the above equations, we get our final closed form expression for the global conformal Carroll blocks in the regime $|u|<1$: 
\begin{align}
    g_{\D,\xi}(p|u,v)= 2^{2 \D _p-2}\, \left(1-u\right)^{-1/2}  \exp{\left(\frac{-\xi_p v}{u\sqrt{1-u}} +2\xi \frac{v}{u} \right)} u^{\D _p-2\D} (1+\sqrt{1-u})^{2-2\D _p}.
\end{align}
More details of the Carroll bootstrap programme can be found in \cite{Bagchi:2017cpu}, where among other calculations, answered obtained intrinsically here were also reproduced by considering the limit from relativistic answers. 

\subsection{Stress Tensors}\label{ccft2-T}

Any quantum field theory with an underlying translation symmetry has a conserved stress tensor. Since Carrollian and Conformal Carrollian theories are translationally invariant, one can readily define stress tensors for these theories. We can construct stress tensors via the Noetherian prescription \cite{Saha:2022gjw} or by the variation of the equivalent of the metric on the Carroll manifold \cite{Dutta:2022vkg}. In this subsection, following the historical route, we will again appeal to the limit \eqref{contr} from the relativistic 2D  CFT to arrive at the stress tensor for the 2D CCFT. Other approaches mentioned above yield identical results. 

\medskip
{\ding{112}} \underline{\em{Constructing the CCFT Stress tensor}}
\medskip

We remind the reader that in a relativistic 2D  CFT, the holomorphic and antiholomorphic stress tensors on the plane are given by 
\begin{align}\label{T-cft-pl}
    T_{plane} (z) = \sum_n \L_n z^{-n-2}, \quad \bar{T}(\z)=  \sum_n \bL_n \z^{-n-2}. 
\end{align}
We wish to perform the contractions on the cylinder. To do this, we map the relativistic 2D  CFT to the cylinder and the stress tensor transforms by the Schwarzian derivative to give:
\begin{align}
    T_{cyl} = z^2 T_{plane} - \frac{c}{24} \Rightarrow T_{cyl} = \sum \L_n e^{in \omega} - \frac{c}{24},
\end{align}
and similarly for $\bar{T}_{cyl}$. Now we follow the limit \eqref{contr-ur} and define:
\begin{align}\label{EMccft}
    & T_1 (\sigma, \t) = \lim_{\e\to0} \left( T_{cyl} - \bar{T}_{cyl} \right), \quad  T_2 (\sigma, \t) = \lim_{\e\to0} \e \left( T_{cyl} + \bar{T}_{cyl} \right).
\end{align}
This leads to
\begin{align}\label{Carr-T}
    T_1 (\sigma, \t) = \sum_n \left(L_n - in \t M_n\right) e^{-in\sigma} + \frac{c_L}{24}, \quad T_2 (\sigma) = \sum_n M_n e^{-in\sigma} + \frac{c_M}{24}. 
\end{align}
These relations can be inverted to get the modes of the algebra in terms of the stress tensors defined above:
\begin{align}
    L_n = \int d\sigma \, (T_1 + in\t T_2) e^{in\sigma}, \quad M_n = \int d\sigma \, T_2 e^{in\sigma}.
\end{align}
The central terms, which are ignored above, lead to a shift of the zero modes $L_0$ and $M_0$. 
In some work, the stress tensors are defined in a way that is a bit different from what we have done above. In particular, instead of $T_1(\sigma, \t)$, a different stress tensor $\hat{T}(\sigma)$ is defined with 
\begin{align}
    \hat{T}(\sigma) = \sum_n L_n e^{-in\sigma}, 
\end{align}
while $T_2(\sigma)$ stays the same. We note here that $\hat{T}(\sigma)$ is just the restriction of $T_1 (\sigma, \t)$ to $\t=0$. To see this, we translate the operator $\hat{T}(\sigma)$ in $\t$ with the Hamiltonian on the Carroll cylinder: 
\begin{align}
    e^{iH_c \t} \hat{T}(\sigma) e^{-iH_c \t} & = e^{iM_0 \t} \hat{T}(\sigma) e^{-iM_0 \t} \quad \text{since} \quad H_c =\partial_\t = M_0\cr 
    = \sum_n e^{iM_0 \t} & L_n e^{-iM_0 \t} e^{-in\sigma} = \sum_n \left(L_n - in \t M_n\right) e^{-in\sigma} = T_1 (\sigma, \t). 
\end{align}

\medskip
{\ding{112}} \underline{\em{Stress tensor correlations}}
\medskip

In a 2D  relativistic CFT, the seminal analysis of Belavin, Polyakov and Zamolodchikov \cite{Belavin:1984vu} showed that there was a recursion relation between the $n$-point and $(n-1)$-point correlation function of stress tensors. The connected part of this identity is what we would be interested in and this reads: 
\begin{align}
    \< T(z_1) T(z_2) \ldots T(z_n)\> = \sum_i \left(\frac{2}{(z_1-z_i)^2} + \frac{1}{z_{1}-z_i}\partial_i\right) \< T(z_2) \ldots T(z_n)\>\,.
\end{align}
Using the properties of the Virasoro algebra and the expansion \eqref{T-cft-pl}, one can find the two-point function of the stress tensor on the plane: 
\begin{align}
    \< T(z_1) T(z_2)\>  = \frac{c/2}{(z_1 -z_2)^4}\,.
\end{align}
Using the above recursion relation and the two-point function
let's compute arbitrary $n$-point functions of the stress tensor in 2D  CFTs. This can further be checked by using just the Virasoro algebra.  

\medskip

In the same spirit, one can use the expressions \eqref{Carr-T} and the infinite dimensional 2D CCA to compute Carroll stress tensor two point functions. For the correlation functions on the cylinder, the answers are the following: 
\begin{subequations}\label{CTT}
   \begin{align}
    \<0|T_1(\sigma_1, \t_1)  T_1(\sigma_2, \t_2)|0\> &= \frac{c_L - 2c_M (\t_1 - \t_2)\cot\left(\frac{\sigma_1 - \sigma_2}{2}\right) }{32 \sin^4\left(\frac{\sigma_1 - \sigma_2}{2}\right)}\,,\\
   \<0|T_1(\sigma_1, \t_1)  T_2(\sigma_2)|0\> &= \frac{c_M}{32 \sin^4\left(\frac{\sigma_1 - \sigma_2}{2}\right)}\,,\\
   \<0|T_2(\sigma_1)  T_2(\sigma_2)|0\> &= 0. 
\end{align} 
\end{subequations}
One can exploit symmetries to figure out higher point correlation functions of the stress tensors as well. It is also possible to find recursion relations analogous to the one for the Virasoro algebra for the 2D CCA. These read
\begin{align}
  \< T_2(\sigma_1) T_1(\t_2, \sigma_2) \ldots T_1(\t_n, \sigma_n)\> = & \nonumber \\
  \sum_{i=2}^n \left(\frac{2}{\sin^2(\sigma_{1i}/2)} \right. &+ \left. \frac{\cot(\sigma_{1i}/2)}{2}\partial_{\sigma_i}\right) \< T_1(\t_2, \sigma_2) \ldots T_1(\t_n, \sigma_n)\>, \nonumber\\
  \< T_1(\t_1, \sigma_1) T_1(\t_2, \sigma_2) \ldots T_1(\t_n, \sigma_n)\> = &\frac{c_L}{c_M} \< T_2(\sigma_1) T_1(\t_2, \sigma_2) \ldots T_1(\t_n, \sigma_n)\> \cr & \qquad + \sum_{i=1}^n \t_i \partial_{\sigma_i} \< T_1(\t_2, \sigma_2) \ldots T_1(\t_n, \sigma_n)\>.
\end{align}
Using the result for the two point functions and the above recursion relations, it is again possible to figure out arbitrary $n$-point correlation functions for Carroll stress tensors. These can be cross-checked against answers coming solely from symmetries and the two methods agree. We refer the reader to \cite{Bagchi:2015wna} for more details.

\newpage

{\ding{112}} \underline{\em{Schwarzian derivative}}

\medskip

We have been exclusively dealing with expressions of stress tensors of the 2D CCFT in the cylinder coordinates in the above. It is also useful to look at the expressions on the null plane. The Carroll stress tensor on the plane is given by
\begin{eqnarray}
    T_1(t,x) = \sum_{n} \Big\{L_n  - (n+2)\, \frac{t}{x}\, M_n \Big\}\, x^{-n-2}\,,\quad T_2 (x) = \sum_{n} M_n x^{-n-2} \,.
\end{eqnarray}
This is related to the cylinder representation discussed in Sec. \ref{ssec:2dreps} by the plane to cylinder map \eqref{p2c} and the corresponding Carroll Schwarzian derivative we define below. Under a Carroll transformation: 
\begin{align}
x \to \tilde{x} = f(x),   \quad t\to \tilde{t} = f'(x)t + g(x)\,,  
\end{align}
the stress tensors change as \cite{Basu:2015evh, Jiang:2017ecm, Hao:2021urq}
\begin{subequations}\label{Tcar-sch}
    \begin{align}
    \widetilde{T_2}(x) &= f'^2\,T_2\bigl(\tilde x\bigr) + \frac{c_M}{12}\,\{f,x\}, \\
\widetilde{T_1}(x,y) &= f'^2\,T_1\bigl(\tilde x,\tilde y\bigr)
   + 2f'\,g'T_2\!\bigl(\tilde x\bigr)
   + (f')^2g\,T'_2\!\bigl(\tilde x\bigr)
   + \frac{c_L}{12}\,\{f,x\}
   + \frac{c_M}{12}\,\left[ \left( {f}, {g} \right), {x} \right].
\end{align}
\end{subequations}
Here $\{~ ,\}$ is the usual Schwarzian derivative 
\begin{eqnarray}
    \{f,x\} = \frac{f'''}{f'} - \frac{3}{2}\left(\frac{f''}{f'}\right)^2
\end{eqnarray}
and the last term in the parentheses above is the Carroll version of it, which involves two functions, and is given by:
\begin{equation}\label{car-sch}
    \left[ \left( {f}, {g} \right), {x} \right] = 
\frac{
3 ({f}'' )^2 - {f}' {f}''' {g}' - 3 {f}' {f}'' {g}'' + ({f}')^2 {g}'''
}{({f}')^3}.
\end{equation}
\medskip

\subsection{Partition function and Carroll modular invariance}\label{part-CCFT}
In this subsection, we focus on the partition function of 2D  CCFTs and the notion of modular invariance in these theories. We will then briefly touch upon characters of the representations that we have discussed above and reconstruct the partition function as a sum of these characters. 

\medskip
{\ding{112}} \underline{\em{Partition function}}
\medskip

Any statistical mechanical system is defined in terms of its partition function
\begin{align}
    Z = \text{Tr} \, e^{-\beta H}, 
\end{align}
where $H$ is the Hamiltonian of the system and $\beta= \frac{1}{T}$ is the inverse of the temperature. In relativistic 2D  CFTs, it is important to define the theory away from the complex plane where holomorphic and antiholomorphic sectors arise and hence on arbitrary genus Riemann surfaces. The most important is the first non-trivial example: the torus. A torus is defined by two linearly independent lattice vectors $v_1, v_2$ on a plane and identifying points on the plane which differ by an integer combination of them. The ratio of these vectors is called the modular parameter $\t = v_2/v_1$. The partition function of a relativistic 2D  CFT on a torus with modular parameters $(\t, \bar{\t})$ is given by 
\begin{align}\label{Z-cft}
    Z_{CFT} = \text{Tr} \left( e^{2\pi i \t \L_0} e^{-2\pi i \bar{\t} \bL_0} \right). 
\end{align}
In direct analogy, we can define the partition function of a 2D  CCFT on a ``torus'' with Carroll modular parameters $(\zeta, \rho)$ as
\begin{align}\label{Z-ccft}
    Z_{CCFT} = \text{Tr} \left( e^{2\pi i \zeta L_0} e^{2\pi i \rho M_0} \right). 
\end{align}
If we follow the ultra-relativistic limit \eqref{contr}, and demand that 
\begin{align}
    Z_{CFT} \to Z_{CCFT}, \quad \text{as} \, \,  \e \to 0
\end{align}
we see that the Carroll modular parameters are related to the relativistic CFT ones by 
\begin{align}\label{modmap}
    \zeta = \frac{1}{2} \left(\t + \bar{\t}\right), \quad \rho = \frac{1}{2\e} (\t -\bar{\t}). 
\end{align}
The above relation would help us in understanding a form of modular invariance that 2D  CCFTs inherit from their relativistic counterparts. 

\medskip

In \eqref{Z-ccft}, the trace is over states in a particular representation. It is thus expected that if this sum is performed over very different representations, the contribution to the partition function would be very different. So at the outset, the expectation is that since the highest weight and the induced representations are very different looking, the partition functions we get in theories governed by these representations also would be very different. Rather miraculously, this does not happen and we will elaborate on this when we discuss the characters of the representations. 

\medskip
{\ding{112}} \underline{\em{Modular invariance}}
\medskip

One of the most powerful tools in a 2D  relativistic CFT is the notion of modular invariance on the torus. The properties of the 2D  CFT should not depend on the overall scale of the lattice which was identified to make the torus or by the absolute orientation of the lattice vectors. If $(v_1, v_2)$ and $(v_1', v_2')$ are two lattice vectors defining the same torus, then 
\begin{align}
    \begin{pmatrix} v_1' \\ v_2' \end{pmatrix} = \begin{pmatrix} a & b \\ c & d \end{pmatrix} \begin{pmatrix} v_1 \\ v_2 \end{pmatrix}, \quad a, b, c, d \in \mathbb{Z}, \quad ad-bc = 1. 
\end{align}
The unit cell has to have the same area and hence the matrices we consider should be unimodular. And of course the inverse has to exist. So we are dealing with the group $SL(2, \mathbb{Z})$. The action of $SL(2, \mathbb{Z})$ on the modular parameter is 
\begin{align}\label{mod-cft}
    \t \to \frac{a \t +b}{c\t +d}, \quad ad-bc = 1.
\end{align}
The group to be considered is actually $SL(2, \mathbb{Z})/\mathbb{Z}_2$ since the signs of all the integers can be changed simultaneously without affecting the transformation. The partition function of a 2D  CFT \eqref{Z-cft} should be invariant under \eqref{mod-cft}. This is the statement of modular invariance of a 2D  CFT on a torus. 

\medskip

For a 2D CCFT, we again need to identify the plane to get to a torus. But here we must remember that the initial plane where the Carroll theory lived was a null plane. So the torus which we will end up with would be a null torus. The notion of modular invariance hence would also change in these CFT which live on null surfaces. 

\medskip

The easiest way to figure out a notion of modular invariance in these Carroll theories is to appeal again to the UR limit \eqref{contr}. We have seen that under this the modular parameters of the theory are related by \eqref{modmap}. We now write the CFT modular transformation in terms of these variables and take the limit:
\begin{align*}
        \zeta + \e \rho \to \frac{a(\zeta + \e \rho) + b}{c(\zeta + \e \rho) + d}. 
\end{align*}
Expanding and comparing $\mathcal{O}(1)$ and $\mathcal{O}(\e)$ terms we get the Carroll version of modular invariance \cite{Bagchi:2012xr, Bagchi:2013qva}: 
\begin{align}\label{Car-mod}
   \zeta \to \frac{a \zeta +b}{c\zeta +d}, \quad \rho \to \frac{\rho}{(c\zeta +d)^2}, \quad ad-bc = 1.
\end{align}
We will demand that the 2D CCFT partition function \eqref{Z-ccft} is invariant under these transformations \eqref{Car-mod}. When we discuss applications of 2D CCFTs as duals to 3D asymptotically flat spacetimes, we will use the Carroll modular invariance to derive a version of the famous Cardy formula for 2D CCFTs.

\medskip
{\ding{112}} \underline{\em{Characters}}
\medskip

As we have seen earlier, all states in the highest weight representation can be created by acting with raising operators $L_{-n}$ and $M_{-n}$ on primaries $|\D, \xi\>$, which raise the eigenvalue $\D$. All set of states created from $|\D, \xi\>$ and their linear combinations form what we will call the Carroll module $\mathcal{C}(\D, \xi, c_L, c_M)$. The Hilbert space of the Carroll CFT ($\mathcal{H}_{CCFT}$) is thus the direct sum of modules for all Carroll primaries of the given theory:
\begin{align}
    \mathcal{H}_{CCFT}(c_L, c_M) = \bigoplus_{(\D_i, \xi_i)} \mathcal{C}(\D_i, \xi_i, c_L, c_M)
\end{align}
We have seen that the partition function of the 2d CCFT is given by \eqref{Z-ccft}. The operator $\hat{\mathcal{O}}= e^{2\pi i(\zeta L_0 +\rho M_0)}$ does not mix states in each Carroll module. Hence the trace over each module can be performed separately. The characters of a Carroll highest weight module are defined by
\begin{align}
    \chi_{\D, \xi} (\zeta,\rho) = \text{Tr}_{\D, \xi} \, e^{2\pi i \{\zeta (L_0 - \frac{c_L}{24}) + \rho (M_0 - \frac{c_M}{24})\}}. 
\end{align}
One can explicitly calculate this \cite{Bagchi:2019unf}, and the character for a non-vacuum primary module is given by: 
\begin{align}
    \chi_{\D,\xi}(\zeta,\rho) = \frac{e^{\frac{2\pi i \zeta}{12}} e^{-2\pi i(\zeta \frac{c_L}{24}+\rho \frac{c_M}{24})}e^{2\pi i(\zeta\Delta+\xi\rho)}}{\eta(\zeta)^2} \, ,
\end{align}
where $\eta(\zeta)$ is the Dedekind eta function. For the vacuum $(\Delta=0,\xi=0)$ the answer is different and the Carroll identity character takes the form:  
\begin{align}
    \chi_{\mathbb{I}}(\zeta,\rho) =  \frac{e^{\frac{2\pi i \zeta}{12}} e^{-2\pi i(\zeta \frac{c_L}{24}+\rho \frac{c_M}{24})}}{\eta(\zeta)^2}(1-e^{2\pi i \zeta})^2\,.
\end{align}
Expressions for the characters can also be understood from the contraction of Virasoro characters \cite{Bagchi:2019unf}. The partition function can be written down now in terms of the characters:
\begin{align}
    & Z_{CCFT}(\zeta,\rho) = \sum_{\D,\xi} D(\D, \xi) \chi_{\D,\xi} (\zeta,\rho) \\
    & \quad \qquad= \frac{e^{\frac{2\pi i \zeta}{12}} e^{-2\pi i(\zeta \frac{c_L}{24}+\rho \frac{c_M}{24})}}{\eta(\zeta)^2} \left[D(0,0) (1-e^{2\pi i \zeta})^2 + \sum_{(\D,\xi) \neq (0,0)} D(\D, \xi) e^{2\pi i(\zeta\Delta+\xi\rho)} \right] .\nonumber
\end{align}
It is of interest to note that one can also calculate the character of the a priori very different induced representations and the characters (up to absolute values) are identical to the above \cite{Oblak:2015sea}. The possible reason for this is an automorphism in the Virasoro algebra 
\begin{align}
    \L_n \to \L_{-n}, \quad c\to -c
\end{align}
which leads to an automorphism in the two copies of the Virasoro algebra
\begin{align}
    \L_n + \bL_n \to \L_n - \bL_{-n}, \quad \L_n - \bL_n \to \L_n + \bL_{-n}, \quad c+\bar{c} \leftrightarrow c-\bar{c}. 
\end{align}
that in turn then descends into the isomorphism of the Galilean and Carrollian conformal algebras in $D=2$ \cite{Bagchi:2019unf}, a phenomenon we have encountered earlier in the review.

\subsection{Carroll via deformations}\label{carrdef}
We have seen how contractions can generate 2D CCFTs from CFTs in the same dimension. 
In this subsection, we will deal with a completely different way to get a Carrollian theory from a relativistic theory, that is inducing flows via current-current deformations. We will argue that composition of $U(1)$ currents can lead to special points in the moduli space of the theory, where we could see a transition from 2D relativistic to Carroll 2D CFTs \cite{Bagchi:2022nvj,Bagchi:2024unl}. As in many cases, our starting point will be the action of free massless 2D  relativistic scalar field, in general, we will consider $D$ such scalar fields:  
\be
\mathcal{L}=\frac{1}{2} \sum_{i=1}^{D}\left(\left(\partial_\tau \phi_i\right)^2-\left(\partial_\sigma \phi_i\right)^2\right),
\ee
which gives rise to the equations of motion for each field: $\left(\partial_{\tau}^2-\partial_{\sigma}^2\right)\phi_i=0$.
In what follows, we will be applying two distinct types of deformations to this theory - symmetric and antisymmetric current-current deformations.  We will then examine the system's behaviour when the deformation parameter is dialled to the extremes of the parameter space. Our focus will be to see how the spacetime symmetry algebra changes under such a flow. Note that we are mainly concerned about the classical theory here, so we will extensively use Poisson brackets, which are extended to commutator brackets in the quantum theory. 

\subsubsection{Symmetric current-current deformation} 
First consider deformations of the parent Lagrangian by a symmetric current-current deformation, by which we mean adding a $\delta_{ab}J^aJ^b$ type term to the Lagrangian, namely:
\be \label{lag.deform1}
\widetilde{\mathcal{L}}^\a=\mathcal{L}+\frac{\a}{2}\sum_i \left({{J}}^2_{i}+\overline{{J}}^2_{i}\right).
\ee
Here $J,  \overline{{J}}$ are conserved currents of the free theory: 
\be
	{J}_i=\frac{1}{\sqrt{2}}(\dot{\phi}_{i}+ \phi_{ i}^{\prime}),\quad \overline{{J}}_{i}=\frac{1}{\sqrt{2}}(\dot{\phi}_{i}- \phi_{i}^{\prime}),
\label{JbarJ.currents}
\ee
where $\dot{\phi}_i=\d_\tau{\phi}_i$ and ${\phi}'_i=\d_\phi{\phi}_i$ are the temporal and spatial derivatives, and $\a$ is a continuous parameter that determines the strength of the marginal deformation.
The currents satisfy the conservation laws $\bar{\d}J_i=\d \overline{J}_i=0$ with $\d, \bar{\d}=\d_\tau\pm\d_\sigma$. Note that we consider the CFT on the cylinder in this case. The deformed Lagrangian reads:
\be\label{JJdef}
	\widetilde{\cL}^\a=\frac{1}{2} 
	\sum_{i=1}^{D}\left((1+\a)\left(\partial_\tau \phi_i\right)^2-(1-\a)\left(\partial_\sigma \phi_i\right)^2\right).
\ee
Depending on whether the sign of $\a$ is selected, the deformation scales down space derivatives and up time derivatives in the action, or vice versa. When $\a = \pm 1$, one of the derivatives drop out. In this case temporal and spatial coordinates can not be rescaled, and full conformal symmetry can not be ``restored''. In this perspective,  
$\a = \pm 1$ serves as the bounds on the parameter space. It can be realised that $\a=1$ is the Carroll point (and $\a=-1$ is the Galilean point) and reaching this point leads to a change in space-time symmetry. At this precise point, Carroll algebra takes control, both from a classical and quantum point of view. 
\medskip

Recall, the stress tensors of the free theory satisfy the following algebra, written in the Poisson Bracket form
\begin{align}
\left\{T(\sigma), T\left(\sigma^{\prime}\right)\right\}_{\mathrm{PB}} 
& =2 T(\sigma) \partial_\sigma \delta\left(\sigma-\sigma^{\prime}\right)
+\left(\partial_\sigma T(\sigma)\right) \delta\left(\sigma-\sigma^{\prime}\right),
\\
\left\{ \overline{T}(\sigma),  \overline{T}\left(\sigma^{\prime}\right)\right\}_{\mathrm{PB}} 
& =-2  \overline{T}(\sigma) \partial_\sigma \delta\left(\sigma-\sigma^{\prime}\right)
-\left(\partial_\sigma  \overline{T}(\sigma)\right) \delta\left(\sigma-\sigma^{\prime}\right),
\\
\left\{T(\sigma), \overline{T}\left(\sigma^{\prime}\right)\right\}_{\mathrm{PB}} 
& =0.
\end{align}
Writing in terms of Fourier modes, we have
\be\label{Lmodes}
	T(\sigma)=\sum_m \cL_m\,e^{-im\sigma},
	\qquad\quad
	\overline{T}(\sigma)=\sum_m \overline{\cL}_m\,e^{+im\sigma}.
\ee
Here the modes $\cL$ and $\overline{\cL}$ correspond to two copies of the Witt algebra:
\be\label{Vir1}
\{\cL_n, \cL_m\}_{\text{PB}} = -i(n-m) \cL_{n+m}, \qquad \{\bL_n, \bL_m\}_{\text{PB}} = -i(n-m)\bL_{n+m}.
\ee
Now if we consider the deformed Lagrangian \eqref{JJdef}, using $T$ and $\overline{T}$ we can define the Hamiltonian and the momentum flux density for the theory
\be\label{hami}
\widetilde{\cH}^\a =T + \overline{T}+2\a \sum_i  {{J}}_{i} \overline{{J}}_{i}\,,\quad \cJ=T - \overline{T}.
\ee
It can be checked they lead to the following brackets:
\be\label{flowalgebra}
\begin{aligned}
& \left\{\mathcal{J}(\sigma), \mathcal{J}\left(\sigma^{\prime}\right)\right\}_{\mathrm{PB}}
=2 \cJ(\sigma) \partial_\sigma \delta\left(\sigma-\sigma^{\prime}\right) 
+\left(\partial_\sigma \cJ(\sigma)\right) \delta\left(\sigma-\sigma^{\prime}\right)\\
& \left\{\mathcal{J}(\sigma), \widetilde{\cH}^\a\left(\sigma^{\prime}\right)\right\}_{\mathrm{PB}}
=2 \widetilde{\cH}^\a(\sigma) \partial_\sigma \delta\left(\sigma-\sigma^{\prime}\right)
+\left(\partial_\sigma \widetilde{\cH}^\a(\sigma)\right) \delta\left(\sigma-\sigma^{\prime}\right) \\
& \left\{\widetilde{\cH}^\a(\sigma), \widetilde{\cH}^\a\left(\sigma^{\prime}\right)\right\}_{\mathrm{PB}}
=\left(1-{\alpha}^2\right)
\left(2 \cJ(\sigma) \partial_\sigma \delta\left(\sigma-\sigma^{\prime}\right)
+\left(\partial_\sigma \cJ(\sigma)\right) \delta\left(\sigma-\sigma^{\prime}\right)\right).
\end{aligned}
\ee
Now this is almost the conformal algebra, without the central terms and with the extra scaling in the last term.
To see this, let us write this in terms of modes, 
\be\label{LMmodes}
	\cJ(\sigma)=\sum_m L_m\,e^{-im\sigma},
	\qquad
	\widetilde{\cH}^\a(\sigma)=\sum_m M^\a_m \, e^{-im\sigma},
\ee
Where we have particularly defined the combinations $L_m$ which is independent of $\alpha$ and $M_m^\alpha$, which is not. Using these combined modes, we will have the algebra
\begin{align}
\{L_n, L_m\}_{\text{PB}} &= -i(n-m) L_{n+m}, \quad \{L_n, M^\a_m\}_{\text{PB}} = -i(n-m)M^\a_{n+m},  \cr
\{M^\a_n, M^\a_m\}_{\text{PB}} &= \left(1-{\alpha}^2\right)\Big[-i(n-m) L_{n+m}\Big].
\end{align}
Note that $\a=0$ is the usual conformal  algebra in two dimensions. For $-1<\a<1$, it is always possible to reinterpret the above as the same by rescaling the $\widetilde{\cH}^\a$ by a constant factor. However, at $\a=\pm 1$, the last bracket vanishes\footnote{Reader can recall here that $\{\cH(\sigma),\cH(\sigma')\}=0$ is the sign of emergence of Carroll symmetries.} and such a rescaling of the generator becomes singular. The spacetime algebra changes from the two copies of the (the classical part of) Virasoro algebra to (the classical part of) the BMS$_3$ algebra. When we refer to the algebra as the BMS algebra, we will be doing to with the understanding that we are not attempting to distinguish Carrollian and Galilean algebras. 

\medskip

One can ask now how to connect this to the usual Carroll limit in the sense of vanishing of speed of light. This can be answered by looking at the equation of motion of $\phi$ of the deformed theory which reads
\be\label{waveeqndef}
	\ddot{\phi}_i-\kappa^2{\phi}''_i=0,
\ee
Here $\kappa=\sqrt{\frac{1-\a}{1+\a}}$ is an extra factor, reinterpreted as a scaled version of speed of light, which goes to zero as $\a \to +1$ (and diverges as $\a \to -1$). This portrays the role of effective speed of wave propagation and exactly at  $\a = +1$ we get the equation for electric scalar field theory, as seen in \eqref{eleceom}.
\medskip

{\ding{112}} \underline{\em{Carroll at infinite boosts}}
\medskip

Our discussion for the symmetric deformation has far-reaching consequences. Following \cite{Bagchi:2022nvj}, it can be shown that this particular deformation can be thought of as a $SL(2,\mathbb{R})$ rotation of the $U(1)$ Kac-Moody currents, which correspond to the stress tensor of the conformal fields, by some angle $\theta$:
\begin{eqnarray}\label{bogo1}
	\begin{pmatrix}
		J^\mu(\ta)\\
		\bar{J}^\mu(\ta)
	\end{pmatrix} = 
	\begin{pmatrix}
		\cosh\theta(\ta) \, \, & \, \, \sinh\theta(\ta)\\
		\sinh\theta(\ta) \, \, & \, \, \cosh\theta(\ta)
	\end{pmatrix}
	\begin{pmatrix}
		J^\mu(0)\\
		\bar{J}^\mu(0)
	\end{pmatrix}
\end{eqnarray}
where,
\be
\cosh\theta(\ta) =\frac{1}{\sqrt{1-\ta^2}} ,~~\sinh\theta(\ta)=\frac{\ta}{\sqrt{1-\ta^2}}, ~~~\ta= \left(\frac{2\a}{1+\a^2}\right).
\ee
Now since $J\sim \partial\phi$ only, this rotation of currents actually tell us that adding such a deformation term is nothing but a boost transformation acting on the (anti)holomorphic coordinates $z_{L/R}$, which mixes contributions/currents from both sectors:  
\be{}
\label{Symm:z_LR}
	\begin{pmatrix}
	z_{\rm L} \\ z_{\rm R}
	\end{pmatrix}
	=\frac{1}{\sqrt{1-\a^2}}
	\begin{pmatrix}
	1 & -\a \\ -\a & 1
	\end{pmatrix}
	\begin{pmatrix}
	z \\ \overline{z}
	\end{pmatrix}
	\qquad
	\quad
	\ee

Now the reader should recall that a timelike observer in $(\t,\sigma)$ coordinates can never reach the lightcones via finite Lorentz boosts. So the CCFT point $\a =1$ is nothing but the manifestation of infinite boosts in our language. This goes hand in hand with our previous discussions of speed of light going to zero limit.
\medskip

One can show that this boosted structure controls the full flow from CFT to CCFT in our setup. In fact this also plays a role in the quantum mechanical regime. Briefly, we can notice the mode expansion for the Kac-Moody currents, $J^a(\sigma)=\sum_m J_m^a e^{-im\sigma}$, where $J_m^a$ are oscillators having the Abelian algebra with a level $k$, written in form of commutators:
\be\label{AffKMrel}
	[J^a_m,J^b_n]=\frac{k}{2}m\delta^{ab}\delta_{m+n}.
\ee
Given this, we can now simplify and write down the transformation \eqref{bogo1} in terms of the oscillators:
\begin{eqnarray}\label{bogodef}
J_n(\a) = \cosh\theta~ J_n(0) + \sinh\theta ~\bar{J}_{-n}(0);\quad \bar{J}_n(\a) = \cosh\theta~ \bar{J}_n(0) + \sinh\theta~ J_{-n}(0).
\end{eqnarray}
This evidently keeps the algebra of modes unchanged, and at the level of states, this imposes a \textit{Bogoliubov transformation}, which relates the undeformed CFT vacuum and the deformed one. One can show that when we take $\a =\pm 1$, and consequently $\tanh\theta = \pm 1$, the transformed vacuum corresponds to a CCFT vacuum in a particular representation. Moreover, the spectrum also smoothly interpolates to the BMS spectrum, details of which can be found in \cite{Bagchi:2024unl}. Those discussions, unfortunately, are beyond the scope of this review. 

\medskip

{\ding{112}} \underline{\em{A related pathway: $\sqrt{T\overline{T}}$ deformation}}

\medskip

Let us also point out an intriguing relation here. The idea of adding marginal deformations to CFT and flowing to CCFTs actually came out of progress in a seemingly unrelated direction, from the advent of $\sqrt{T\overline{T}}$ deformations\footnote{This is, however, very different from its well known cousin the $T\overline{T}$ type irrelevant deformation. For details please see \cite{Jiang:2019epa, He:2025ppz} and references thereof for comprehensive reviews on the subject.}. To understand this, consider a particular nonlinear class of deformation for the CFT Hamiltonian and momentum density:
\begin{subequations}
\begin{align}
	\cJ(\sigma)&=T(\sigma)-\overline{T}(\sigma)\\
	\cH(\sigma )&=T(\sigma)+\overline{T}(\sigma)+2\tilde{\a}(T(\sigma)\overline{T}(\sigma))^\b.
\end{align}
\end{subequations}
It can be easily seen these generators automatically satisfy the classical BMS$_3$ algebra when we put the values $\tilde{\a}=1, \b=1/2$,
\begin{subequations}
    \begin{eqnarray}
&\{\cJ(\sigma),\cJ(\sigma')\}_{\rm PB}	=\{\partial_{\sigma},\cJ(\sigma)\}\delta(\sigma-\sigma')
	\\
	&\{\cJ(\sigma),\cH(\sigma')\}_{\rm PB}
	=\{\partial_{\sigma},\cP(\sigma)\}\delta(\sigma-\sigma')
	\\
	&\{\cH(\sigma),\cH(\sigma')\}_{\rm PB}
	=0.
          \end{eqnarray}
\end{subequations}
where we have used a shorthand notation: $\{\partial_\sigma,X\} = 2X\partial_\sigma+ \partial_\sigma X$. So now we have a set of exact generators deformed by the $\sqrt{T\overline{T}}$ operator, which again is a classically marginal operator, but for a special value of the coupling it leads to the BMS algebra. This is a non-trivial statement since this map is non-linear in the stress tensors, and has been discussed in a set of papers \cite{sqrt1,sqrt2,sqrt3}. This avenue is obviously different from the boost perspective and current-current deformations we discussed earlier, but clearly produces the same algebra at special points of the moduli space. However, except for the classical equivalence, not much is understood about these flows.

\subsubsection{Asymmetric current-current deformation} 
We will now turn our attention to current-current deformations of the form $\epsilon_{ab}J^a\bar J^{b}$. These are exactly marginal operators and have been studied in the context of string theory in curved spaces \cite{Apolo:2019zai}. In contrast to the symmetric $J\bar J$ scenario where spacetime indices simply decouple from each another, a notable characteristic of such deformations is that they mix spacetime indices on the fields. Discussions on such marginal deformations of CFT have become very important, in conjunction with their irrelevant cousins (see, for example \cite{Anous:2019osb, Guica:2020eab, Guica:2021fkv}). 
\medskip

For simplicity, we will limit our analysis to the two-dimensional Lagrangian density of two separate bosonic scalar fields $\phi_{1,2}$ deformed by a $J^1\wedge \bar{J}^2$ operator. Here, we consider the same $U(1)$ currents as in the previous section. The deformed Lagrangian takes the form
\be\label{lagwedge}
	\mathcal{\tilde{L}}^{(g)}=\frac{1}{1+g^2}\left[\frac{1}{2} 
	\sum_{i=1,2}\left(\left(\partial_\tau \phi_i\right)^2-\left(\partial_\sigma \phi_i\right)^2\right)
	-g\left(\partial_\tau \phi_1 \partial_\sigma \phi_2-\partial_\sigma \phi_1 \partial_\tau \phi_2\right)\right].
\ee
Compared to the symmetric situation, the action of this deformation is very different\footnote{Note that this can be thought to be equivalent to as turning on a constant B-field.}. The theory is a relativistic CFT for all finite values of the coupling $g$, since the perturbation can be regarded as a boundary contribution under the integral.  The effective speed of light in this $g$-deformed line of CFTs, however, remain unchanged since the propagating modes' equation of motion stays the same.  Only at $g\to \infty$ does this boundary term become important, indicating the change in the spacetime algebra.  We shall briefly demonstrate how the spacetime algebra abruptly changes when the boundary conditions for the fields in this limit become non-invertible, implying an emergent zero velocity of mode propagation. Starting point is the Hamiltonian density 
\begin{align}
	\mathcal{H}^{(g)}
	=\frac{1}{2}\left(\left(1+g^2\right)\left(\Pi_1^2+\Pi_2^2\right)+{\phi_1'}^2+{\phi_2'}^2\right)
	+ g\left(\Pi_1 \phi'_2-\Pi_2 \phi'_1\right)\,,
	\label{ham.in.phi12}
\end{align}
which can be represented in a matrix form
\be
	\mathcal{H}^{(g)}
	=
	\,\begin{pmatrix}
	\Pi_1 \\ \Pi_2 \\ \phi_1' \\ \phi_2'
	\end{pmatrix}^{\rm T}
	M^{(g)}
	\begin{pmatrix}
	\Pi_1 \\ \Pi_2 \\ \phi_1' \\ \phi_2'
	\end{pmatrix}\ ,
	\qquad \text{where} \qquad
	M^{(g)}=\begin{pmatrix}
	\frac{1+g^2}{2} & 0 & 0 & \frac{g}{2} \\
	0 & \frac{1+g^2}{2} & -\frac{g}{2} & 0 \\
	0 & -\frac{g}{2} & \frac{1}{2} & 0 \\
	\frac{g}{2} & 0 & 0 & \frac{1}{2}
	\end{pmatrix}\ .
\ee
We now want to diagonalize the Hamiltonian for all values of $g$. The matrix (kernel) $M^{(g)}$ has two sets of eigenvalues given by,
\be
	\lambda^{(g)}_\pm=\frac{2+g^2\pm g\sqrt{g^2+4}}{4}=\frac{\gamma_\pm^2}{2}\ ,
	\qquad\text{where}\qquad
	\gamma_\pm=\frac{\sqrt{g^2+4}\pm g}{2}=\frac{1}{\gamma_\mp}\ .
\ee
and the corresponding orthonormal eigenvectors:
\be
	u^{+}_1=c^+
	\begin{pmatrix}
	1 \\ 0 \\ 0 \\ \gamma_-
	\end{pmatrix}
	\quad
	u^{+}_2=c^+
	\begin{pmatrix}
	0 \\ 1 \\ -\gamma_- \\ 0
	\end{pmatrix}
	\quad
	u^{-}_1=c^-
	\begin{pmatrix}
	1 \\ 0 \\ 0 \\ -\gamma_+
	\end{pmatrix}
	\quad
	u^{-}_2=c^-
	\begin{pmatrix}
	0 \\ 1 \\ \gamma_+ \\ 0
	\end{pmatrix}	
\ee
where we have the parameters
\be{}
c^\pm=\frac{1}{\sqrt{1+2\lambda_\mp}}.
\ee
$M^{(g)}$ can be then diagonalised as
\be
	\Lambda={\rm diag}(\lambda^{(g)}_+,\lambda^{(g)}_+,\lambda^{(g)}_-,\lambda^{(g)}_-)
	=S^{\rm T}M^{(g)}S\,,
\ee
by an element in SO(4)given by :$S=\begin{pmatrix}u^{+}_1 & u^{+}_2 & u^{-}_1 & u^{-}_2\end{pmatrix}$. The Hamiltonian density can then be recast in a bilinear form in terms of new field variables $\Xi$:
\be\label{hamil.in.Xi}
	\mathcal{H}^{(g)}
	=\,\Xi^{\rm T}\Lambda \Xi
	=\,\lambda^{(g)}_+\left\{\Pi_+^2+(\chi'_-)^2\right\}
	+\,\lambda^{(g)}_-\left\{\Pi_-^2+(\chi'_+)^2\right\}\,,
\ee
where $\Xi$ is defined by
\be\label{def.Xi}
	\Xi=
	\begin{pmatrix}
	\Pi_+ \\ \chi'_- \\ -\Pi_- \\ \chi'_+
	\end{pmatrix}
	:=S^{\rm T}
	\begin{pmatrix}
	\Pi_1 \\ \Pi_2 \\ \phi_1' \\ \phi_2'
	\end{pmatrix}
	=
	\begin{pmatrix}
	c^+(\Pi_1+\gamma_-\phi_2') \\
	c^+(\Pi_2-\gamma_-\phi_1') \\
	c^-(\Pi_1-\gamma_+\phi_2') \\
	c^-(\Pi_2+\gamma_+\phi_1')
	\end{pmatrix}.
\ee
Now let's check out the Poisson brackets for these new fields, 
\begin{eqnarray}
    &&\{\Pi_+(\sigma),\Pi_-(\sigma')\}_{\rm PB}=\{\chi'_+(\sigma),\chi'_-(\sigma')\}_{\rm PB}=0, \\
    &&\{\Pi_+(\sigma),\chi_-(\sigma')\}_{\rm PB} = \{\Pi_-(\sigma),\chi_+(\sigma')\}_{\rm PB} = 0\\
    &&\{\chi_+(\sigma),\Pi_+(\sigma')\}_{\rm PB} = \{\chi_-(\sigma),\Pi_-(\sigma')\}_{\rm PB} =\delta(\sigma - \sigma').
\end{eqnarray}
These show that $(\chi_+, \Pi_+)$ and $(\chi_-, \Pi_-)$ are two sets of pairs of canonically conjugate variables. Now pay attention to our Hamiltonian $\mathcal{H}^{(g)}$ in the \eqref{hamil.in.Xi}. When we dial the coupling $g \to \infty$, it is clear that $\lambda_{+}^{(g)} \to \infty$, while $\lambda_{-}^{(g)}\to 0$. In terms of canonical variables, one can easily see
\be
 \mathcal{H}^{(g\to \infty)}\sim \left\{\Pi_+^2+(\chi'_-)^2\right\}
 \ee
which commutes with itself at two different spatial points, thus Carroll dynamics emerging in this system. Now the standard Kac-Moody currents in this diagonalised system are
\begin{eqnarray}
    J^\pm=\frac{1}{\sqrt{4\lambda^{(g)}_\pm}}\left(2\lambda^{(g)}_\pm\Pi_\pm+{\chi}'_\pm\right),
	\qquad\quad
	\overline{J}^\pm=\frac{1}{\sqrt{4\lambda^{(g)}_\pm}}\left(2\lambda^{(g)}_\pm\Pi_\pm-{\chi}'_\pm\right).
\end{eqnarray}
Using Sugawara construction, one can now write down the relevant stress-energy tensors, in terms of the undeformed ones. Using the above currents we define
\be\label{contract}
	L^{(g)}=T-\overline{T}
	\qquad\qquad
	M^{(g)}=\frac{1}{\lambda^{(g)}_+}(T+\overline{T}).
\ee
They satisfy the following algebra
\begin{eqnarray}
	\left\{L^{(g)}(\sigma),L^{(g)}(\sigma')\right\}_{\rm PB}
	&&=2L^{(g)}(\sigma)\d_\sigma\delta(\sigma-\sigma')+\left(\d_\sigma L^{(g)}(\sigma)\right) \delta(\sigma-\sigma')\\
    \left\{L^{(g)}(\sigma),M^{(g)}(\sigma')\right\}_{\rm PB}
	&&=2M^{(g)}(\sigma)\d_\sigma\delta(\sigma-\sigma')+\left(\d_\sigma M^{(g)}(\sigma)\right) \delta(\sigma-\sigma')
	\\
	\left\{M^{(g)}(\sigma),M^{(g)}(\sigma')\right\}_{\rm PB}
	&&=\frac{1}{(\lambda^{(g)}_+)^2}
	\left[2L^{(g)}(\sigma)\d_\sigma\delta(\sigma-\sigma')+\left(\d_\sigma L^{(g)}(\sigma)\right) \delta(\sigma-\sigma')\right]\nonumber
	\\
	&& \overset{g\to\infty}{\longrightarrow}\ 0
\end{eqnarray}

which shows the transition from a pair of Virasoro algebras to BMS$_3$ algebra as $g$ is cranked up to infinity.
\medskip

So in general, we can have different kinds of marginal deformations inducing different flows in CFTs. Clearly in the antisymmetric case, as the deformation is just a boundary term, the space-time symmetry algebra remains intact at finite $g$. In fact, this kind of deformation is different from the symmetric one in the quantum sense as well, since we can show that \cite{Bagchi:2024unl} the canonical CFT vacua does not change in this case. But intriguingly, at special points in the parameter space, the symmetry algebra does change to BMS$_3$, since all of these cases boil down to a zero effective propagation speed. We have only shown this for the symmetry algebra, but in principle this works out for the spectrum and correlation functions as well. 
\medskip

{\ding{112}} \underline{\em{Remarks on Wilsonian paradigm}}
\medskip

As mentioned before, in \cite{Bagchi:2024unl} the flows from relativistic CFTs to BMS invariant field theories were closely discussed, and some intriguing structures were noted regarding such flows. In standard Renormalisation Group (RG) flow \cite{Wilson:1974mb} lore, we talk about integrating out high energy modes of a theory to arrive at a low energy effective field theory. Now revisiting the discussion in the introduction, we have already pointed out that ultra-relativistic modes are also the ultra-high energy modes of the theory, which become important in the Carrollian regime\footnote{For a simple explanation, recall that a Carrollian contraction of the relativistic algebra requires that energies become much larger than spatial moment i.e. $E \to E/\epsilon,~~~p^i\to p^i.$}. It is also clear from the usual paradigm that marginal deformations keep the flow on a line of CFTs, as opposed to the (ir)relevant ones. So the appearance of BMSFTs as some kind of fixed points in such marginal flows are indeed astounding.
\medskip

To understand this situation better, one may focus on the energy scales a bit more. In Wilsonian RG paradigm, when one integrates out high-energy modes of a relativistic QFT to arrive at a Galilean QFT, the resultant theories, like the ones important in condensed matter physics, are still relativistic theories defined at some lower velocity scale. This relies on the quintessential duality between so called `slow-modes' and `fast-modes' as propounded by Wilson \cite{Wilson:1974mb}. This is blind to the change in spacetime algebra. This becomes even more interesting once we only focus on ultra-high energy excitations, the `fast-modes' in this case. 
Going to such regimes requires one to systematically throw out all the other lower energy modes to keep the ones which are travelling at the speed of light (or any characteristic velocity setting the energy scale). Clearly this is not exactly compatible with the Wilsonian paradigm. This actually makes sense in terms of locality in the relativistic sense, which completely goes away in the Carroll regime, where lightcones close and we get ultra-locality setting in\footnote{This may still be made compatible with Wilsonian paradigm by considering observer dependence of states, which may be able to redistribute what we call `fast' and `slow' modes. This might need a better understanding of quantum equivalence principle and observer dependent Hilbert spaces, some recent discussions towards which can be found in \cite{Bagchi:2024tyq, Dutta:2024gkc, Sheikh-Jabbari:2025tkh}.}. 
\medskip

\subsection{Pointers to literature}
A lot of the discussions of 2D CCFTs are with holographic applications to 3D asymptotically flat spacetimes in mind. For the works that focus on applications to flat holography, we refer the reader to the analogous subsection of Sec.~\ref{sec:3dafs} (viz. Sec.~\ref{pt-3dAFS}). Below we list work that is mainly field theoretic in nature. 

\begin{itemize}
    \item{\em Generic discussions of 2D CCFTs:} Since 2D Carrollian and Galilean CFTs are isomorphic with an interchange of space and time directions, a lot of the older literature on 2D GCFTs is very relevant. A particularly useful reference in this context is \cite{Bagchi:2009pe}. More recently, a detailed discussion of various aspects of 2D CCFTs from intrinsic Carrollian symmetries appears in \cite{Saha:2022gjw}. In \cite{Chen:2019hbj}, a generalised 2D non-Lorentzian CFT with anisotropic scaling is presented. 

    \item{\em Explicit examples:} \cite{Hao:2021urq} considers details of the 2D massless scalar theory and associated conformal structures. \cite{Hao:2022xhq} generalises this discussion to fermions. Representation theory aspects are dealt with carefully and in a lot of detail. See also \cite{Yu:2022bcp} for a discussion of Carroll fermions and conformal structure. Also see the analogous discussions in the examples mentioned in Sec.~\ref{pt-3} for more references. 

    \item{\em Correlation functions:} The first examples of 2D Conformal Carrollian correlation functions arose masquerading in their Galilean avatars in \cite{Bagchi:2009ca, Bagchi:2009pe}. 

    \item  {\em{CCFT blocks:}} These were further discussed from a holographic perspective using irreps of the algebra in \cite{Hijano:2018nhq}.  A construction of Carroll blocks in the large central charge $c_M \to \infty$ limit was done in \cite{Ammon:2020wem} using a oscillator construction of the highest weight representation.

    \item {\em Boundary Carroll CFTs:} One can consider boundaries on the Carroll manifold and how Carroll CFTs would perceive this. This was recently addressed in  \cite{Bagchi:2024qsb}, where a new algebra was shown to emerge. This also arises as a non-trivial contraction of a single copy of the Virasoro algebra and is the symmetry for the open null string. 

     \item  {\em{Classical Integrability:}} From CFT perspective, the Classical Integrability of any theory can be realised in terms of integrable hierarchies associated to the Poisson structure thereof. The infinite charges corresponding to relativistic CFT in 2D comes from the Kortewig de-Vries (KdV) hierarchy derived from the stress tensor OPE. The similar construction in the case of dynamical systems with BMS$_3$ symmetry algebra was discussed in \cite{Fuentealba:2017omf} and the commuting charges were constructed following the Drinfeld-Sokolov formulation.
\end{itemize}
\newpage

\section{Carrollian Conformal Field Theories: Higher dimensions}
\label{sec:carrfields4}

%----------------------------------------------------------------------------------------

\newcommand{\ttil}{\tilde{t}}
\newcommand{\ktil}{\widetilde{\kappa}}
\newcommand{\Htil}{\widetilde{\mathcal{H}}}
\newcommand{\Mtil}{\widetilde{M}}
\newcommand{\Ytil}{\widetilde{Y}}
%-----------------------------------------------------------------------------------------

We now look beyond $D=2$ and at CCFTs in general dimensions. This section would outline a general approach to CCFTs based on the representations focusing on the analogue of the 2D global sub-algebra generated by $L_{0, \pm1}, M_{0,\pm1}$. As stated earlier, this amounts to looking at the version of the Conformal Carroll algebra we met at the beginning of the review, given by Carroll generators \eqref{Car-gen} and their conformal extensions \eqref{CCar-gen}. As mentioned in Sec. \ref{ssec:carpoincare}, the underlying geometry means that there are infinite dimensional enhancements in any dimension. For the start of this section, we will not focus on these infinite enhancements. However from Sec., we will turn our attention specifically to $D=3$ and here we will talk about the infinite algebra as well. Predominantly, though, this section would be devoted to understanding the Conformal Carroll algebra and its consequences for Ward identities and correlation functions in $D=3$.

\subsection{Conformal Carroll in generic dimensions}

For the sake of completeness, here is again the full Carroll conformal algebra in arbitrary dimensions. The generators on the $D$-dimensional Carroll plane $\mathbb{R}_t \times \mathbb{R}^{D-1}$ (where $\mathbb{R}_t$ represents the null direction) are given by \footnote{We have instated factors of $i$ in the generators when compared to the ones we wrote down in Sec. \ref{section2}.}: 
\begin{subequations}\label{CCA-gen}
 \begin{align}
& H=-i\partial_{t} , \quad P_i=-i\partial_{i}, \quad C_i=-ix_i\partial_{t}, \quad J_{ij}=i(x_i\partial_{j}-x_j\partial_{i}), \\
& D=-i(t\partial_{t} + x_i\partial_{i}) ,\quad K_0 = -ix_i^2\partial_{t} ,\quad K_i = 2ix_i(t\partial_{t} + x_j\partial_{j}) - ix_j^2\partial_{i}. 
\end{align}   
\end{subequations}
The non-zero commutation relations are: 
\begin{subequations}\label{ccft-d}
\begin{align}
        [J_{ij},J_{kl}]&=so(D-1),\quad [C_i,P_j]=i\delta_{ij}H, \quad [J_{ij},X_{k}]=-i\delta_{ik}X_{j} + i\delta_{jk}X_{i},  \\
        [D,P_i] &= iP_i, \quad [D,H] = iH, \quad [D,K_i] = -iK_i, \quad [D,K_0] = -iK_0, \\
        [K_0,P_i] &= 2iC_i, \quad [K_i,H]=-2iC_i, \quad [K_i,P_j] = -2i\delta_{ij}D+2iJ_{ij}\,,\quad [K_i,C_j]=-i\delta_{ij}K_0.
    \end{align}     
\end{subequations}
In the above, $X_i\equiv (P_i, C_i, K_i)$. 

\medskip

We now identify the Little group of the conformal Carroll group that leaves the origin, $\textbf{x}=0$ invariant. This is generated by rotations, Carroll boosts, dilatations and Carroll SCTs. Let us denote the values of these generators at $\textbf{x}=0$ by $\mathcal{S}_{ij}, \mathscr{C}_{i}, \mathbf{\Delta}, k_i, k_0$ respectively. The non-zero commutations of this reduced algebra are given by  
\begin{subequations}\label{ccar-lit}
\begin{align}
        &[\mathcal{S}_{ij},\mathcal{S}_{kl}]=so(D-1), \quad [\mathcal{S}_{ij},\mathscr{C}_{k}]=-i\delta_{ik}\mathscr{C}_{j} + i\delta_{jk}\mathscr{C}_{i}, \\
        &[\mathbf{\Delta},k_i] = -ik_i, \quad [\mathbf{\Delta},k_0] = -ik_0, \quad [\mathcal{S}_{ij},k_{k}]=-i\delta_{ik}k_{j} + i\delta_{jk}k_{i}. 
    \end{align}     
\end{subequations}
We consider a generic field $\Phi(t,\vec{x})$ on this $d$-dimensional Carroll manifold. At the origin the field would transform as: 
\begin{subequations}\label{little}
  \begin{align} 
	&[J_{ij},\Phi(0)]=\mathcal{S}_{ij}\Phi(0), \quad [C_i,\Phi(0)]=\mathscr{C}_i \Phi(0), \quad [D,\Phi(0)]=-i\mathbf{{\Delta}}\Phi(0)  \\
	&[K_0,\Phi(0)]=\mathit{k}\Phi (0), \quad [K_i,\Phi(0)]=\mathit{k}_i \Phi(0).
\end{align}  
\end{subequations}
If we demand $\Phi(t,\vec{x})$ to transform in an irreducible representation of the Carroll group, then a matrix that commutes with 
$\mathcal{S}_{ij}$ and $\mathscr{C}_{i}$ must be, by Schur's lemma, a multiple of identity. Thus $\mathbf{\Delta}$ is a multiple of identity and manifestly $i \D$ where $\D$ is the scaling dimension of the field $\Phi$. 
This forces $\mathit{k}$ and $\mathit{k}_i$ to zero as a consequence of the algebra. 

\medskip

The transformation rules for the field $\Phi$ at an arbitrary spacetime point $(\textbf{x})$ can  be figured out by translating the relations  at the origin (\ref{little}) to this arbitrary point by means of the Baker-Campbell-Hausdorff formula. The relations then become: 
\begin{subequations}\label{Carrpri}
  \begin{align} 
	&[H,\Phi(t,x^i)]=-i\partial_t \Phi (t,x^i), \quad [P_i,\Phi(t,x^i)]=-i\partial_i \Phi (t,x^i)  \\  
	&[J_{ij},\Phi(t,x^i)]=-i(i\mathcal{S}_{ij}-x_i\partial_j+x_j\partial_i) \Phi (t,x^i)  \\ 
    &[C_i,\Phi(t,x^i)]=(-ix_i\partial_t+\mathscr{C}_i) \Phi (t,x^i)  \\ 
	&[D,\Phi(t,x^i)]=-i(\Delta+t\partial_t+x^i\partial_i) \Phi (t,x^i)  \\ 
	&[K,\Phi(t,x^i)]=(-ix^2\partial_t+2x^i\mathscr{C}_i) \Phi (t,x^i)  \\ 
&[K_i,\Phi(t,x^i)]=-i(-2x_i\Delta +2ix^j\mathcal{S}_{ij}-2it\mathscr{C}_i-2tx_i\partial_t-2x_ix^j\partial_j+x^2\partial_j) \Phi (t,x^i) 
\end{align}  
\end{subequations}
We notice from the algebra \eqref{ccft-d} that $K_i$ and $K_0$ act as lowering operators and $H, P_i$ act as raising operators in the spectrum labelled by dilatation eigenvalues $\D$. The notion of Carroll primary operators, keeping with our discussions in $D=2$ is thus given by: 
\begin{align}
    [K_i, \Phi_p(0)] = 0 = [K_0, \Phi_p(0)]   
\end{align}
This means that these form operators whose conformal dimension cannot be reduced by acting with lowering operators. This aligns well with the discussion of the representations we have constructed above and is in close analogy with relativistic CFTs. 

\medskip

We finally draw our attention to the Carroll subalgebra and in particular remind the reader of the reduced Carroll algebra \eqref{ccar-lit}. This means that when we are looking at generic spin primaries $[J_{ij},\Phi(0)]=\mathcal{S}_{ij}\Phi(0)$, one needs to take care of boost matrices $\mathscr{C}_i$ such that the algebra is satisfied. For scalar primaries, this is not an issue since $\mathcal{S}_{ij}=\delta_{ij}$ means that $\mathscr{C}_i$ are identically zero. But for non-trivial spins, there is the option of non-trivial boost matrices. This is reminiscent of our discussion of the non-diagonalisability of the $M_0$ generator in the $D=2$ case. $M_0$ was also the boost in the $D=2$ case. However, $\mathscr{C}_i=0$ remains a viable option for which the Carroll algebra closes even for non-trivial spins. 

\subsection{3D Carroll CFTs}
\label{ssec:3dccfts}

We now focus our attention on Carroll CFTs in 3D. This is particularly important as 3D CCFTs would play a pivotal role in understanding holography for asymptotically flat spacetime in 4D and hence to a holographic description of the real world. We will come back to this in Part B. In this section, we outline some of the salient features of 3D CCFTs. 

We begin with the infinitely extended conformal Carroll algebra in 3D: 
\begin{subequations}\label{eq:bms4alg}
  \begin{align}
    & [L_n, L_m] = i(n-m) L_{n+m} \, , \quad [\bar{L}_n, \bar{L}_m] = i(n-m) \bar{L}_{n+m}\, , \\
    & [L_n, M_{r,s}] = i\left(\frac{n+1}{2} -r \right) M_{n+r, s}\, , \quad  [\bar{L}_n, M_{r,s}] = i\left(\frac{n+1}{2} -s \right) M_{r, n+s}\, . 
\end{align}  
\end{subequations}
In the above, we have only displayed non-zero commutation relations between generators. The algebra can be obtained from the conformal isometry equations of 3D Carroll manifolds \eqref{eq:carkilling}. The global part of the algebra is spanned by 10 generators: 
\begin{align}
  \mathfrak{G} = \{L_{0,\pm 1}, \bar{L}_{0,\pm 1}, M_{00}, M_{01}, M_{10}, M_{11} \}.   
\end{align}
This is the algebra that is isomorphic to the one obtained as a limit of the relativistic conformal algebra in $D=3$. 

\medskip

It is particularly useful to write these generators in the following coordinate basis:
\begin{align}
L_n = - iz^{n+1}\partial_z - \frac{i}{2}(n+1)z^n t\partial_t, \, \, \bar{L}_n = - i\z^{n+1}\partial_\z - \frac{i}{2}(n+1)\z^n t\partial_t, \, \, M_{rs} = -iz^r \z^s \partial_t. 
\end{align}
Here we start on the plane $\mathbb{R}_t\times \mathbb{R}^2$ parametrized by $(x,y, t)$ and move to complexified coordinates 
\begin{align}
   z, \z = x\pm iy.
\end{align}
$(z, \z)$ can also be thought about as stereographic coordinates for the sphere where now the Carroll manifold is $\mathbb{R}_t\times \mathbb{S}^2$. With the identification $z, \z = x\pm iy$, it is easy to see that the global subalgebra $\mathfrak{G}$ is the same as \eqref{ccft-d}, with the identifications:
\begin{align}
    & L_{-1}= \frac{1}{2}(P_x + i P_y), \quad \bar{L}_{-1} = \frac{1}{2}(P_x - i P_y), \quad M_{00} = H,  \quad L_0 = \frac{1}{2}(D + iJ), \\ &\bar{L}_0 = \frac{1}{2}(D- iJ), \quad
     M_{10} = C_x - i C_y, \quad M_{01} = C_x + iC_y, \nonumber \\ &L_1 = -\frac{1}{2}(K_x -i K_y), \quad \bar{L}_1 = -\frac{1}{2}(K_x + iK_y), \quad M_{11} = K_0. \nonumber
\end{align}

We now focus on representations. In keeping with our discussions for 2D CCFTs, we will label our fields $\Phi$ with Dilatations. This amounts to 
\begin{equation}\label{eq:l0act}
    [L_0,\Phi(0)]=-ih \,\Phi(0) \, , ~~~~ \quad [\bar{L}_0,\Phi(0)]=-i\bar{h}\,\Phi(0) \, .
\end{equation}
Notice however that unlike the 2D case \eqref{carr2}, we have $[L_0,M_{00}] \neq 0$ and $[\bar{L}_0,M_{00}] \neq 0$ here and hence the representations that carry a label under $L_0, \bar{L}_0$ cannot carry a label under $M_{00}$. This would mean that the 3D representations are of a different flavour compared to the 2D representations. Also 
\begin{align}
L_0+\bar{L}_0=D, \quad L_0-\bar{L}_0=iJ.    
\end{align}
So, we can immediately see that the labels $h,\bar{h}$ are related to the scaling dimension $\Delta$ and spin $\sigma$ through
\begin{equation}
    h + \bar{h} = \Delta \, , ~~~~~ h - \bar{h} = \sigma \, .
\end{equation}
We now define Carroll primaries as 
\begin{subequations}
   \begin{align}
   & [L_n,\Phi(0)]=0,\quad [\bar{L}_0,\Phi(0)]=0, \quad \forall \, n>0, \\
   & [M_{r,s}, \Phi(0)] = 0, \quad \forall \, r,s>0.
\end{align} 
\end{subequations}
The representations look very similar to what one would have for a 2D relativistic CFT. But notice that here in a 3D CCFT, in addition to Virasoro positive modes,  half of the supertranslation generators also annihilate the primary field at the origin. This is a crucial difference from a 2D CFT.  The transformation rules for primary fields at an arbitrary point at the Carrollian manifold are then given by
\begin{subequations} \label{BMS-primary}
	\begin{align}  \label{Primary}
         [L_n,\Phi(t,z,\bar{z})]&=-iz^{n+1}\partial_z\Phi(t,z,\bar{z})-i(n+1)(h+\frac{t}{2}\partial_t)\Phi(t,z,\bar{z})z^{n},  \\    [\bar{L}_n,\Phi(t,z,\bar{z})]&=-i\bar{z}^{n+1}\partial_{\bar{z}}\Phi(t,z,\bar{z})-i(n+1)(\bar{h}+\frac{t}{2}\partial_t)\Phi(t,z,\z)\bar{z}^n   \\
[M_{r,s},\Phi(t,z,\bar{z})]&=-iz^r\bar{z}^s\partial_t\Phi(t,z,\bar{z}).
	\end{align}
\end{subequations}
One can check that when restricted to $n\in (0, \pm1)$ and $r,s \in (0,1)$ i.e. when one considers the global subalgebra $\mathfrak{G}$, the primary transformation rules are identical to the quasi-primary transformations defined in general dimensions in \eqref{Carrpri}. 

\medskip

We discussed subtleties with respect to the boost matrices for non-trivial spins earlier for generic dimensions. We now focus $D=3$ and explicitly construct transformation laws of spinless fields and fields with spin, focusing on quasiprimaries in $D=3$. In $D=3$, the Carroll group is generated by one rotation  and two Carroll boosts along with the spatial translations. The algebra of the spin matrices associated to rotations ($\mathcal{J}$) and Carroll boosts ($\mathscr{C}_x, \mathscr{C}_y$)  in this case reduces to
\begin{align}  \label{spin algebra}
	[\mathcal{J},\mathscr{C}_x]=-i\mathscr{C}_y, \quad [\mathcal{J},\mathscr{C}_y]=i\mathscr{C}_x, \quad [\mathscr{C}_x,\mathscr{C}_y]=0.
\end{align}

\paragraph{Spin 0:} The spin 0 or the scalar representation of the conformal Carroll fields can be obtained by trivially setting 
\begin{equation}
	\mathcal{J}=\mathscr{C}_x=\mathscr{C}_y=0
\end{equation}
For this case, the transformation rules \eqref{Carrpri} of the scalar primaries become
\begin{subequations}
\begin{align}
	&	[L_n,\Phi(t,z,\bar{z})]=-i\left[z^{n+1}\partial_z+\left(\frac{\Delta}{2}+\frac{t}{2}\partial_t\right)z^n\right]\Phi(t,z,\bar{z})  \quad    \forall n  \in 0, \pm 1 \\ 
	&	[\bar{L}_n,\Phi(t,z,\bar{z})]=-i\left[\bar{z}^{n+1}\partial_{\bar{z}}+\left(\frac{\Delta}{2}+\frac{t}{2}\partial_t\right)\bar{z}^n\right]\Phi(t,z,\bar{z})     \quad    \forall n  \in 0, \pm 1  \\  
	&\text{and} \quad
	[M_{r,s},\Phi(t,z,\bar{z})]=-iz^r\bar{z}^s\partial_t \Phi(t,z,\bar{z}) \quad   \forall r,s \in 0,1
\end{align}     
\end{subequations}
These transformation properties expectedly match up with the previously defined Carroll primaries with $h=\bar{h}=\frac{\Delta}{2}$.

\paragraph{Higher spins:} Representation of spinning primary fields have interesting features, which we will mostly omit in this review. More details can be found in \cite{Bagchi:2023cen}. For illustration, let us briefly comment on spin 1. The spin 1 representation of the rotation generator is given by
\begin{align} \label{eq:spin1rot}
	\mathcal{J}=
	\begin{bmatrix}
		0 & 0 & 0 \\
		0 & 0 & i  \\
		0 & -i & 0
	\end{bmatrix}
\end{align}
We should choose the boost matrices such a way that is consistent with the commutation relations given in (\ref{spin algebra}). One  choice would be to take the matrices equal to zero, i.e.
\begin{equation} \label{trivial}
	\mathscr{C}_x=\mathscr{C}_y=0.
\end{equation}
Clearly this choice is consistent with the Carroll algebra in \eqref{spin algebra}. In \cite{Bagchi:2023cen}, we focussed on this choice for holographic purposes. However, more generic choices are allowed, and should lead to interesting holographic features as well. With this choice, the primary transformation rules are given by
\begin{subequations}
\begin{align}
	&	[L_n,\Phi^i(t,z,\bar{z})]=-i[z^{n+1}\partial_z+(h_i+\frac{t}{2}\partial_t)z^n]\Phi^i(t,z,\bar{z})   \\ 
		&	[\bar{L}_n,\Phi^i(u,z,\bar{z})]=-i[\bar{z}^{n+1}\partial_{\bar{z}}+(\bar{h}_i+\frac{t}{2}\partial_t)\bar{z}^n]\Phi^i(t,z,\bar{z})     \quad    \forall n  \in 0, \pm 1  \\ 
		&[M_{r,s},\Phi^i(t,z,\bar{z})]=-iz^r\bar{z}^s\partial_t \Phi^i(u,z,\bar{z})
\end{align}    
\end{subequations}
Here the vector primary $\Phi^i(t,z,\bar{z})$ contains $t, z, \bar{z}$ components and we have defined $\Phi^z=\Phi^x+i\Phi^y$ and $\Phi^{\bar{z}}=\Phi^x-i\Phi^y$. Looking at the above equations it can be immediately figured out that  each component of the vector primary again transform like a scalar Carroll primary. The information about the spin is manifested in holomorphic and anti-holomorphic weights ($h$ and $\bar{h}$ ) only. As a consequence of this choice the spin of this 3D Carroll theories effectively reduces to a 2D spin. The weights of the components organise themselves according to the transformation rules above. This is given in the table below.  
\begin{table}[h!]
	\centering
	\begin{tabular}{||c | c c c||} 
		\hline
		weights &  $\Phi_t$ & $\Phi_z$ & $\Phi_{\bar{z}}$ \\ [0.75ex] 
		\hline\hline
		$h$ &  ${\Delta}/{2}$ & $\frac{\Delta+1}{2}$ & $\frac{\Delta-1}{2}$ \\ 
		\hline
		$\bar{h}$ & ${\Delta}/{2}$ & $\frac{\Delta-1}{2}$ & $\frac{\Delta+1}{2}$ \\
		\hline
	\end{tabular}
	\caption{holomorphic and anti-holomorphic weights of vector primaries}
	\label{t1}
\end{table}

Higher spin primaries can be similarly constructed. We refer the interested reader to \cite{Bagchi:2023cen} for more details on the higher spin primaries. 

\subsection{Correlation functions}
\label{ssec:carcorfunctions}

As we have remarked in Sec.~\ref{ssec:2dccfts}, like in relativistic CFTs in generic dimensions, for $D$-dimensional CCFTs, the underlying symmetry is powerful enough to fix the lower point correlation functions. The two point function is entirely fixed and the three point functions can be fixed up to some overall coefficients. Here we are interested in $D=3$, but the analysis can be extended to all $D$.  

\medskip

We shall demand that the vacuum state invariant under the global subgroup of the 3D Conformal Carroll generators. Ward identities admits two classes of solutions, usually designated as the ``CFT'' \cite{Bagchi:2016bcd} and ``delta function'' branches. We will see in Part 2, specifically in Sec.~\ref{sec:4dafs} that it is the delta branch that would be related to scattering amplitudes in the bulk 4D asymptotically flat spacetime{\footnote{One could ponder here why we have not come across a similar situation in $D=2$. The delta-function branch was historically not considered in this case and hence is not included in this review. But there are ongoing efforts aimed at resolving this issue.}}.

\medskip
In this subsection, we will treat the two-point function in detail and also sketch the construction of three-point functions. 

\medskip
{\ding{112}} \underline{\em{Two Point function}}
\medskip

We denote two point function as
\begin{equation}
	G^{(2)}(t,z,\bar{z},t',z',\bar{z}')=\langle 0|\Phi(t,z,\bar{z})\Phi'(t',z',\bar{z}')|0 \rangle. 
\end{equation}
Here $\Phi(t,z,\bar{z})$ and $\Phi'(t',z',\bar{z}')$ are primaries with weight $(h,h')$ and $(\bar{h},\bar{h}')$ respectively. With respect to the supertranslations $M_{r,s}$, $G^{(2)}$ varies as 
\begin{equation}
	\delta_{M_{r,s}}G^{(2)}(t,z,\bar{z},t',z',\bar{z}')=\big( z^r\bar{z}^s\partial_t + z'^r \bar{z}'^s \partial_{t'}\big)G(t,z,\bar{z},t',z',\bar{z}')=0.
\end{equation}
The above  equation has two independent classes of solutions that corresponds to two branches of correlators we mentioned. In one case, the $t$ dependence from the correlation function can be eliminated. In this case, upon further imposing the constraints from global superrotations,  $G^{(2)}$ becomes a 2D CFT primary correlator \cite{Bagchi:2016bcd}: 
\begin{equation}\label{eq:cftbranch2pt}
	G^{(2)}(t,z,\bar{z},t',z',\bar{z}')=\frac{1}{(z-z')^{2h}(\bar{z}-\bar{z}')^{2\bar{h}}}. 
\end{equation}
The other option is keeping time dependence at the expense a contact term in the spatial part \cite{Bagchi:2022emh}. This choice leads to  
\begin{equation} \label{f}
		G^{(2)}(t,z,\bar{z},t',z',\bar{z}')=f(t-t')\delta^{2}(z-z',\bar{z}-\bar{z}')
\end{equation}
where $f(t-t')$ remains arbitrary for now. We now impose superrotation Ward identities on this correlation function. The variation of $G^{(2)}$ with respect to holomorphic superrotations leads to: 
\begin{align}
	\delta_{L_n}G^{(2)}(t,z,\bar{z},t',z',\bar{z}')&=\big[ (z^{n+1}\partial_z+z'^{n+1}\partial_{z'})\\ \nonumber 
	&+(n+1)\big((hz^n+h'z'^n)+\frac{1}{2}(tz^n\partial_t+t'z'^n\partial_{t'})\big)\big]G(t,z,\bar{z},t',z',\bar{z}').
\end{align}
Similarly for $\bar{L}_n$s we have 
\begin{align}
	\delta_{\bar{L_n}}G^{(2)}(t,z,\bar{z},t',z',\bar{z}')&=\big[ (\bar{z}^{n+1}\partial_{\bar{z}}+\bar{z}'^{n+1}\partial_{\bar{z}'}) \\ \nonumber
	&+ (n+1)\big((\bar{h}\bar{z}^n+\bar{h}'\bar{z}'^n)+\frac{1}{2}(t\bar{z}^n\partial_t+t'\bar{z}'^n\partial_{t'})\big)\big]G(t,z,\bar{z},t',z',\bar{z}'). 
\end{align}
$n=-1$ imposes spatial translational invariance but does not add anything nontrivial as these expressions are already translationally invariant. For $n=0$, we get 
\begin{subequations}
  \begin{align}
& [(z\partial_z+z'\partial_{z'})+(h+h')+\frac{1}{2}(t\partial_t+t'\partial_{t'})]f(t-t')\delta^{2}(z-z',\bar{z}-\bar{z}')=0, \\
& [(\bar{z}\partial_{\bar{z}}+\bar{z}'\partial_{\bar{z'}})+(\bar{h}+\bar{h'})+\frac{1}{2}(t\partial_t+t'\partial_{t'})]f(t-t')\delta^{2}(z-z',\bar{z}-\bar{z}')=0.
\end{align}  
\end{subequations}
Using properties of delta functions, these two equations become
\begin{align}
	(\Delta+\Delta'-2)f(t-t')+(t-t')\partial_t f(t-t')=0, \quad   
	(\sigma+\sigma')f(t-t')&=0.
\end{align}
Here $\Delta=(h+\bar{h})$ , is the scaling dimension and $\sigma=(h-\bar{h})$, is 2D spin. The solution of the above equations is 
\begin{equation}
	f(t-t')=\delta_{\sigma+\sigma',0}(t-t')^{-(\Delta+\Delta'-2)}.
\end{equation}
Hence
\begin{equation} \label{Sym-cor}
	G^{(2)}(t,z,\bar{z},t',z',\bar{z}')=\delta_{\sigma+\sigma', 0}\frac{\delta^{(2)}(z-z',\bar{z}-\bar{z}')}{(t-t')^{\Delta+\Delta'-2}}.
\end{equation}
Once this correlator has this form, the equations for $n=1$, which impose the special conformal invariance, are also trivially satisfied. 

\medskip

The delta function in the spatial part of the correlation function reflects ultra-locality of Carrollian theories. When the light cones close up, the causal particles are not allowed to move in the spatial directions any more, thus the transition amplitude between different spatial points vanish. Also unlike relativistic CFTs, two-point function is non-zero even when primaries have unequal scaling dimensions. The restriction is only determined by the spin. 

\medskip
{\ding{112}} \underline{\em{Three Point function}}

\medskip

Even for three-point correlators of Carroll primary fields, 
\begin{equation}
	G^{(3)}(t_i,z_i,\bar{z}_i)=\langle 0|\Phi_1(t_1,z_1,\bar{z}_1)\Phi_2(t_2,z_2,\bar{z}_2)\Phi_3(t_3,z_3,\bar{z}_3)|0 \rangle, 
\end{equation}
the Ward identities admit two classes of solutions. The solution associated with the CFT branch is again the 3-point 2D CFT correlator: 
\begin{align}\label{eq:cftbranch3pt}
	G^{(3)}(t_i,z_i,\bar{z}_i)=\frac{C_{123}}{z_{12}^{h_1+h_2-h_3}z_{23}^{h_2+h_3-h_1}z_{31}^{h_1+h_3-h_2}}.
\end{align}
 There are several subtleties associated with the solution in the other branch and there are substructures that emerge. This categorisation mainly depends on whether we treat  $z$ and $\bar{z}$ independently{\footnote{These different cases would compute scattering amplitudes in (1,3) and (2,2) signature spacetimes when we connect it holographically to 4D AFS.}}. We will give a glimpse of one of the possibilities and point the reader to \cite{Bagchi:2023cen} for further details. 

\medskip

We consider the case where Carrollian manifold has a topology of $\mathbb{R}\times \mathbb{S}^2$ (or $\mathbb{R}\times \mathbb{R}^2$) and $z, \bar{z}$ are dependent variables. Generic three point functions in this case are zero, but there are non trivial ones for a few specific cases.  {\footnote{In the bulk, the generic answers are zero simply because of momentum conservation in 4D AFS. The non-zero solutions correspond to either collinear scattering of massless particles, where all 3 particles are parallel to each other or when one of the particles has zero energy.}}

\medskip

$\bullet$ {\textbf{Case I:}}  We work with the ansatz
\begin{align}
	G^{(3)}(t_i,z_i,\bar{z}_i)=F(t_i)\delta^2(z_{12})\delta^2(z_{13}).
\end{align}
The delta function in $z_{12}$ and $z_{13}$ only allows the momenta that are parallel to each other. Global generators of the algebra will further fix the form of $F(u_i)$. Supertranslation invariance implies
\begin{align}
	\sum_{i=1}^3z^{r}_i{z}^s_i\partial_{t_i}G^{(3)}(t_i,z_i,\bar{z}_i)=0  \qquad \forall r,s =0,1.
\end{align}
Because of the delta functions in the ansatz, the equations reduce to
\begin{align}
	\sum_{i=0}^3 \partial_{t_i} F(t_i)=0.
\end{align}
This equation implies translation invariance along $u$ direction
\begin{align} \label{series}
	F(t_i)\equiv F(t_{12},t_{23},t_{31})=\sum_{a,b,c} f_{abc}t_{12}^a t_{23}^b t_{31}^c.
\end{align}
There are still six more equations that would impose global superrotation invariance, which is isomorphic to bulk Lorentz group. Variations with respect to the holomorphic and anti-holomorphic generators are given by 
\begin{subequations}
    \begin{align}
	\delta_{L_n} G^{(3)}(t_i,z_i,\bar{z}_i)=\sum_{i=1}^{3} \Big[z_{i}^{n+1}\partial_{z_{i}}+(n+1)z_{i}^{n}(h_i+\frac{1}{2}t_i\partial_{t_i})\Big]G^{(3)}(t_i,z_i,\bar{z}_i), \\
	\delta_{\bar{L}_{n}} G^{(3)}(t_i,z_i,\bar{z}_i)=\sum_{i=1}^{3} \Big [\bar{z_i}^{n+1}\partial_{\bar{z}_{i}}+(n+1)\bar{z_i}^{n}(\bar{h}_{i}+\frac{1}{2}t_{i}\partial_{t_i})\Big]G^{(3)}(t_i,z_i,\bar{z}_i).
\end{align}
\end{subequations}
We need to solve these equations for $n=0,\pm 1$. We shall do that for an arbitrary term in the series expansion given by \eqref{series}. i.e.
\begin{equation}
	G^{(3)}(t_i,z_i,\bar{z}_i)= t_{12}^a t_{23}^b t_{31}^c \delta^2(z_{12})\delta^2(z_{13}).
\end{equation}
$n=-1$ equations only impose translational invariance along $z$ and $\bar{z}$. The ansatz we work with already satisfy these conditions. The equation for $n=0$ give
\begin{align}
	&\sum_{i=1}^{3} \Big[z_{i}\partial_{z_{i}}+z_{i}(h_i+\frac{1}{2}t_i\partial_{t_i})\Big] F(t_i)\delta^2(z_{12})\delta^2(z_{13})=0  \Rightarrow a+b+c= 4-2\sum_{i}h_i.
\end{align}
Similarly the anti-holomorphic equation fixes the other weights: 
\begin{align}
	a+b+c=4-2\sum_i\bar{h}_i.
\end{align}
It can shown the equations imposed by special conformal generators $L_1$ and $\bar{L}_1$ don't further add any constraints on the solutions.
Together the above two equations become
\begin{align}
	a+b+c=4-\sum_{i}\Delta_i, \qquad \sum_{i}\sigma_i=0
\end{align}
The final result is thus
\begin{align}\label{eq:carrollthreepointcollinear}
   G^{(3)}(t_i,z_i,\bar{z}_i)= t_{12}^a t_{23}^b t_{31}^c \delta^2(z_{12})\delta^2(z_{13}) ~~ \text{with} ~~
		a+b+c=4-\sum_{i}\Delta_i, ~~  \sum_{i}\sigma_i=0 
\end{align}

$\bullet$ {\textbf{Case II:}} For this case we assume only one delta function which imposes collinearity in two particle momenta unlike the previous case where all three particles are collinear. The ansatz we start off with is
\begin{align}
	G^3(t_i,z_i,\bar{z}_i)=F(t_i) \, {z_{12}}^p \, {\bar{z}_{12}}^q \, \delta^2(z_{13})
\end{align}
Following similar steps we find that the final answer is of the form
\begin{align}
	G^3(t_i,z_i,\bar{z}_i)= t_{13}^c z_{12}^p\bar{z}_{12}^q\delta^2(z_{13})
\end{align}
subjected to the following conditions
	\begin{align}\label{Noncollinear}
		\Delta_1+\Delta_3=(2-c)-\frac{p+q}{2}, \, \Delta_2=-\frac{p+q}{2}, 
		\quad \sigma_1+\sigma_3=-\frac{p-q}{2}, \, \sigma_2=-\frac{p-q}{2}
	\end{align}

$\bullet$ {\textbf{Case III:}} We finally consider a Carroll manifold $R \times C
^2$. On this manifold, $z$ and $\bar{z}$ are independent, allowing a more general ansatz. For e.g one can start by considering
\begin{align}
    G^{3}(t_i,z_i,\bar{z}_i)=F(t_i,z_i)\delta (\bar{z}_{12})\delta (\bar{z}_{23}).
\end{align}
Subsequently symmetry arguments could be used to fix $F(t_i,z_i)$. This yields
\begin{align}
F(t_i,z_i)=U^{-2(\sum_{i=i}^3 \bar{h}_i-2)}z_{12}^a z_{23}^b z_{31}^{c}, \quad \text{where} \quad U=z_1t_{23}+z_2t_{31}+z_3t_{12}
\end{align}
Furthermore $a,b,c$ can also be determined to be
\begin{align}
	a=\Delta_3+\sigma_1+\sigma_2-2, \quad  
	b= \Delta_1+\sigma_2+\sigma_3-2, \quad
	c=\Delta_2+\sigma_3+\sigma_1-2
\end{align}
Notice that unlike the previous cases here we don't have any restrictions on the conformal weights ($h,\h$) in order for the correlation functions to be nontrivial{\footnote{In the 4D AFS bulk dual theory, this would correspond to scattering of different helicity particles in the MHV sector.}}. 

\newpage

\subsection{Pointers to the literature}
A lot of the relevant work in higher dimensional CCFTs is focussed around applications to holography in AFS, especially 4D AFS. We direct the reader to Sec.~\ref{ssec:pointersotherholo} for papers in this direction. Below we mention a few more works which are mainly field theoretic in nature. 
\begin{itemize}
\item \textit{Symmetries of EOM}: Early work on higher dimensional CCFTs include \cite{Bagchi:2016bcd, Bagchi:2019xfx, Bagchi:2019clu}, which considered various examples and showed the emergence of Carrollian conformal structure in equations of motion. 

\item\textit{Intrinsic analyses}: Stress tensors for 3D CCFTs were constructed from Carroll symmetries in \cite{Dutta:2022vkg}. A more detailed intrinsic analysis of 3D CCFTs was done in \cite{Saha:2023hsl} following analogous 2D computations in \cite{Saha:2022gjw}. Generic higher dimensional Carroll field theories were addressed in  \cite{Chen:2021xkw}.

\item \textit{OPE}: Carrollian OPEs were further discussed in \cite{Nguyen:2025sqk}, following up on earlier work related to Celestial CFTs and BMS in \cite{Banerjee:2020kaa}. Carroll OPEs were also discussed in \cite{Dutta:2022vkg, Saha:2023hsl, Bagchi:2024gnn} earlier from slightly different perspectives.

\item \textit{Extensions}: Supersymmetric extensions of higher dimensional CCFTs were studied in \cite{Bagchi:2022owq} and. In addition,  off-shell supersymmetric field theories in four dimensions were studied in \cite{Koutrolikos:2023evq}. Carrollian field theories involving higher spins were discussed in \cite{Campoleoni:2021blr}.

\item \textit{Quantum effects}: Some efforts in understanding quantum effects in Carroll field theories and CCFT have been made in the literature. 
\begin{itemize}
    \item [$\star$] \textit{One-loop effects}: The loop corrections to the two point and four point correlators of an interacting electric scalar theory were studied in \cite{Banerjee:2023jpi}. An RG flow tailored to Carrollian field theories was proposed and a novel fixed point (which is not Gaussian) was discovered. For the fermionic version of this, see \cite{Ekiz:2025hdn}.
    \item [$\star$] \textit{UV/IR mixing}: In \cite{Cotler:2024xhb}, it was found that the electric theories exhibited UV sensitivity while the magnetic theories exhibited IR sensitivity. If $V$ is the volume of the hypercubic spatial lattice with lattice scale $a$ on which the electric scalar theory lives, then the free energy scales as $\frac{V}{a^d}$ (number of sites) which demonstrates the UV/IR mixing in the continuum limit $a \to 0$. It was noted that the magnetic theory has fields whose insertions result in soft theorems.
    \item [$\star$] \textit{Finite effective central charge}: The analysis \cite{Cotler:2024xhb} indicated that the canonical stress tensor correlators diverge in the continuum limit (lattice scale tending to zero). Thus, the central charge of the CCFT defined through the stress tensor two point function diverges. In order to remedy this, \cite{Cotler:2025dau} worked with a Carrollian vector model with $N$ components and proposed a singular $N \to 0$ limit in order to obtain a finite ``effective central charge'' in the continuum limit and hence finite quantum corrections owing to non-Gaussianities. 
    \item [$\star$] Other relevant works on quantum aspects include \cite{Hao:2021urq,Rivera-Betancour:2022lkc,Hao:2022xhq,Yu:2022bcp,Mehra:2023rmm,Chen:2023pqf,Figueroa-OFarrill:2023qty,Chen:2024voz,Vassilevich:2024vei,Sharma:2025rug,Bekaert:2024itn}.
\end{itemize}
\end{itemize}

\newpage

\part{Flatspace Holography: The Carroll way}

\bigskip

\bigskip

\bigskip

\bigskip

\section*{Outline of Part II}

In this part of our review of Carroll symmetries and their role in various aspects of theoretical physics, we focus on the most popular application of Carroll and Conformal Carroll symmetries, viz.~in the context of holography of asymptotically flat spacetimes (AFS). 

\medskip

We will organise this part into five sections. The first one lays out the basics of building a holographic dual for AFS. Sec.~\ref{sec:3dafs} addresses 3D AFS and 2D Carroll CFTs. AFS$_3$/CCFT$_2$ has seen a lot of work in the past decade and we try to summarise what we believe are the most important developments in this section. 

\medskip

The more relevant physically relevant 4D  AFS and its Carrollian dual has been the cynosure of recent interest and has seen multiple approaches. We devote three sections to this. The connection between 4D scattering matrix elements and correlations of 3D CCFTs, established in \cite{Bagchi:2022emh}, forms the basis of a lot of recent work. We give details of this correspondence in  Sec.~\ref{sec:4dafs}. The next section is devoted to the details of the prescription to get Carrollian flat space duals as a limit from AdS/CFT via the embedding space. Sec.~\ref{sec:otherapproaches} briefly discusses other Carrollian approaches to 4D  AFS. This also contains a run through of existing literature not dealt with in detail in the review.

\bigskip \bigskip

\newpage

\section{How to build a Flatspace Hologram}
Over the last two decades, the most promising route to formulation of the elusive quantum theory of gravity has been the a priori unconventional route of the Holographic Principle \cite{tHooft:1993dmi, Susskind:1994vu} that boldly asserts that a theory of quantum gravity is equivalent to a theory without gravity or an ordinary quantum field theory in a lower dimension. The explosion in activity in the field was triggered by Maldacena's path-breaking AdS/CFT correspondence \cite{Maldacena:1997re} relating type IIB superstring theory in AdS$_5 \times$S$^5$ to $\mathcal{N}=4$ Super Yang-Mills theory living on the $D=4$ boundary of AdS$_5$. One of the outstanding questions in the field is how to build a hologram from a general spacetime and a lot of effort has been devoted to this over the years. Below we give a prescription for holography in general taking inspiration from AdS/CFT. 

\medskip

\ding{112} {\underline{\em{Asymptotic symmetries and holography}}}

\smallskip

Symmetries of a certain spacetime at times enhance at its boundaries. One of the important ideas in gravitational physics is the definition of asymptotic symmetries. The asymptotic symmetry group of a spacetime given a particular set of boundary conditions is the quotient of the set of allowed diffeomorphisms by the set of trivial diffeomorphisms: 
\begin{align}
    \text{Asymptotic Symmetry Group} = \frac{\text{Allowed diffeomorphisms}}{\text{Trivial diffeomorphisms}} \, . \nonumber
\end{align}
The Asymptotic symmetry group and its associated algebra, the Asymptotic Symmetry Algebra (ASA), often is given by the isometry of the spacetime. But there are very important exceptions. The seminal analysis of Brown and Henneaux \cite{Brown:1986nw} showed that for AdS$_3$, the ASA was given by two copies of the infinite dimensional Virasoro algebra \eqref{Vira} with central terms $c=\bar{c}=\frac{3\ell}{2G}$, where $\ell$ is the radius of AdS and $G$ is the Newton's constant. To many this is seen as the principle precursor of the AdS/CFT correspondence. Holography for higher spin theories in AdS$_3$ has also followed this route with the discovery on infinite dimensional W-algebras \cite{Campoleoni:2010zq, Henneaux:2010xg, Gaberdiel:2010pz}. The potential co-dimension one dual field theories inherit the  infinite dimensional ASA as their underlying symmetries. 

\medskip

Many years before the analysis of Brown and Henneaux, Bondi, van der Burgh, Metzner \cite{Bondi:1962px} and independently Sachs \cite{Sachs:1962wk}, found that in 4D  asymptotically flat spacetimes (AFS) the symmetries at the null boundary enhance from the Poincar\'{e} algebra to what is now called the BMS$_4$ algebra. 
It was later discovered that in 3D AFS, there is also an infinite ASA, the BMS$_3$ algebra \cite{Ashtekar:1996cd, Barnich:2006av}. If we wish to build holography in general spacetimes borrowing insight from AdS/CFT, it is then very natural to assume that the hologram of a general spacetime is a quantum field theory in one lower dimension which has as its global symmetry the ASA of the gravitational theory. A hologram of AFS should then be given by a co-dimension one field theory with BMS symmetries. 

\medskip

\ding{112} {\underline{\em{Flat holography and Carroll}}}

\smallskip

Following similar observations in lower dimensions in \cite{Bagchi:2010zz}, it was shown that Conformal Carrollian symmetries are isomorphic to BMS symmetries in one higher dimension \cite{Duval:2014uva}:
\begin{align}
    \mathfrak{Conf Carr}_d = \mathfrak{bms}_{d+1}.
\end{align}

\begin{figure}[t]
	\centering	
	\includegraphics[width=0.7\textwidth]{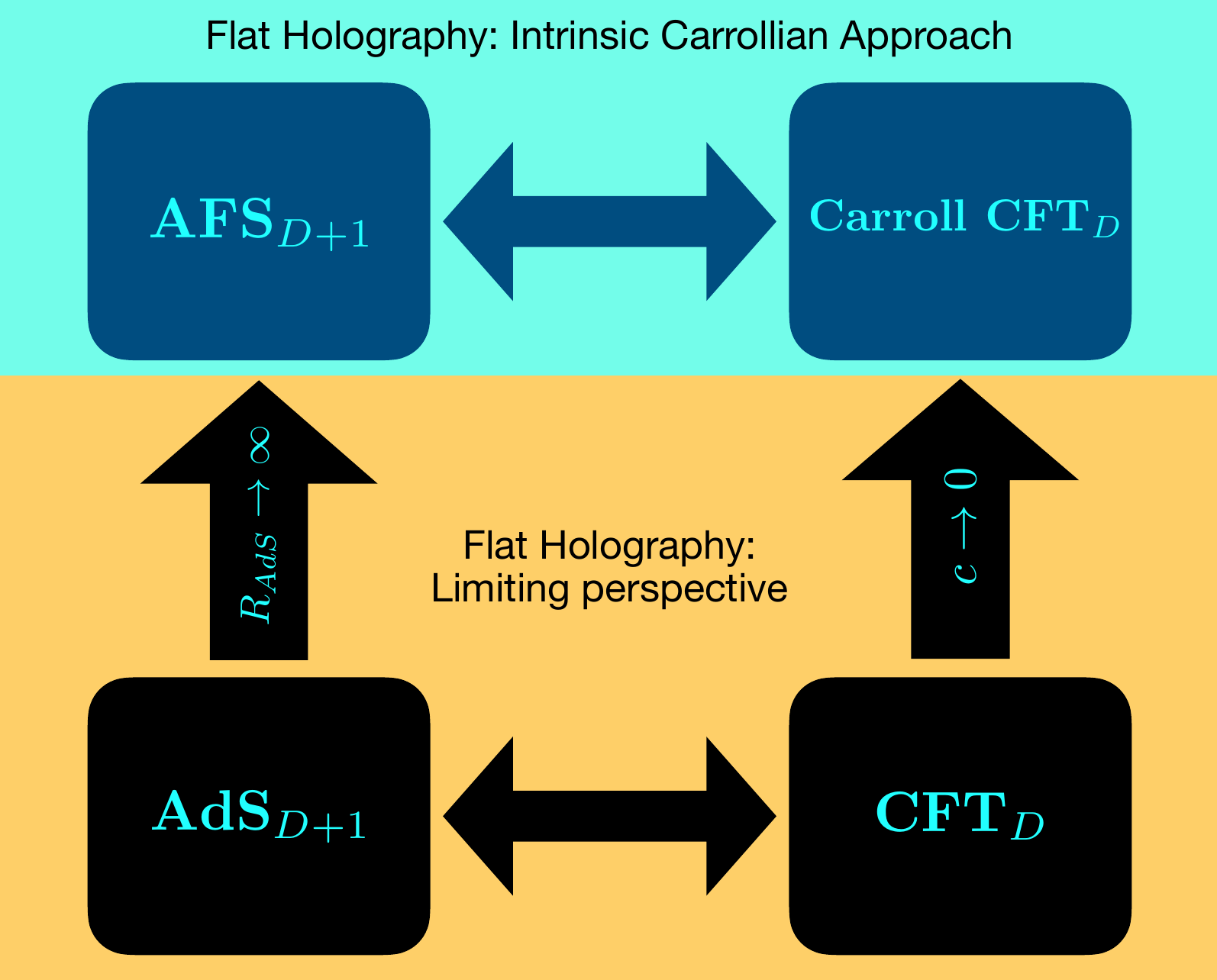}
	\caption{Carrollian holography.}
    \label{fig:flatholo}
\end{figure}

If we adopt the prescription for holography in generic spacetimes outlined above, then this is the statement for holography in asymptotically flat spacetimes:

\medskip

\noindent\fbox{%
    \parbox{\textwidth}{%
        {\em {The holographic dual of a theory of gravity in asymptotically flat spacetimes is a Carrollian conformal field theory (CCFT) in one lower dimension.}}
    }%
}

\medskip

This is the main message of what is nowadays known as the Carrollian version of Flatspace Holography, or in short Carrollian Holography. We summarize the main idea in Fig.~\ref{fig:flatholo}. The light blue box in the figure is what we have outlined as Carrollian holography above. 

\medskip

Let us elucidate a bit more and in a simplified setting \cite{Bagchi:2016bcd}. The first requirement of a holographically duality is that the symmetries of the bulk gravitational theory and the putative dual field theory should match. If we, for now, forget about asymptotic symmetries and assume just the isometries of the bulk theory, then this isometry group has to be realised as global symmetries of the lower dimensional field theory that lives on the boundary. Let us focus on $D=4$. In the case of the AdS/CFT correspondence, the isometry of AdS$_4$ is $SO(3,2)$. This also turns out to be the symmetry of the conformal group in $D=3$, hence providing the first rudimentary check of the  AdS$_4$/CFT$_3$ duality. Now, if we wish to construct a dual to 4D AFS, the isometry group in this case is $ISO(3,1)$. The dual theory, following the lead of holography in AdS, should be a co-dimension one field theory with the isometry group realised as global symmetries. Thus we should have a quantum field theory in $D=3$ with $ISO(3,1)$ as its symmetry. Notice that in 3D, a relativistic QFT has symmetries $ISO(2,1)$. So this is an exotic QFT. As we have discussed, the finite dimensional Carroll Conformal group in $D=3$ is $ISO(3,1)$. So this makes a Carroll CFT a possible candidate for the dual theory to 4D AFS, since symmetries match. Of course the matching between the symmetries in the boundary and bulk goes beyond the isometry group to the entire infinite dimensional asymptotic symmetry algebra.   

\medskip

The other and very obvious route to Flatspace holography is through a limit of usual AdS/CFT. The infinite radius limit of AdS leads to AFS. In terms of the isometry algebra, the AdS$_{D+1}$ isometry algebra $so(D,2)$ contracts to $iso(D,1)$. This singular limit on the bulk side should correspond to a similar singular limit on the boundary. It turns out that this is the limit where the speed of light goes to zero \cite{Bagchi:2012cy}. The timelike boundary of AdS is boosted to the mostly null boundary of AFS and this limit of infinite boost is the Carroll limit{\footnote{We saw how one can get a 2D CCFT from relativistic 2D CFT by performing an infinite boost in Sec.~\ref{ssec:2dccfts}. This is the holographic manifestation of the same phenomenon.}}. This can also be viewed as zooming into the centre of AdS where the AdS radius is imperceivable. We will explain all of this in more detail in what follows.

\medskip

Historically, the non-Lorentzian road to holography in asymptotically flat spacetimes started in \cite{Bagchi:2010zz} with the observation that the BMS$_3$ algebra as isomorphic to the 2D Galilean conformal algebra, drawing inspiration from the BMS/CFT correspondence \cite{Barnich:2010eb}, but pointing out that the CFT to look at was not the usual relativistic one since that only formed a sub-algebra of the entire ASA. As stressed earlier, in $D=2$, the Galilean and Carrollian Conformal algebras are identical and the connections to Carroll emerged. In \cite{Bagchi:2012cy}, it was understood that the flatspace limit of AdS can be understood an ultra-relativistic $c\to0$ contraction on the dual field theory. 

\medskip

After the isomorphism between conformal Carroll algebras and BMS symmetries were formally established in \cite{Duval:2014uva}, the general philosophy of Carrollian holography was first clearly mentioned in \cite{Bagchi:2016bcd}. The initial successes of the field were in the lower dimensional AFS$_3$/CCFT$_2$. This is what we discuss in the next section (Sec.~\ref{sec:3dafs}). More recently, there has been a lot of activity in the context of AFS$_4$/CCFT$_3$ and connections to scattering in 4D  AFS has been understood in terms of correlations of 3D CCFTs \cite{Bagchi:2022emh}. We will discuss this in detail in Sec.~\ref{sec:4dafs}. In Sec.~\ref{sec:flatlimit}, we focus on the limit of AdS to AFS in general dimensions and obtain Carroll correlations from the flat limit of Witten diagrams in AdS, following \cite{Bagchi:2023fbj, Bagchi:2023cen}. Although we primarily focus on the approach that some of the authors in this review have worked on, we will attempt to give a wider perspective for the uninitiated reader and present alternate proposals in Sec.~\ref{sec:otherapproaches}. However, these will not be as comprehensive as the review of our own work. We also give a more detailed list of pointers to the literature at the end of Sec.~\ref{sec:otherapproaches} (cf.~Sec.~\ref{ssec:pointersotherholo}).

\bigskip

\newpage
 
\section{The dual of 3D AFS}
\label{sec:3dafs}

Gravity in three dimensions has been a fertile playground for understanding quantum gravity. Due to the absence of propagating degrees of freedom in $D=3$ in Einstein gravity, the theory was initially thought to be trivial. But the discovery of BTZ black holes \cite{Banados:1992wn, Banados:1992gq} in asymptotically AdS spacetimes dramatically changed this outlook. As was mentioned above, the discovery of two copies of the Virasoro algebra in the asymptotic symmetries of AdS$_3$ \cite{Brown:1986nw} is seen as many as the primary precursor to the AdS/CFT correspondence. Many of the first checks of AdS/CFT were constructed in terms of AdS$_3$/CFT$_2$. 

\medskip

In this section we discuss gravity in three dimensional asymptotically flat spacetimes and its putative dual theory, a 2D Conformal Carrollian theory. We will see that in direct analogy with gravity in AdS$_3$, we will have non-trivial zero mode solutions, the flat cousins of BTZ black holes, which morph into cosmological solutions called Flat Space Cosmologies. We will understand how to extract bulk physics by doing computations in the putative dual 2D CCFT. We have already presented some details of 2D CCFTs in Sec. \ref{ssec:2dccfts}. We will use many of these results in the subsections below.

\subsection{Symmetries}
The asymptotic symmetry algebra of 3D AFS at its null boundary is given by the BMS$_3$ algebra \cite{Ashtekar:1996cd, Barnich:2006av}
\begin{subequations}\label{bms3}
    \begin{align}
        [L_n, L_m] &= (n-m) L_{m+n} + {c_L}\delta_{n+m,0} (n^3-n). \\
        [L_n, M_m] &= (n-m) M_{n+m} + {c_M}\delta_{n+m,0} (n^3-n). \quad [M_n, M_m] = 0. 
    \end{align}
\end{subequations}
The structure of $\mathscr{I}^\pm$ is $\mathbb{R\times S}^1$ where $\mathbb{R}$ represents a null direction. In the above, $L_n$'s are the diffeomorphisms of the circle at infinity called superrotations. $M_n$'s are angle dependent translations of the null direction. The central terms for Einstein gravity are given by \cite{Barnich:2006av}
\begin{align}\label{bmsc}
    c_L = 0, \quad c_M = \frac{1}{4G}.
\end{align}
where $G$ is the Newton's constant. 

\medskip

It is obvious that the BMS$_3$ algebra is isomorphic to the 2D Carrollian conformal algebra \eqref{carr2}. The Brown-Henneaux analysis \cite{Brown:1986nw} of the ASA of AdS$_3$ gives two copies of the Virasoro algebra with central terms 
\begin{align}\label{bhc}
    c=\bar{c}= \frac{3 \ell}{2G}.
\end{align}
Here $\ell$ is the radius of AdS$_3$. In order to go from AdS to flatspace, one needs to take the AdS radius $\ell \to \infty$. The two copies of Virasoro of AdS$_3$ contracts to form \eqref{bms3}. This is the ultra-relativistic contraction mentioned in Sec.~3 : 
\begin{align}
    L_n = \L_n - \bL_{-n}, \quad M_n = \frac{1}{\ell} (\L_n + \bL_{-n}), \quad \ell \to \infty. 
\end{align}
It is easy to check that under this limit the central charges become
\begin{align}\label{BC-central}
    c_L = \frac{1}{12} (c - \bar{c})=0, \quad c_M = \frac{1}{12\ell} (c+ \bar{c})= \frac{1}{4G}.
\end{align}
when one uses the Brown-Henneaux central charges \eqref{bhc} for AdS$_3$, matching up with the intrinsic analysis results \eqref{bmsc}. {\footnote{We note that in this section we will use scale the central charges by 12 with respect to Sec.~\ref{ssec:2dccfts}.}}

\subsection{Solution space of gravity in 3D AFS: Flat Space Cosmologies}
The most general solution to 3D Einstein gravity in asymptotically flat spacetimes with Barnich-Comperé boundary conditions \cite{Barnich:2006av} is given by
\begin{align}\label{eq:gen3dmetric}
 ds^2 = \Theta(\psi) du^2 - 2 du dr - 2 \left( \Xi(\psi) + \frac{u}{2} \partial_\psi \Theta(\psi) \right) du d\psi + r^2 d\psi^2.    
\end{align}
In the above, $\Theta(\psi)$ is the mass aspect and $\Xi(\psi)$ is called the angular momentum aspect. The asymptotic structure at $\mathscr{I}^\pm$ is preserved by two arbitrary functions of $\psi$ whose modes are given by \cite{Bagchi:2012yk, Bagchi:2012xr}
\begin{align}
    l_n = i e^{in\psi} \left( (1+n^2\frac{u}{r}) \partial_\psi + inu\partial_u - in r\partial_r \right), \quad m_n = i e^{in\psi} \partial_u. 
\end{align}
These generate the centreless BMS$_3$ algebra and the zero modes $l_{0, \pm1}, m_{0, \pm1}$ coincide with the exact Killing vectors of Minkowski spacetime (up to a trivial diffeomorphism). The corresponding conserved charges can also be Fourier decomposed into modes $L_n$ and $M_n$, which are then related to the mass and angular momentum aspect as
\begin{align}\label{MJ}
    \Theta(\psi) = -1 + 8G \sum M_n e^{in\psi}, \quad \Xi(\psi) = 4G\sum L_n e^{in\psi}. 
\end{align}
For $\Theta = -1$ and $\Xi=0$, we get back usual flat spacetime:
\begin{align}
    ds^2 = -du^2 -2 dudr + r^2 d\psi^2. 
\end{align}
We will be interested in the most general zero mode solutions with 
\begin{align}
    \Theta(\psi)  = 8GM, \quad \Xi(\psi) = 4GJ
\end{align}
where $M, J$ are the mass and angular momentum of the solution. For $M>0$ these are called {\em Flat Space Cosmologies} (FSC):
\begin{align}\label{eq:fscmetric}
  ds^2 = 8 M du^2 - 2 du dr - 8 J du d\psi + r^2 d\psi^2.  
\end{align}
In this parametrization, flat space has
\begin{align}
    M = -\frac{1}{8G}, \quad J=0. 
\end{align}

\begin{figure}[h]

\begin{subfigure}{0.5\textwidth}
\includegraphics[width=0.8\linewidth, height=5.5cm]{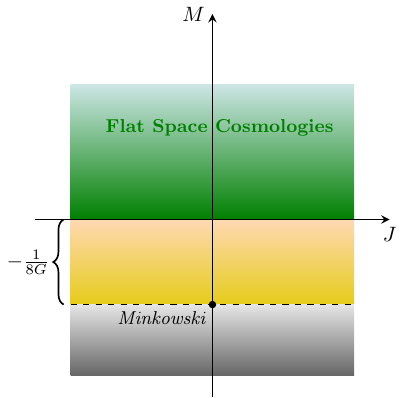} 
\end{subfigure}
\begin{subfigure}{0.5\textwidth}
\includegraphics[width=0.8\linewidth, height=5.5cm]{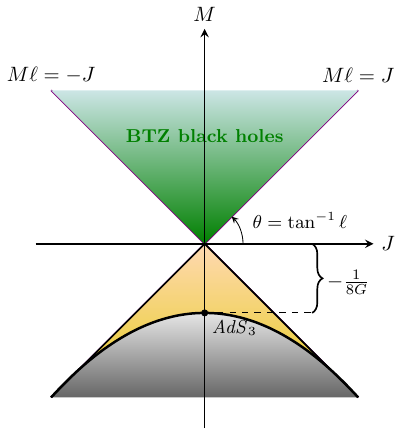}

\end{subfigure}

\vspace{0.3cm}
\centering
\begin{tabular}{ll}
\textcolor{yellow}{\rule{1em}{1em}}  Conical deficit \quad \quad 
\textcolor{gray}{\rule{1em}{1em}} & Conical excess \\
\end{tabular}
\caption{Solution space of AFS (left) and AdS (right) in 3D.}
\end{figure}

\bigskip

In order to understand how these zero mode solutions are related to the zero mode solutions in AdS$_3$, i.e. the BTZ black holes \cite{Banados:1992wn}, let us begin from the metric of the non-extremal BTZ in Schwarzschild like coordinates:
\begin{align}
    ds^2 = - \frac{(r^2 - r^2_+)(r^2 - r^2_-)}{r^2 \ell^2} dt^2 + \frac{r^2 \ell^2}{(r^2 - r^2_+)(r^2 - r^2_-)} dr^2 + r^2 \left( d\phi - \frac{r_+ r_-}{\ell r^2} dt \right)^2. 
\end{align}
Here $r_\pm$ are the outer and inner horizons of the rotating black hole related to its mass and angular momentum by
\begin{align}\label{r+-}
   r_\pm = \sqrt{2G \ell (\ell M + J)} \pm \sqrt{2G \ell(\ell M - J)} \, .
\end{align}
In the above, $\ell$ is the AdS radius. We now take the infinite radius limit $\ell \to \infty$ limit on these solutions to find the corresponding flatspace solutions. Very interestingly, in this limit, the outer horizon is pushed all the way to infinity while the inner horizon remains at a finite distance
\begin{align}
    r_+ \to \ell \sqrt{8GM}= \ell \hat{r}_+, \quad r_0 = \sqrt{\frac{2G}{M}} J = r_0.
\end{align}
So the inside of the original black hole becomes the entire spacetime and since this entails the flipping of the radial and temporal directions, the solution becomes time dependent and hence is a cosmology. In Schwarzschild like coordinates, this is given by
\begin{align}\label{fsc-sch}
    ds^2 = \hat{r}_+^2 dt^2 - \frac{r^2 dr^2}{\hat{r}_+^2(r^2-r_0^2)} - 2 \hat{r}_+ r_0 dt d\phi+ r^2 d\phi^2. 
\end{align}
It is clear that there is a cosmological horizon at $r=r_0$. To connect with the previous discussion about the general solutions to 3D AFS, one needs to make the following coordinate transformation:
\begin{align}
    d\psi = d\phi + \frac{r_0 dr}{\hat{r}_+ (r^2-r_0^2)}, \quad du = dt + \frac{r^2 dr}{\hat{r}_+^2 (r^2-r_0^2)}.
\end{align}
This brings the above metric to the following form: 
\begin{align}
    ds^2 = \frac{\hat{r}_+^2(r^2 - r^2_0)}{r^2} du^2 - 2 du dr + r^2 \left( d\phi - \frac{\hat{r}_+ r_0}{r^2} du \right)^2. 
\end{align}
In Figure \ref{pdfsc} below, we have drawn the Penrose diagram for the FSC to elucidate its causal structure. 
\begin{figure}[h] 
\centering
\includegraphics[scale=1]{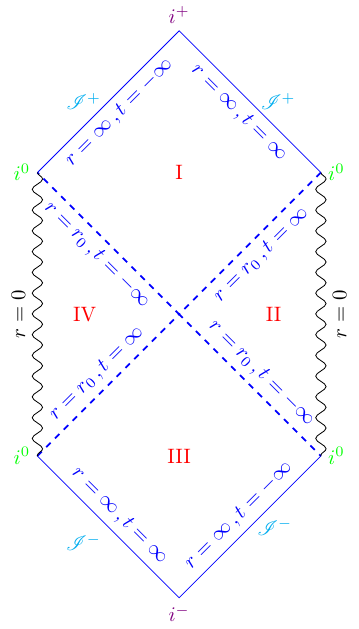}
\captionof{figure}{Penrose diagram of the FSC solution. Dashed lines represent cosmological horizons. Timeline singularities and past-future null infinities are represented by solid blue lines and black wiggled lines.}
\label{pdfsc}
\end{figure}

\medskip

{\ding{112}} \underline{\em{Orbifolds}}

\smallskip

It is well established that the BTZ black hole solutions can be constructed as quotients of AdS$_3$ \cite{Banados:1992gq}. Let us quickly review this: AdS$_3$ can be represented as a hyperboloid embedded within a four-dimensional spacetime $\mathbb{R}^{2,2}$, governed by the equation
\begin{align}
  -u^2 - v^2 + x^2 + y^2 = -\ell^2,  
\end{align}
with the corresponding metric given by
\begin{align}
ds^2 = -du^2 - dv^2 + dx^2 + dy^2.    
\end{align}
BTZ black holes are obtained through identifications of discrete subgroups of the isometry group of AdS$_3$. The identification that gives rise to the non-extremal BTZ is 
\begin{equation}
    \xi = \frac{r_+}{\ell}J_{ux} - \frac{r_-}{\ell}J_{vy}
\end{equation}
where $J_{\mu\nu}$ are the Lorentz generators of the embedding spacetime.

\medskip

In a similar fashion, one can define FSCs as quotients of 3D Minkowski spacetime. This identification involves a combination of a translation (shift) and a Lorentz boost, leading to the interpretation of FSCs as \textit{shifted-boost orbifolds} of flat spacetime \cite{Cornalba:2007zb}. 

\medskip

Following \cite{Cornalba:2007zb}, we now recall how the AdS orbifold goes over to the flatspace orbifold in the $\ell\to \infty$ limit. Now, before we take the flat limit, we relabel the coordinates: 
\begin{align}
 x \to X, \quad y \to Y, \quad u \to T, \quad  \frac{v}{\ell}\to -1.    
\end{align}
The Killing vector then becomes 
\begin{eqnarray}
    \xi_{FSC}=\hat{r}_+(X\partial_T+T\partial_X)+r_0\partial_Y.
\end{eqnarray}
Here, the first term corresponds to a boost in the $X$-direction with velocity $v = \tanh(2\pi \hat{r}_+)$, while the second term represents a translation by $2\pi r_0$ along the $Y$-direction. Hence the name \textit{shifted-boost orbifolds} of 3D Minkowski space. To connect with the coordinates we have been using above \eqref{fsc-sch}, we do the following coordinate transformation for the region $r>r_0$:
\begin{eqnarray}
    X^2 = \frac{r^2 - r_0^2}{\hat{r}_+^2}  \sinh^2(\hat{r}_+ \phi)\,, \quad
T^2 = \frac{r^2 - r_0^2}{\hat{r}_+^2}\cosh^2(\hat{r}_+ \phi)\,, \quad
Y = r_0 \phi - \hat{r}_+ t\,.
\end{eqnarray}
One can find similar transformations valid for region $r<r_0$. In these coordinates \eqref{fsc-sch}, identification of flatspace that gives rise to the FSC simply becomes 
\begin{align}
    \xi_{FSC} = \d_{\phi}.
\end{align}
This is similar to the BTZ identification in AdS$_3$. 

\medskip

{\ding{112}} \underline{\em{Thermodynamics of FSC}}

\smallskip

These FSC solutions have a horizon and hence thermodynamical quantities can be ascribed to them. $r=r_0$ is a Killing horizon with normal 
\begin{align}
    \chi = \partial_u + \frac{\hat{r}_+}{r_0} \partial_\phi
\end{align}
The surface gravity is given by
\begin{align}
    \kappa^2 = - \frac{1}{2} \nabla^\mu \xi^\nu \nabla_\mu \xi_\nu = \frac{\hat{r}_+^4}{r_0^2} \Rightarrow \kappa = \frac{\hat{r}_+^2}{r_0}. 
\end{align}
The associate Hawking temperature of the FSC is given by 
\begin{align}
    T_{FSC} = \frac{\kappa}{2 \pi} = \frac{\hat{r}_+^2}{ 2\pi r_0}.
\end{align}
One can then associate a Bekenstein-Hawking entropy to the FSC by means of the famous area law:
\begin{align}\label{fsc-en}
    S_{FSC} = \frac{2 \pi r_0}{4G}. 
\end{align}
The thermodynamical quantities enables one to write down a first law for the FSC
\begin{align}\label{FLfsc}
    dM = -T_{FSC} \, dS_{FSC} + \Omega_{FSC} \, dJ
\end{align}
where $\Omega_{FSC}$ is the angular velocity of the FSC and $\Omega_{FSC} = \frac{\hat{r}_+}{r_0}$. We notice that the entropy of the FSC is a remnant of one which could have been attributed to the inner horizon of the BTZ \cite{Bagchi:2013lma}. The first law of the FSC \eqref{FLfsc} has an unfamiliar minus sign in front of the TdS term. This too is a reflection of a first law that arises out of a limit from the inner horizon first law which is known to have ``wrong'' signs. 

\medskip

In 3D AdS spacetimes, there is a phase transition between thermal AdS and the BTZ black hole known as the Hawking-Page transition \cite{Hawking:1982dh}, where above a certain critical temperature, the BTZ black hole becomes the dominant saddle of the Euclidean gravity partition function. {\footnote{These transitions also exist in higher dimension, where large AdS black holes emerge as the stable phase at higher temperatures.}} In \cite{Bagchi:2013lma}, it was discovered that there is an analogue of this transition even in flatspace where now the transition is between hot flatspace (HFS) and the FSC solutions. HFS transitions into a time-evolving geometry when the temperature exceeds a critical value giving rise to a cosmological solution with a horizon from apparently boring flat space!  

\subsection{Cardy counting in dual Carroll theory}
In this subsection, we recount the first major success of the Carrollian holography programme following \cite{Bagchi:2012xr}, which provided a matching between a counting of states in the boundary 2D CCFT and the entropy of the cosmological horizon of 3D FSCs we have discussed above. 

\medskip

{\ding{112}} \underline{\em{Cardy counting in AdS/CFT}}

\smallskip

One of the first checks of the AdS$_3$/CFT$_2$ duality was the matching \cite{Strominger:1997eq} of the Bekenstein-Hawking entropy of the BTZ black hole to a counting of states in the dual 2D CFT via the famous Cardy formula:
\begin{align}
    S_{Cardy} =  2\pi \left( \sqrt{\frac{c h}{6}} + \sqrt{\frac{\bar{c} \h}{6}} \right)
\end{align}
where $h, \h$ are the eigenvalues of $\L_0, \bL_0$ that label the states of a 2D CFT, and $c = \bar{c}$ are the Brown-Henneaux central terms \eqref{bhc}. BTZ black holes are thermal states in the dual theory with Virasoro weights
\begin{align}
    h= \ell M + J, \quad \h = \ell M - J,
\end{align}
where $M, J$ are the mass and angular momentum of the BTZ. It is straight forward to see that plugging this into the Cardy formula results in 
\begin{align}
    S_{Cardy} = \frac{\pi r_+}{2G} = S_{BTZ},
\end{align}
where we have used the relation between the horizons and mass and angular momentum \eqref{r+-}. 

\medskip

{\ding{112}} \underline{\em{BMS-Cardy formula}}

\smallskip

The holographic dictionary between 3D AFS and 2D CCFTs hinges on the matching of partition functions between the bulk and boundary theories:
\begin{align}
    Z_{AFS_3} = Z_{CCFT_2}.
\end{align}
We remind the reader of the partition function of the 2D CCFT defined in Sec.~\ref{ssec:2dccfts} (cf. \eqref{Z-ccft}):
\begin{align}\label{PFcar}
    Z_{CCFT}(\zeta, \rho) = \text{Tr} \, e^{2\pi i \zeta L_0} e^{2\pi i \rho M_0} = \sum d(\D, \xi) e^{2\pi i \zeta \D} e^{2\pi i \rho \xi}.
\end{align}
We now recall that transforming the stress tensor from the plane to the cylinder involved Schwarzian derivatives for usual 2D CFTs and analogous ones for 2D CCFTs which led to additional central term contributions. This affects the zero modes and hence the partition function that we would be interested in for the torus would involve shifts in $L_0$ and $M_0$ by central terms. The partition function we will be interested in is thus
\begin{align}
    Z^{(0)}_{CCFT}(\zeta, \rho) = \text{Tr} \, e^{2\pi i \zeta(L_0 - \frac{c_L}{2})} e^{2\pi i \rho(M_0 - \frac{c_M}{2})} \, = e^{-{\pi i}(\zeta c_L + \rho c_M)}Z_{CCFT}(\zeta, \rho).
\end{align}
Under Carroll modular transformations \eqref{Car-mod}, we will demand this partition function to be invariant:
\begin{align}
    Z^{(0)}_{CCFT}\left( \frac{a \zeta + b}{c \zeta + d}, \frac{\rho}{(c \zeta + d)^2} \right) = Z^{(0)}_{CCFT}(\zeta, \rho).
\end{align}
In particular, under the Carroll S-transformation $(\zeta, \rho) \to (- \frac{1}{\zeta},  \frac{\rho}{\zeta^2})$ , we have
\begin{align}
     Z_{CCFT}(\zeta, \rho)  =& e^{\pi i(\zeta c_L + \rho c_M)} Z^{(0)}_{CCFT}(\zeta, \rho) = e^{\pi i(\zeta c_L + \rho c_M)} Z^{(0)}_{CCFT}\left( - \frac{1}{\zeta},  \frac{\rho}{\zeta^2}\right) \cr
      & = e^{\pi i(\zeta c_L + \rho c_M)} e^{- \pi i((-\frac{1}{\zeta}) c_L +  (\frac{\rho}{\zeta^2}) c_M)}Z_{CCFT}\left( - \frac{1}{\zeta},  \frac{\rho}{\zeta^2}\right). 
\end{align}
To find the density of states $d(\D, \xi)$ defined above in \eqref{PFcar}, we do an inverse Laplace transformation:
\begin{align}
   d(\D, \xi) = \int d\D d\xi \, \, e^{\pi i \left\{c_L\left(\zeta + \frac{1}{\zeta}\right) + c_M \left(\rho  - \frac{\rho}{\zeta^2}\right) + \zeta \D + \rho \xi \right\}} \, Z_{CCFT}\left( - \frac{1}{\zeta},  \frac{\rho}{\zeta^2}\right). 
\end{align}
In the limit of large $\D, \xi$, we can perform the integral in the saddle point approximation and the logarithm of the density of states yields the entropy of the high energy states of the field theory:
\begin{align}
    S_{BMS-Cardy}= \ln d(\D, \xi) = 2 \pi \left(c_L\sqrt{\frac{\xi}{2c_M}} + \D \sqrt{\frac{c_M}{2\xi}}\right)
\end{align}
The above formula is nowadays referred to as the BMS-Cardy formula \cite{Bagchi:2012xr} and hence the subscript on the entropy. 

\medskip

{\ding{112}} \underline{\em{FSC entropy from state counting}}

\smallskip

Now let us go back to the holographic dictionary. The dual to thermal states in a 2D CCFT are 3D Flat Space Cosmologies. So a counting of states in the field theory, in keeping with generic holographic expectations, should lead to the entropy of FSCs that we reviewed above in \eqref{fsc-en}. Let us remember that $L_0, M_0$ generate angular momentum and time translations in the boundary theory that lives on $\mathbb{R} \times \mathbb{S}^1$ and states are labelled by their eigenvalues $\D, \xi$. These quantities are related to the angular momentum and mass of the zero mode solutions in 3D AFS. Hence remembering \eqref{MJ}
\begin{align}\label{xid}
    \xi = M_{bulk} + \frac{1}{8G}, \quad \D = J_{bulk}.
\end{align}
We will drop the bulk subscripts on the mass and angular momentum in what follows. We note that in going from the plane to the cylinder, there is a shift in the zero modes as
\begin{align}
    L_0 \to L_0 + \frac{c_L}{2}, \quad M_0 \to M_0 + \frac{c_M}{2}.
\end{align}
In the case of Einstein gravity, the central charges are given by \eqref{BC-central}. This explains the shift in the $\xi$ and no shift in $\D$ in \eqref{xid}. We now put back the values of $\D, \xi, c_L, c_M$ into the BMS-Cardy formula and work in the limit $\xi\gg c_M$. This means, for Einstein gravity, we have
\begin{align}
    S= 2 \pi J \sqrt{\frac{1/4G}{2M}} = \frac{\pi J}{\sqrt{2GM}} = \frac{2\pi r_0}{4G} = S_{FSC}. 
\end{align} 
We thus have a matching between the entropy calculated by a counting of states from modular invariance of the partition function in a 2D Carroll CFT and the Bekenstein Hawking entropy of the FSC, in a manner reminiscent of the seminar matching of the BTZ entropy with Cardy counting in a 2D CFT in \cite{Strominger:1997eq}. This is the first and one of most important checks of the Carrollian approach to flatspace holography. 

\medskip

Logarithmic corrections to the BMS-Cardy formula were first computed in \cite{Bagchi:2013qva}. In \cite{Riegler:2014bia}, it was understood that this entropy formula could be obtained as a limit of the {\em inner-horizon} Cardy formula \cite{Riegler:2014bia, Fareghbal:2014qga}: 
\begin{align}
    S_{inner} = 2\pi \left( \sqrt{\frac{c h}{6}} - \sqrt{\frac{\bar{c} \h}{6}} \right)
\end{align}
To see this, remember that in the flatspace limit, the map from the dual theory is the UR map: 
\begin{align}
    \D = h - \h, \quad \xi = \e (h+\h), \quad c_L = \frac{1}{12}(c-\bar{c}), \quad c_M = \frac{\e}{12}(c+\bar{c}).
\end{align}
It is then straightforward to follow the $\e\to0$ limit to extract $S_{BMS-Cardy}$ from $S_{inner}$. FSC entropies have also been calculated from different perspectives in \cite{Barnich:2012xq, Bhattacharjee:2023sfd}. 

\subsection{Flatspace Chiral Gravity}
One of the goals of a holographic duality is to find an explicit dual pair between a gravitational theory and a lower dimensional field theory. Following \cite{Bagchi:2012yk}, we now review the first holographic dual pair established for the Carroll holography programme in terms of a gravitational theory in 3D AFS and a dual 2D field theory. 

\medskip

{\ding{112}} \underline{\em{Topologically Massive Gravity in 3D AFS}}

\smallskip

As emphasized earlier, the central charges for the asymptotic BMS$_3$ algebra for Einstein gravity have a vanishing Virasoro central charge $c_L=0$. This is hence not the most generic scenario one could envision for a theory of gravity with asymptotically flat boundary conditions. In order to generate a non-zero $c_L$, one can turn on a gravitational Chern-Simons term. The gravitational action is then given by 
\begin{align}
    I_{TMG} = I_{EH} + \frac{1}{2\mu} I_{GCS} = \frac{1}{16\pi G}\int d^3x \sqrt{-g} \left[ R \, + \frac{1}{2\mu}\e^{\lambda \mu \nu} \Gamma^\rho_{\lambda\sigma} \left(\partial_\mu \Gamma^\sigma_{\rho \nu} + \frac{2}{3}\Gamma^\rho_{\mu \tau} \Gamma^\t_{\nu \rho}\right)\right].
\end{align}
This theory is called Topologically Massive Gravity (TMG). In the above, $I_{EH}$ is the Einstein Hilbert action, $I_{GCS}$ is the action for the gravitational Chern-Simons term. $R$ is the Ricci scalar, $\Gamma$ are Christoffel connections. A canonical analysis of TMG with asymptotically flat boundary conditions \cite{Bagchi:2012yk} yields the BMS$_3$ algebra \eqref{bms3}, now with two non-zero central terms:
\begin{align}
    c_L = \frac{1}{4\mu G}, \quad c_M = \frac{1}{4G}. 
\end{align}
A holographic dual for TMG in 3D AFS would thus be a 2D CCFT, now with both central terms turned on. 

\medskip

{\ding{112}} \underline{\em{Chern-Simons Gravity and the conjecture}}

One can now perform a rather intriguing limit on TMG. In this, we do the following double scaling limit:
\begin{align}
    G\to \infty, \, \mu\to0, \quad \mu G = \frac{1}{8k}.
\end{align}
In this limit, one can neglect the Einstein-Hilbert term in the TMG action and the gravitational action becomes
\begin{align}
    I_{CGS} = \frac{k}{4\pi}\int d^3x \, \, \varepsilon^{\lambda \mu \nu} \,\Gamma^\rho_{\lambda\sigma} \left(\partial_\mu \Gamma^\sigma_{\rho \nu} + \frac{2}{3}\Gamma^\rho_{\mu \tau} \Gamma^\t_{\nu \rho}\right).
\end{align}
Here the subscript CGS stands for Chern-Simons gravity. The central charges of the asymptotic symmetry algebra have now scaled and become
\begin{align}
    c_L \to 2k, \quad c_M \to 0. 
\end{align}
One can calculate the charges of the asymptotic symmetries in this scaling limit and it turns out that all $M_0$ charges vanish due to the scaling. We are thus left with a situation where 
\begin{align}
    c_M = 0, \quad \xi_i = 0.
\end{align}
where $\xi_i$ are the $M_0$ weights of all states in the theory. This particular subsector of the 2D CCFT is known to admit a chiral truncation which can be seen by considering the highest weight representations and the structure of null vectors in this highest weight module \cite{Bagchi:2009pe}. The symmetry algebra that rules this sector is thus
\begin{align}
    [L_n, L_m] = (n-m)L_{n+m} + 2k \, \delta_{n+m,0}(n^3-n). 
\end{align}
This is thus a single copy of the Virasoro algebra with a central charge 
\begin{align}
    c_{vir} = 24k. 
\end{align}
in canonical CFT normalisation. The ``Flatspace Chiral Gravity (F$\chi$G) conjecture'' thus is: 

\smallskip

{\em{The dual to Chern-Simons gravity in 3D asymptotically flat spacetime with appropriate boundary conditions is a single copy of the Virasoro algebra with central charge $c_{vir} = 24k$}.}
\smallskip

This was the first example of an explicit dual pair for holography in AFS. In particular the $k=1$ case was conjectured to be a dual connection between CSG with flat boundary conditions and the Monster CFT with $c=24$ \cite{Bagchi:2012yk}. 

\medskip

One can go further with the F$\chi$G correspondence and look at solutions to CSG. The solution space admits FSC as the generic zero modes and it can be shown that the entropy of these FSCs now in CSG is given by a chiral Cardy formula \cite{Bagchi:2013hja}. The F$\chi$G conjecture has also been extended to a supersymmetric version in \cite{Bagchi:2018ryy}.  

\subsection{Carroll Liouville theory}
\label{ssec:carlioville}
In the previous subsection, we presented the first example of a concrete correspondence between a theory of 3D gravity with asymptotically flat boundary conditions and a putative dual 2D field theory. Although the 2D field theory arises as a sector of a 2D Carrollian CFT, it is a bit of a stretch to call it an example of a 2D Carroll CFT that is holographic. In this subsection, we present the first example of a 2D Carroll CFT which can be regarded as holographic. Following \cite{Barnich:2012rz}, we will provide a Carroll version of Liouville theory. In  direct analogy with our discussions in the Carroll field theory section (Sec.~\ref{sec:carfieldtheories}), we will find that there are two versions of Carroll Liouville we can write down, which in modern parlance, are an electric theory and a magnetic Carroll theory. We elucidate this below. 

\medskip

{\ding{112}} \underline{\em{Standard Liouville theory}}

\medskip

Relativistic Liouville theory plays an important role in the AdS$_3$/CFT$_2$ correspondence, where the classical theory can be thought of as describing the boundary dynamics of asymptotically AdS$_3$ spacetimes. The theory is defined by the action:
\begin{eqnarray}\label{Liouville}
    \mathcal{S}_{L}[\varphi] = \frac{1}{4\pi}\int\,d^2x \sqrt{-\hat{g}}\left(\frac{1}{2}\hat{g}^{ab}\d_a\varphi\d_b\varphi + \frac{1}{\gamma}\varphi R(\hat{g}) + \frac{\mu}{2\gamma^2}e^{\gamma\varphi}\right).
\end{eqnarray}
Here $\varphi$ is the Liouville field which is also related to the conformal factor of the metric - $g_{ab} = e^{\gamma \varphi}\hat{g}_{ab}$, $R$ is the Ricci scalar of the background metric, $\gamma$ is the coupling constant and $\mu$ is the cosmological constant. When moving to a Hamiltonian form on the Minkowski cylinder with coordinates $(u,\phi)$, the Liouville action becomes, 
\begin{equation}\label{LiouvilleHamil}
    \mathcal{S}_H[\varphi, \pi] = \int du\, d\phi \left[ \pi \dot{\varphi} - \frac{1}{2} \pi^2 - \frac{1}{2l^2} (\varphi')^2 - \frac{\mu}{2\gamma^2} e^{\gamma \varphi} \right],
\end{equation}
where $u$ and $\phi$ are the time and angular coordinate, respectively and we have used $\hat{g}_{ab} = \eta_{ab} = \text{diag}(-1,\ell^2)$, $\ell$ being the AdS radius. 
\medskip

The theory \eqref{LiouvilleHamil} is invariant under conformal transformations on the cylinder parametrised by $x^{\pm} = \frac{u}{\ell} \pm \phi$, generated by the vector fields $\xi = f(u,\phi)\d_u + Y(u,\phi)\d_{\phi}$ that preserve the flat boundary structure. The conserved charge associated with such a transformation is given by
\begin{equation}
    \mathcal{Q}_{\xi} = \int d\phi\,\left[f\mathcal{H} + Y\mathcal{P}\right] 
\end{equation}
where $\mathcal{H},\mathcal{P}$ are the Hamiltonian and momentum densities, respectively. The charges obey the Dirac bracket algebra: 
\begin{eqnarray}
    \{ \mathcal{Q}_{\xi_1}, \mathcal{Q}_{\xi_2} \} = \mathcal{Q}_{[\xi_1, \xi_2]_H} + K_{\xi_1, \xi_2}.
\end{eqnarray}
Here $K_{\xi_1, \xi_2}$ is the central extension term. The conformal Killing vectors in terms of the Fourier modes, on the cylinder are written as, 
\begin{subequations}
\begin{align}
    m_n &= e^{i n \phi} \frac{1}{2l} \left[ \left( e^{i n \frac{u}{l}} + e^{-i n \frac{ u}{l}} \right) l \partial_u + \left( e^{i n \frac{u}{l}} - e^{-in\frac{ u}{l}} \right) \partial_\phi \right], \\
l_n &= e^{i n \phi} \frac{1}{2} \left[ \left( e^{in\frac{u}{l}} - e^{-in\frac{u}{l}} \right) l \partial_u + \left( e^{in\frac{u}{l}} + e^{-in\frac{ u}{l}} \right) \partial_\phi \right].
\end{align}    
\end{subequations}
Now, if one denotes the corresponding charges by $M_m = \mathcal{Q}_{p_m}$ and $L_m = \mathcal{Q}_{j_m}$ 
the resulting algebra becomes, 
\begin{align}\label{PJalge}
    i \{ M_m, M_n \} &= \frac{1}{l^2}(m - n) L_{m+n}, \nonumber\\
i \{ L_m, L_n \} &= (m - n) L_{m+n}, \\
i \{ L_m, M_n \} &= (m - n) M_{m+n} + \frac{8\pi}{\gamma^2 l^2} m(m^2-1) \delta_{m+n, 0}\nonumber.
\end{align}
If one change the basis to $\mathcal{L}_m^+ = \frac{1}{2}\left(\ell M_m + L_m\right)$ and $\mathcal{L}_m^- = \frac{1}{2}\left(\ell M_{-m} - L_{-m}\right)$, the \eqref{PJalge} is precisely the two copies of Virasoro algebra 
\begin{eqnarray}
    i \{ \L^{\pm}_m, \L^{\pm}_n \} = (m - n) \L^{\pm}_{m+n} + \frac{c^{\pm}}{12} m(m^2-1)  \delta_{m+n, 0}\,,\quad\,i \{ \L^{\pm}_m, \L^{\mp}_n \} = 0
\end{eqnarray}
with $c^{\pm} = \frac{48\pi}{\g^2\ell} = \frac{3\ell}{2G}$. This is precisely the Brown-Henneaux central charge \cite{Brown:1986nw}, where $G$ is the Newton's constant given by $G = \frac{\g^2\ell^2}{32\pi}$.

\medskip

{\ding{112}} \underline{\em{Carroll Liouville theory: Electric version}}

\medskip

In the flat limit, the relativistic conformal symmetry changes to the Carrollian version or equivalently to the BMS$_3$ algebra, generated by the vector field $\xi = f\d_u + Y\d_{\phi}$ where  
\begin{eqnarray}
    Y = Y(\phi), \quad f = T(\phi) + u Y'(\phi).
\end{eqnarray}
The algebra of Noether charges close to form the 2D CCA or equivalently the BMS$_3$ algebra with with $c_L = 0$ and $c_M = 3/G$ for Einstein gravity. 

\medskip 

As mentioned above, there exist two distinct flat limits of the above action for relativistic Liouville theory which give rise to two different 2D CCFTs. In the first case, one naively takes the limit $\ell \to \infty$ with $\gamma,\mu$ fixed. The action \eqref{LiouvilleHamil} reduces to 
\begin{eqnarray}\label{elh}
    \tilde{\mathcal{S}}_H[\varphi, \pi] = \int du\, d\phi \left[ \pi \dot{\varphi} - \frac{1}{2} \pi^2 - \frac{\mu}{2\gamma^2} e^{\gamma \varphi} \right].
\end{eqnarray}
The spatial derivative term in the action vanishes. The resulting theory is ultralocal in space and lacks a central extension in its symmetry algebra. This is the electric Carroll theory and this can readily be checked by moving back to the Lagrangian description. 

\medskip

Carrying on with the Hamiltonian formulation we find the expression for Hamiltonian and momentum densities are given by,
\begin{eqnarray}
    \mathcal{H} = \frac{1}{2} \pi^2 + \frac{\mu}{2\gamma^2} e^{\gamma \varphi},\quad \mathcal{P} = \pi \varphi' - \frac{2}{\g}\pi'.
\end{eqnarray}
One can compute the Noether charges and check their algebras. Here the central extension $K_{\xi_1, \xi_2}$ vanishes. In terms of modes, the charges are expressed as,
\begin{eqnarray}\label{modes}
    M_m = \mathcal{Q}_{e^{i m \phi}  \partial_u}\,, \quad L_m = \mathcal{Q}_{e^{i m \phi} \left( i m u\, \partial_u + \partial_\phi \right)}
\end{eqnarray}
For this case, one finds the central charges $c_L = c_M = 0$. What can be concluded from this is that although the theory defined by \eqref{elh} is invariant under 2D Conformal Carroll transformations, the algebra of its associated Noether charges lacks a central extension. As a result, this theory does not correspond to Einstein gravity in 3D AFS, whose symmetry algebra, as we have mentioned several times before has a non-zero central extension $c_M=1/4G$.

\medskip

{\ding{112}} \underline{\em{Carroll Liouville theory: Magnetic version}}

\medskip

In the second case, before taking the flat limit, the fields and parameters are rescaled: 
\begin{align}
  \varphi = \ell \Phi, \quad \pi = \Pi/\ell, \quad \beta = \gamma \ell, \quad \nu = \mu\ell^2.
\end{align}
Then, taking $\ell \to \infty$ yields the Magnetic version of the Carrollian field theory whose action is given by: 
\begin{eqnarray}\label{magh}
    \tilde{\mathcal{S}}_H[\Phi, \Pi] = \int du\, d\phi \left[ \Pi \dot{\Phi} - \frac{1}{2} (\Phi')^2 - \frac{\nu}{2\beta^2} e^{\beta \Phi} \right].
\end{eqnarray}
Now, expression for Hamiltonian and momentum densities are given by
\begin{eqnarray}
    \mathcal{H} = \frac{1}{2} (\Phi')^2 + \frac{\nu}{2\beta^2} e^{\beta \Phi} - \frac{2}{\beta} \Phi'', \quad 
\mathcal{P} = \Phi' \Pi - \frac{2}{\beta} \Pi'
\end{eqnarray}
Similarly here also the Noether charges can be calculated. However, in this case the central extension term $K_{\xi_1, \xi_2}$ doesn't vanish. In terms of modes \eqref{modes}, one gets the centrally extended CCA$_2$ or BMS$_3$ algebra with $c_L = 0$ and $c_M = 8\pi/\beta^2 = 1/4G$, where $\beta = \sqrt{32 \pi G}$ has been held fixed in the limit. This exactly matches with the gravitational result \cite{Barnich:2006av} and supports the identification of this field theory as a candidate holographic dual for Einstein gravity in 3D AFS.

\medskip

We have thus provided two distinct Carroll Liouville theories, one electric and one magnetic and from this above analysis, it seems that the magnetic one is more relevant from the holographic point of view because of the non-trivial central extensions. It is interesting to ponder here about the meaning of this. We will see in the next section that in 4D AFS, the electric or the leading Carroll branch is the one which is connected to scattering matrices in the bulk and not the magnetic one. Perhaps the lack of propagating degrees of freedom in Einstein gravity 3D AFS forces the putative dual theories to behave differently and make the magnetic branch more important here. 

\subsection{Holographic Entanglement Entropy in AFS$_3$/CCFT$_2$}
\label{ssec:hee}
In this subsection, we focus on entanglement entropy and its calculation in 2D CCFTs. We will then provide a very quick review of bulk computations that reproduce the field theory answers.

\medskip

{\ding{112}} \underline{\em{A quick recap of Entanglement Entropy in 2D CFT}}

\smallskip

Entanglement Entropy (EE) is a fundamental measure of quantum correlations between subsystems. In relativistic 2D CFT, for a single spatial interval of length $\ell$ in the vacuum state of an infinite system, EE is given by the von-Neumann entropy of the reduced density matrix $\rho_A$ of the interval:
\begin{eqnarray}\label{vne}
    S_A = - \mbox{Tr} \big(\rho_A \ln \rho_A\big)\, = \frac{c+\bar{c}}{6}\, \ln{\frac{\ell}{a}}, 
\end{eqnarray}
where $c, \bar{c}$ are the central charges of the CFT and $a$ is a UV cutoff regulating short-distance divergences. This result is derived using the replica trick, which involves computing the Rényi entropy of order $n$, defined as 
\begin{align}
    S_A^{(n)} = \frac{1}{1-n} \ln \mbox{Tr} \left(\rho_A^n\right),
\end{align}
and then analytically continuing $n\to 1$ to recover EE. 

\medskip

More generally, in a 2D quantum field theory defined on a lattice, when the subsystem A consists of $m$ disjoint intervals, the computation of the $n$th Rényi entropy via this replica trick involves considering $n$ non-interacting copies of the original system and analytically continuing the trace. The copies are sewn together creating an $n$-sheeted Riemann surface with branch points at the endpoints of the intervals. The Rényi entropy is then related to the partition function $Z_n$ on this branched cover by 
\begin{eqnarray}
    S_A^{(n)} = \frac{1}{1 - n} \log\left( \frac{Z_n}{Z_1^n} \right) = -\frac{\d}{\d n} \frac{Z_n(A)}{Z_1(A)^n}\,.
\end{eqnarray}

EE for a single spatial interval $ A = [u, v] $ of length $\ell = |u - v|$ in a 2D relativistic CFT is computed in the replica method as a path integral on an $n$-sheeted Riemann surface $\mathcal{R}_n$, constructed by gluing $n$ copies of the original complex plane cyclically along the interval A. This branched surface $ \mathcal{R}_n$ is conformally mapped to the complex plane $\mathbb{C}$, where computations become tractable.
This map is given by 
\begin{align}
    z= \zeta^{1/n} = \left(\frac{\omega-u}{\omega-v}\right)^{1/n}, 
\end{align}
where $\omega$ are coordinates on $ \mathcal{R}_n$. The conformal stress tensor transforms by a Schwarzian derivative: 
\begin{eqnarray}\label{Ttrans}
T(w) = \left( \frac{dz}{dw} \right)^2 T(z) + \frac{c}{12} \left[\frac{z'''(w)}{z'(w)} - \frac{3}{2} \left( \frac{z''(w)}{z'(w)} \right)^2 \right].
\end{eqnarray}
Now, if we take the expectation on both sides of \eqref{Ttrans}, and use $\langle T(z)\rangle_{\mathbb{C}} = 0$, we get
\begin{eqnarray}
    \langle T(w) \rangle_{\mathcal{R}_n} = \frac{c}{24} \,\Big(1 - \frac{1}{n^2}\Big) \frac{(v-u)^2}{(w-u)^2 (w - v)^2}\,.
\end{eqnarray}

From conformal Ward identities, we can show that 
\begin{eqnarray}\label{Tcw}
\langle T(w) \rangle_{\mathcal{R}_n} = \frac{\langle T(w) \tilde{\Phi}_n (u) \tilde{\Phi}_{-n} (v) \rangle_{\mathbb{C}}}{\langle\tilde{\Phi}_n (u) \tilde{\Phi}_{-n} (v) \rangle_{\mathbb{C}}}
\end{eqnarray}
where $\tilde{\Phi}_n, \tilde{\Phi}_{-n}$ are twist fields in the 2D CFT, i.e., primaries with conformal weights 
\begin{align}
  \Delta =  \frac{c}{24}\left(1 - \frac{1}{n^2}\right), \quad 
  \bar{\Delta} = \frac{\bar{c}}{24}\left(1 - \frac{1}{n^2}\right). 
  \end{align}
 It can be argued from the above that the renormalized partition function $\mbox{Tr} \rho^n_A$ behaves like the $n$-th power of the 2-point function of twist fields under a generic conformal transformation. Hence 
\begin{align}
    \mbox{Tr} \rho^n_A \propto \left(\frac{v-u}{a}\right)^{-c (n-1/n)/6},
\end{align}
and the Rényi entropies are given by
\begin{eqnarray}
    S_A^{(n)} = \frac{c}{6} \Big( 1+ \frac{1}{n} \Big) \ln \frac{\ell}{a} \,.
\end{eqnarray}
In the limit $n\to 1$, we recover \eqref{vne} (by adding an identical piece with $c$ replaced by $\bar{c}$ for the antiholomorphic piece). 

\medskip

{\ding{112}} \underline{\em{Entanglement Entropy in 2D CCFT}}

\smallskip

We now wish to build toward an understanding of EE in CCFTs. Here we will be using the plane representation of conformal Carroll algebra given by \eqref{CCA-pl}. As we saw earlier in Sec.~{\ref{ccft2-T}}, the stress tensor on the plane is given by
\begin{eqnarray}
    T_1(t,x) = \sum_{n} \Big\{L_n  - (n+2)\, \frac{t}{x}\, M_n \Big\}\, x^{-n-2}\,,\quad T_2 (x) = \sum_{n} M_n x^{-n-2} \,.
\end{eqnarray}
This is related to the cylinder representation by the plane to cylinder map \eqref{p2c} and the corresponding Carroll Schwarzian derivative \eqref{Tcar-sch}. 

\medskip

We already know the two-point function of two primary operators satisfy \eqref{bms32pt}. Using this and Carroll conformal Ward identities, the three-point function of the stress tensor with two primary fields of weight $(\Delta,\xi)$ gives: 
\begin{eqnarray}\label{T3pt}
    &&\langle T_1(x_w,t_w) \mathcal{O}(x_1, t_1) \mathcal{O}(x_2, t_2)\rangle = \Big(\frac{x_{12}}{x_{w1} x_{w2}}\Big)^2 
\Big[ \Delta - 2 \xi \Big(\frac{t_{12}}{x_{12}} - \frac{t_{w1}}{x_{w1}} - \frac{t_{w2}}{x_{w2}} \Big) \Big]\,  x_{12}^{-2\Delta} \, e^ {2\xi \frac{t_{12}}{x_{12}}}, \nonumber\\
&&\langle T_{(2)}(x_w,t_w) \mathcal{O}(x_1, t_1) \mathcal{O}(x_2, t_2)\rangle = -\Big(\frac{x_{12}}{x_{w1} x_{w2}}\Big)^2  \xi x_{12}^{-2\Delta}\, e^ {2\xi \frac{t_{12}}{x_{12}}}. 
\end{eqnarray}
 We proceed by first mapping the $n$-sheeted Riemann surface to the Carroll plane and then using Carroll Schwarzian derivatives to finally arrive at
\begin{eqnarray}
    \langle T_{1}(x_w, t_w)\rangle _{\mathcal{R}_n} &=& \Big(1-\frac{1}{n^2}\Big) \Big(\frac{x_{12}}{x_{w1} x_{w2}}\Big)^2 \Big[ \frac{c_L}{24} - \frac{c_M}{12} \Big(\frac{t_{12}}{x_{12}}-\frac{t_{w1}}{x_{w1}}-\frac{t_{w2}}{x_{w2}}\Big)\Big] \nonumber \\
\langle T_{2}(x_w, t_w)\rangle _{\mathcal{R}_n}&=& \Big(1-\frac{1}{n^2}\Big)\Big(\frac{x_{12}}{x_{w1} x_{w2}}\Big)^2 \, \frac{c_M}{24}\,.
\end{eqnarray}
Here the interval is at $[(x_1, t_1), (x_2, t_2)]$. The Carroll analogue of \eqref{Tcw} then becomes
\begin{eqnarray}
    \langle T_{i}(x, t) \rangle_{\mathcal{R}_n} = \frac{\langle T_{i}(x,y) \Phi_n (x_1, t_1) \Phi_{-n} (x_2, t_2) \rangle_{\mathbb{C}}}{\langle\Phi_n (x_1, t_1) \Phi_{-n} (x_2, t_2) \rangle_{\mathbb{C}}}, 
\end{eqnarray}
where $\Phi_n$'s are the 2D CCFT twist fields with weights 
\begin{align}
    \Delta_\Phi = \frac{c_L}{24} \left( 1- \frac{1}{n^2}\right),\quad \xi_\Phi = \frac{c_M}{24} \left( 1- \frac{1}{n^2}\right).
\end{align}
Following an analysis that mirrors the relativistic 2D CFT one, it can be shown that
\begin{eqnarray}
    \mbox{Tr} \, \rho_A^n = k_n \langle \Phi _{\Delta , \xi}(x_1, t_1) \Phi _{\Delta , \xi}(x_2, t_2)\rangle ^n _{\mathbb{C}}
= k_n \, x_{12}^{-\frac{c_L}{12}(n-\frac{1}{n})} \exp \Big[\frac{c_M}{12} (n-\frac{1}{n}) \frac{t_{12}}{x_{12}} \Big]
\end{eqnarray}
for some arbitrary constants $k_n$. From this, using $S_{A} = -\lim_{ n \rightarrow 1} \frac{\partial}{\partial n} \mbox{Tr} \, \rho_A^n $, we can get our final expression for EE in 2D CCFT as
\begin{eqnarray}\label{eq:ccft2vacee}
S^{\mbox{\tiny{CCFT}}_2}_{\mbox{\tiny{EE}}}  = \frac{c_L}{6}\, \ln \frac{\ell_x}{a} + \frac{c_M}{6} \,\frac{\ell_t}{\ell_x},
\end{eqnarray}
where $\ell_x=a x_{12}$, $\ell_t=a t_{12}$ and $a$ is earlier UV cutoff. 

\medskip

\newpage

{\ding{112}} \underline{\em{Entanglement Entropy in 3D AFS}}

\smallskip

Having reviewed the computation of \cite{Bagchi:2014iea} which used twist operators to compute the EE in CCFT$_2$, we now move towards a bulk understanding. The initial holographic interpretation of the same was in terms of bulk Chern-Simons formulation. To be precise, \cite{Basu:2015evh} used the Wilson line techniques of \cite{Ammon:2013hba,deBoer:2013vca,Castro:2014tta} to match the CCFT$_2$ result of \eqref{eq:ccft2vacee}. 

\medskip

The holographic interpretation in terms of the metric formalism analogous to the RT (HRT) proposal \cite{Ryu:2006bv,Ryu:2006ef,Hubeny:2007xt} was worked in \cite{Jiang:2017ecm,Apolo:2020bld}. For spherical entangling surfaces, the idea of \cite{Jiang:2017ecm} was to generalize the Rindler method of \cite{Casini:2011kv} for CCFTs. In the traditional Rindler method, one can use conformal transformations to map the vacuum entanglement entropy to a thermodynamic entropy on a Rindler spacetime. These transformations map the vacuum AdS to a black hole with a horizon. Thus, the AdS/CFT dictionary enables us to use the Bekenstein-Hawking entropy to compute the thermal entropy on the black hole spacetime. 

\medskip

\cite{Apolo:2020bld} generalized the methods of \cite{Jiang:2017ecm} for generic entangling surfaces consistent with \cite{Lewkowycz:2013nqa,Dong:2016hjy}. The proposal considers a generic holographic dual between a $D-1$ dimensional quantum field theory in the boundary and a $D$ dimensional gravitational theory which reduces to Einstein gravity in the semiclassical limit. The EE of a codimension one subregion $\mathcal{C}$ in the boundary field theory is then holographically computed from an RT like formula:
\begin{equation}\label{eq:swingee}
    S_{\mathcal{C}} = \text{min} \, \text{ext}_{X_{\mathcal{C}} \, \sim \, \mathcal{C}} \,\dfrac{\text{Area}(X_{\mathcal{C}})}{4G} \, , ~~~~~~ X_{\mathcal{C}} = X \, \cup_{p \in \partial \mathcal{C}} \gamma_{(p)} \, ,
\end{equation}
where $X_{\mathcal{C}}$ is a surface homologous to $\mathcal{C}$ and it contains a space-like surface $X$ along with a set of null geodesics $\gamma_{(p)}$. Let us now unpack \eqref{eq:swingee}. The null geodesics arise from the points $p$ located at the boundary of the entangling region $\partial \mathcal{C}$ and end at the spacelike surface. The configuration of $\gamma_{\mathcal{C}} = \gamma \, \cup_{p \in \partial \mathcal{C}} \gamma_{(p)}$ is called a ``swing surface'' because $\gamma$ is a spacelike ``bench'' connecting the null ``ropes'' $\gamma_{(p)}$. Thus, formula \eqref{eq:swingee} instructs us to first extremize the area of every possible configuration $X$ and then choose the minimal area. In vanilla AdS/CFT, \eqref{eq:swingee} reduces to the RT/HRT prescription because the null geodesics $\gamma_{(p)}$ shrink to the boundary and the swing surface becomes the usual RT/HRT surface.

\medskip

We now test the formula \eqref{eq:swingee}. Suppose if one considers the Poincar\'e vacuum in the bulk with metric
\begin{equation}\label{eq:poincarevacmetric}
    ds^2 = - 2 du \, dr + r^2 dz^2 \, ,
\end{equation}
which describes a dual CCFT on the plane as $z \in (-\infty,\infty)$. The central terms for Einstein's gravity are \eqref{bmsc}. Thus, the EE \eqref{eq:ccft2vacee} is proportional to $\frac{\ell_t}{\ell_x}$. The formula \eqref{eq:swingee} matches the field theoretic EE \eqref{eq:ccft2vacee}. For the Flat space Cosmologies described by \eqref{eq:fscmetric}, one can parametrize the interval $\mathcal{C}$ along $u$ and $\psi$ as $\ell_u$ and $\ell_\psi$ respectively. Thus, the holographic EE given by swing surface formula \eqref{eq:swingee} gives
\begin{equation}
    S_{\mathcal{C}} = \dfrac{1}{4G} \Bigg| 2\sqrt{2M} \left(\ell_u + \dfrac{J \ell_\psi}{2M} \right) \coth \left( \sqrt{2M} \ell_\psi \right) - \dfrac{J}{M} \bigg| \, .
\end{equation}
This agrees with the EE obtained from generalized Rindler methods \cite{Jiang:2017ecm}.

\subsection{Stress tensor correlation functions}
In this subsection, we outline a procedure to reproduce correlation functions of stress tensors in a 2D CCFT as discussed in Sec. \ref{ccft2-T} by a suitable calculation in Einstein gravity in 3D AFS. We will use the Chern-Simons formulation of 3D gravity below following the seminal works \cite{Achucarro:1986uwr, Witten:1988hc}.

\medskip

{\ding{112}} \underline{\em{Chern-Simons formulation of 3D AFS}}

\smallskip

The Einstein Hilbert action for gravity in 3D AFS can be written as a Chern-Simons action:
\begin{align}
    I_{EH} = I_{CS}[A] = \frac{k}{4\pi} \int \, \text{Tr} \left( A \wedge dA + \frac{2}{3}A\wedge A \wedge A \right), 
\end{align}
where $A$ is a $isl(2)$ valued connection
\begin{align}
    A= A^{\, n}_L L_n + A^{\, n}_M M_n, \quad n=0, \pm1.
\end{align}
and $L_{0,\pm1}, M_{0,\pm1}$ generate the global 2D CCA, as in our usual notation. The trace in the Chern-Simons action is defined with the non-degenerate bilinear form
\begin{align}
    \text{Tr} (L_n M_m) = \gamma_{mn}, \quad \text{with} \quad \gamma = \text{antidiag}(-2, 1, -2)
\end{align}
and the CS level is related to the Newton's constant as $k = \frac{1}{4G}$. Also below we call $L_{-1} = L_-, \, L_{+1} = L_+$ etc. We will now make our computations considerably simpler by choosing a gauge where
\begin{align}
    A = b^{-1} (d+a) b, \quad\text{where} \quad {b=e^{rM_-/2}}, \quad a = a_u(u, \varphi) du + a_\varphi(u, \varphi) d\varphi. 
\end{align}
The boundary conditions \cite{Barnich:2006av} in the CS language correspond to 
\begin{align}
    a_u = M_+ - \frac{\mathcal{M}}{4} M_, \quad a_\varphi = L_+ - \frac{\mathcal{M}}{4} L_- - \frac{\mathcal{N}}{2} N_-.
\end{align}
It is straightforward to check that these lead to a metric 
\begin{align}
    ds^2 = g_{\mu\nu} dx^\mu dx^\nu = \eta_{mn} A^{\, m}_M A^{\, n}_M= \mathcal{M} du^2 - 2 dudr -2{\mathcal{N}} dud\varphi + r^2 d\varphi^2. 
\end{align}
This is exactly the form we found earlier \eqref{eq:gen3dmetric} for generic zero mode solutions in 3d AFS following \cite{Barnich:2006av}. 
The Chern-Simons connection $A$ solves the equations of motion
\begin{align}
    F= dA + [A,A] =0,
\end{align}
provided we have 
\begin{align}
    \partial_u \mathcal{M} =0, \quad \partial_u \mathcal{N} = - \frac{1}{2} \partial_\varphi\mathcal{M}.  
\end{align}
Using the expressions of the Carroll stress tensors \eqref{Carr-T}, we recognise the above equations to be Carroll conservation laws, with the identification of the functions $\mathcal{M, N}$ with the Carroll stress-tensor components as follows:
\begin{align}
    \mathcal{M} = T_2, \quad \mathcal{N} =T_1. 
\end{align}

\medskip

{\ding{112}} \underline{\em{Calculating correlations in CS formulation}}

\smallskip

We will now figure out how to compute correlation functions in the Chern-Simons formulation. To this end, consider the deformation of a free CFT by a source term for the stress tensor:
\begin{align}
    S_\mu = S_0 + \int d^2z \, \mu(z, \z) T(z)
\end{align}
and use a localize the source at some point $(z_2, \z_2)$: $\mu(z, \z) = \e \de^2(z-z_2, \z-\z_2)$. The one point function of the deformed theory thus gives the two-point function of the undeformed theory:
\begin{align}
    \< T(z_1) \>_\mu = \<T(z_1)\>_0 + \e \<T(z_1) T(z_2)\>_0 + \mathscr{O}(e^2). 
\end{align}
For a 2D CCFT, one needs to do this for the stress tensors $T_1(u, \varphi), T_2(\varphi)$. In the dual 3D AFS, in the Chern-Simons language, turning on sources is the same as turning on chemical potentials. In this case, $a_u$ changes but $a_\varphi$ does not \cite{Gary:2014ppa}:  
\begin{align}
    & a_u \to a_u -\mu_M M_- - \mu_L L_- + a_0 M_0 + a_1 L_0 + a_2 M_+ + a_3 L_+, \quad a_\varphi \to a_\varphi.
\end{align}
Equations of motion now determine the state dependent functions as
\begin{subequations}\label{MNeom}
  \begin{align}
&    \partial_u \mathcal{M} = k \partial_\varphi^3 \mu_L + \mu_L \partial_\varphi \mathcal{M}  + 2 \mathcal{M} \partial_\varphi \mu_L \\
& \partial_u \mathcal{N} = k \partial_\varphi^3 \mu_M + \mu_L \partial_\varphi \mathcal{N}  + 2 \mathcal{M} \partial_\varphi \mu_M + (1- \mu_M) \partial_\varphi \mathcal{M} + 2 \mathcal{N} \partial_\varphi \mu_L
\end{align}  
\end{subequations}
These equations are analogous to Ward identities of the dual 2D CCFT. 

\medskip

To show an explicit matching to the CCFT two-point functions we derived earlier in \eqref{CTT}, we choose the global Minkowski background such that $\mathcal{M}^{(0)} = k/2, \, \mathcal{N}^{(0)} =0$. We localize sources at $(u_2, \varphi_2)$ by
\begin{align}\label{source}
      \mu_M(u, \varphi) = \e_M \de^2(u-u_2, \varphi- \varphi_2),  \quad \mu_L(u, \varphi) = \e_L \de^2(u-u_2, \varphi- \varphi_2). 
\end{align}
Plugging back into the equations of motion \eqref{MNeom} with 
\begin{align}
  \mathcal{M} =   \frac{k}{2} + \mathcal{M}^{(1)}, \quad \mathcal{N} = \mathcal{N}^{(1)}, 
\end{align}
we get 
\begin{subequations}
  \begin{align}
&    \partial_u \mathcal{M}^{(1)} = - k \e_L ( \partial_\varphi^3 + \partial_\varphi) \de^2(u-u_2, \varphi- \varphi_2) \\
& \partial_u \mathcal{N}^{(1)} = - k \e_M ( \partial_\varphi^3 + \partial_\varphi) \de^2(u-u_2, \varphi- \varphi_2) + \partial_\varphi \mathcal{M}^{(1)}
\end{align}  
\end{subequations}
To solve these equations, we use the Green's function
\begin{align}
   \partial_u  \partial_\varphi G(u-u_2, \varphi- \varphi_2) = \de^2(u-u_2, \varphi- \varphi_2),
\end{align}
which can be solved by the method of images to give
\begin{align}
   G(u-u_2, \varphi- \varphi_2) = \ln \left\{(u-u_2) \sin \left(\frac{\varphi- \varphi_2}{2}\right) \right\},
\end{align}
So we find
\begin{align}
   \mathcal{M}^{(1)} = \frac{6k \e_L}{32 \sin^4 \left(\frac{\varphi_1- \varphi_2}{2}\right)}, \quad \mathcal{N}^{(1)} = \frac{6k\left[\e_M - 2\e_L(u_1-u_2)\cot \left(\frac{\varphi_1- \varphi_2}{2}\right)\right]}{32 \sin^4 \left(\frac{\varphi_1- \varphi_2}{2}\right)} 
\end{align}
We are now in a position to read of the correlation functions of the Carroll stress tensors from our bulk computation. To do this, we would focus on the coefficients associated with the sources as indicated by \eqref{source}. 

\medskip

The coefficient of $\e_L$ in $\mathcal{M}^{(1)}$ (which is the whole answer) and the coefficient of $\e_M$ in the expression of $\mathcal{N}^{(1)}$ are identical and give the two point correlation function $\<T_1 T_2\>$ in the undeformed 2D CCFT.  This is identical to the answer that was obtained from the field theory side in Sec.~\ref{ccft2-T} and specifically to the second equation of \eqref{CTT}. Similarly, the coefficient of $\e_L$ of $\mathcal{N}^{(1)}$ gives the correlator $\<T_1 T_1\>$.  This is to be checked with the first equation of \eqref{CTT}. The vanishing of the correlator $\<T_2 T_2\>$ is reflected by the fact that there is no $\e_M$ piece in 
$\mathcal{M}^{(1)}$. 

\medskip

The higher point correlation functions can also be obtained similarly. For more details the reader is pointed to \cite{Bagchi:2015wna}. 

\bigskip \bigskip

\subsection{Pointers to literature}\label{pt-3dAFS}
We now provide a list of important works in the context of AFS$_3$/CCFT$_2$. A lot of these are papers written in the earlier decade and there is some newer work, which we try to point out below. 

\begin{itemize}

\item \textit{Representation theory:} Unitary Irreducible Representations of BMS$_3$ were discussed in \cite{Barnich:2014kra, Barnich:2015uva} and in details in the thesis \cite{oblak2017bms}. These UIRs also include superrotations.

\item{\em Variational Principle:} The variational principle and one-point functions for 3D AFS were addressed in \cite{Bagchi:2013lma, Detournay:2014fva}. 

\item{\em BMS Blocks and Flat Holography:} We have discussed Carroll blocks in Sec.4 in some detail. In \cite{Hijano:2017eii, Hijano:2018nhq}, holographic computations of these blocks as well as an extrapolate dictionary for flat holography were proposed and correlation functions were computed. 

\item{\em {Characters and 1-loop partition function:}} We have had an encounter with characters of the 2D CCFTs in Sec.~\ref{part-CCFT}. As mentioned there, the characters for induced representations were computed in \cite{Oblak:2015sea} and the ones for the highest weight representations, which we reviewed was done in \cite{Bagchi:2019unf}. The expression for these characters \cite{Oblak:2015sea} were matched with the computation of a 1-loop partition function for Einstein gravity in 3D AFS in \cite{Barnich:2015mui}.

\item{\em {Higher point functions on the torus:}} We have seen earlier in this section that the BMS-Cardy formula was derived from the modular invariance of the partition function, which can be thought of as a zero-point function on the torus. Similar considerations for the torus one-point functions in a 2D relativistic CFT leads to a formula for asymptotic structure constants \cite{Kraus:2016nwo}. The Carroll equivalent was addressed in \cite{Bagchi:2020rwb}, where a bulk calculation of a one-point function in the FSC background reproduced the field theory results. 
Two-point functions on the torus were addressed in \cite{Bagchi:2023uqm}. 

\item{\em {Anomalies:}} A calculation for Weyl anomaly in Carrollian CFTs was performed in \cite{Bagchi:2021gai}. 

\item{\em {Supersymmetric extensions:}} We will briefly encounter supersymmetric extensions of 2D CCFTs in our penultimate section Sec. \ref{penult}. Here we give a list of papers addressing this in the context of AFS$_3$/CCFT$_2$. 

\begin{itemize}

\item [$\star$] Asymptotic symmetries in 3D flat supergravities was first discussed in \cite{Barnich:2014cwa}.
 
\item [$\star$] In \cite{Barnich:2015sca}, the authors construct a 2D super-BMS$_3$ invariant theory dual to 3D flat $\mathcal{N} = 1$ supergravity. The theory is described by a gauged chiral Wess-Zumino-Witten model.

\item [$\star$] In \cite{Lodato:2016alv}, the above analysis is extended to find realisations of both the homogeneous super-BMS$_3$ algebra (the one constructed in \cite{Barnich:2015sca}) and the inhomogeneous one \cite{Bagchi:2017cte} in terms of 3D flat supergravities. We will discuss the two algebras briefly in Appendix B. 
        
\item [$\star$] Some more explorations of Super BMS algebras can be found in \cite{Banerjee:2015kcx, Banerjee:2016nio, Banerjee:2017gzj}. 

\item[$\star$] Asymptotic structure of 3D $\mathcal{N} = (2,0)$ supergravity have been studied in \cite{Fuentealba:2017fck}, where the asymptotic symmetries are found to be governed by an extended super-BMS$_3$ algebra.  

\end{itemize}

\item{\em{Kac-Moody extensions:}} Non-Lorentzian Kac-Moody algebras were addressed in detail in \cite{Bagchi:2023dzx}. Modular aspects of the $U(1)$ case and the bulk duals were investigated in \cite{Bagchi:2022xug}. 

\item{\em Spectral flows:} Spectral flows in 3D AFS as well as in the dual field theory were considered in \cite{Basu:2017aqn}, where the duals considered had as symmetries additional $U(1)$s or supersymmetry on top of the usual 2D CCA.  

\item{\em Soft hairs on FSC:} Soft hairs on FSC horizons in 3D gravity were discussed in \cite{Afshar:2016kjj}. 

\item{\em Higher spins:} Higher spin algebras and their bulk duals were discussed first in \cite{Afshar:2013vka} followed closely by \cite{Gonzalez:2013oaa}. FSCs with higher spins were investigated in \cite{Gary:2014ppa, Ammon:2017vwt, Matulich:2014hea}. We will briefly encounter the algebra in Sec.14. Higher spin extensions of BMS$_3$, with fermionic contributions have been studied in \cite{Fuentealba:2015jma}.

\item{\em {Limit and Grassmann:}} An interesting implementation of the flat limit from 3D AdS to 3D AFS was found in \cite{Krishnan:2013wta}. Here the AdS radius was treated as a Grassmann variable. Generalisations to higher spin were made simple in this formulation.

\item{\em Holographic renormalisation:} Aspects of holographic renormalization in terms of the associated geometric structures of the underlying Carroll manifold were addressed in \cite{Hartong:2015usd}. 

\item{\em Aspects of entanglement:} We have discussed EE in 2D CCFTs and their holographic realisation following \cite{Bagchi:2014iea, Jiang:2017ecm} earlier in this section.  

\begin{itemize}

\item[$\star$] {\em{More EE:}} There have been other attempts at understanding holographic EE in flatspace including \cite{Hijano:2017eii, Godet:2019wje}. It is also important to mention earlier work \cite{Li:2010dr}. 

\item[$\star$] \textit{Swing surfaces, Modular Hamiltonians}: Following up on  \cite{Jiang:2017ecm}, generalised Swing surfaces were constructed in \cite{Apolo:2020bld}. Modular Hamiltonians in the context of flat holography were introduced in \cite{Apolo:2020qjm}. 

\item[$\star$] \textit{Holographic RG flows} In \cite{Grumiller:2023rzn}, the authors holographically construct $c$-functions for 2D CCFTs. They establish that for every holographic flow in AdS$_3$, there is a corresponding holographic flow in flat space.

\item[$\star$] \textit{Symmetry resolved entanglement}: In \cite{Banerjee:2024ldl} authors calculate EE in CCFTs considering a boosted setup by introducing an Aharonov-Bohm flux on the replica manifold. Starting from such a boosted interval in a 2D CFT, they calculate RE and EE of the ground state, and take the boost parameter to infinity. The Symmetry Resolved Entanglement Entropy (SREE) has also been computed for the Carroll case. Possible connections with flatspace holography computation were explored.
\end{itemize}

\item{\em Chaos and OTOCs:} An important marker of chaos in quantum systems is the calculation of Out-of-time-ordered correlators (OTOC) and Lyapunov exponents. For a relativistic CFT, the calculation of Lyapunov exponents can be done in a holographic setting. While the CFT calculation relies on the block expansion of the OTOC of primary operators \cite{Shenker:2013pqa,kitaev2014hidden}, the bulk gravity calculation can be done using a backreacted `shockwave' geometry where the OTOC can be encoded as a shift on the future (past) horizon \cite{Shenker:2013yza,Maldacena:2015waa}. 

\begin{itemize}
    \item[$\star$] {\em{Chaos in AFS$_3$/CCFT$_2$:}} Analogous calculations to the above in the Carrollian (and Galilean) case were performed in \cite{Bagchi:2021qfe}. The Carroll case here is interesting since the diagnostic here has to be a Out-of-space-ordered correlator, however the gravity calculation goes through seamlessly for a shockwave FSC background. The holographic Lyapunov exponents from both calculations match in this case. 
    
\item[$\star$] {\em{OTOC:}} For the CFT to CCFT flows (see Sec.~\ref{carrdef}), OTOCs can be calculated as well, as in \cite{Banerjee:2022ime}, by going to the oscillator representation of operators. A holographic picture for the same was outlined in the same work. 
\end{itemize}

\item{\em Quasi Normal Modes for FSC:} QNM for FSC solutions have been difficult to compute due to issues with boundary conditions. From the point of view of the boundary theory, the first attempt was made in \cite{Bagchi:2023uqm}. Thereafter, 
\cite{Bagchi:2023ilg} reproduced the answers in \cite{Bagchi:2023uqm} partially. A purely bulk computation has been recently attempted in \cite{Bagchi:2025erx}.

\item{\em {More on Carroll Partition functions:}} 

\begin{itemize}

        \item [$\star$] In \cite{Merbis:2019wgk}, the Chern-Simons formulation was used to do a dimensional reduction of gravity in 3D AFS and 1-loop partition function was computed using this. The paper further computed corrections to entanglement entropy in 2D CCFTs, which indicated that $c_L$ changes due to quantum corrections, but $c_M$ does not.  
        
        \item [$\star$] In \cite{Bagchi:2023ilg}, the Selberg trick was used to compute one-loop partition function and this matched with results in \cite{Barnich:2015mui}. 
    
        \item [$\star$] In \cite{Cotler:2024cia}, the authors extend the infra-red triangle (4D bulk version is reviewed briefly in Appendix \ref{ap:celreview}) to 3D bulk. The boundary Carrollian description is a magnetic theory. However, this is not a dual in the traditional AdS/CFT sense, but rather should be considered as akin to a Schwarzian description of JT gravity \cite{Jensen:2016pah,Maldacena:2016upp}. 
        
     \item [$\star$]  \cite{Poulias:2025eck} revisits older work and comments on possible subtleties analogous to \cite{deBoer:2023fnj}.

\end{itemize}

\item{\em Other recent developments:} More recent works on AFS$_3$/CCFT$_2$ include 
\cite{Adami:2024rkr, Hao:2025btl}.

\end{itemize}

\medskip

\newpage

\section{Dual of 4D AFS: S-Matrices to Carroll Correlators}
\label{sec:4dafs}

We will now focus our attention on the physically relevant of $D=4$ AFS. The fundamental difference between the 3D and the 4D  case is that there would be propagating degrees of freedom in $D=4$ as opposed to Einstein gravity in $D=3$. In AFS, radiation would leak out through the null boundaries and hence the dual field theory should have a way to accommodate this. The observable which will encode this radiation data is the $S$-matrix and hence any holographic dual theory has to reproduce $S$-matrix elements. In this section, we will summarize the main ideas of \cite{Bagchi:2022emh} which argue how the Carrollian structures of the boundary dual could holographically encode the $S$-matrix of the bulk, via the so-called {\em modified Mellin transformation} \cite{Banerjee:2018gce}. We will then briefly recall the results of \cite{Bagchi:2024gnn} which relates a 3D boundary stress tensor to soft theorems in the 4D bulk.

\begin{figure}[h]
	\centering	
	\includegraphics[width=0.5\textwidth]{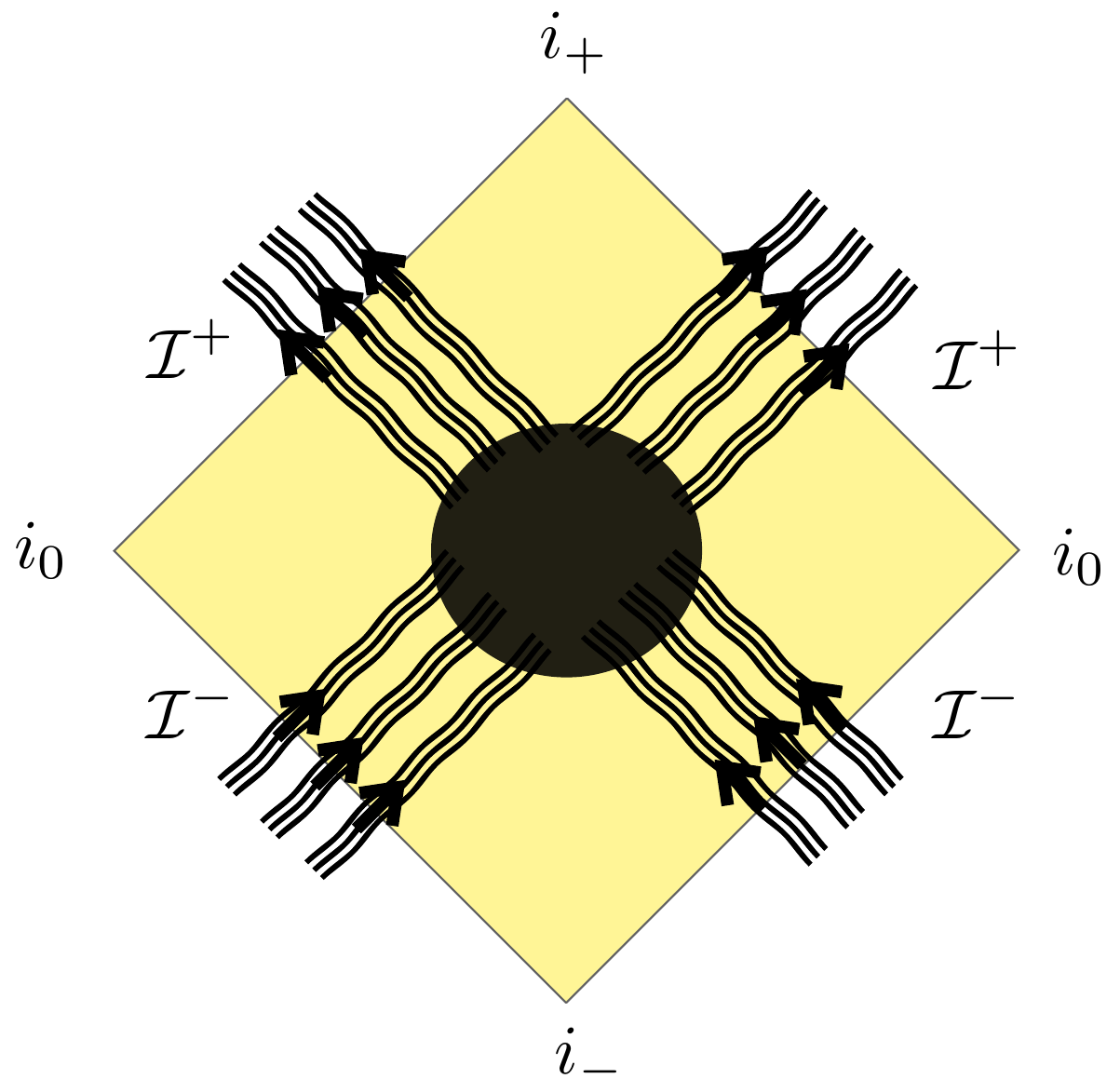}
	\caption{Penrose diagram and scattering in AFS.}
    \label{fig:smatrix}
\end{figure}

\medskip

\subsection{Symmetries}

The asymptotic symmetries, as in the 3D case, will play a pivotal role and will dictate the symmetries of the boundary field theory in the Carrollian formulation. In $D=4$, at $\mathscr{I}^\pm$ which are topologically $\mathbb{R} \times \mathbb{S}^2$ where $\mathbb{R}$ is a null direction, the ASA is given by
\begin{subequations}\label{bms4}
    \begin{align}
        & [L_n, L_m] = (n-m) L_{n+m}\, , \quad ~~ [\bar{L}_n, \bar{L}_m] = (n-m) \bar{L}_{n+m} \, , \\
        & [L_n, M_{r,s}] = \left(\frac{n+1}{2} -r \right) M_{n+r, s} \, , \quad ~~  [\bar{L}_n, M_{r,s}] = \left(\frac{n+1}{2} -s \right) M_{r, n+s} \, . 
    \end{align}
\end{subequations}
In the above, we have only displayed non-zero commutation relations. This algebra is called the BMS$_4$ algebra after the original authors Bondi, van der Burgh, Metzner \cite{Bondi:1962px} and Sachs \cite{Sachs:1962wk}. Here, the $M_{r,s}$ are called supertranslations which are angle dependent translations of the null direction, while $L_n$ are the super-rotations that generalise the usual Lorentz group to the conformal group on the sphere. The usual Lorentz group acts as the global conformal group on the celestial sphere. Barnich and Troessart \cite{Barnich:2010eb} considered an enhancement of this global conformal group to the infinite dimensional (two copies of) Virasoro algebra. This is the version of superrotations we would be interested in for the purposes of this review. One can also have other extensions, which e.g. consider enhancements to the full diffeomorphism of the sphere \cite{Campiglia:2014yka}. 

\medskip

The algebra \eqref{bms4}, in the Carrollian perspective, forms the global symmetries of the dual theory. This is of course the same as the 3D Carroll Conformal Algebra \cite{Duval:2014uva}, which was dealt with in some details in Sec.~\ref{sec:3dafs}. We will compare our holographic answers with the ones obtained in that section throughout our present section. 

\medskip

There exists a different and, at least at the time of writing, more popular approach to flatspace holography called Celestial holography. Here the dual theory is conjectured to be a co-dimension two relativistic CFT that lives entirely on the celestial sphere. The power of 2D CFTs is a very useful tool to have in this case, although the 2D Celestial CFT has many features that make it very different from usual relativistic 2D CFTs. This approach has led to the discovery of many new results in the context of asymptotic symmetries and scattering amplitudes and continues to do so. We present a short summary of Celestial holography in Appendix \ref{ap:celreview}. Since our review focuses on Carrollian symmetries, our discussion of 4D  flatspace holography would also mainly tell the Carrollian story.

\bigskip

\subsection{A quick recap of Celestial elements}
Our focus in this part of the review would be the formulation initiated by \cite{Bagchi:2022emh} which related, for the first time, scattering matrix elements in the 4D  bulk AFS and correlation functions of a 3D Carroll CFT. The formulation of \cite{Bagchi:2022emh} could be thought of as a bridge between the Celestial and Carrollian approaches since \cite{Bagchi:2022emh} generalizes the pre-existing structures of the Celestial CFT. So, to begin with, we will summarize the salient features of Celestial CFTs. A more detailed account is presented in Appendix \ref{ap:celreview}, which also contains some pointers to the literature. 

\medskip

Let us consider single particle states (with momentum $p^{\mu}$ and helicity $\sigma$) which are constructed from Fock space creation and annihilation operators $a(p^{\mu},\sigma)/a^{\dagger}(p^{\mu},\sigma)$:
\begin{equation}
    |p^\mu,\sigma\rangle= a^\dagger(p^\mu,\sigma) |0 \rangle .
\end{equation}
Any massless particle can be parametrized in terms of it's energy $\omega$ and the local coordinates of the Celestial sphere it hits $(z,\z)$:
\begin{equation}\label{eq:3dpara}
    p^{\mu}=\omega \Big( 1+z\bar{z},z+\bar{z},i(z-\bar{z}),1-z\bar{z}\Big).
\end{equation}
Thus, given a bulk operator $a(p^{\mu},\sigma) = a(\epsilon \omega,z,\z,\sigma)$, one can define a local operator on the Celstial sphere through the Mellin transform \cite{Pasterski:2016qvg,Pasterski:2017kqt}
\begin{equation}\label{eq:mellindef}
    \mathcal{O}_{h,\bar{h}}(z,\bar{z})=\int_{0}^{\infty}d\omega \, \omega^{\Delta-1}  \,a(\epsilon \omega,z,\bar{z},\sigma).
\end{equation}
One can show that the action of the bulk Lorentz group on the fock space creation/annihilation operator manifests itself as a global conformal transformation (SL$(2,\mathbb{C})$) on the Celestial sphere \cite{Pasterski:2017kqt}. Thus, a consequence of this is that $\mathcal{O}_{h,\bar{h}}$ is a conformal primary with weights determined from
\begin{equation}\label{eq:mellinweights}
    h + \bar{h} = \Delta , ~~~~~ h-\bar{h} = \sigma.
\end{equation}
Since the $S$-matrix is made up of asymptotic in/out creation and annihilation operators, one can use \eqref{eq:mellindef} to propose a holographic correspondence between the bulk 4D  $S$-matrix and the correlation functions of the 2D Celestial CFT \cite{Pasterski:2017kqt}
\begin{equation}\label{eq:mellinsmatrix}
    \<\mathcal{O}_{1}(z_1,\bar{z}_1)\mathcal{O}_2(z_2,\bar{z}_2)...\mathcal{O}_n(z_n,\bar{z}_n)\>\sim \int_{0}^{\infty} \left( \prod_{i=1}^n d \omega_i \, \omega_{i}^{\Delta_i-1} \right)\mathcal{S}_n(\omega_i,z_i,\bar{z}_i,\sigma_i) .
\end{equation}

\medskip

We have remarked above \eqref{eq:mellinweights} that $\mathcal{O}_{h,\bar{h}}$ of \eqref{eq:mellindef} transforms as a SL$(2,\mathbb{C})$ primary under bulk Lorentz transformations. However, the action of bulk translations $p^{\mu} \to p^{\mu} + l^{\mu}$ is non-covariant. For instance, if you consider an infinitesimal time translation generated by a Hamiltonian $H$, then
\begin{align}
    \delta_H \mathcal{O}_{h,\bar{h}}(z,\bar{z})&=\int_0^{\infty} d\omega \, \omega^{\Delta-1} \, [H,a(\epsilon\omega,z,\bar{z},\sigma)]= \int_0^{\infty} d\omega \, \omega^{\Delta-1} \, [-\epsilon\omega(1+z\bar{z}) a(\epsilon\omega,z,\bar{z},\sigma)] \nonumber\\
	&=-\epsilon (1+z\bar{z}) \, \mathcal{O}_{h+\frac{1}{2},\bar{h}+\frac{1}{2}}(z,\bar{z}) .
\end{align}
Notice the unusual feature of how the weights \eqref{eq:mellinweights} shift under the action of a simple bulk time translation. This is expected from the transformation \eqref{eq:mellindef} which essentially diagonalizes the action of boosts and boost eigenstates are not eigenstates under translations in the 2D picture. The other translations act in a similar manner
\begin{equation}
    \begin{split}
        [P^1,\mathcal{O}_{h,\bar{h}}(z,\bar{z})]&=-\epsilon (z+\bar{z}) \,\mathcal{O}_{h+\frac{1}{2},\bar{h}+\frac{1}{2}}(z,\bar{z}) ,  \\
	[P^2,\mathcal{O}_{h,\bar{h}}(z,\bar{z})]&=i\epsilon(z-\bar{z})\, \mathcal{O}_{h+\frac{1}{2},\bar{h}+\frac{1}{2}}(z,\bar{z}) ,  \\
	[P^3,\mathcal{O}_{h,\bar{h}}(z,\bar{z})]&=-\epsilon(1-z\bar{z})\, \mathcal{O}_{h+\frac{1}{2},\bar{h}+\frac{1}{2}}(z,\bar{z}) .
    \end{split}
\end{equation}
This feature is undesirable from the point of view of holography because the bulk isometry group is the full Poincar\'e group (and not just the Lorentz group). For instance, the $S$-matrix on the RHS of \eqref{eq:mellinsmatrix} is covariant under translations but the correlation function in the LHS of \eqref{eq:mellinsmatrix} isn't. We will see below how there is a natural way to incorporate the covariance under bulk translations. The idea is to evolve from a time-independent 2D picture to a time-dependent 3D picture.

\subsection{Modified Mellin transformation and scattering amplitudes}

A natural way to build a 3D picture from \eqref{eq:mellindef} is consider a null time evolution through the Hamiltonian $H$:
\begin{equation}\label{eq:modmellindef}
    \begin{split}
        \Phi^{\epsilon}_{h,\bar{h}}(u,z,\bar{z})&=e^{-iHU}\mathcal{O}_{h,\bar{h}}(z,\bar{z})e^{iHU} =\int_{0}^{\infty} d\omega \, \, \omega^{\Delta-1}e^{-iHU}a(\epsilon\omega,z,\bar{z})e^{iHU} , \\ 
	&=\int_{0}^{\infty}d\omega \, \, \omega^{\Delta-1}e^{-i\epsilon\omega u}a(\epsilon\omega,z,\bar{z}) ,
    \end{split}
\end{equation}
where $u = U(1+z\z)$ .\footnote{The $z$ factors arise from the fact that the boundary of 4D  Minkowski space has topology $\mathbb{R} \times S^2$ and the sphere metric is given by $ds^2 = 4\dfrac{dz \, d\z}{(1+z\z)^2}$.} The action of the SL$(2,\mathbb{C})$ transformation (bulk Lorentz transformation) is
\begin{equation}
    \Phi^{\epsilon}_{h,\bar{h}}(u,z,\bar{z}) \to \Phi^{\epsilon}_{h,\bar{h}}(u',z',\bar{z}')=\frac{1}{(cz+d)^{2h}}\frac{1}{(\bar{c}\bar{z}+\bar{d})^{2\bar{h}}}
	\Phi^{\epsilon}\Big(\frac{u}{|cz+d|^2},\frac{az+b}{cz+d},\frac{\bar{a}\bar{z}+\bar{b}}{\bar{c}\bar{z}+\bar{d}}\Big) ,
\end{equation}
which is expected and most importantly, under bulk translations, the weights don't shift:
\begin{equation}
    \Phi^{\epsilon}_{h,\bar{h}}(u,z,\bar{z}) \to \Phi^{\epsilon}_{h,\bar{h}}(u',z',\bar{z}')=\Phi^{\epsilon}_{h,\bar{h}}(u+p+qz+r\bar{z}+sz\bar{z},z,\bar{z}) ,
\end{equation}
which is evidently covariant. The infinitesimal version of these transformations are precisely the ones we worked out from the Primary conditions in \eqref{BMS-primary} for $n=0,\pm1$ and $r,s=0,1$. The transformation in \eqref{eq:modmellindef} is called the \textit{modified Mellin transformation} \cite{Banerjee:2018gce}. Let us pause to comment on what we have achieved here. The Celestial primaries defined through the Mellin transform \eqref{eq:mellindef} were only defined on the Celestial sphere but the \textit{Carrollian} primaries of \eqref{eq:modmellindef} are now defined across the entire future/past null infinity $\mathscr{I}^{\pm}$. We know that the BMS algebra in 4D  is isomorphic to the conformal Carroll algebra in $D=3$. We have now been able to define primaries that transform covariantly under the full BMS group and we have a way to do so from operators in momentum space through \eqref{eq:modmellindef}. This exemplifies an explicit construction of a primary that satisfies the primary conditions \eqref{BMS-primary}. 

\medskip

We now have the following correspondence between the scattering amplitudes in the 4D  bulk and correlation functions of 3D Carrollian primaries
\begin{equation}\label{eq:modmellinsmatrix}
    \langle \Phi_1^{\epsilon_1}(u_1,z_1,\bar{z}_1)\dots \Phi_n^{\epsilon_n}(u_n,z_n,\bar{z}_n)\rangle=\int_{0}^{\infty} \left( \prod_{i=1}^{n} d\omega_i \, e^{-i\epsilon_i\omega_iu_i} \, \omega_i^{\Delta_i-1} \right) \mathcal{S}_n .
\end{equation}
Here $\epsilon_i = \pm 1$ for ingoing/outgoing particles. There are several advantages of \eqref{eq:modmellinsmatrix} over \eqref{eq:mellinsmatrix}.  First of all, both the sides are now covariant under bulk translations and the weights don't shift under translations. The interpretation of holographic correspondence is thus much cleaner. The second advantage is that the $e^{i \omega u}$ factors render the integral transform UV finite. In particular, graviton amplitudes that are generically UV divergent in the transform of \eqref{eq:mellinsmatrix} are now UV finite in the transform of \eqref{eq:modmellinsmatrix} \cite{Banerjee:2019prz}. As we will elucidate below in section \ref{ssec:softstresstensor}, the extraction of soft graviton theorems is cleaner in the modified Mellin basis where the soft factor is extracted from a finite amplitude as opposed to a UV divergent amplitude suggested by the Mellin transform \eqref{eq:mellinsmatrix}.

\medskip

Finally, in Sec. \ref{sec:carrfields4}, we have seen that symmetry arguments lead to two branches for the correlation functions of Carrollian primaries (\eqref{eq:cftbranch2pt} and \eqref{Sym-cor}). The main proposition of \cite{Bagchi:2022emh} is to identify the ``time-dependent branch'' \eqref{Sym-cor} with the correlation functions in the LHS of the correspondence \eqref{eq:modmellinsmatrix}. The time dependent branch is the 3D Carrollian CFT branch whose correlation functions have the $\delta^{2}(z-z')$ factors signifying ultra-locality. Thus, it is natural to identify \eqref{Sym-cor} (and \eqref{eq:carrollthreepointcollinear}) as holographic correlator encoding the two point (and three point) scattering amplitude. In section \ref{ssec:examplemodmellin}, we will perform the modified Mellin map of \eqref{eq:modmellinsmatrix} to establish this claim.  

\subsection{Examples}
\label{ssec:examplemodmellin}

{\ding{112}} \underline{\em{Two point function}}

\medskip

The two point free particle scattering amplitude is simply given by the inner product between two one-particle states
\begin{equation}
    \langle{p_1,\sigma_1}|{p_2,\sigma_2}\rangle = (2\pi)^3 2E_{p_1} \delta^3(\vec p_1 - \vec p_2) \delta_{\sigma_1+\sigma_2,0} .
\end{equation}
In the parametrization of \eqref{eq:3dpara}, we have
\begin{equation}
    \langle{p_1,\sigma_1}|{p_2,\sigma_2}\rangle = 4\pi^3 \frac{\delta(\omega_1 - \omega_2)\delta^2(z_1 - z_2)}{\omega_1}\delta_{\sigma_1 + \sigma_2, 0} .
\end{equation}
We get a $\delta_{\sigma_1+\sigma_2,0}$ simply because we have labelled the external particles as if they were outgoing. Performing the modified Mellin transformation \eqref{eq:modmellinsmatrix},
\begin{equation}
    \begin{split}
        &\mathcal{\tilde {M}}(u_1, z_1, \bar z_1, u_2, z_2, \bar z_2, h_1, \bar h_1, h_2, \bar h_2, \epsilon_1 = 1, \epsilon_2 = -1)  \\  
	& = 4\pi^3 \delta_{\sigma_1 + \sigma_2,0} \int_{0}^{\infty} d\omega_1 \int_{0}^{\infty} d\omega_2 \omega_1^{\Delta_1 -1}\omega_2^{\Delta_2-1} e^{- i\omega_1 u_1}e^{i\omega_2 u_2} \frac{\delta(\omega_1 - \omega_2)\delta^2(z_1 - z_2)}{\omega_1} \\
	&= 4\pi^3 \Gamma(\Delta_1 + \Delta_2 -2) \frac{\delta^2(z_1 - z_2)}{(i ( u_1 - u_2))^{\Delta_1 + \Delta_2 -2}} \delta_{\sigma_1 + \sigma_2, 0} ,
    \end{split}
\end{equation}
which up to an overall normalization factor, is identical to \eqref{Sym-cor}. Here $\mathcal{\tilde{M}}$ is a label we define for the modified Mellin transformation of the scattering amplitude (RHS of \eqref{eq:modmellinsmatrix}).

\medskip

{\ding{112}} \underline{\em{Three point function}}

\medskip

The three point scattering amplitude involving three massless particles has a momentum conserving delta function of the form
\begin{equation}\label{eq:threepointdelta}
	\delta^{(4)}(\omega_3 \Tilde{q}_3 - \omega_1\Tilde{q}_1 - \omega_2 \Tilde{q}_2) .
\end{equation}
This essentially evaluates to zero for generic momenta. It is impossible to have momentum conservation in $\mathbb{R}^{1,3}$ with just three null momenta. One can see that from
\begin{equation}\label{eq:momcons}
	(\omega_1 \Tilde{q}_1 + \omega_2 \Tilde{q}_2)^2 = 2\omega_1 \omega_2 \Tilde{q}_1 \cdot \Tilde{q}_2 ~ \bm{\neq} ~ \omega^2_3 \Tilde{q}^2_3.
\end{equation}
Hence, the argument of the delta function \refb{eq:threepointdelta} is not satisfied and always gives zero. From \eqref{eq:momcons}, it is evident that there is another class of solutions when $\Tilde{q}_1 =\Tilde{q}_2$ \cite{Chang:2022seh}. We have a non-zero result because now the three particles become collinear. Thus, the split of the momentum conserving delta function becomes \footnote{See Appendix C.4 of \cite{Bagchi:2023fbj} for the proof of the Lorentz invariance of the split.}
\begin{equation}\label{eq:momcons2}
	\delta^{4}(\omega_3 \Tilde{q}_3 -\omega_1\tilde{q}_1 -\omega_2\Tilde{q}_2) = \dfrac{1}{\omega^3_3}\delta(\omega_3 - \omega_1 - \omega_2)\,\delta(z_{12})\, \delta(z_{13})\,\delta(\bar{z}_{12})\,\delta(\bar{z}_{13}) .
\end{equation}
\begin{equation}
    \begin{split}
        \tilde{\mathcal{M}}^{(3)}(u_i,z_i,\bar{z}_i)
		&=\int_{0}^{\infty}\prod_{n=1}^3d\omega_i e^{i\epsilon_i\omega_iu_i}\omega_i^{\Delta_i-1}
            \frac{1}{\omega_3^2}\delta(\omega_1+\omega_2-\omega_3)\delta^2(z_{12})\delta^2(z_{13})   \\ 
            &= \sum_{k=0}^{\Delta_3 -4} \dfrac{{}^{\Delta_3-4}C_k \, \Gamma(k+\Delta_1) \, \Gamma(\Delta_2+\Delta_3-k-4)}{(i(u_3-u_1))^{\Delta_1+k} \, (i(u_3-u_2))^{\Delta_2+\Delta_3-4-k}}\delta^{2}(z_{12})\delta^{2}(z_{13}) .
    \end{split}
\end{equation}
This expression agrees with the correlation function in \eqref{eq:carrollthreepointcollinear}.

\medskip

Another channel in which \eqref{eq:threepointdelta} is non-zero is when one of the incoming particles is soft. In this case, one has the following split of the momentum conserving delta function
\begin{equation}
    \delta^4(	\omega_1q_1+\omega_2q_2-\omega_3q_3 )=\frac{1}{\omega_3^2 z_{12}\bar{z}_{12}}\delta(\omega_2)\delta(\omega_1-\omega_3)\delta^2(z_{13}) .
\end{equation}
The associated modified Mellin amplitude would be 
\begin{align}
		\tilde{M}^{(3)}(u_i,z_i,\bar{z}_i)=\delta_{\Delta_2,1}\frac{\Gamma(\Delta_1+\Delta_3-3)\delta^2(z_{13})}{u_{13}^{\Delta_1+\Delta_3-3}z_{12}\bar{z}_{12}} .
	\end{align}
This agrees with \eqref{Noncollinear} with $p=q=-1$ and $c=\Delta_1+\Delta_3-3$.

\medskip

Through these examples, we learn that even though the symmetry-based arguments give rise to two branches of correlation functions, the modified Mellin transformation picks up only the time-dependent branch. This is precisely the dynamical input that enters into the correlation functions themselves. For instance, one could consider a free electric scalar theory and explicitly work out the correlation function \cite{Bagchi:2022emh} to verify that it indeed matches with \eqref{Sym-cor} for specific values of the weights $h,\h$.

\subsection{3D Carroll stress tensors and soft theorems}
\label{ssec:softstresstensor}

Noether's theorem guarantees that any local translationally invariant quantum field theory has a stress tensor. In this section, we will briefly elucidate the role of stress tensors in the holography of asymptotically flat spacetimes. It is now well understood that soft theorems (which are universal statements about scattering amplitudes \cite{Low:1954kd,PhysRev.110.974,PhysRev.140.B516,PhysRevLett.20.86,Cachazo:2014fwa}) are Ward identities corresponding to the asymptotic BMS symmetries \cite{Strominger:2014pwa,He:2014laa,Kapec:2016jld}. In traditional AdS/CFT, the stress tensor correlators encode graviton scattering processes in the bulk. This is because the stress tensor in the boundary CFT is dual to the bulk metric fluctuations \cite{Witten:1998qj,Balasubramanian:1999re}. 

\medskip

We have seen that in AFS$_3$/CCFT$_2$, one can match stress tensor correlation functions from the bulk and the boundary. We will also see in the next section that the limit from AdS/CFT would results in structures that encode scattering processes in AFS. Thus, it is entirely reasonable to expect that features of graviton scattering in flat space might potentially be encoded in the stress tensors of the dual Carrollian field theory. This question was answered in the affirmative in \cite{Bagchi:2024gnn}, where it was argued that the leading and sub-leading soft graviton theorems can be encoded into Carrollian stress tensor Ward identities. The relevant literature on boundary Carroll stress tensors includes \cite{Mann:2005yr,Fareghbal:2013ifa, Bagchi:2015wna,Ciambelli:2018ojf,Ciambelli:2018wre,Jafari:2019bpw,Chandrasekaran:2021hxc,deBoer:2021jej,Adami:2021nnf,Freidel:2022vjq,Saha:2023hsl,Campoleoni:2023fug,Adami:2024rkr,Bhambure:2024ftz,Ruzziconi:2024kzo,Ciambelli:2025mex,Hartong:2025jpp}. We briefly discuss the main elements of \cite{Bagchi:2024gnn} below.

\medskip

{\ding{112}} \underline{\em{3D Carroll Stress Tensor}}

\medskip

Our previous construction of the 2D CCFT stress tensor (in Sec.~\ref{ccft2-T}) was based on the contraction of the two copies of the Virasoro algebra. Now that the 3D CCFT is infinite dimensional as opposed to the 3D relativistic conformal algebra, this method is no longer as useful. We will thus resort to a more geometric way of constructing the Carroll stress tensor. {\footnote{For a geometric derivation of the 2D CCFT stress tensor, the reader is referred to \cite{Bagchi:2021gai, Bagchi:2022eav}.}} 

\medskip

The Carroll stress tensor is one that is constructed on null surfaces. Hence, the variation must be performed with respect to the intrinsic geometric elements characterizing the null surface \cite{Henneaux:1979vn}. For instance, in terms of the time and spacelike vielbeins $\tau^{\alpha}$ and $e^{\alpha}_{~\,a}$ of the null surfaces, the stress tensor can be defined through the following variation \cite{Baiguera:2022lsw,Dutta:2022vkg}
\begin{equation}
    \delta S = \int du\ d^2z\ e \left[ \tau^\beta \delta \tau_\alpha + e^\beta_{~\,a} \delta e^a_{~\,\alpha} \right] T^\alpha_{~\,\beta}  .
\end{equation}
From the variation of the vielbeins under local Carroll transformations, we have the following restrictions
\begin{equation}
T^i_{~\,u}=0 \, , \qquad T^i_{~\,j}=T^j_{~\,i}  .
\end{equation}
From the stress tensor conservation equations, one can arrive at the following quantities that are conserved
\begin{equation}\label{eq:stresscomp}
    T_u(z, \bar{z}) \equiv T^u_{~\,u}(z,\bar{z})\, , \quad T_i(z, \bar{z}) \equiv T^u_{~\,i} - \frac{u}{2} \partial_i T^u_{~\,u} = C_i(z, \bar{z}) - \int du\, \partial_j \theta^j_{~\, i}(u, z, \bar{z}) .
\end{equation}
These components go into the definition of conserved charges and the modes of the components close to the BMS$_4$ algebra \cite{Dutta:2022vkg}. Suppose if you consider supertranslations parametrized by $\alpha(z,\z)$, they act on \eqref{eq:stresscomp} as
\begin{equation}
    \begin{split}
        \delta_{\alpha(z,\bar{z})}T_z&= \frac{1}{2}\alpha(z,\bar{z}) \partial_z T_u(z,\bar{z})+\frac{3}{2}T_u(z,\bar{z})\partial_z\alpha(z,\bar{z}) , \\ 
\delta_{\alpha(z,\bar{z})}T_{\bar{z}}&=\frac{1}{2}\alpha(z,\bar{z}) \partial_{\bar{z}} T_u(z,\bar{z})+\frac{3}{2}T_u(z,\bar{z})\partial_{\bar{z}}\alpha(z,\bar{z}) , \\
\delta_{\alpha(z,\bar{z})}T_u&= 0 .
    \end{split}
\end{equation}
One can check that these stress tensor components transform as primaries under $L_n, \bar{L}_n$ and the conformal weights of these primaries are summarized in Table \ref{tab:stress}. From the transformation properties, one can derive the Ward identities associated with these stress energy tensor components \cite{Bagchi:2024gnn}:
\begin{subequations}\label{eq:wardid}
  \begin{align} 
	\langle
	&T_u(z,\bar{z})\prod_{i=1}^n \Phi^{\epsilon_i}_{h_i,\bar{h}_i}(u_i,z_i,\bar{z}_i)\rangle=-\Big(\sum_{i} \delta^2(z-\omega_i)\partial_{u_i}\Big)\langle \prod_{i=1}^{n} \Phi^{\epsilon_i}_{h_i,\bar{h}_i}(u_i,\omega_i,\bar{\omega}_i) \, \rangle\, , \\ 
	&\langle T_z(z,\bar{z})\prod_{i=1}^n \Phi^{\epsilon_i}_{h_i,\bar{h}_i}(u_i,z_i,\bar{z}_i))\rangle \nonumber \\ &~~~~ =\Big(\sum_{i}(h_i+\frac{u_i}{2}\partial_{u_i})\partial_{z_i}\delta^2(z-z_i) +\delta^2(z-z_i)\partial_{z_i}\Big) \langle\prod_{i=1}^{n} \Phi^{\epsilon_i}_{h_i,\bar{h}_i}(u_i,z_i,\bar{z}_i)\rangle .
\end{align}  
\end{subequations}
These stress-tensor Ward identities would be pivotal in connecting to bulk physics in 4D  AFS below. We would like to point out that the only information that goes into the construction of \eqref{eq:stresscomp} is the geometric elements of null hypersurfaces and the input of symmetries. This is a universal sector of a field theory that inherits BMS$_4$ symmetries. In particular, we are not making any claim about the flux because it involves dynamics.

\begin{table}[t!]
\centering
\begin{tabular}{||c | c c c||} 
 \hline
 weights &  $T_u$ & $T_z$ & $T_{\bar{z}}$ \\ [0.75ex] 
 \hline\hline
 $h$ &  ${3}/{2}$ & 2 & 1 \\ 
 \hline
 $\bar{h}$ & ${3}/{2}$ & 1 & 2 \\
 \hline
\end{tabular}
\caption{Weights of Carroll stress tensor components}
\label{tab:stress}
\end{table}

\medskip
\medskip

{\ding{112}} \underline{\em{Connections with Soft Theorems}}

\medskip

The statement of the soft theorem for a single graviton is given by \footnote{See Appendix \ref{apssec:softsmatrix} for a lightning introduction to soft theorems.}
\begin{equation}\label{eq:gravsoft}
    \mathcal{M}_{n+1}(p_i,k)=\left[\frac{1}{\omega}\sum_{i=1}^{n}\frac{\epsilon_{\mu\nu}p_i^\mu p_i^\nu}{p_i.\hat{k}}+\sum_{i=1}^{n}\frac{\epsilon_{\mu\nu}p_i^\mu\hat{J}^{\nu\lambda}k_{\lambda}}{p_i.\hat{k}}+\mathcal{O}(\omega)\right]\mathcal{M}_n(p_i) .
\end{equation}
Here $\mathcal{M}_{n+1}$ is the scattering amplitude involving $n$ particles and one graviton which goes soft. In \eqref{eq:gravsoft}, $p_i$ and $k$, respectively, denote the momentum of hard and soft particles. $e_i$ denotes the charge of the particles. $\omega$  and $\epsilon_{\mu\nu}$ are the energy and polarisation vector of the photon. $\hat{J}^{\nu\lambda}$ denotes the angular momentum. The first and the second terms in \eqref{eq:gravsoft} are the leading and sub-leading soft theorems. In the modified Mellin basis of \eqref{eq:modmellinsmatrix}, the leading soft graviton theorem takes the following form
\begin{equation}
    \lim_{\omega \to 0} \omega \mathcal{M}_{n+1}(\omega,z,\bar{z},\omega_i,z_i,\bar{z}_i)=-\Big(\sum_{i}\frac{\epsilon_i\,\omega_i\,(z-z_i)}{\bar{z}-\bar{z}_i}\Big) \, \mathcal{M}_n(\omega_i,z_i,\bar{z}_i,\sigma_i) .
\end{equation}
If we perform a modified Mellin transformation on the hard particles \cite{Banerjee:2020zlg},
\begin{equation}
    \langle S_0^-(z,\bar{z})\prod_{i=1}^{n} \Phi^{\epsilon_i}_{ h_i,\bar{h}_i}(u_i,z_i,\bar{z}_i)\rangle =-\Big(\sum_{i}\frac{\epsilon_i\,(z-z_i)}{\bar{z}-\bar{z}_i}\partial_{u_i}\Big)\langle \prod_{i=1}^{n} \Phi^{\epsilon_i}_{h_i,\bar{h}_i}(u_i,z_i,\bar{z}_i)\rangle ,
\end{equation}
where $S^-_0$ corresponds to a soft insertion of a graviton of negative helicity. Acting twice with $\partial_{z}$ on either side of the above equation, we have
\begin{equation}\label{eq:s0wardid}
	\langle \partial^2_{z}S_0^-(z,\bar{z})\prod_{i=1}^{n} \Phi^{\epsilon_i}_{h_i,\bar{h}_i}(u_i,z_i,\bar{z}_i)\rangle=-\sum_{i} \delta^2(z-z_i) \epsilon_i\partial_{u_i}\langle\prod_{i=1}^{n} \Phi^{\epsilon_i}_{h_i,\bar{h}_i}(u_i,z_i,\bar{z}_i)\rangle .
\end{equation}
Comparing \eqref{eq:s0wardid} with \eqref{eq:wardid}, we see that we can make the following identification of the soft operators with the stress tensor component
\begin{equation}\label{eq:tuurel}
    T_u(z,\z)=\partial^2_{z}S_0^-(z,\bar{z}) .
\end{equation}
If $S^-_1$ was the negative helicity soft insertion that incorporated the sub-leading soft graviton theorem of \eqref{eq:gravsoft} in modified Mellin basis, then through analogous arguments one can argue
\begin{equation}\label{eq:tzzrel}
    T_z(z, \z)=\partial^3_zS_1^{-}(z,\bar{z}) .
\end{equation}
Effectively, through \eqref{eq:tuurel} and \eqref{eq:tzzrel}, we have argued that one could create quantities out of the stress tensor components obtained from the variation of intrinsic geometric elements to encode soft theorems in the bulk. Further, one could relate the quantities $\{T_u,T_z,T_{\z}\}$ to the soft sectors in Celestial CFT \cite{Strominger:2013jfa,Barnich:2013axa,He:2014laa,Strominger:2014pwa,Donnay:2018neh}.

\medskip
\medskip

{\ding{112}} \underline{\em{Connection to the Celestial soft operators}}

\medskip

The soft sectors of Celestial CFT are reviewed in Appendix \ref{Celestial CFT}. From \eqref{eq:celsoftprimaries}, one can construct soft operators in the 2D dual Celestial CFT \eqref{softop} which encode the leading \eqref{celestialkac} and sub-leading soft theorems \eqref{Shadow}. For convenience, we will re-write \eqref{softop} in terms of the soft operators associated with the negative helicity graviton $S^-_0$ and $S^-_1$ defined above in \eqref{eq:tuurel} and \eqref{eq:tzzrel}:
\begin{eqnarray}
    & P_{(2d)}(z,\z) = \partial_z S^-_0(z,\z), \label{eq:2dkacmoody} \\
    & T_{(2d)}(z) = \int d^2w \dfrac{1}{(z-w)} \partial^3_w S^-_1(w,\bar{w}). \label{eq:2dshadow}
\end{eqnarray}
From the Ward identities and from counting the weights of operators, one can immediately identify the following map between Celestial and Carrollian soft operators \cite{Bagchi:2024gnn}:
\begin{equation}\label{eq:carcelmapstress}
    T_u(z,\z) = \partial_z P_{(2d)}(z,\z), ~~~~ T_z(z,\z) = \partial_z \bar{T}_{(2d)}(\z), ~~~~ T_\z(z,\z) = \partial_\z T_{(2d)}(z).
\end{equation}
It is clear that the asymptotic symmetries are packaged into operators differently for Celestial and Carrollian approaches. The integral transformation in \eqref{eq:2dshadow} renders the operator $T_{(2d)}$ non-local on the celestial sphere. In addition to such a non-local operator, one needs to introduce another $U(1)$ Kac-Moody current of the form $P_{(2d)}$ \eqref{eq:2dkacmoody} to encode the leading soft theorem. However, we have shown that from the 3D Carrollian perspective, we can encode both the leading and sub-leading soft theorems using \textit{local} operators $T_u,T_z,T_\z$ constructed from an intrinsic local geometric stress tensor of the Carrollian CFT. In fact, the partial derivatives in \eqref{eq:carcelmapstress} localize the non-local Virasoro stress tensor $T_{(2d)}$ in the following way
\begin{equation}
    \begin{split}
        T_\z(z,\z) &= \partial_\z T_{(2d)}(z) = \partial_\z \int d^2w \dfrac{1}{(z-w)} \partial^3_w S^-_1(w,\bar{w}) \\
        &= \int d^2w \, \delta^2(z-w) \, \partial^3_w S^-_1 = \partial^3_z S^-_1(z,\z).
    \end{split}
\end{equation}
The 3D Carroll stress tensor thus overcomes one of the principle issues of the 2D Celestial stress tensor by rendering it local. We have given an in-principle argument based on how the bulk isometries organize themselves into boundary degrees of freedom through \eqref{eq:tuurel} and \eqref{eq:tzzrel}. Further, we do not make claims about this being the unique way of encoding bulk soft theorems and there certainly exist other proposals \cite{Donnay:2021wrk,Saha:2023hsl,Ruzziconi:2024kzo}. Our proposal is guided purely by the zeroth order requirement of the holographic principle: match between the asymptotic symmetries of the bulk and the boundary isometries. The difference lies in how the boundary isometries organize themselves into boundary structures. We will return to the Carroll stress tensor construction briefly at the end of the next section when we discuss the limit from AdS/CFT to AFS/CCFT.

\newpage

\section{Flat space limit of AdS/CFT: View from embedding space}
\label{sec:flatlimit}

One of the ways to arrive at flat space is by taking the large radius limit of AdS. This large radius (equivalently the zero cosmological constant) limit in the bulk manifests itself as a Carrollian limit in the boundary CFT \cite{Bagchi:2012cy}. We have also seen that this limiting approach towards constructing a dual theory for flat space works very well in lower dimensions. However, due to the presence of gravitational radiation and its leaky behaviour, this approach is not very straightforward in higher dimensions. Nonetheless, in section \ref{sec:4dafs}, we have showcased hints of holography through Carrollian boundary correlation functions defined via the modified Mellin transformations of scattering amplitudes.  We can expect to arrive at this framework from some large radius limit of AdS/CFT as well. In fact, this approach has been taken by the vast majority of the literature in the past \cite{Polchinski:1999ry,Susskind:1998vk,Giddings:1999jq,Balasubramanian:1999ri,Giddings:1999qu,PhysRevD.94.065017,Gary:2009ae,Penedones:2010ue,Fitzpatrick:2011jn,Fitzpatrick:2011hu,Fitzpatrick:2011dm,Raju:2012zr,Paulos:2016fap,Hijano:2019qmi,Hijano:2020szl,Li:2021snj} whose history we briefly recapitulate below. 

\subsection{A bit of history}

Polchinski \cite{Polchinski:1999ry} and Susskind \cite{Susskind:1998vk} were the first to sketch a prescription that gives rise to a flat space S-matrix in the large AdS radius limit. \cite{Giddings:1999jq} introduced the notion of scattering amplitudes in AdS \cite{Balasubramanian:1999ri,Giddings:1999qu} by elaborating on the prescription of \cite{Polchinski:1999ry,Susskind:1998vk}. Defining such amplitudes can be subtle \cite{PhysRevD.94.065017} because AdS has a time-like boundary and one cannot construct traditional \textit{in} or \textit{out} states. \cite{Gary:2009ae} pointed out that the bulk S-matrices can only be extracted if the boundary correlators exhibited a particular singularity structure. These works were then generalized in the seminal work of Penedones \cite{Penedones:2010ue} which showed that AdS scattering amplitudes are precisely given by the Mellin space representation \cite{Mack:2009gy,Mack:2009mi} of boundary CFT correlators. The large AdS radius $R$ limit with the momenta (Mandelstam variables) scaled with suitable powers of $R$ results in a bulk flat space massless \footnote{The massive case was addressed in \cite{Paulos:2016fap}.} S-matrix at the centre of AdS \footnote{If we zoom into the centre of global AdS in the large AdS radius limit, then curvature corrections become negligible and we are effectively left with a Minkowski diamond at the centre of global AdS.}. The conjecture of \cite{Penedones:2010ue} was proven in \cite{Fitzpatrick:2011jn,Fitzpatrick:2011hu,Fitzpatrick:2011dm} which basically constructed smeared scattering states in AdS that reduced to plane waves in the centre of AdS.

\medskip

The use of HKLL bulk reconstruction techniques \cite{Hamilton:2005ju,Hamilton:2006az,Hamilton:2006fh} in the flat space limit was pioneered in \cite{Hijano:2019qmi,Hijano:2020szl,Li:2021snj}. The idea here is to first identify suitable bulk observables that encode the S-matrix within the Minkowski diamond at the centre of AdS, also known as the ``scattering region''. One then reconstructs these bulk fields in the CFT using appropriate smearing kernels of the HKLL technique. In the large AdS radius limit, these smearing kernels will localize and have support over certain ``bands'' of the CFT. These smearing kernels construct states that are generalizations of \cite{Fitzpatrick:2011jn}. Thus, the prescription of \cite{Hijano:2019qmi} reduces to \cite{Penedones:2010ue} when the particles are massless and reduces to \cite{Paulos:2016fap} when the particles are massive. \cite{Li:2021snj} pointed out how HKLL helps in unravelling the flat space limit of momentum space correlators \cite{Raju:2012zr} \footnote{See \cite{Raju:2012zs,Farrow:2018yni,Gadde:2022ghy,Marotta:2024sce} for works addressing the flat limit in Momentum space. The recent work of \cite{Marotta:2024sce} shows how to obtain the flat space limit of a CFT correlator expressed in momentum space. CFT three point functions in momentum space can be expressed non-perturbatively in terms of triple-$K$ integrals \cite{Bzowski:2013sza} and depending on whether the particles are massive or massless, one can appropriately take a flat limit in the deep interior of AdS to recover flat space scattering amplitudes in one higher dimension as shown in general in \cite{Marotta:2024sce}.}. In global AdS, the smearing kernel localizes to regions around the following global time slices
\begin{equation}
    \tau = \pm \dfrac{\pi}{2} + \mathcal{O}(R^{-1}) .
\end{equation}
These slices will turn out to be important for our analysis.

\medskip
 
In the following, we consider elements of AdS Witten diagrams \cite{Witten:1998qj} and consider its flat space limit. Recent attempts to link flat holography to AdS/CFT using Witten diagrams are \cite{Lam:2017ofc,Casali:2022fro,deGioia:2022fcn,Iacobacci:2022yjo,Sleight:2023ojm,Bagchi:2023fbj,deGioia:2023cbd,Bagchi:2023cen,Alday:2024yyj,Marotta:2024sce,Chakrabortty:2024bvm,Banados:2024kza,Surubaru:2025fmg,Lipstein:2025jfj}. The original work of \cite{Lam:2017ofc} showed how the bulk point singularity \cite{Gary:2009ae,Penedones:2010ue,Maldacena:2015iua} in AdS Witten diagrams is manifested in Celestial amplitudes in two lower dimensions. By foliating the Minkowski spacetime analogous to \cite{deBoer:2003vf}, \cite{Casali:2022fro} were able to relate 4D Celestial amplitudes to AdS$_3$ Witten diagrams. In the next section, we summarize the main ideas of \cite{Bagchi:2023fbj,Bagchi:2023cen}. The works of \cite{Bagchi:2023fbj,Bagchi:2023cen} are inspired from \cite{deGioia:2022fcn} which showed how AdS Witten diagrams with boundary insertions at specific time slices reduce to Celestial amplitudes (in two dimensions lower) in the large AdS radius limit. \cite{Bagchi:2023fbj,Bagchi:2023cen} considers a ``band'' of insertion points around the fixed time slices of \cite{deGioia:2022fcn} which at first glance are seemingly sub-leading in the large $R$ limit. However, we will see that this band turns out to be crucial in ensuring the boundary dimension does not arbitrarily reduce in dimensions in the large $R$ limit.

 \medskip

 One of the motivations of \cite{Bagchi:2023fbj} in the investigation of the rather well-developed literature has to do with the boundary CFT interpretation of the S-matrices in flat spacetimes. In the large AdS radius limit, one should also account for the symmetries of the dual of Minkowski space. The literature mentioned above mostly considers this aspect in terms of a CFT dual to some parent AdS. This interpretation precisely misses the emergence of the null direction in the dual of Minkowski space. Finally, there is no dimensional reduction in the bulk when one goes from AdS to flat space and hence there should be no dimensional reduction in the dual description as well. This has to be contrasted with the Celestial picture where the null direction was discarded.

\subsection{Prescription for flat limit }
\label{ssec:flatpres}

\medskip

We are interested in implementing the flat limit in AdS Witten diagrams which have bulk to boundary propagators, bulk to bulk propagators and bulk interaction vertices. In order to obtain a clean limit, we will follow \cite{Giddings:1999jq} and consider global AdS. Due to the nature of reflecting boundary conditions in the boundary of AdS, the scattering processes occur indefinitely in repeating patterns. To alleviate this, one constructs AdS wavepackets \cite{Gary:2009ae} (or analogous ``scattering'' states \cite{Fitzpatrick:2011jn,Fitzpatrick:2011hu,Fitzpatrick:2011dm}) that scatter in the centre of the global AdS cylinder and extract one scattering process. As shown in the seminal work of Penedones \cite{Penedones:2010ue}, this scattering process can be encoded in AdS Witten diagrams. As we explained the motivations of \cite{Bagchi:2023fbj}, if we are interested in the boundary interpretation, one must carefully track the boundary insertions of the AdS Witten diagrams. For the bulk to boundary propagators, \cite{Bagchi:2023fbj} considers the following assumption ($R$ is the AdS radius): \textit{insertion points of boundary operators are at global time slices $\tau=\pm \frac{\pi}{2}+\frac{u}{R}$.}
As a consequence, we will see that there is an anti-podal identification between the two spheres at $\tau= +\frac{\pi}{2} + \frac{u}{R}$ and $\tau= -\frac{\pi}{2} + \frac{u}{R}$ on the boundary. Since the AdS radius explicitly enters into the boundary insertions, we will see that the boundary interpretation becomes transparent in the large AdS radius limit.

\medskip

To make progress, one considers the embedding space formalism \cite{Dirac:1936fq,Penedones:2007ns} where the computation of Witten diagrams is simplified. 
In an appropriate slicing of embedding space coordinates, the  AdS metric is given by the usual global AdS metric:
\begin{equation}\label{eq:adsmetric}
	ds^2 =\frac{R^2}{\cos^2\rho}\left(-d\tau^2+d\rho^2+\sin^2\rho \, d\Omega_{S^{d-1}}^2\right) ,
\end{equation}
where $\tau \in (-\infty , \infty)$ \footnote{As usual, we unwrap the AdS (and go to the covering space) by extending the time like coordinate from the original range $\tau \in [-\pi,\pi]$.} and $\rho \in [0,\frac{\pi}{2})$. The boundary limit is given by $\rho\rightarrow\frac{\pi}{2}$.

\medskip

Following \cite{Giddings:1999jq}, we implement the flat limit on the AdS solution  \eqref{eq:adsmetric} through the rescaling:
\begin{equation}\label{eq:largerlimit}
	\tau = \dfrac{t}{R} , ~~~~~~ \rho = \dfrac{r}{R} ,
\end{equation}
and then we let $R \to \infty$.  \eqref{eq:largerlimit} also implies that  \eqref{eq:adsmetric} becomes
\begin{equation}\label{mflat}
	ds^2 \xrightarrow{R \to \infty} -dt^2 + dr^2 + r^2 \, d\Omega^2_{S^{d-1}} .
\end{equation}
Effectively, one zooms into the centre of AdS where curvature scale becomes negligible and the geometry becomes flat. As mentioned above, we consider the following insertion point 
\begin{equation} \label{tau}
\tau=\pm \frac{\pi}{2}+\frac{u}{R} .
\end{equation}
One can motivate this as follows. The boundary metric (reached by $\rho \to \frac{\pi}{2}$) is flat:
\begin{equation}\label{eq:bdymetricads}
ds^2_{\text{CFT}} = -d{\tau}^2 + d\Omega_{d-1}^2 .
\end{equation}
We now specifically keep track of the null boundary of flat spacetime that emerges at the centre of AdS, when we implement the large AdS radius limit on  \eqref{eq:adsmetric}. Consider $\mathscr{I}^+$ ($\mathscr{I}^-$ follows naturally). The standard Minkowski retarded time can be expressed in terms of the AdS coordinates  \eqref{eq:largerlimit}
\begin{equation}\label{eq:minku}
u = t - r = R(\tau-\rho)  . 
\end{equation}
We want to relate this $u$ with the boundary time of AdS. If $\tau = \tau_p$ is the boundary time we reach as $\rho \to \frac{\pi}{2}$, we have 
\begin{equation}\label{eq:bdytime}
u = R(\tau_p - \frac{\pi}{2} ) \Rightarrow \tau_p = \frac{\pi}{2} + \frac{u}{R} .
\end{equation}
Thus, the insertion point we considered is precisely the point which gets related to the boundary retarded time of the Minkowski diamond at the center of AdS. One can see this at the level of symmetries by looking at the boundary metric in the large $R$ limit. If we substitute  \eqref{eq:bdytime} in  \eqref{eq:bdymetricads}, we get
\begin{equation}\label{eq:cftbdyrmet}
ds^2_{\text{bdy}} = -\frac{1}{R^2} du^2 + d\Omega_{d-1}^2 .
\end{equation}
Clearly, in the limit of $R\to \infty$ the metric degenerates to a null or Carrollian metric at $\mathscr{I}^+$:
\begin{equation}
ds^2_{\text{Carroll}} = 0 . du^2 + d\Omega_{d-1}^2 .
\end{equation}
To reach $\mathscr{I}^-$, one must consider $\tau= - \frac{\pi}{2}+\frac{v}{R}$ where $v$ is the advanced time $v=t+r$. This justifies our choice  \eqref{tau}.

\medskip

The above approach is schematically represented in figure \ref{fig:carrolliancase}. Since there is no dimensional reduction in the AdS boundary in the large AdS radius limit, we see the emergence of the null boundary of Minkowski space in the dark green patches around $\tau = \pm \frac{\pi}{2}$. The red lines indicate massless particle scattering. It is clear through \eqref{eq:cftbdyrmet}, the large $R$ limit is realised as an infinite boost limit which takes the timelike boundary to a null boundary if we identify the speed of light $c = \frac{1}{R}$. Similar conclusions were reached in the recent works of \cite{Alday:2024yyj,Surubaru:2025fmg}. We see that if we had set $u=0$, the dark green patches reduce to a single time slice at $\tau = \frac{\pi}{2}$. This reproduces the Celestial CFT picture of \cite{deGioia:2022fcn,deGioia:2023cbd}. However, in the large $R$ limit, the subleading terms in $\mathcal{O} \left( \frac{1}{R} \right)$ become crucial to account for the null boundary of the flat Minkowski space that emerges in the centre of AdS.

\begin{figure}[tbp]
	\centering
    \hspace{1.5cm}
	\includegraphics[width=0.5\textwidth]{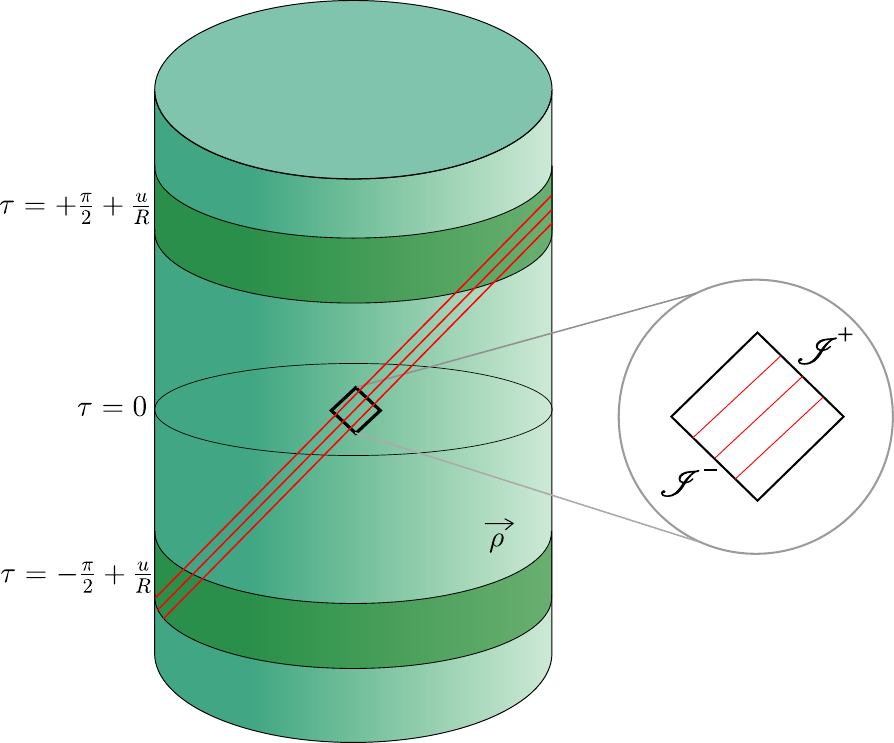}
	\caption{The Carrollian approach to flat limit of AdS.}
    \label{fig:carrolliancase}
\end{figure}

\subsection{Witten diagrams in the flat space limit}
\label{ssec:wittendiag}

The main result of \cite{Bagchi:2023fbj} is that the leading order result of AdS Witten diagrams in a large $R$ expansion precisely correspond to Carrollian correlation functions which encode bulk scattering processes. One of the crucial elements of Witten diagrams is the bulk to boundary propagator. The scalar bulk to boundary propagator $K_\Delta(\vec{p},\vec{x})$ in the embedding space is given by \cite{Penedones:2010ue,Penedones:2007ns}: 
\begin{equation}\label{eq:bulkbdyprop}
	K_\Delta(\vec{p},\vec{x})=\frac{C_\Delta^d}{(-2\vec{p}\cdot \vec{x}+i\epsilon)^\Delta} ,
\end{equation}
where $C_{\Delta}^d$ is a constant and we have parametrized the bulk and boundary points by $\vec{x}\in(\tau,\rho,\Omega)$ and $\vec{p}\in(\tau_p,\Omega_{\vec{p}})$ respectively. In the large $R$ limit (substituting \eqref{eq:largerlimit} and \eqref{eq:bdytime} in \eqref{eq:bulkbdyprop}), we get
\begin{equation}\label{eq:outgoingfn}
	K_\Delta(\vec{p},\vec{x})= C^d_\Delta \left(\frac{1}{(-u-\tilde{q} \cdot x+i\epsilon)^\Delta}+ \mathcal{O}(R^{-1})\right) ,
\end{equation}
where $x=(t,r\Omega)\in\mathbb{R}^{1,d}$ is a bulk point and the vector $\tilde{q}=(1,\Omega_{\vec{p}})\in\mathbb{R}^{1,d}$  is a null vector in the direction of the boundary point $\vec{p} \in (\tau_p,\Omega_{\vec{p}})$. For $\mathscr{I}^-$, one uses $\tau_p= -\frac{\pi}{2}+\frac{u}{R}$ which implies \eqref{eq:bulkbdyprop} in the large $R$ limit becomes:
\begin{equation}\label{eq:incomingfn}
	K_\Delta(\vec{p},\vec{x})= C^d_\Delta \left(\frac{1}{(u+\tilde{q} \cdot x+i\epsilon)^\Delta}+ \mathcal{O}(R^{-1})\right) ,
\end{equation}
where $x=(t,r\Omega)$ but the $\tilde{q}=(1,\Omega_{\vec{p}}^A)$, where crucially, we have
\begin{equation}\label{eq:antipodal}
	\Omega_{\vec{p}}^A= -\Omega_{\vec{p}} ,
\end{equation}
which is the antipodal point of $\Omega_{\vec{p}}$. The spheres at $\tau_p = \frac{\pi}{2}+\frac{u}{R}$ and $\tau_p =- \frac{\pi}{2}+\frac{u}{R}$ are antipodally matched. 

\medskip

We now identify \eqref{eq:incomingfn} and \eqref{eq:outgoingfn} with the modified Mellin transform \cite{Banerjee:2018gce} of plane waves:
\begin{equation}\label{eq:modmellin}
	\int^{\infty}_{0} d\omega \, \omega^{\Delta -1} \, e^{\mp i\omega u} e^{\mp i\omega \, \tilde{q} \cdot x} e^{-\epsilon \omega} = \dfrac{(\pm i)^{\Delta}\Gamma(\Delta)}{(\mp u\mp \tilde{q} \cdot x+i\epsilon)^{\Delta}} .
\end{equation}
Thus, the large $R$ expansion of bulk to boundary propagators reduce to Carrollian primary basis of plane waves. Rewriting the bulk to boundary propagators in terms of modified Mellin transforms, we get:
\begin{equation}\label{eq:bbdypropmodmellin}
	K_{\Delta}(\vec{p},\vec{x}) = N_{\Delta}^{d}\psi_{\Delta,\tilde{q},u}^{\pm}(x) +O(R^{-1}),
\end{equation}
for $N_{\Delta}^{d}$ being a normalization constant and
\begin{equation}
	\psi_{\Delta,\tilde{q},u}^{\pm}(x)=\int_{0}^{\infty}d\omega \, \omega^{\Delta_1-1}e^{\mp i \omega (\tilde{q}\cdot x+u)}e^{-\epsilon\omega}.
\end{equation}
The superscript $+(-)$ indicates outgoing (incoming) wave basis respectively. 

\medskip

If we consider non-derivative couplings in the bulk, the vertex factors in the large $R$ limit can be worked out to be
\begin{equation}\label{eq:vertexfactor}
	i \mu \int_{\text{AdS}_{d+1}} d^{d+1}\vec{x} = i \mu \int_{\mathbb{R}^{1,d}}\left(d^{d+1}x + \mathcal{O}(R^{-2})\right) .
\end{equation}
Finally, an internal AdS propagator satisfying
\begin{equation}
	\left(\Box_{\text{AdS}_{d+1}}-\frac{\Delta(\Delta-d)}{R^2}\right)\Pi_\Delta(\vec{x}_1,\vec{x}_2)= i\delta_{\text{AdS}_{d+1}}(\vec{x}_1,\vec{x}_2) ,
\end{equation}
goes over to the Feynman propagator in the large $R$ limit:
\begin{equation}\label{eq:propagator}
	\Pi_\Delta(\vec{x}_1,\vec{x}_2)= G(x_1,x_2)+\mathcal{O}(R^{-2}) ,
\end{equation}
where $G(x_1,x_2)$ obeys the equation 
\begin{equation}\label{eq:flatspacekg}
	\left(\Box_{\mathbb{R}^{1,d}}-m^2\right)G(x_1,x_2)= i\delta_{\mathbb{R}^{1,d}}(x_1,x_2) , \quad~~~ m\equiv \lim_{R\rightarrow\infty}\frac{\Delta}{R}.
\end{equation}

\medskip

To illustrate the limit, consider the example of a four point function with the $\phi^3$ interaction:
\begin{equation}\label{eq:fourpointeg}
	\begin{split}
		\langle O_{\Delta_1}(\vec{p}_1)O_{\Delta_2}(\vec{p}_2)O_{\Delta_3}(\vec{p}_3)O_{\Delta_4}(\vec{p}_4)\rangle= &(i\mu)^2\int_{AdS_{d+1}}d^{d+1}\vec{x} \, d^{d+1}\vec{y} \, \Pi_\Delta(\vec{x},\vec{y})\\ &~~ K_{\Delta_{1}}(\vec{p}_1,\vec{x}) K_{\Delta_{2}}(\vec{p}_2,\vec{x}) K_{\Delta_{3}}(\vec{p}_3,\vec{y}) K_{\Delta_{4}}(\vec{p}_4,\vec{y}) .
	\end{split}
\end{equation}
If $\vec{p}_1,\vec{p}_2$ are inserted in $\tau = -\frac{\pi}{2}+\frac{u_1}{R}$ and $\tau = -\frac{\pi}{2}+\frac{u_2}{R}$ respectively, and $\vec{p}_3,\vec{p}_4$ are inserted in $\tau = \frac{\pi}{2}+\frac{u_3}{R}$ and $\tau = \frac{\pi}{2}+\frac{u_4}{R}$ respectively, one can use the results of \eqref{eq:bbdypropmodmellin}, \eqref{eq:vertexfactor} and \eqref{eq:propagator} we get
\begin{equation}
	\begin{split}
		\begin{split}
			\langle O_{\Delta_1}(\vec{p}_1)O_{\Delta_2}(\vec{p}_2)O_{\Delta_3}(\vec{p}_3)O_{\Delta_4}(\vec{p}_4)\rangle= &(i\mu)^2\int_{\mathbb{R}^{1,d}}d^{d+1}x \, d^{d+1}y \, G(x,y)\\ &~~ \psi_{\Delta_1,q_1,u_1}^{-}(x) \psi_{\Delta_2,q_2,u_2}^{-}(x) \psi_{\Delta_3,q_3,u_3}^{+}(y) \psi_{\Delta_4,q_4,u_4}^{+}(y) .
		\end{split}
	\end{split}
\end{equation}

\subsection{Explicit examples}
\label{ssec:examplesdiag}

In this subsection, we work out some explicit examples in detail to match with the field theory analysis of Carrollian correlation functions. Working in $3+1$ dimensions, we will use the following parametrization for the null vector in the direction of $\Omega_{\vec{p}}$:
\begin{equation}\label{eq:4dparamq}
	\tilde{q}^{\mu} = \left[1,\dfrac{z+\bar{z}}{1+z\bar{z}},\dfrac{-i(z-\bar{z})}{1+z\bar{z}},\dfrac{1-z\bar{z}}{1+ z\bar{z}} \right] .
\end{equation}
The topology of the null boundary is $\mathbb{R}\times \mathbb{S}^2$ and $(z,\bar{z})$ characterize the coordinates on the celestial sphere. 

\medskip

{\ding{112}} \underline{\em{Two point function}}

\medskip

The two-point scalar Witten diagram is given by
\begin{equation}\label{eq:tpf}
	\langle O_{\Delta_1}(\vec{p}_1)O_{\Delta_2}(\vec{p}_2)\rangle= \int_{\text{AdS}_{4}}d^{4}\vec{x} \, K_{\Delta_{1}}(\vec{p}_1,\vec{x}) \, K_{\Delta_{2}}(\vec{p}_2,\vec{x}) ,
\end{equation}
where $K_{\Delta_{1}}(\vec{p}_1,\vec{x})$ is inserted at $\tau= -\frac{\pi}{2}+\frac{u_1}{R}$ (ingoing) and $K_{\Delta_{2}}(\vec{p}_2,\vec{x})$ is inserted at $\tau= \frac{\pi}{2}+\frac{u_2}{R}$ (outgoing). Thus, in the large $R$ limit, the leading term is given by
\begin{equation}
	\langle O_{\Delta_1}(\vec{p}_1)O_{\Delta_2}(\vec{p}_2) \rangle \simeq N^3_{\Delta_1} N^3_{\Delta_2} \int_{\mathbb{R}^{1,3}} d^{4}x \, \psi^-_{\Delta_1,\Tilde{q}_1,u_1}(x) \, \psi^+_{\Delta_2,\Tilde{q}_2,u_2}(x) .
\end{equation}
This evaluates to
\begin{equation}\label{result1}
	\langle O_{\Delta_1}(\vec{p}_1)O_{\Delta_2}(\vec{p}_2)\rangle = \mathcal{A} \frac{\delta^2(z_2-z_1)}{(i(u_2-u_1))^{\Delta_1+\Delta_2-2}} ,
\end{equation}
where $\mathcal{A}$ is a normalization constant. The result \eqref{result1} matches with the  Carrollian CFT two-point correlation function discussed in  \eqref{Sym-cor} up to normalisation.

\medskip

{\ding{112}} \underline{\em{Three point function}}

\medskip

Consider the contact AdS Witten diagram corresponding to a $\phi^3$ interaction in the bulk
\begin{equation}\label{eq:threepointflat}
	\begin{split}
		\langle O_{\Delta_1}(\vec{p}_1)O_{\Delta_2}(\vec{p}_2)O_{\Delta_3}(\vec{p}_3) \rangle= (i\mu)\int_{\text{AdS}_{4}}d^{4}\vec{x} \, K_{\Delta_{1}}(\vec{p}_1,\vec{x}) \, K_{\Delta_{2}}(\vec{p}_2,\vec{x}) \, K_{\Delta_{3}}(\vec{p}_3,\vec{x}) .
	\end{split}
\end{equation}
The insertion points are at $\vec{p}_1$ at $\tau=-\frac{\pi}{2}+ \frac{u_1}{R}$, $\vec{p}_2$ at $\tau=-\frac{\pi}{2}+ \frac{u_2}{R}$ and $\vec{p}_3$ at $\tau=\frac{\pi}{2}+\frac{u_3}{R}$. In the large $R$ limit,
\begin{equation}\label{eq:threepointfirst}
	\begin{split}
		\langle O_{\Delta_1}(\vec{p}_1)O_{\Delta_2}(\vec{p}_2)O_{\Delta_3}(\vec{p}_3) \rangle \simeq ~ &(i\mu) N^d_{\Delta_1}N^d_{\Delta_2}N^d_{\Delta_3}\int_{\mathbb{R}^{1,3}} \, d^{4}x \, \\ &\hspace{1cm}\times \psi^-_{\Delta_1,\Tilde{q}_1,u_1}(x) \,\psi^-_{\Delta_2,\Tilde{q}_2,u_2}(x)\,\psi^+_{\Delta_3,\Tilde{q}_3,u_3}(x) .
	\end{split}
\end{equation}
The result is proportional to $\delta^3(\omega_3 \tilde{q}_3 - \omega_1 \tilde{q}_1 - \omega_2 \tilde{q}_2)$. Thus following the discussion of \eqref{eq:threepointdelta}, we have
\begin{equation}
	\langle O_{\Delta_1}(\vec{p}_1)O_{\Delta_2}(\vec{p}_2)O_{\Delta_3}(\vec{p}_3) \rangle \simeq 0.
\end{equation}
This is because $\Tilde{q}_i \in \mathbb{R}^{1,3}$ and $\tilde{q}^2=0$. The three point function in the large $R$ limit evaluates to zero for generic momentum. This matches with the expected result \cite{Banerjee:2018gce}, where we also have a vanishing three point function.

\medskip

\eqref{eq:momcons2} will ensure that  \eqref{eq:threepointfirst} is non-zero. In fact, substituting  \eqref{eq:momcons2} in  \eqref{eq:threepointfirst}, we get the following result
\begin{equation}\label{eq:threepointcollinear}
	\begin{split}
	    \langle O_{\Delta_1}(\vec{p}_1)&O_{\Delta_2}(\vec{p}_2)O_{\Delta_3}(\vec{p}_3) \rangle \\ &= \mathcal{A}_{(3)}\delta^2(z_{12})\delta^2(z_{13}) \sum_{k=0}^{\Delta_3 -4} \dfrac{{}^{\Delta_3-4}C_k \, \Gamma(k+\Delta_1) \, \Gamma(\Delta_2+\Delta_3-k-4)}{(i(u_3-u_1))^{\Delta_1+k} \, (i(u_3-u_2))^{\Delta_2+\Delta_3-4-k}} .
	\end{split}
\end{equation}
Here, $\mathcal{A}_{(3)}$ is a normalization constant. The result is valid for $\Delta_3 \in \mathbb{N}$ and $\Delta_3 \geq 4$. This analysis can be extended to spinning particles, especially the scattering processes that involve photons and gravitons \cite{Bagchi:2023cen}. In \cite{Bagchi:2023cen}, it was also noticed that there is another channel where the three point delta function has a non-zero split (analogous to \eqref{eq:momcons2}): soft channel when one of the particles become soft.

\subsection{Comparison to other approaches}

In this subsection, our goal is to make a brief note comparing the approach prescribed in Sec.~\ref{ssec:flatpres} to the approach described in \cite{Alday:2024yyj}. \eqref{eq:largerlimit} basically followed from the argument of \cite{Giddings:1999jq}. Any point in AdS can be moved to the center of AdS by using the AdS isometries. In the global AdS picture, it is natural to rescale the global time and global radial coordinate by factors of the AdS radius to identify the center of AdS as a region where the curvature scale is negligible. This denotes the ``scattering region'' depicted by the causal diamond in fig.~\ref{fig:carrolliancase}. 

\medskip

The restriction of the insertion of the operators into two tiny dark green strips as shown in fig.~\ref{fig:carrolliancase} is a consequence of \cite{Penedones:2010ue} and the well established HKLL bulk reconstruction techniques \cite{Fitzpatrick:2011jn,Fitzpatrick:2011hu,Fitzpatrick:2011dm,Hijano:2019qmi,Hijano:2020szl,Li:2021snj}. The setup is clearly translationally invariant when the strips are at $\Delta \tau \sim \pi$. Since we have argued that the causal diamond exists in the center of AdS purely due to the symmetries, we can ask if the bulk operators within that ``scattering region'' could be reconstructed from the boundary CFT following \cite{Hamilton:2005ju,Hamilton:2006az,Hamilton:2006fh}. Even though the smearing region of the boundary operators is given by the light crossing time of the bulk point, that smearing region localizes to the tiny dark green strips around global time $\tau = \pm\frac{\pi}{2}$ \cite{Fitzpatrick:2011jn,Fitzpatrick:2011hu,Fitzpatrick:2011dm,Hijano:2019qmi,Hijano:2020szl,Li:2021snj}. Such a restriction is merely a consequence of fact that the smearing function of the HKLL bulk reconstruction localizes to the time bands around $\tau = \pm \frac{\pi}{2}$. There is significant compelling evidence that this is the case because one can build wavepackets in AdS smeared around $\tau = \pm \frac{\pi}{2}$ that scatter in the center of AdS \cite{Gary:2009ae} to reproduce the results of \cite{Penedones:2010ue}. The information around $\tau = \pm \frac{\pi}{2}$ is sufficient to ensure that the bulk point singularity \cite{Maldacena:2015iua} appears within the scattering region of $\mathcal{O}\left( \frac{1}{R} \right)$.

\medskip

The approach of \cite{Alday:2024yyj} basically blows up the casual diamond at the center of global AdS into two causal Poincar\'e patches. We will precisely argue this below. \cite{Alday:2024yyj} considers planar Bondi coordinates in AdS given by
\begin{equation}\label{eq:adsbondi}
	ds^2_{\text{AdS}} = -\dfrac{r^2_b}{R^2}du_b^2 - 2du_b \, dr_b + 2r^2_b dw \, d\bar{w} \, ,
\end{equation}
where $u_b$ and $r_b$ denote the planar AdS bondi time and radial coordinate respectively. The coordinate transformation relating \eqref{eq:adsbondi} to \eqref{eq:adsmetric} can be straightforwardly worked out to be
\begin{equation}\label{eq:coordtransf}
	\begin{split}
		r_b &= \dfrac{R \, \cos\tau}{\cos \rho} - R \tan \rho \, \Omega_3 \, , ~~~~~~~~
		u_b = \dfrac{R}{\left( \dfrac{ \cos\tau}{\cos \rho} - \tan \rho \, \Omega_3 \right)} \left[1 + \dfrac{\sin \tau}{\cos \rho} \right] \, ,\\
		w &= \dfrac{\tan \rho}{\left( \dfrac{ \cos\tau}{\cos \rho} - \tan \rho \, \Omega_3 \right)} \dfrac{\sqrt{2} z}{1+z\bar{z}} \, , ~~~~~
		\bar{w} = \dfrac{\tan \rho}{\left( \dfrac{ \cos\tau}{\cos \rho} - \tan \rho \, \Omega_3 \right)} \dfrac{\sqrt{2} \bar{z}}{1+z\bar{z}} \, , ~~~~~
		\Omega_3 = \dfrac{1-z \bar{z}}{1+z\bar{z}} \, .
	\end{split}
\end{equation}
To relate these coordinates to the one in the causal diamond, we implement the rescaling of \eqref{eq:largerlimit} in \eqref{eq:coordtransf}. This gives us
\begin{equation}\label{eq:blowup}
	r_b = R - r \dfrac{1-z\z}{1+z\z} \, , ~~~~~ u_b = R + u + \dfrac{2 r}{(1+z\z)} \, ,
\end{equation}
where $u$ is the Minkowski retarded time coordinate \eqref{eq:minku}. Clearly $R \to \infty$ in $\eqref{eq:blowup}$ implements $r_b \to \infty$ boundary limit in \eqref{eq:adsbondi}. From the explicit appearance of factors of AdS radius $R$, it is clear that Minkowski causal diamond in the center of global AdS \eqref{eq:adsmetric} is blown up in the planar coordinates \eqref{eq:adsbondi}. 

\medskip

From \eqref{eq:blowup}, it is clear that the dark green strips of fig.~\ref{fig:carrolliancase} or equivalently \eqref{tau} are simply blown up to extend to the whole Poincar\'e patch as $u_b \to \infty$. One can confirm this expectation by directly implementing the boundary limit $\rho \to \frac{\pi}{2}$ in \eqref{eq:coordtransf} to get 
\begin{equation}\label{eq:bdyurel}
	u_b(\text{bdy}) = \dfrac{R \, \sin \tau_{\text{bdy}}}{\cos \tau_{\text{bdy}} - \Omega_3} \, .
\end{equation}
Here $\text{bdy}$ is used to indicate that \eqref{eq:bdyurel} is a relation between the boundary coordinates of planar Bondi \eqref{eq:adsbondi} and spherical global AdS \eqref{eq:adsmetric}. To see the precise relation between the strips of \eqref{tau} and $u_b(\text{bdy})$, we substitute \eqref{tau} in \eqref{eq:bdyurel} to get
\begin{equation}
	u_b(\text{bdy}) = \dfrac{R}{- \frac{u}{R} - \Omega_3} \, .
\end{equation}
which generically diverges (see \eqref{eq:poincarerel} for a rescaled version). This essentially means that the strip around $\tau = \frac{\pi}{2}$ is blown up to cover the entire Poincar\'e patch. Thus, in particular, one is not losing information due to the restriction of \eqref{tau}. \eqref{eq:adsbondi} simply entails a different slicing of the global AdS.

\medskip

Indeed, this is a generic feature of the Poincar\'e coordinates which are given by
\begin{equation}\label{eq:adspoincare}
	ds^2_{\text{AdS}} = \dfrac{R^2}{\rho^2} \left( d\rho^2 - dx^2_0 + dx^2_1 + dx^2_2\right) \, .
\end{equation}
The relation between the boundary $x_0$ and $u_b(\text{bdy})$ is given by 
\begin{equation}\label{eq:poincarerel}
	x_0(\text{bdy}) = \dfrac{u_b(\text{bdy})}{R} = \dfrac{\sin\tau}{\cos\tau - \Omega_3} \, .
\end{equation}
For a finite global time $\tau$, one already can reach the boundary of the Poincar\'e coordinates. It is clear that if one has to get to flat Minkowski space from \eqref{eq:adspoincare}, one has to also rescale the Poincar\'e bulk coordinate $\rho$ as
\begin{equation}
	\rho = \exp \left( \dfrac{x_3}{R} \right) \, ,
\end{equation}
when taking $R \to \infty$. If one does not account for this exponential rescaling, the Minkowski diamond present in the deep interior of AdS gets blown up by factors of the AdS radius via relations like \eqref{eq:blowup} but the picture is equivalent and in fact conformally invariant. This completes the comparison of Sec.~\ref{ssec:flatpres} to \cite{Alday:2024yyj}.

\subsection{A note on stress tensors}
In \eqref{eq:tuurel} and \eqref{eq:tzzrel}, we saw how the stress tensor components encoded information about the soft operators in the bulk. To connect this discussion with the limit of AdS/CFT we have reviewed earlier in this section, it is useful to note that the flat limit could be considered as a high energy limit in which one zooms in on the flat region at the centre of AdS through \eqref{eq:largerlimit}. Thus, the boundary limit focuses on only the high energy sector of the original theory. In this limit, all finite energy excitations scale to zero and become soft. For instance, as in Penedones' formula \cite{Penedones:2010ue}, if we do not scale the energy (Mandelstam variables) with factors of AdS radius, the particles will end up becoming soft in this limit. One can understand this from the fig.~\ref{fig:graviton} which shows how the wavelengths get stretched out by a factor of the AdS radius.
\begin{figure}[tbp]
	\centering
    \hspace{1.5cm}
	\includegraphics[width=0.5\textwidth]{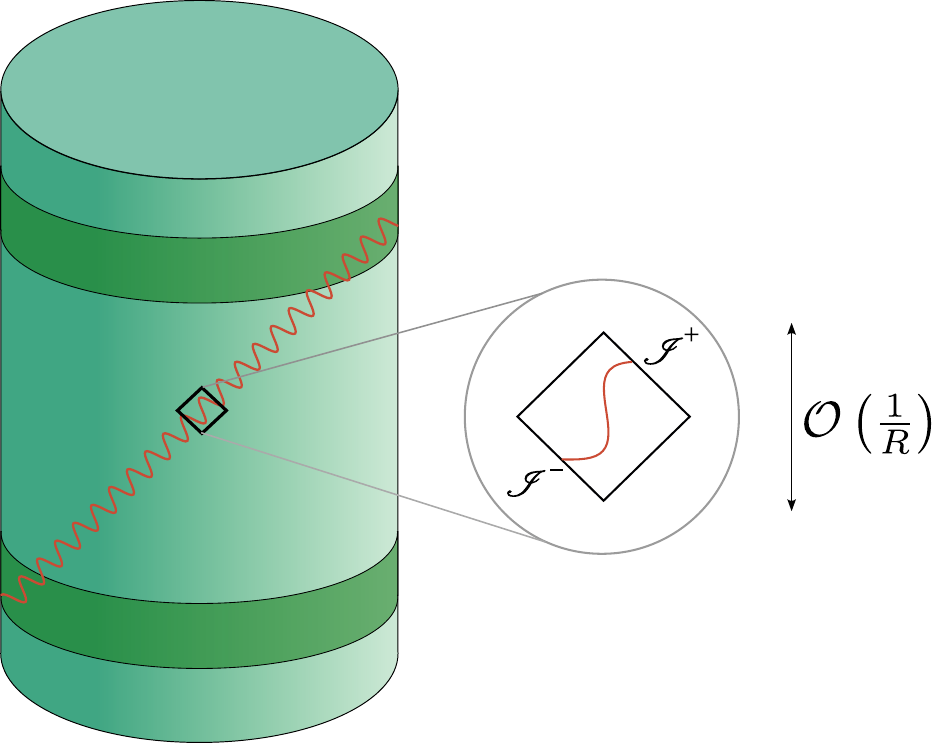}
	\caption{Soft scaling in the flat limit of AdS.}
    \label{fig:graviton}
\end{figure}

\medskip

When it concerns the stress tensor components of \eqref{eq:stresscomp}, there is a hierarchy of scales because $T^u_{\,~u}$ becomes soft faster than $T^u_{\,~i}$ in the $c \to 0$ limit. One can see this from the relativistic stress tensor of a massless scalar given by
\begin{equation}
    T^{\mu}_{\,\,~\nu} = \partial^{\mu} \phi \partial_{\nu} \phi - \dfrac{1}{2} \delta^{\mu}_{\,~\nu} \left( \partial_{\rho} \phi \, \partial^{\rho} \phi \right) .
\end{equation}
Thus, the usual Carroll limit $t \to \epsilon \, t$, $x^i \to x^i$ with $\epsilon \to 0$ implies $T^u_{\,~u} \sim \mathcal{O}(\epsilon^2)$ and $T^{u}_{\,~i} \sim \mathcal{O}(\epsilon)$. From the point of view of the large AdS radius limit (as $c = \frac{1}{R}$), it is natural that $T_u$ \eqref{eq:tuurel} encodes the leading soft theorem while $T_z$ \eqref{eq:tzzrel} encodes the sub-leading soft theorem. In AdS/CFT, the boundary stress tensor is dual to graviton fluctuations in the bulk. We see that this feature survives in the large AdS radius limit. Since all finite energy fluctuations become soft in this limit, the limiting Carroll boundary stress tensor precisely encodes these soft graviton fluctuations through \eqref{eq:tuurel} and \eqref{eq:tzzrel}.

\section{Other approaches to Carrollian holography}
\label{sec:otherapproaches}

As we reviewed in the previous section, historically, the approach to holography in asymptotically flat spacetimes has been from the point of view of large AdS radius limit of AdS/CFT. Though this ``top-down approach'' has been fruitful, it has so far not led us to an explicit dual description of the boundary degrees of freedom in the singular limit. 

\medskip

The recent literature has focused on the so-called ``bottom-up'' approach, mainly driven by symmetry principles. This approach was first initiated by de Boer and Solodukhin \cite{deBoer:2003vf}, who considered a hyperbolic slicing of Minkowski space in terms of Euclidean AdS and de-Sitter slices and applied AdS/CFT techniques to those slices. This resulted in a holographic reduction of the bulk fields along the slices in terms of an infinite family of conformal primaries located in the co-dimension two sphere. This served as a harbinger for the celestial CFT co-dimension two approach to flat space holography. This led to a host of interesting results concerning the infrared structure of asymptotically flat spacetimes. As mentioned before, we provide a brief summary of Celestial holography in Appendix \ref{ap:celreview}.  

\medskip

So far, our review has focussed on the Modified Mellin route to Carrollian holography based on \cite{Bagchi:2022emh, Bagchi:2023fbj, Bagchi:2023cen}. Keeping in tune with the major theme of our review article, in this section, we will provide an overview of alternate approaches within the Carrollian perspective on flat space holography. Some of other approaches we will review here in a bit of detail are 
\begin{itemize}
    \item Sourced Carrollian field theory \cite{Donnay:2022aba,Donnay:2022wvx},
    \item Source-less intrinsic field theory \cite{Saha:2022gjw,Saha:2023hsl,Saha:2023abr,Ruzziconi:2024kzo}, 
    \item Path integral and Carrollian partition function \cite{Kim:2023qbl,Kraus:2024gso,Kraus:2025wgi}. 
\end{itemize}
We will provide a more detailed list of references to other works in this direction in Sec.~{\ref{ssec:pointersotherholo}}. 

\subsection{Sourced Carrollian field theory}

We begin with the formulation of \cite{Donnay:2022aba,Donnay:2022wvx}, which along with \cite{Bagchi:2022emh}, has initiated a great flurry of activity in Carrollian holography. 

The goal in these works is to prescribe a boundary description that can suitably encode the leaky nature of gravity in AFS. For instance, it is well known in 4D  AFS that the $\mathfrak{bms}$ charges are not conserved owing to radiation reaching null infinity \cite{Bondi:1962px,Sachs:1962wk,Sachs:1962zza,Trautman:1958zdi,Wald:1999wa,Barnich:2011mi}, due to the null nature of $\mathscr{I}$. This is unlike the situation in AdS/CFT where one has standard conservative Dirichlet boundary conditions \cite{Witten:1998qj} \footnote{For an implementation of ``leaky'' boundary conditions in AdS, the reader is referred to \cite{Compere:2019bua,Compere:2020lrt}.}. The works \cite{Donnay:2022aba,Donnay:2022wvx} propose that one way to account for the radiation leaking through $\mathscr{I}$ is to couple the dual Carrollian CFT at $\mathscr{I}$ to external sources $\sigma_{(k,\bar{k})}$. These source operators can be related to Celestial conformal primaries $\mathcal{O}_{\Delta,J}$ and thereby providing a complementary view-point of the Celestial approach to the boundary dual description. 

\medskip

Since one of the main observables in flat space is the S-matrix, it is useful to quickly review the three different bases \cite{Donnay:2022sdg} in which we can express the S-matrix ($s$ denotes the spin):
\begin{enumerate}
    \item Position space basis: $\phi^{(s)\, \text{bdy}}(u,z,\z)$,
    \item Fourier space basis: $a^{(s)\, \text{out/in}}_{\pm} (\omega,z,\z)$,
    \item Mellin space basis: $a^{(s)\, \text{out/in}}_{\Delta,\pm}(z,\z)$.
\end{enumerate}
A massless, transverse spin $s$ field in the De-Donder gauge satisfying the Klein-Gordon equation can be mode expanded as
\begin{equation}\label{eq:fieldexppos}
    \phi^{(s)}_I(X) = \mathcal{K}_{(s)} \sum_{\alpha = \pm} \int \omega d\omega \, d^2z \,  \left( a^{(s)}_{\alpha}(\omega,z,\z) \varphi^{*\alpha}_I (\omega,z,\z|X) + a^{(s)}_{\alpha}(\omega,z,\z)^{\dagger} \varphi^{\alpha}_I (\omega,z,\z|X) \right) ,
\end{equation}
where $X$ is some arbitrary bulk point, $\mathcal{K}_{(s)}$ is a constant, $I$ denotes the symmetrized index $(\mu_1 \dots \mu_s)$, $\alpha = \pm$ denotes the polarization and 
\begin{equation}
    \varphi^{*\alpha}_I(\omega,z,\z) = \varepsilon^\alpha_{(\mu_1} \dots \varepsilon^{\alpha}_{\mu_s)}(z,\z) e^{i \omega q^{\mu} X_{\mu}} , 
\end{equation}
where we have used the parametrization  \eqref{eq:4dparamq} and the polarization vectors $\varepsilon^\alpha_{\mu}$ are given by
\begin{equation}
    \varepsilon^+_{\mu} = \partial_{z} q_{\mu} =\frac{1}{\sqrt{2}}(-\z,1,-i,-\z) , ~~~~~ \varepsilon^-_{\mu} = \partial_{\z} q_{\mu} = \frac{1}{\sqrt{2}}(-z,1,i,-z) .
\end{equation}
One can do the following Mellin transform of plane waves \cite{deBoer:2003vf,Pasterski:2017kqt,Pasterski:2017ylz}
\begin{equation}
    U^{*\alpha}_I(\Delta,z,\z|X) = \lim_{\epsilon \to 0^+} \varepsilon^{*\alpha}_I(z,\z) \int_0^{\infty} d\omega \, \omega^{\Delta-1} e^{i \omega q^{\mu}X_{\mu} - \epsilon \omega} = \lim_{\epsilon \to 0^+} \varepsilon^{*\alpha}_I(z,\z) \frac{i^{\Delta} \Gamma(\Delta) }{(q^{\mu}X_{\mu} + i \epsilon)^{\Delta}} .
\end{equation}
The demand of a delta-function normalizable basis requires $\Delta= 1 + i \, \nu$, i.e., $\Delta$ lies on the principal continuous series \cite{Pasterski:2017kqt}. The field expansion in the Mellin basis takes the following form
\begin{equation}
    \phi^{(s)}_I(X) = \mathcal{K}_{(s)} \int d\nu \, d^2 z \, \left( a^{(s)}_{2-\Delta,\alpha}(z,\z) \, U^{*\alpha}_I(\Delta,z,\z|X) + a^{(s)}_{2-\Delta,\alpha}(z,\z)^{\dagger} \, U^{\alpha}_I(\Delta,z,\z|X) \right) ,
\end{equation}
where the creation/annihilation operators are in the Mellin basis
\begin{equation}
    a^{(s)}_{\Delta,\alpha}(z,\z) = \int^{\infty}_0 d\omega \, \omega^{\Delta-1} \, a^{(s)}_{\alpha}(\omega,z,\z) .
\end{equation}
Given these two bases, we can construct boundary operators by doing a large-$r$ expansion by expressing the bulk flat spacetime in Bondi coordinates. One can show that the asymptotic behaviour near $\mathscr{I}^+$ is encoded in the following boundary value field
\begin{equation}\label{eq:bdyfieldfourier}
    \phi^{(s)\, \text{bdy}}_{z \dots z}(u,z,\z) = - 2i \mathcal{K}_{(s)} \int_0^{\infty} d\omega \, \left( a^{(s)}_+(\omega,z,\z) \, e^{-i \omega u} - a^{(s)}_-(\omega,z,\z)^{\dagger} \, e^{i \omega u} \right) ,
\end{equation} 
and its conjugate $\phi^{(s) \, \text{bdy}}_{\z \dots \z}(u,z,\z)$. This gives us the standard relation between position space basis and Fourier space basis. One can show that this extrapolated boundary field transforms as a conformal Carrollian primary under the $\mathfrak{bms}$ group with weights
\begin{equation}\label{eq:extrapolatek}
    k = \frac{1+J}{2} , ~~~~ \bar{k} = \frac{1-J}{2} .
\end{equation}
The above weights are purely derived for single-particle states and in general, we expect a different relation for multi-particle states \cite{Kulp:2024scx}. Now, the relation between the position space operators $\phi^{(s) \, \text{bdy}}(u,z,\z)$ and the Mellin space basis operators $a^{s}_{\Delta,\alpha}(z,\z)$ is clear:
\begin{equation}\label{eq:amellinbasis}
    a^{(s)}_{\Delta,+}(z,\z) = \frac{i}{4\pi^2 \mathcal{K}_{s}} \lim_{\epsilon \to 0^+} \int_0^{\infty} d\omega \, \omega^{\Delta-1} \int_{-\infty}^{+\infty} du \, e^{i\omega u - \epsilon \omega} \phi^{(s) \, \text{bdy}}_{z \dots z}(u,z,\z) . 
\end{equation}
The above relation is simply the Mellin transform of the inverse Fourier transform of  \eqref{eq:bdyfieldfourier}:
\begin{equation}
    a^{(s)}_+(\omega,z,\z) = \frac{i}{4\pi \mathcal{K}_{(s)}} \int_{-\infty}^{+\infty} du \, e^{i\omega u} \, \phi^{(s) \, \text{bdy}}_{z \dots z}(u,z,\z) .
\end{equation}
Thus, the map from position basis to Mellin basis is clear from  \eqref{eq:amellinbasis}:
\begin{equation}\label{eq:btransform}
    a^{(s)}_{\Delta,+}(z,\z) = \frac{1}{4\pi \mathcal{K}_{(s)}} i^{\Delta+1} \Gamma(\Delta) \, \lim_{\epsilon \to 0^+} \int_{-\infty}^{+\infty} du \, (u+i\epsilon)^{-\Delta} \, \phi^{(s) \, \text{bdy}}_{z \dots z}(u,z,\z) .
\end{equation}

\medskip

The maps described above are useful in translating between bulk and boundary statements. However, the bulk allows for non-trivial radiation to reach $\mathscr{I}$. For instance, the BMS charges given by \cite{Compere:2020lrt,Campiglia:2020qvc,Freidel:2021dfs,Donnay:2021wrk}
\begin{equation}\label{eq:bmscharges}
    Q_{\xi} = \frac{1}{16 \pi G} \int d^2z \, \left( 4 f(z,\z) \, \bar{M} + 2 Y^A(z/ \z) \, \bar{N}_A \right) ,
\end{equation}
where $\bar{M}$ and $\bar{N}_A$ are expressed in terms of the shear $C_{AB}$ and the news tensor $N_{AB}$
\begin{equation}
    \begin{split}
        \bar{M} &= M + \dfrac{1}{8} N_{AB} C^{AB} , \\
        \bar{N}_A &= N_A - u \partial_A \bar{M} + \dfrac{1}{4} C_{AB} \partial_C C^{BC} + \dfrac{3}{32} \partial_A \left( C_{BC} C^{BC} \right) \\
        & ~~~~~~ + \dfrac{u}{4} \partial^B \left[ (\partial_B \partial_C - \frac{1}{2} N_{BC} ) C_A^{~C} \right] - \dfrac{u}{4} \partial^B \left[ (\partial_A \partial_C - \frac{1}{2} N_{AC} ) C_B^{~C} \right] ,
    \end{split}
\end{equation}
are not conserved. This is because, one can use Einstein's equations coupled with some null matter stress tensor to show that the charges satisfy the following flux equation:
\begin{equation}\label{eq:bmsflux}
    \dfrac{d Q_{\xi}}{d u} = \int d^2 z \, F_{\xi(f,Y)},
\end{equation}
with
\begin{equation}
    \begin{split}
        F_{\xi(f,0)} &= \dfrac{1}{16\pi G} f(z,\z) \left[ \partial^2_z N_{\z\z}  + \frac{1}{2} C_{\z\z} \partial_u N_{zz} + \text{c.c.} \right] - f(z,\z) \,T^{m(2)}_{uu} , \\
        F_{\xi(0,Y)} &= \dfrac{1}{16\pi G} Y^z(z) \left[-u \partial^3_z N_{\z\z} + C_{zz} \partial_z N_{\z\z} - \frac{u}{2} \partial_z C_{zz} \partial_u N_{\z\z} - \frac{u}{2} C_{zz} \partial_z \partial_u N_{\z\z} \right] \\
        & ~~~~~ - Y^z(z) \, T^{m(2)}_{uz} + \frac{u}{2} Y^z \partial_z T^{m(2)}_{uu} + \text{c.c.} ,
    \end{split}
\end{equation}
where the matter stress tensor components arise from the large $r$ expansion near $\mathscr{I}$ as follows
\begin{equation}
    \begin{split}
        T^m_{uu}(u,r,z,\z) &= \dfrac{T^{m(2)}_{uu}(u,z,\z)}{r^2} + \mathcal{O}(r^{-3}) , \\
        T^m_{uA}(u,r,z,\z) &= \dfrac{T^{m(2)}_{uA}(u,z,\z)}{r^2} + \mathcal{O}(r^{-3}) .
    \end{split}
\end{equation}
One can do a careful radiative phase analysis of the flux equation  \eqref{eq:bmsflux} to split the hard and soft degrees of freedom \cite{Campiglia:2020qvc,Campiglia:2021bap} to show that the soft sector encodes the leading and sub-leading soft graviton theorems \cite{He:2014cra,Kapec:2014opa} as reviewed in Appendix \ref{ap:celreview}. Thus, the issue lies in the question of how to incorporate external radiation from an intrinsic field theoretic perspective of the dual boundary theory at $\mathscr{I}$.

\medskip

The central theme of \cite{Donnay:2022aba,Donnay:2022wvx} is to consider a sourced quantum Carrollian field theory at $\mathscr{I}$ to adequately incorporate radiation leaking to $\mathscr{I}$. Consider a field theory with fields $\Phi^j$ having global symmetries of the form $\delta_K \Phi^i = K^i[\Phi]$ and corresponding conserved currents $j^c_K$. If we have external sources $\sigma^m(x)$ (additional non-dynamical fields), then the symmetry of the action is lost. However, following \cite{Donnay:2022wvx}, we take the following action 
\begin{equation}\label{eq:gensym}
    \delta_K \Phi^i = K^i[\Phi|\sigma] , ~~~~~~ \delta_K \sigma^m = K^m[\sigma] ,
\end{equation}
to be a symmetry of the fully sourced equations of motion
\begin{equation}
    \dfrac{\delta S}{\delta \Phi^i} = 0 \implies \delta_K \left( \dfrac{\delta S}{\delta \Phi^i} \right) = 0 .
\end{equation}
Since the action is not preserved, the usual Noether current conservation equation can be expressed as
\begin{equation}\label{eq:sourceclasscons}
    \partial_c j^c_K = K^i \dfrac{\delta S}{\delta \Phi^i} + F_K \implies \partial_c j^c_K = F_K ~~(\text{on-shell}) ,
\end{equation}
where $F_K$ is the flux proportional to the terms in the Lagrangian that change under \eqref{eq:gensym}. The path integral for the fully sourced quantum field theory is \footnote{The formalism is valid in general, and we simply consider a three-dimensional field theory at $\mathscr{I}$.}
\begin{equation}
    Z[J_i,J_m] = \int [\mathcal{D} \Phi]_{\sigma} [\mathcal{D} \sigma] \, \exp \frac{i}{\hbar} \left( S[\Phi|\sigma] + \int du \, d^2z \, (J_i \Phi^i + J_m \sigma^m ) \right) .
\end{equation}
Thus, one can straightforwardly see that the conservation equation  \eqref{eq:sourceclasscons} inside correlation functions would lead to Ward identities of the form
\begin{equation}\label{eq:sourcedward}
    \begin{split}
        \dfrac{\partial}{\partial x^c} \langle j^c_K(x) \, X^{\Phi}_n  \rangle &= \dfrac{\hbar}{i} \sum_{k=1}^n \delta^{(3)}(x-x_k) \delta_{K^{i_k}} \langle X^{\Phi}_n \rangle + \langle F_K(x) \, X^{\Phi}_n \rangle , \\
        \dfrac{\partial}{\partial x^c} \langle j^c_K(x) \, X^{\sigma}_n  \rangle &= \langle F_K(x) \, X^{\sigma}_n \rangle ,
    \end{split}
\end{equation}
where $X^{\Phi}_n \equiv \Phi^{i_1}(x_1) \dots \Phi^{i_n}(x_n)$ and $X^{\sigma}_n$ denoting a string of $\sigma$ fields. For a conformal Carrollian field theory, the fields $\Phi^i$ and $\sigma^m$ transform as conformal Carroll primaries  \eqref{BMS-primary}. Then, the classical expression of the conserved Noether current is given by
\begin{equation}\label{eq:carstresstensor}
    j^c_{\xi} = \mathcal{C}^c_{~\,d} \xi^d , \quad \text{where} \quad \mathcal{C}^c_{~\,d} = \begin{pmatrix}
        \mathcal{M} & \mathcal{N}_D \\
        \mathcal{B}^C & \mathcal{A}^C_{~\,D}
    \end{pmatrix} .
\end{equation}
where $\mathcal{C}^c_{~\, d}$ is the Carrollian stress tensor.  
Insertion of these components into correlators would lead to Ward identities of the form  \eqref{eq:sourcedward}. If we now demand that the $\mathfrak{bms}$ charges of  \eqref{eq:bmscharges} correspond to Noether currents  \eqref{eq:carstresstensor} integrated on a constant $u$ slice, we get the following map between Carrollian momenta and gravitational data:
\begin{equation}\label{eq:gravstresstensor}
    \begin{split}
        \langle \mathcal{M} \rangle &= \dfrac{M}{4 \pi G} , ~~~~~~~~~~~~ \langle \mathcal{A}^C_{~\,D} \rangle + \dfrac{1}{2} \delta^C_{~\, D} \langle \mathcal{M} \rangle = 0 , \\
        \langle \mathcal{N}_A \rangle &= \dfrac{1}{8 \pi G} \Bigg( N_A + \dfrac{1}{4} C_{AB} \partial_C C^{BC} + \dfrac{3}{32} \partial_A \left( C_{BC} C^{BC} \right) \\
        & ~~~~~~ + \dfrac{u}{4} \partial^B \partial_B \partial_C C_A^{~C}  - \dfrac{u}{4} \partial^B \partial_A \partial_C C_B^{~C} \Bigg) .
    \end{split}
\end{equation}
In this case, the Bondi news tensor $N_{AB}$ is the free data that encodes the gravitational radiation reaching $\mathscr{I}$ which would suggest the following identification of the sources in  \eqref{eq:gensym}
\begin{equation}
    \sigma_{AB} = C_{AB} ,
\end{equation}
which is the asymptotic shear.

\medskip

In standard AdS/CFT, the normalizable mode will source a boundary operator through extrapolation \cite{Witten:1998qj}.  \eqref{eq:gravstresstensor} is reminiscent of this feature but for asymptotically flat spacetimes. Taking a cue from  \eqref{eq:gravstresstensor}, it is now natural to identify the boundary value of the fields  \eqref{eq:bdyfieldfourier} with the sources $\sigma^m$ of the dual Carrollian field theory i.e., $\sigma_{(k_i,\bar{k}_i)} \equiv \phi^{(s) \, \text{bdy}}_{z \dots z}$ \footnote{Depending on whether we consider $\mathscr{I}^+$ or $\mathscr{I}^-$, we get ``$\text{out}$'' or ``$\text{in}$'' conformal primaries.}. Thus, if we invert the map of  \eqref{eq:bdyfieldfourier}, we can define a map between the position space correlators of $\sigma^m$ and the S-matrix elements constructed from in/out ladder operators:
\begin{equation}\label{eq:delta1dictionary}
    \begin{split}
        &\langle \sigma^{\text{out}}_{k_1,\bar{k}_1}(x_1) \dots \sigma^{\text{out}}_{k_n,\bar{k}_n}(x_n) \sigma^{\text{in}}_{k_{n+1},\bar{k}_{n+1}}(x_{n+1}) \dots \sigma^{\text{in}}_{k_N,\bar{k}_N}(x_N) \rangle \\
        &= \dfrac{1}{(2\pi)^N} \prod_{k=1}^n \int d \omega_k \, e^{-i\omega_k u_k} \prod_{l=n+1}^N \int_{0}^{\infty} d\omega_l \, e^{i\omega_l v_l} \mathcal{S}_N(p_1,\dots,p_N) .
    \end{split}
\end{equation}
For example, the two-point free particle scattering amplitude from $\mathscr{I}^-$ to $\mathscr{I}^+$ is
\begin{equation}
    \mathcal{S}_2(p_1,p_2) \propto \dfrac{\delta(\omega_1 - \omega_2)}{\omega_1} \delta^{(2)}(z_1-z_2) \delta_{\alpha_1,\alpha_2} ,
\end{equation}
where the $\alpha_i$s denote the helicities. After suitably regulating the $\omega$ integral with an IR cut-off, we find
\begin{equation}
    \begin{split}
        \langle &\sigma^{\text{out}}_{k_1,\bar{k}_1}(u,z_1,\z_1) \sigma^{\text{in}}_{k_2,\bar{k}_2}(v,z_2,\z_2)  \rangle \\ & \propto \lim_{\beta \to 0^+}\left( \dfrac{1}{\beta} -\gamma - \ln|u-v|- \dfrac{i \pi}{2}\text{sign}(u-v) \right) \delta^{(2)}(z_1-z_2) \delta_{k^+_{12},2} \, \delta_{k^-_{12},0}, 
    \end{split}
\end{equation}
where $k^{\pm}_{12} \equiv \sum_i (k_i \pm \bar{k}_i)$. This expression is consistent with the conformal Carroll Ward identities once we integrate the sourced Ward identities  \eqref{eq:sourcedward} on all of $\mathscr{I}$ \footnote{See \cite{Donnay:2022wvx} for details of an antipodal identification of spheres $\mathscr{I}^+_-$ and $\mathscr{I}^-_+$ which glues $\mathscr{I}^+$ and $\mathscr{I}^-$ to a single $\mathscr{I} = \mathscr{I}^+ \cup \mathscr{I}^-$. This identification crucially kills the contribution of $\mathscr{I}^+_-$ and $\mathscr{I}^-_+$ to the integrated Ward identity.}:
\begin{equation}
    \Bigg\langle \left( \int_{\mathscr{I}}du \, d^2z \,F_{\bar{\xi}}(x) - \int_{\mathscr{I}^+_+} d^2z \, j^u_{\bar{\xi}}(u \to +\infty,z,\z) + \int_{\mathscr{I}^-_-} d^2z \, j^u_{\bar{\xi}}(u \to -\infty,z,\z)  \right) X^{\sigma}_n\Bigg\rangle = 0 .
\end{equation}
$\bar{\xi}$ is some arbitrary conformal Carroll vector of the form \eqref{eq:carkilling}
\begin{equation}
    \bar{\xi} = \left( f(z,\z)+ \dfrac{u}{2}(\partial_z Y^z + \partial_\z Y^\z) \right) \partial_u + Y^z \partial_z + Y^\z \partial_\z .
\end{equation}
If we assume that the scattering process involves only massless particles, then the contribution from $\mathscr{I}^+_+$ and $\mathscr{I}^-_-$ drops out to imply \footnote{The $u$ integral is explicitly written just for the sake of clarity.The integration over $\mathscr{I}$ would involve the integration over $\mathscr{I}^+$ ($u$) and $\mathscr{I}^-$ ($v$).}
\begin{equation}\label{eq:fluxnomassive}
    \langle \mathcal{F}_{\bar{\xi}} X^{\sigma}_n \rangle = 0 \, , ~~~~~ \mathcal{F}_{\bar{\xi}} = \int_{\mathscr{I}} du \, d^2z \, F_{\bar{\xi}}(x).
\end{equation}

\medskip

One can show that the Flux operator in  \eqref{eq:fluxnomassive} generates a transformation under $\bar{\xi}$ in the asymptotic phase space \cite{Donnay:2022wvx}. In fact, the hard and soft contributions of the fluxes act independently in the phase space. We thus have,
\begin{equation}
    \langle \mathcal{F}_{\bar{\xi}} X^{\sigma}_n \rangle = 0 \iff \delta_{\bar{\xi}} \langle X^{\sigma}_n\rangle = 0,
\end{equation}
which is nothing but the invariance of the correlators under the conformal Carroll symmetries studied in \cite{Bagchi:2009ca,Chen:2021xkw,Bagchi:2019xfx,Bagchi:2022emh}. As expected \cite{Chen:2021xkw,Bagchi:2022emh}, the demand of conformal Carroll invariance for a two-point correlator would lead to two branches of solutions to the Ward identities: time-independent CFT branch correlator
\begin{equation}
    \langle X_2 \rangle \propto \dfrac{1}{z^{k_1+k_2}_{12} \z^{\bar{k}_1 + \bar{k}_2}_{12}} \delta_{k_1,k_2} \delta_{\bar{k}_1,\bar{k}_2},
\end{equation}
and the time-dependent branch $(2+1)D$ correlator
\begin{equation}\label{eq:tdep2ptgen}
    \langle X_2 \rangle \propto \dfrac{\delta^2(z_{12})}{u^{k^+_{12}-2}_{12}} \delta_{k^-_{12},0}.
\end{equation}
In the limit $k^+_{12} \to 2$,  \eqref{eq:tdep2ptgen} becomes
\begin{equation}\label{eq:tdep2pt1}
    \langle X_2 \rangle \propto \left( \dfrac{1}{k^+_{12} - 2} - \gamma - \ln|u_{12}| \right) \delta^{(2)}(z_{12}) \, \delta_{k^-_{12},0} ,
\end{equation}
which is clearly IR divergent. If $k^+_{12} = 2 + n$ for $n \in \mathbb{N}$, then one finds a discrete set of distributional solutions to the Ward identities of the form
\begin{equation}\label{eq:tdep2pt2}
    \langle X_2 \rangle  \propto \dfrac{d^n}{du^n_1} \text{sign}(u_{12}) \delta^{(2)}(z_{12}) \delta_{k^-_{12},0} \, \delta_{k^+_{12},2+n} , ~~~~~ \forall n \in \mathbb{N} .
\end{equation}
Thus, it has been proposed that the dictionary  \eqref{eq:delta1dictionary} for the extrapolated bulk fields in general is a linear combination of  \eqref{eq:tdep2pt1} and  \eqref{eq:tdep2pt2}:
\begin{equation}
    \langle X_2 \rangle \propto \left( \dfrac{1}{k^+_{12} - 2} - \gamma - \ln|u_{12}| - \dfrac{i\pi}{2} \text{sign}(u_{12}) \right) \delta^{(2)}(z_{12}).
\end{equation}
This analysis indeed hints that the sourced sector of the Carrollian CFT encodes the hard scattering in the bulk.

\medskip

Finally, the sourced sector of the Carrollian CFT can be related to the Celestial CFT through the integral transform  \eqref{eq:btransform} dubbed $\mathcal{B}-$transform. Since we have related the boundary value of the extrapolated bulk field to the sources of the dual Carrollian field theory, we have the following relation between the Celestial operator insertions and Carrollian source operators via  \eqref{eq:btransform}:
\begin{equation}
    \mathcal{O}^{\text{out}}_{(\Delta_i,J_i)}(z_i,\z_i) \propto \lim_{\epsilon \to 0^+} \int_{-\infty}^{+\infty} \dfrac{d u_i}{(u_i+i\epsilon)^{\Delta_i}} \sigma^{\text{out}}_{k_i,\bar{k}_i}(u_i,z_i,\z_i) .
\end{equation}
Using the above transform in the sourced Carrollian ward identities  \eqref{eq:sourcedward} for Electromagnetic theory and gravity, one can derive the Celestial Ward identities for the soft photon and soft graviton theorems.

\subsection{Source-less intrinsic field theoretic approach}

Unlike the above approach, which relied on the input from the bulk theory, this intrinsic approach \cite{Saha:2022gjw,Saha:2023hsl,Saha:2023abr,Ruzziconi:2024kzo} purely works with constraints imposed by the symmetries on the quantum fields without recourse to coupling the field theory to arbitrary sources motivated from the bulk. In particular, \cite{Saha:2023hsl} showed that the leading and sub-leading soft graviton theorems in 4D  AFS arise from stress tensor Ward identities of a 3D pure, source-less Carrollian conformal field theory. To see this, consider the action of a Carrollian field theory in 3 dimensions:
\begin{equation}
    S[\Phi] = \int du \, d^2z \, \mathcal{L}(\Phi, \partial_{\mu} \Phi) .
\end{equation}
One can obtain the stress tensor from the standard Noether procedure. From the action of Carroll boosts, rotations, and Weyl invariance, the resulting canonical stress tensor can be Belinfante-improved to $T^{\mu}_{~\nu}$ satisfying
\begin{equation}
    T^{i}_{~u} = 0 , ~~~~~~~ T^{ij} = T^{ji}, ~~~~~~~ T^{\mu}_{~\mu} = 0.
\end{equation}
If we can define Carrollian primary fields purely based on $\mathscr{I}^+$ or $\mathscr{I}^-$ alone, then one can define correlation functions for a string of primaries $X = \Phi_1(x_1) \cdots \Phi_n(x_n)$ in the usual way:
\begin{equation}
    \langle X \rangle \equiv \langle \Phi_1(x_1) \cdots \Phi_n(x_n)\rangle = \dfrac{\int [\mathcal{D} \Phi] \, \Phi_1(x_1) \cdots \Phi_n(x_n) \, e^{i S[\Phi]} }{\int [\mathcal{D} \Phi] \, e^{i S[\Phi]}} .
\end{equation}
For such correlation functions, one can derive the following Ward identity corresponding to global translations as follows:
\begin{equation}\label{eq:transward}
    \partial_{\mu} \langle T^{\mu}_{~\nu}(x) \, X \rangle = -i \sum_{m=1}^{n} \partial_{\nu_m} \langle X \rangle \, \delta(u-u_m) \, \delta^2(z-z_m).
\end{equation}
For the sake of simplicity, we assume that the fields don't transform under the Carroll boosts, which implies $\langle T^i_{~u}(x) \, X\rangle = 0$ (the general case is worked out in \cite{Saha:2023hsl}). Thus, from  \eqref{eq:transward}, 
\begin{equation}\label{eq:supertransward}
    \partial_u \langle T^u_{~u}(x)\, X \rangle = -i \sum_{m=1}^n \delta(u-u_m) \, \delta^2(z-z_m) \partial_{u_m} \langle X \rangle .
\end{equation}

\medskip

To solve  \eqref{eq:supertransward}, \cite{Saha:2023hsl,Ruzziconi:2024kzo} use the following initial condition
\begin{equation}
    \lim_{u \to -\infty} \langle T^u_{~u}(x) \, X \rangle = 0 ,
\end{equation}
which would imply a solution of the form
\begin{equation}\label{eq:tuuward}
    \langle T^u_{~u}(x) \, X \rangle = - i \sum_{m=1}^n \theta(u-u_m) \, \delta^2(z-z_m) \partial_{u_m} \langle X \rangle.
\end{equation}
The $\theta(u-u_m)$ is the temporal step function that arises from integrating out  \eqref{eq:transward}, which corresponds to the memory effect of the super-translated vacua. Using $T^u_{~u}$, we can define a new operator \footnote{$+$ denotes a positive helicity graviton. If we choose to treat $S^+_0$ as a local field, then $S^-_0$ would be a non-local shadow transformation of $S^+_0$. Thus, one can only treat one of the helicities as a local field \cite{Banerjee:2022wht}.}
\begin{equation}\label{eq:gravopsoft1}
    S^+_0 (u,z,\z) = \int d^2z' \dfrac{\z-\z'}{z-z'} \, T^u_{~u}(u,z',\z'),  ~~~~~ \partial_u S^+_0 = 0 .
\end{equation}
This operator in the Ward identity would imply
\begin{equation}\label{eq:softward1}
    \langle S^+_0 (u,z,\z) \, X \rangle = -i \sum_{m=1}^n \theta(u-u_m) \dfrac{\z-\z_m}{z-z_m} \partial_{u_m} \langle X \rangle .
\end{equation}
This precisely corresponds to Weinberg's leading soft graviton theorem when $u \to \infty$ \footnote{As $\lim_{u \to \infty} \theta(u-u_m) = 1$.} written in a null parametrization for the momenta of the scattering particles. Even though $S^+_0$ is the bulk soft graviton operator, from  \eqref{eq:gravopsoft1}, it is clear that it is constructed purely from the intrinsic field theoretic stress tensor. The vanishing of the RHS of  \eqref{eq:tuuward} whenever $x \neq x_m$ would imply that $T^u_{~u}$ is a null state from the point of view of the 3D Carrollian CFT. It is unclear what a component of the stress tensor being a null state would imply for a quantum field theory.  

\medskip

Analogous to  \eqref{eq:softward1}, one can show that the super-rotation Ward identity is equivalent to the sub-leading soft graviton theorem with the sub-leading soft operator given by
\begin{equation}\label{eq:gravopsoft2}
    S^+_1(u,z,\z) = \dfrac{1}{2} \int d^2z' \, \dfrac{(\z-\z')^2}{z-z'} \left[ T^u_{~\z}(u,z',\z') + \int_{u_0}^u du' \, \partial_{z'} T^z_{~\z}(u',z',\z') \right] .
\end{equation}
It is useful to note that simply inserting $S^+_0$ and $S^+_1$ (\eqref{eq:gravopsoft1} and  \eqref{eq:gravopsoft2}) into the a correlator and performing a Fourier transform wouldn't be enough; one has to perform a $\omega \to 0$ limit as well. The Ward identity corresponding to $S^+_1$ will also have temporal step functions analogous to  \eqref{eq:softward1}. Due to the presence of these temporal step functions, one has a 2D CFT-like relation between the operator commutation relations and OPEs. Since there is no notion of radial quantisation, one has to introduce a ``$j\epsilon$-prescription'' to incorporate the temporal step function in the OPE statements \cite{Saha:2022gjw}. Given these OPE statements, \cite{Saha:2023abr} showed that it is possible to encode the subsubleading soft graviton theorem through an insertion of an operator $S^+_2$ provided the following OPE is satisfied
\begin{equation}\label{eq:s2consis}
    (\partial_u S^+_2 - S^+_1)(u,z,\z) \Phi(u',z',\z') \sim 0 .
\end{equation}
Given $S^+_2$, the consistency of $S^+_2 S^+_2$ OPE would demand the existence of another field $S^+_3$ satisfying  \eqref{eq:s2consis} with $S^+_2$ replaced by $S^+_3$ and $S^+_1$ replaced by $S^+_2$. One can repeat the argument to generate a whole tower of fields $S^+_k$ for $k \geq 3$ \cite{Saha:2023abr}. From the OPEs of the tower of fields, one can extract a current algebra precisely given by the (wedge sub-algebra) of $w_{1+\infty}$ algebra \cite{Pope:1991ig} consistent with the 2D Celestial result \cite{Strominger:2021mtt,Ball:2021tmb,Banerjee:2023zip}. 

\medskip

In fact, from the $S^+_k S^+_l$ OPE, one could obtain the Celestial OPE of two conformally soft primary gravitons in the bulk linearized Einstein gravity \cite{Pate:2019lpp,Guevara:2021abz}. This $u \to \infty$ limit was investigated further in \cite{Ruzziconi:2024kzo}, where they showed that the $u \to \infty$ limit of Carrollian CFT stress tensor Ward identities precisely match the leading and sub-leading soft graviton theorems in the bulk. As the stress tensor components in \cite{Ruzziconi:2024kzo} were related to the radiative modes of the graviton in the bulk, the dual Carrollian description should be thought of as describing only the radiative sector of the bulk theory. In contrast, in the sourced approach \cite{Donnay:2022aba,Donnay:2022wvx} described above, the stress tensor was constructed to describe the non-radiative sector in the bulk. The relation between stress tensor components and the radiative modes was non-local in $u$, and in order to make them local, one has to introduce twistor potentials in twistor space. Using this construction, it was shown that the infinite-dimensional current algebra derived in \cite{Saha:2023abr} encodes the collinear scattering of hard gravitons in the bulk. Basically, one demands the OPE statement of  \eqref{eq:s2consis} be satisfied exactly for all $k \geq 0$:
\begin{equation}
    \partial_u S^+_{k+1}(u,z,\z) = S^+_k(u,z,\z) .
\end{equation}
This would imply that the infinite tower of currents are ``primary-descendants'' (in the language of \cite{Banerjee:2020kaa}) of $S^+_0$ which encodes the leading soft graviton theorem through  \eqref{eq:softward1}. One could then argue that the OPE of $S^+_0$ at finite $u$ encodes the collinear splitting property of hard gravitons.

\subsection{Path integral approach to the S-matrix}

In this approach \cite{Kim:2023qbl,Jain:2023fxc,Kraus:2024gso,Kraus:2025wgi}, the main idea is to consider the S-matrix as a Path integral over field configurations that asymptote to plane waves as we reach $\mathscr{I}^{\pm}$. This was first conceived by Arefeva, Faddeev, and Slavnov \cite{Arefeva:1974jv}. Suppose we consider a scalar field action
\begin{equation}\label{eq:fullactphi}
    S[\phi,\tilde{\phi}] = \int d^4x \, \left( \frac{1}{2} \partial_{\mu} \phi \partial^{\mu} \phi - V(\phi) \right) + S_{\text{bdy}}[\phi,\tilde{\phi}],
\end{equation}
where $\tilde{\phi}$ encodes the appropriate boundary conditions, then near $\mathscr{I}^{\pm}$ in the standard Bondi gauge, the field $\phi \sim \frac{1}{r} \phi_1$ takes the following form:
\begin{equation}\label{eq:phiasymptotics}
    \phi(x) \approx \begin{cases}
        \frac{1}{r} \tilde{\phi}^-_1(u,z,\z) +\text{positive frequency} ~~~ \text{on} ~ \mathscr{I}^+ , \\
        \frac{1}{r}\tilde{\phi}^+_1(v,z,\z) +\text{negative frequency} ~~~ \text{on} ~ \mathscr{I}^- ,
    \end{cases}
\end{equation}
where ``positive frequency'' means the Fourier expansion contains $e^{-i \omega t}$ with $\omega>0$. Thus, the $\pm$ of the asymptotics  \eqref{eq:phiasymptotics} denotes the appropriate frequency. The boundary term in the full action  \eqref{eq:fullactphi} can be deduced from demanding the variations that satisfy $\delta \tilde{\phi}^-_1|_{\mathscr{I}^+} = \delta \tilde{\phi}^+_1|_{\mathscr{I}^-} = 0$:
\begin{equation}\label{eq:bdyterm}
    S_{\text{bdy}}[\phi,\tilde{\phi}] = (\tilde{\phi}^-,\phi)_{\mathscr{I}^+} - (\tilde{\phi}^+,\phi)_{\mathscr{I}^-},
\end{equation}
where,
\begin{equation}\label{eq:bdytermspec}
    \begin{split}
        (\tilde{\phi}^-,\phi)_{\mathscr{I}^+} &= \dfrac{1}{2} \int_{\mathscr{I}^+} du \, d^2z \, \left( \tilde{\phi}^-_1 \partial_u \phi_1 - \partial_u \tilde{\phi}^-_1 \, \phi_1 \right) , \\
        (\tilde{\phi}^+,\phi)_{\mathscr{I}^-} &= \dfrac{1}{2} \int_{\mathscr{I}^-} dv \, d^2z \, \left( \tilde{\phi}^+_1 \partial_v \phi_1 - \partial_v \tilde{\phi}^+_1 \, \phi_1 \right) .
    \end{split}
\end{equation}
After performing the standard mode expansion of the field $\phi$ of the form
\begin{equation}\label{eq:modeexpphi}
    \tilde{\phi}(x) = \int \dfrac{d^3p}{(2\pi)^3} \dfrac{1}{2 \omega_p} \left( a(\vec{p}) e^{i p \cdot x} + a^{\dagger}(\vec{p}) \, e^{-i p \cdot x} \right) , ~~~~~~~ \omega_p = |\vec{p}| ,
\end{equation}
the large $r$ expansion can be worked out from saddle point approximation
\begin{equation}
    \int \dfrac{d^2 \hat{p}}{(2\pi)^2} \omega_p \, f(\hat{p}) \, e^{i p \cdot x} = \begin{cases}
        -\frac{i}{2\pi r} f(\hat{x}) \, e^{-i\omega u} ~~~~~ \text{on} ~ \mathscr{I}^+ , \\
        \frac{i}{2 \pi r} f(-\hat{x}) \, e^{-i \omega v} ~~~~~ \text{on} ~ \mathscr{I}^- ,
    \end{cases}
\end{equation}
where $d^2 \hat{p}$ is the measure of the unit sphere in the $\vec{p}$ plane and $\hat{x}$ denotes the unit vector that points in the direction of the point $(z,\z)$ on the Celestial sphere. Hence, the field $\tilde{\phi}$, which encodes the boundary condition, can be mode expanded near $\mathscr{I}^{\pm}$ as  \eqref{eq:phiasymptotics}:
\begin{equation}\label{eq:modedata}
    \begin{split}
        \tilde{\phi}(x) \approx \begin{cases}
            \frac{-i}{8 \pi^2 r} \int_0^{\infty} d \omega \, \left( a(\omega \hat{x}) e^{-i \omega u} - a^{\dagger}(\omega \hat{x}) e^{i \omega u} \right) ~~~~~~~~ \text{on} ~ \mathscr{I}^+,  \\
            \frac{i}{8 \pi r} \int_0^{\infty} d \omega \, \left( a(-\omega \hat{x}) e^{-i \omega v} - a^{\dagger}(-\omega \hat{x}) e^{i \omega v} \right) ~~~~~ \text{on} ~ \mathscr{I}^- ,
        \end{cases}
    \end{split}
\end{equation}
where $a,a^{\dagger}$ are the standard ladder operators. The boundary term  \eqref{eq:bdyterm} can now be 
expressed in terms of the conditions
\begin{equation}\label{eq:bdycondnsphi}
    \tilde{\phi}^-_1(u, \hat{x}) = \dfrac{i}{8 \pi^2} \int_0^{\infty} d\omega \, a^{\dagger}(\omega \hat{x}) \, e^{i \omega u}, ~~~~~~~ \tilde{\phi}^+_1(v,\hat{x}) = \dfrac{i}{8 \pi^2} \int_0^{\infty} d\omega \, a(-\omega \hat{x}) e^{-i\omega v} . 
\end{equation}
This is related to  \eqref{eq:bdyfieldfourier} except that we consider the field expansions at both $\mathscr{I}^+$ and $\mathscr{I}^-$ on the same footing to describe the S-matrix through a path integral which is a functional of the boundary data:
\begin{equation}\label{eq:pathintsmatrix}
    Z[\tilde{\phi}] = \int_{\tilde{\phi}} \mathcal{D}\phi \, e^{i S[\phi,\tilde{\phi}]} .
\end{equation}

\medskip

One can now express the S-matrix in terms of  \eqref{eq:modedata} and  \eqref{eq:pathintsmatrix} as:
\begin{equation}
    \langle q_1, \dots , q_M |\mathcal{S}| p_1 , \dots , p_N\rangle = \left[ \prod_{k=1}^N \left( 2\omega_{p_k} (2\pi)^3 \dfrac{\delta}{\delta a_k(\vec{p}_k)} \right) \prod_{l=1}^M \left( 2\omega_{q_l} (2\pi)^3 \dfrac{\delta}{\delta a^{\dagger}_l(\vec{q}_l)} \right) Z[\tilde{\phi}] \right]_{\tilde{\phi} = 0}.
\end{equation}
To make this relation more transparent and to connect it to the usual AdS/CFT partition function of the GKPW dictionary \cite{Gubser:1998bc,Witten:1998qj}, we will use the following solution obeying the boundary conditions  \eqref{eq:bdycondnsphi}:
\begin{equation}\label{eq:phiintermsofk}
    \phi(x) = \int_{\mathscr{I}^+} d^3x'\, K_{\mathscr{I}^+}(x;x') \, \tilde{\phi}^-_1(x') + \int_{\mathscr{I}^-} d^3x' \, K_{\mathscr{I}^-}(x;x') \, \tilde{\phi}^+_1(x') ,
\end{equation}
where
\begin{equation}\label{eq:bbdypropspath}
    \begin{split}
        K_{\mathscr{I}^+}(x;x') &= \dfrac{i}{(2\pi)^2} \dfrac{1}{(u'+ q(\hat{x}') \cdot x - i \epsilon )^2} , \\
        K_{\mathscr{I}^-}(x;x') &= \dfrac{i}{(2\pi)^2} \dfrac{1}{(v'+ q(-\hat{x}') \cdot x + i \epsilon )^2},
    \end{split}
\end{equation}
with $x' = (u',\hat{x}')$ or $x'=(v',\hat{x}')$ and $q^{\mu}(\hat{x}') = (1,\hat{x}')$. The bulk to boundary propagators  \eqref{eq:bbdypropspath} are clearly the bulk to boundary propagators of  \eqref{eq:bbdypropmodmellin} for $\Delta =2$. Since the extrapolated field  \eqref{eq:modedata} has $\Delta =1$ (cf.~\eqref{eq:bdyfieldfourier}), the bulk-to-boundary propagators of the form  \eqref{eq:bbdypropspath} in the path integral would lead to a differentiated boundary correlator through $\partial_u$ derivatives (analogous to  \eqref{eq:tdep2pt2}) as we will argue below.

\medskip

One can work out the path integral in  \eqref{eq:pathintsmatrix} by expanding around a free solution $\tilde{\phi}$ as $\phi = \tilde{\phi} + \phi_G$ and doing the path integral over $\phi_G$. Due to the specific nature of the boundary terms in the AFS path integral  \eqref{eq:bdytermspec}, $Z[\tilde{\phi}]$ becomes a generating functional of boundary correlators. For instance, the $2 \to 2$ scattering is described by
\begin{equation}\label{eq:pathintbdy}
    \begin{split}
        Z^{\text{bdy}}_{2,2} = &\dfrac{1}{(2!)^2} \int_{\mathscr{I}^-} dv_1 \, d^2 \hat{x}_1 \, dv_2 \, d^2 \hat{x}_2 \int_{\mathscr{I}^+} du_3 \, d^2 \hat{x}_3 \, du_4 \, d^2 \hat{x}_4 \, G_{2,2}(v_1,\hat{x}_1;v_2,\hat{x}_2;u_3,\hat{x}_3; u_4, \hat{x}_4) \\
        &~~~~\times \partial_{v_1} \tilde{\phi}^+_1(v_1,\hat{x}_1) \partial_{v_2} \tilde{\phi}^+_1(v_2,\hat{x}_2) \left( -\partial_{u_3} \tilde{\phi}^-_1(u_3,\hat{x}_3) \right) \left( -\partial_{u_4} \tilde{\phi}^-_1(u_4,\hat{x}_4) \right).
    \end{split}
\end{equation}
Alternatively, one could also view $Z[\tilde{\phi}]$ as computing conventional off-shell amplitudes via Feynman diagrams. If we express the amplitude through a Fourier transform
\begin{equation}
    \mathcal{A}_n(x_1,\dots, x_n) = \int \left( \prod_{i=1}^n \dfrac{d^4p_i}{(2\pi)^4} \right) \tilde{A}_n(p_1,\dots,p_n) \, e^{i \sum_i p_i \cdot x_i},
\end{equation}
which leads to the following ``bulk definition'' of  \eqref{eq:pathintsmatrix}:
\begin{equation}\label{eq:pathintbulk}
    Z^{\text{bulk}}_n[\tilde{\phi}] = \dfrac{1}{n!} \int \left( \prod_{i=1}^n d^4 x_i \right) \,\mathcal{A}_n(x_1,\dots,x_n) \, \tilde{\phi}(x_1) \dots \tilde{\phi}(x_n) .
\end{equation}
 \eqref{eq:pathintbulk} computes the on-shell amplitude via the LSZ procedure where $\tilde{\phi}$ carry the on-shell wavefunctions. Thus, the equivalence between ``bulk''  \eqref{eq:pathintbulk} and ``boundary''  \eqref{eq:pathintbdy} descriptions can be viewed as an equivalence between AFS \cite{Arefeva:1974jv} and LSZ formulations. This has been showed within perturbation theory \cite{Arefeva:1974jv,Kim:2023qbl}. Equating  \eqref{eq:pathintbulk} and  \eqref{eq:pathintbdy} for an $n$ point amplitude, we find the following relation between position space boundary correlators and momentum space amplitudes:
\begin{equation}
    \begin{split}
        G&(v_1, \hat{x}_1 ; \dots ; v_m, \hat{x}_m ; u_{m+1}, \hat{x}_{m+1} ; \dots ; u_n, \hat{x}_n ) \\
        &= \dfrac{1}{(2\pi)^n} \left( \prod_{i=1}^m \int_{-\infty}^0 \dfrac{d\omega_i}{2\pi} \right) \left( \prod_{j=m+1}^n \int_{0}^{\infty} \dfrac{d\omega_i}{2\pi} \right) \mathcal{A}_n(\omega_1,|\omega_1| \hat{x}_1; \dots ; \omega_n,|\omega_n|\hat{x}_n) \\
        & \hspace{2cm} \times e^{-i \sum_{i=1}^m \omega_i v_i -i \sum_{j=m+1}^n \omega_j u_j}
    \end{split}
\end{equation}
This is the conjectured relation between boundary conformal Carrollian correlators and the bulk S-matrix via the integral transform in \eqref{eq:modmellindef} (with $\Delta=1$) and  \eqref{eq:modmellinsmatrix} \cite{Banerjee:2018gce,Bagchi:2022emh,Donnay:2022aba,Donnay:2022wvx}. We see that the relation arises as a natural consequence of equivalence between bulk and boundary description of the path integral in  \eqref{eq:pathintsmatrix}.

\medskip

The AFS path integral specified through appropriate boundary terms in  \eqref{eq:bdyterm} is entirely analogous to the GKPW dictionary \cite{Gubser:1998bc,Witten:1998qj} in AdS/CFT. In the GKPW dictionary, one defines the boundary correlators of operators in terms of a bulk path integral, which is done over configurations with appropriate boundary conditions (which source these boundary operators). These CFT boundary correlators can then be expressed through AdS Witten diagrams. From the bulk to boundary propagators  \eqref{eq:bbdypropspath}, it is clear that one could now have a flat space analogue of the Witten diagrams through the AFS path integral in  \eqref{eq:pathintbulk}. If one substitutes  \eqref{eq:phiintermsofk} in  \eqref{eq:pathintbulk}, one gets
\begin{equation}\label{eq:afspathintres}
    \begin{split}
        Z^{\text{bulk}}_n = \dfrac{1}{n!} \int_{\mathscr{I}^-} \left( \prod_{i=1}^m d^3x'_i \right) \int_{\mathscr{I}^+} \left( \prod_{j=m+1}^n d^3x'_j \right) \int_{\mathcal{M}_4} \left( \prod_{k=1}^n d^4x_k\right) \mathcal{A}_n(x_1,\dots,x_n) \\ \times K_{\mathscr{I}^-}(x_1;x'_1) \dots K_{\mathscr{I}^-}(x_m;x'_m)
        K_{\mathscr{I}^+}(x_{m+1};x'_{m+1}) \dots K_{\mathscr{I}^+}(x_n;x'_n) \\ \times \tilde{\phi}^+_1(x'_1) \dots \tilde{\phi}^+_1(x'_m) \tilde{\phi}^-_1(x'_{m+1}) \dots \tilde{\phi}^-_1(x'_n) .
    \end{split}
\end{equation}
Since $\mathcal{A}_n$ is a sum over Witten diagrams, we basically get the fact that the sum of Witten diagrams computes a differentiated Carrollian correlator:
\begin{equation}
    \begin{split}
        (-\partial_{v_1}) \dots (-\partial_{v_m}) \partial_{u_{m+1}} \dots \partial_{u_n} G_n(x'_1,\dots,x'_n) = \int_{\mathcal{M}_4} \left( \prod_{k=1}^n d^4x_k\right) \mathcal{A}_n(x_1,\dots,x_n) \\
        \times K_{\mathscr{I}^-}(x_1;x'_1) \dots K_{\mathscr{I}^-}(x_m;x'_m)
        K_{\mathscr{I}^+}(x_{m+1};x'_{m+1}) \dots K_{\mathscr{I}^+}(x_n;x'_n) .
    \end{split}
\end{equation}

\medskip

The invariance of the partition function  \eqref{eq:pathintbdy} under large $U(1)$ gauge transformations would lead to the leading soft photon theorem. However, for the sub-leading soft photon theorem, one has to extract the symmetry transformations that leave the partition function invariant by working backward from the amplitude expressions. The symmetry would be a sub-leading large $U(1)$ gauge transformation, and one cannot see this directly from the scalar QED action. Thus, the interpretation of the sub-leading symmetry within the partition function framework is subtle. 

\medskip

Finally, one could show that the AdS partition function reduces to the AFS path integral \eqref{eq:afspathintres} in a careful limit prescribed in \cite{Giddings:1999jq,Giddings:1999qu}. We know that the AdS partition function can be expressed as a sum of Witten diagrams smeared with boundary sources \cite{Witten:1998qj}. Thus, the argument in terms of Witten diagrams is entirely analogous to Sec. \ref{ssec:wittendiag}. The wavefunctions in AdS given by
\begin{equation}
    \psi(x) = \int_{\partial\text{AdS}} \sqrt{-h} d^dy \, K(x;y) J(y) ,
\end{equation}
where $K$ is the bulk to boundary propagator and $J$ denotes the source. These wavefunctions will be reduced to flat space wavefunctions if the sources are collimated enough to localize them in the flat space ``scattering region''. This can be chosen when the sources' support is in a thin strip around $\tau = \pm \frac{\pi}{2}$
\begin{equation}\label{eq:wavepackets}
    J(\tau,\hat{x}) = \int_{0}^{\infty} \dfrac{d\omega}{2\pi} \tilde{J}(\omega,\hat{x}) e^{-i\omega R \left( \tau + \frac{\pi}{2} \right)}.
\end{equation}
To see the same argument in a formal path integral setting, it would be useful to first define a time evolution operator between initial and final time slices $\Sigma_i$ and $\Sigma_f$
\begin{equation}\label{eq:timeevo}
    \hat{U}(\Sigma_f,\Sigma_i) = T \, \text{exp} \left( - i \int_{\Sigma_i}^{\Sigma_f} \hat{H} d\tau + i \int_{\Sigma_i}^{\Sigma_f} J \hat{O} d\tau  \right) ,
\end{equation}
where $J$ denotes the sources  \eqref{eq:wavepackets}. Since the wavepackets  \eqref{eq:wavepackets} have periodicity $2\pi$, there will obviously be multiple collisions in the AdS box. In order to extract an S-matrix with a single collision, one must truncate the AdS box with ``caps'' that project the state onto the vacuum at early and late times \cite{Skenderis:2008dg,Hijano:2020szl}. Thus, the AdS partition function with these Euclidean caps can be expressed in terms of the time evolution operator  \eqref{eq:timeevo} as
\begin{equation}
    Z[J] = \langle 0| \hat{U}(\Sigma_{\pi},\Sigma_{-\pi})|0\rangle.
\end{equation}
One can now incorporate the slices $\Sigma_i$ and $\Sigma_f$, which bound the scattering region at the center of AdS
\begin{equation}\label{eq:adspathint}
    \begin{split}
        Z[J] &= \langle 0| \hat{U}(\Sigma_{\pi},\Sigma_f) \, e^{-i \hat{H}_0 \tau_f} \, e^{i \hat{H}_0 \tau_f} \, \hat{U}(\Sigma_f,\Sigma_i) \, e^{-i \hat{H}_0 \tau_i} \, e^{i\hat{H}_0 \tau_i} \, \hat{U}(\Sigma_i,\Sigma_{-\pi}) |0\rangle \\
        &= \int [ \mathcal{D} \phi^{\alpha}_+ \mathcal{D} \phi^{\alpha}_- ] [ \mathcal{D} \phi^{\beta}_+ \mathcal{D} \phi^{\beta}_- ] e^{-2i (\phi^{\alpha}_-,\phi^{\alpha}_+)_{\Sigma_f}} e^{-2i (\phi^{\beta}_-,\phi^{\beta}_+)_{\Sigma_i}} \\
        &~~~~\times \langle 0| \hat{U}(\Sigma_{\pi},\Sigma_f) | \phi^{\alpha}_+(\tau_f) \rangle \langle \phi^{\alpha}_-(\tau_f)| \hat{U}(\Sigma_f,\Sigma_i) | \phi^{\beta}_+(\tau_i) \rangle \langle \phi^{\beta}_-(\tau_i) | \hat{U}(\Sigma_i,\Sigma_{-\pi})|0 \rangle,
    \end{split}
\end{equation}
where we have introduced a complete basis in the second step, and we used the inner product notation in  \eqref{eq:bdytermspec}. In  \eqref{eq:adspathint}, we have the generating functional
\begin{equation}
    S[\phi^{\alpha}_-,\phi^{\beta}_+] = \langle \phi^{\alpha}_-(\tau_f) | \hat{U}(\Sigma_f,\Sigma_i) | \phi^{\beta}_+(\tau_i) \rangle,
\end{equation}
which reduces to the AFS expression  \eqref{eq:pathintbulk} only when $\phi^{\alpha}_-$ and $\phi^{\beta}_+$ have support only in the boundary of the flat region. This can be implemented by considering the expression
\begin{equation}
    \langle \phi^{\beta}_-(\tau_i) | \hat{U}(\Sigma_i,\Sigma_{-\pi})|0 \rangle = \int [\mathcal{D} \phi] \, e^{i I[\phi] + i I_{\partial}[\phi]},
\end{equation}
and evaluating the integral over field configurations satisfying the usual AdS boundary conditions in addition to becoming the vacuum state at the Euclidean cap:
\begin{equation}
    \begin{split}
        \phi_-(\tau_i) &= \phi^{\beta}_-(\tau_i) \\
        \phi(\rho,\tau,\hat{x}) &\to \cos^{d-\Delta}\rho \, J(\tau,\hat{x}) \\
        \phi_+(\Sigma_{-\pi}) &=0
    \end{split}
\end{equation}
This construction can be carried out for scalar, $U(1)$ \cite{Kraus:2024gso} and non-Abelian gauge fields \cite{Kraus:2025wgi}.

\subsection{Pointers to literature}
\label{ssec:pointersotherholo}

Keeping in tune with the review, we provide some pointers to the literature of Carrollian holography and we re-emphasize that we are not being exhaustive:
\begin{itemize}

    \item \textit{Representation theory}
    \begin{itemize}

\item[$\star$] We have mentioned some of the earliest references to BMS representation theory already in Sec.~{\ref{section2}}. The reader is referred back to Sec.~\ref{pt:sec2}. 
\item[$\star$] From a quantum gravity perspective, representations of BMS group were perhaps first investigated by \cite{girardello1974continuous}.

 \item[$\star$] In more recent work \cite{Bekaert:2025kjb}, McCarthy's classic work has been extended to include generic supermomentum, involving both a hard and a soft piece. The role of the choice of gravitational vacua in defining the BMS state has also been discussed in the same work. Also look at \cite{AliAhmad:2025hdl}. 

 \item[$\star$] In \cite{Bekaert:2024uuy}, authors showed we can think of BMS particles as quantum superposition of multiple Poincaré particles moving on different gravitational backgrounds. 
\end{itemize}
    
    \item \textit{Flat limit of AdS/CFT:} 
    \begin{itemize}
        \item [$\star$] \textit{Flat sector in CFT} \cite{deGioia:2023cbd}: The authors explicitly work out how the Killing vectors of the conformal algebra, when restricted to the strip \eqref{tau} go over to the symmetry algebra of BMS$_4$ in the large AdS radius limit. The antipodal matching condition \eqref{eq:antipodal} naturally arises a consequence of comparing the Killing vectors restricted to the strips around $\tau = \pm \frac{\pi}{2}$.
        \item [$\star$] \textit{Witten diagrams} \cite{Alday:2024yyj}: The authors consider the flat limit of Witten diagrams discussed in \S\ref{sec:flatlimit} (especially \S\ref{ssec:wittendiag}) in the Bondi coordinates. They then relate it to boundary correlation functions by explicitly computing the $c \to 0$ limit of CFT correlators. In \cite{Surubaru:2025fmg}, the authors extend the analysis to the case of AdS$_3$/CFT$_2$.
        \item [$\star$] \textit{Flat limit of ABJM} \cite{Lipstein:2025jfj}: The authors try to work out an explicit top-down construction of a flat space hologram from the limit of AdS/CFT. They consider correlation functions of protected scalar operators in ABJM theory \cite{Aharony:2008ug} and considered a flat space limit in which the seven sphere decompactifies. Since such operators are dual to the Kaluza-Klein (KK) modes on the seven sphere in the bulk, the flat limit was implemented by fixing the KK mode number. The $c \to 0$ limit of the dual ABJM correlators was then matched to the bulk scattering amplitudes of $\mathcal{N} = 8$ supergravity. 
    \end{itemize}
    
    \item \textit{Embedding space approach} \cite{Salzer:2023jqv}: The Carrollian primaries defined by \eqref{BMS-primary} are lifted to the embedding space. Correlation functions (derived in \S\ref{ssec:carcorfunctions}) are obtained directly from the embedding space formalism. 
    
    \item \textit{Extrapolate dictionary:} 
    \begin{itemize}
        \item [$\star$] In \cite{Nguyen:2023vfz}, the authors work out the representation theory for the conformal Carrollian algebra and propose an extrapolate dictionary. The work of \cite{Nguyen:2023miw} then extends this by computing the correlators of \S\ref{ssec:carcorfunctions} and relating it to the extrapolate dictionary of \cite{Nguyen:2023vfz}.
        \item [$\star$] In \cite{Kulp:2024scx}, the authors detail the role of multiparticle states within the extrapolate dictionary. In particular such states have conformal weight $\Delta \neq 1$ (compared to \eqref{eq:extrapolatek}). They also carefully study the flat space limit (from AdS/CFT) of the CFT primary conditions.
    \end{itemize}

    \item \textit{Bulk extrapolation to a boundary theory at $\mathscr{I}$}: In a series of works \cite{Liu:2022mne,Liu:2023qtr,Liu:2023gwa,Liu:2024nkc}, the authors advocate for a different approach to a dual boundary Carrollian description on $\mathscr{I}$ by reducing (through an asymptotic expansion) the bulk theories in Minkowski space. In this approach, it was shown in \cite{Liu:2024nfc} that one could study loop corrections to Carrollian amplitudes.
    
    \item \textit{Bulk reconstruction}: \cite{Chen:2023naw} approaches bulk reconstruction in flat space from an algebraic representation theory point of view. 
    
    \item \textit{Carrollian amplitudes}: In \cite{Mason:2023mti}, the authors derive closed form expressions for the Carrollian amplitudes \eqref{eq:modmellinsmatrix} for $n$-point MHV amplitudes. From the collinear limit, they then extract Carrollian OPEs which need to be carefully determined owing to the ultra-local nature of the theory. Finally, they initiate a connection between the Carrollian operators at null infinity and twistor space, since twistor space is a natural geometric realisation of Celestial symmetries \cite{Adamo:2021lrv,Bu:2022iak,Kmec:2024nmu}.

    \item {\em{Light transformations}}: \cite{Banerjee:2024hvb} initiates a discussion on Light transformation and its connections to the Modified Mellin transform and consequences for Celestial and Carrollian CFTs.

    \item{\em{Differential representations}}: Differential representations for scalar Carroll correlators were introduced in \cite{Chakrabortty:2024bvm}.
    
    \item \textit{Carrollian string amplitudes}: In \cite{Stieberger:2024shv}, the authors computed the four point Carrollian amplitude of gauge bosons and gravitons in type I open superstring theory and heterotic closed superstring theory. Unlike the Celestial case \cite{Stieberger:2018edy}, the resulting $\alpha'$ dependence doesn't factor out. The $\alpha'$ expansion can be recast into a tower of UV and IR finite $\partial_u$-descendants of the underlying Carrollian field theory amplitudes.

    \item \textit{Massive Carrollian fields}: \cite{Have:2024dff} explores massive representations of Carrollian fields that naturally live in time-like infinity. The details of the intrinsic geometric description of the blowup of timelike infinity are worked out and there is a proposal an extrapolate dictionary between the Carrollian fields that live there and the asymptotic late time limit of bulk massive fields.
    
    \item \textit{Carrollian Chern-Simons theory \& Null reduction}: In \cite{Bagchi:2024efs}, an interesting connection between the dual 3D Carrollian theory and a 2D Celestial CFT was explored in the context of 4D AFS. A 2D Yang-Mills theory was obtained from a particular null reduction of a 3D Carrollian Chern-Simons theory. The theory under consideration was a Chern-Simons theory with bifundamental matter and with the gauge group being $SU(N) \times SU(M)$. This is nontrivial because as we have seen in \S \ref{sec:carfieldtheories}, one arrives at a Carroll field theory by an ultra-relativistic limit of a relativistic field theory. The procedure of \cite{Bagchi:2024efs} shows that the null reduction of the resulting Carroll field theory leads us back to a relativistic field theory in one lower dimension.
    
    \item \textit{Soft physics}: \cite{Jorstad:2024yzm} argues for the existence of a soft contribution that is typically missed in the standard extrapolate dictionary \cite{He:2014laa,Strominger:2017zoo,Donnay:2022aba}. This argument suggests that the extrapolate dictionary should have both electric and magnetic Carroll correlators.

    \item \textit{Anomalies}: \cite{Baulieu:2025itt} studies the anomaly structure of the boundary field theory that inherits the extended BMS symmetries. 
\end{itemize}

\newpage

\part{Hydrodynamics and Condensed Matter}

\bigskip

\bigskip

\bigskip

\section*{Outline}

In this part of the review, we turn our attention to the applications of Carrollian physics to hydrodynamics and condensed matter. Sec.~\ref{carroll_hydro} reviews the so far most popular approach of arriving at equations of Carroll hydrodynamics, by taking the $c\to 0$ limit of the equations of relativistic hydrodynamics. Sec.~\ref{hydro_II} then looks at the symmetry based approach to Carroll hydrodynamics, which carefully incorporates the dynamics of the Goldstone fields associated with the spontaneous breaking of Carroll boost invariance, leading to two classes of Carroll fluids, of which one encompasses the Carroll fluids obtained from the $c\to 0$ limit of relativistic hydrodynamics. Sec.~\ref{heavy_ion} then delves into elaborating the mapping between Carroll fluids and Bjorken/Gubser flow, which are boost-invariant models for the spacetime evolution of the quark-gluon plasma produced in ultrarelativistic heavy-ion collisions.

\medskip

Sec.~\ref{carroll_cond_matt} is devoted to something novel, namely, the emergence and applications of Carrollian symmetries in condensed matter systems. It is genuinely intriguing that Carroll symmetries, thought to be important at ultra-relativistic and/or ultra-high-energy regimes could have anything to do with low energy description of condensed matter systems. But if one recalls the discussion in the introduction for example,\footnote{We don't blame the reader if they have forgotten by now.} we have stressed the fact that in different physical situations, different quantities can act as an effective speed of light, and whenever there is a null hypersurface associated to that speed (in an effective Lorentzian frame), we can always expect Carroll symmetries to emerge. This subtle prowess of Carroll seems to be highly universal in some sense, and it seems there's more often than not some example of beautiful and exotic physics which appears at the zero characteristic velocity limit in real systems. In this section, we will discuss how Carroll physics emerges in theories of exotic quasiparticles called Fractons, takes an important part in strongly coupled electronic theories via the emergence of Flat Bands, acts as the guiding principle for Phase separation in Luttinger Liquids, and last but not the least, facilitates a gauge theory description of shallow water waves. In the first few cases, the characteristic velocities are either particle velocities or Fermi velocities, while in the last case, it points to the speed of the wave (or sound speed).

\newpage

\section{Carroll hydrodynamics I: $c\to 0$ limit of relativistic hydrodynamics}
\label{carroll_hydro}
So far, we have been discussing Carrollian physics focusing on Carroll field theories which provide a microscopic realisation of the Carroll symmetry algebra. But what about the physics on macroscopic scales? At such scales, it is the collective dynamics of the underlying degrees of freedom that holds relevance in determining the properties and response of the system to external perturbations. For systems described by relativistic quantum fields respecting the Poincar\'{e} symmetry, the collective dynamics at long time and length scales compared to the microscopic scales, such as the mean free time and mean free path, is given by \emph{relativistic hydrodynamics.} When perturbed away from a state of global thermal equilibrium, it is the relaxation of conserved charges associated with the symmetries of the system that dictates its macroscopic dynamics, since the non-conserved quantities relax back on much smaller time and length scales. This is so because the conserved charges require physical transport to relax back to equilibrium, unlike the non-conserved quantities, and are thus the relevant degrees of freedom on macroscopic scales. One thus focuses on conserved charges and their conservation equations to extract dynamical information about the system on such scales. For non-relativistic fluids of everyday experience, such as flowing water, the macroscopic dynamics is governed by the Euler (inviscid flow) or the Navier-Stokes (viscous flow) equations, expressing the conservation of mass and momentum. More generally, the same conservation principles can also be used to describe the macroscopic dynamics of relativistic systems in or near thermal equilibrium, albeit with suitable modifications to bring in Poincar\'{e} covariance rather than Galilean, thereby giving the subject its name of relativistic hydrodynamics \cite{Landau-Lifshitz, Kovtun:2012rj, Jeon:2015dfa, Romatschke:2017ejr}.  

\medskip

One may wonder whether Carrollian systems admit a macroscopic description akin to relativistic hydrodynamics. The answer turns out to be a resounding yes! \emph{Carroll hydrodynamics}, as the subject has come to be known, utilizes symmetry principles like its relativistic cousin to arrive at equations governing macroscopic dynamics of Carrollian systems. There are two principal approaches to arrive at the equations of Carroll hydrodynamics. The first approach is to start with relativistic hydrodynamics and carefully impose the $c\to 0$ limit. This is the approach advocated in \cite{deBoer:2017ing, Ciambelli:2018xat, Petkou:2022bmz, Bagchi:2023ysc, Bagchi:2023rwd, Kolekar:2024cfg}. Alternatively, one can start from the symmetries themselves and carefully perform a bottom-up construction of Carroll hydrodynamics, incorporating the dynamics of the Goldstone fields for the spontaneously broken Carroll boost invariance, as discussed in \cite{Armas:2023dcz}. The two approaches are complementary, though it has been argued that the second approach leads to additional classes of Carroll fluids that can not be obtained from a $c\to 0$ limit of relativistic hydrodynamics, and is thus seemingly more general \cite{Armas:2023dcz}.

\medskip 

We will now provide an overview of some of the main ideas and results in Carroll hydrodynamics. The present section focuses on the first approach i.e.~constructing Carroll hydrodynamics as the $c\to 0$ limit of relativistic hydrodynamics. We begin in subsection \ref{PR_para} by introducing the Papapetrou-Randers (PR) parametrization for the background geometry and the hydrodynamic variables. This parametrization makes the $c\to 0$ limit transparent. Subsequently, in subsection \ref{Eqs_Carroll_Hydro}, we derive the equations for Carroll hydrodynamics from the $c\to 0$ limit of relativistic hydrodynamics in the PR parametrization, while in subsection \ref{subleading_hydro} we discuss the subleading terms that arise in the small-$c$ expansion. Our discussion throughout will focus on neutral fluids i.e.~there are no additional symmetries beyond diffeomorphism invariance, for which the conserved charges are energy and momentum.\footnote{In the literature on relativistic hydrodynamics, it is common to also consider an internal global $U(1)$ symmetry and an associated conserved $U(1)$ charge density for the fluid. This could for instance be the global baryon number conservation symmetry for quantum chromodynamics, the theory that underlies the dynamics of quarks and gluons that form the QGP. Such fluids are referred to as ``charged,'' as opposed to the neutral fluids we discuss in the present review. See for instance the references \cite{Erdmenger:2008rm, Banerjee:2008th, Banerjee:2012iz, Jensen:2012jh, Bhattacharyya:2013lha, Bhattacharyya:2014bha, Megias:2014mba, Buzzegoli:2017cqy, Kovtun:2018dvd, Shukla:2019shf, Grieninger:2021rxd} for discussions on charged relativistic hydrodynamics, including thermodynamic and holographic aspects.}  

\subsection{The Papapetrou-Randers parametrization}
\label{PR_para}
The Papapetrou-Randers (PR) parametrization (also termed as the PR gauge) \cite{Ciambelli:2018xat, Petkou:2022bmz, Bagchi:2023ysc, Bagchi:2023rwd} for the line element of a $d+1$-dimensional pseudo-Riemannian manifold is given by
\begin{equation}
\label{PRmetric}
 ds^2 = g_{\mu\nu} dx^{\mu} dx^{\nu} = -c^2 (\Omega \,dt - b_i dx^i)^2 + a_{ij} dx^i dx^j,
\end{equation}
implying that the metric components are
\beq
\label{PRmetric-components}
\begin{split}
g_{tt} &= -c^2 \Omega^2, \quad g_{ti} = c^2 \Omega b_i, \quad g_{ij} = a_{ij}-c^2 b_i b_j,  \\
g^{tt} &= -\frac{1}{c^2 \Omega^2} + \frac{b^2}{\Omega^2}, \quad g^{ti} = \frac{b^i}{\Omega}, \quad g^{ij} = a^{ij}\, ,
\end{split}
\eeq
and
\beq
\label{detgPR}
\sqrt{-g} = c\,\Omega\sqrt{a}\, .
\eeq
The quantities $(\Omega, b_i, a_{ij})$ are all functions of the coordinates $(t,x^i)$. Further, $g\equiv\det(g_{\mu\nu})$ and $a\equiv\det(a_{ij})$. 

\medskip

An important feature of the PR parametrization is the invariance of the form of the line element in \eqref{PR_para} under the following set of coordinate reparameterizations,
\beq
\label{Carroll_diffeos}
t \rightarrow t'(t,\vec{x})\, , \quad \vec{x}\rightarrow \vec{x}{\,'}(\vec{x}). 
\eeq
These form a subset of the most general coordinate reparametrization possible i.e.~the diffeomorphism $t \rightarrow t'(t,\vec{x})\, ,\vec{x}\rightarrow \vec{x}{\,'}(t,\vec{x})$, and by analogy are sometimes referred to as ``Carroll diffeomorphisms'' in the literature \cite{Ciambelli:2018xat, Petkou:2022bmz}. In particular, under the PR gauge preserving coordinate reparametrization \eqref{Carroll_diffeos}, one has
\beq
\label{Carr_diff_met}
\Omega \rightarrow \Omega' = \frac{\Omega}{J}\, , \quad a_{ij}\rightarrow a'_{ij} = (J^{-1})_i^{~k} (J^{-1})_j^{~l} a_{kl}\, , \quad b_i \rightarrow b'_{i} = \left(b_k + \frac{\Omega}{J} J_k \right) (J^{-1})_i^{~k},
\eeq
where we have used the following notation for the elements of the Jacobian of \eqref{Carroll_diffeos},
\beq
J \equiv \frac{\d t'}{\d t}\, , \quad J_k \equiv \frac{\d t'}{\d x^k}\, , \quad J_i^{~k} = \frac{\d x'^{k}}{\d x^i}.
\eeq
The quantity $a_{ij}$ which transforms like a rank-2 covariant tensor under the PR gauge preserving coordinate transformations \eqref{Carroll_diffeos} is termed the ``spatial metric.'' It can be used to raise the index on $b_i$, i.e.~$b^i \equiv a^{ij} b_j$, with $a^{ij}$ being the inverse spatial metric.

\medskip

It is also convenient to introduce a parametrization for the four-velocity $u^\mu$ of the fluid, which satisfies the timelike normalization condition $u^\mu u_\mu = - c^2$, in terms of the geometric quantities entering the PR parametrization of the underlying pseudo-Riemannian manifold. Defining $u \equiv \gamma \d_t + \gamma v^i \d_i$, we can parametrize $(\gamma, v^i)$ in terms of the field $\beta^i(t,\vec{x})$ via
\beq
\label{velo_para}
\gamma = \frac{1+c^2\vec{\beta}\cdot\vec{b}}{\Omega\sqrt{1-c^2\beta^2}} \, , \quad v^i = \frac{c^2\Omega\beta^i}{1+c^2 \vec{\beta}\cdot\vec{b}}\, .
\eeq
It turns out that under the PR gauge preserving coordinate transformations \eqref{Carroll_diffeos}, the field $\beta^i$ transforms like a vector, ${\beta'^{i}} = J^i_{~j} \beta^j$, whose indices can thus be lowered by using the spatial metric $a_{ij}$, thereby leading to $\beta_i = a_{ij} \beta^j, \vec{\beta}\cdot\vec{b} = \beta^i b_i = \beta_i b^i$ and $\beta^2 = \beta_i \beta^i$.
The components of $u^{\mu}$ are then explicitly given by 
\begin{equation}
u^{t} = \frac{1+ c^2 \vec{\beta}\cdot\vec{b}}{\Omega\sqrt{1-c^2 \beta^2}}\, , \quad u_{t} = -\frac{c^2\Omega}{\sqrt{1-c^2 \beta^2}}\, , \quad u^i = \frac{c^2 \beta^i}{\sqrt{1-c^2 \beta^2}}\, , \quad u_i = \frac{c^2 (b_i+\beta_i)}{\sqrt{1-c^2 \beta^2}}\, . 
\label{PR-velo-comps}
\end{equation}

\medskip

{\ding{112}} \underline{\emph{The $c\to 0$ limit in the PR gauge}}

\medskip
Consider taking the $c\to 0$ limit now. In this limit, the metric \eqref{PRmetric} becomes degenerate, and the pseudo-Riemannian manifold becomes a Carrollian manifold $\mathcal{C}$, which has the structure of a fibre bundle, with time fibred over a base spatial manifold. The fibre bundle structure is mathematically described in terms of a degenerate metric $h_{\mu\nu}$ on the Carroll manifold $\mathcal{C}$, and its kernel $k^\mu$ i.e.~$h_{\mu\nu} k^\nu = 0$. In terms of the coordinates $(t,\vec{x})$ and the geometric quantities appearing in the PR parametrization of the original pseudo-Riemannian manifold, the line element on the Carroll manifold and the kernel are given by
\be
\label{Carroll-structure}
d\ell^2 \equiv h_{\mu\nu} dx^\mu dx^\nu = a_{ij} dx^i dx^j \, , \quad k = \frac{1}{\Omega} \d_{\,t}\, .
\ee
This splitting is preserved by the PR gauge preserving coordinate transformations \eqref{Carroll_diffeos}. The field $b_i$ now serves as a connection on the Carroll manifold $\mathcal{C}$, termed the \emph{Ehresmann connection}, and appears in the dual form of the kernel, $\vartheta = - \Omega \, dt + b_i dx^i$, with $k^\mu \vartheta_\mu = - 1$.

\medskip

We can now introduce several quantities on the Carroll manifold $\mathcal{C}$ which will turn out to be useful for the subsequent discussion \cite{Ciambelli:2018xat, Petkou:2022bmz}.  
In particular, one can construct objects that transform covariantly under the PR gauge preserving coordinate transformation \eqref{Carroll_diffeos}. For instance, the Carroll covariant temporal and spatial derivatives are given by
\begin{equation}
    \hat{\partial}_{t} \equiv \frac{1}{\Omega}\partial_{t}\, , \quad \hat{\partial}_i \equiv \partial_i + \frac{b_i}{\Omega}\partial_{t}\, .
\end{equation}
One can also define the components of a temporal and a spatial \emph{Levi-Civita-Carroll} connection via
\begin{equation}
\label{LCC_Conns}
    \hat{\gamma}_{ij} \equiv \frac{1}{2\Omega}\partial_{t} a_{ij}, \quad \hat{\gamma}^i_{jk} \equiv \frac{a^{il}}{2}(\hat{\partial}_j a_{kl} + \hat{\partial}_k a_{jl} - \hat{\partial}_l a_{jk}).
\end{equation}
The indices on $\hat{\gamma}_{ij}$ can be raised using $a^{ij}$, since it too transforms like a Carrollian tensor under the PR gauge preserving coordinate transformations \eqref{Carroll_diffeos}. Further, we can utilize the above definitions for the Levi-Civita-Carroll connections to define temporal and spatial Levi-Civita-Carroll covariant derivatives, $\hat{\nabla}_{t}$ and $\hat{\nabla}_i$, respectively, whose action on Carrollian vectors takes the form 
\be
\begin{split}
&\hat{\nabla}_{t} V^j = \hat{\partial}_{t} V^j + \hat{\gamma}^j_k V^k, \quad \hat{\nabla}_{t} V_j = \hat{\partial}_{t} V_j - \hat{\gamma}_j^k V_k, \\
& \hat{\nabla}_i V^j = \hat{\partial}_i V^j + \hat{\gamma}^j_{ik}V^k, \quad \hat{\nabla}_i V_j = \hat{\partial}_i V_j - \hat{\gamma}^k_{ij}V_k .
\end{split}
\ee
The Levi-Civita-Carroll connections defined above are metric compatible i.e., $\hat{\nabla}_{t} a_{jk} = 0$, $\hat{\nabla}_i a_{jk} = 0$. Further, one can also define a Carrollian expansion $\theta$, a Carrollian acceleration $\varphi_i$ and an anti-symmetric Carrollian tensor $f_{ij}$ via
\begin{equation}
\theta \equiv \hat{\gamma}^i_i = \frac{1}{\Omega}\partial_{t} \operatorname{log} \sqrt{a}, \quad \varphi_i \equiv \frac{1}{\Omega}(\partial_i \Omega+\partial_{t} b_i), \quad f_{ij}\equiv 2(\partial_{[i}b_{j]}+b_{[i}\varphi_{j]}).
\label{SomeDefs}
\end{equation}

\medskip

It is important to note that the temporal and spatial Levi-Civita-Carroll connections $\hat{\gamma}_{ij}$ and $\hat{\gamma}^i_{jk}$ appear naturally in the Christoffel connection for the metric of the original pseudo-Riemannian manifold, \eqref{PRmetric}. To be more explicit, the components of the Christoffel connection 
\begin{equation}
    \Gamma^{\mu}_{\nu\rho} = \frac{g^{\mu\sigma}}{2}(\partial_{\nu}g_{\rho\sigma} + \partial_{\rho}g_{\nu\sigma} - \partial_{\sigma}g_{\nu\rho})
\end{equation}
for the metric \eqref{PRmetric} are
{\allowdisplaybreaks
\begin{align}
& \Gamma^{t}_{tt}=\hat{\partial}_{t}\Omega+ c^2 \Omega \vec{b}\cdot\vec{\varphi}\, ,
\quad \Gamma^{t}_{tj}= \varphi_j-\hat{\nabla}_{t} b_j+c^2 \left( -\frac{1}{2} f_{ij}b^i -\vec{b}\cdot\vec{\varphi} \, b_j\right), \nonumber \\
& \Gamma^{t}_{ij} = \frac{1}{c^2}\frac{\hat{\gamma}_{ij}}{\Omega} - \frac{1}{\Omega}\Big(\hat{\nabla}_{(i}b_{j)}-b_{(i}\hat{\nabla}_{t} b_{j)}+b_{(i}\varphi_{j)}+b_{(i}\hat{\gamma}^l_{j)}b_l\Big)+c^2\frac{b^k}{\Omega}(\varphi_k b_ib_j-b_{(i}f_{j)k}), \nonumber \\
& \Gamma^k_{tt}=c^2 \Omega^2 \varphi^k, \quad \Gamma^k_{tj}=\Omega \hat{\gamma}^k_j+c^2\left(\frac{\Omega}{2}f^{~k}_j-\Omega b_j\varphi^k\right), \nonumber \\
&\Gamma^k_{ij}=\hat{\gamma}^k_{ij}-2\hat{\gamma}^k_{(i}b_{j)}+c^2 \Big(\varphi^k b_ib_j-b_{(i}f_{j)}^{~k}\Big).
\label{PR-Christoffel}
\end{align}}
Note that the expressions above are exact and not truncated in a small-$c$ expansion. In particular, the leading order terms in the small-$c$ expansion depend upon the Levi-Civita-Carroll connections $(\hat{\gamma}_{ij}, \hat{\gamma}^i_{jk})$ introduced in \eqref{LCC_Conns}.

Finally, the fluid velocity in the PR parametrization, \eqref{PR-velo-comps}, reduces to the following in the $c\to 0$ limit,
\beq
\begin{split}
&u^t = \frac{1}{\Omega} + \mathcal{O}(c^2)\, ,\qquad\, u^i = c^2 \beta^i + \mathcal{O}(c^4)\, ,\\
&u_t = - c^2 \Omega + \mathcal{O}(c^4) \, , \quad u_i = c^2 (b_i + \beta_i) + \mathcal{O}(c^4).
\end{split}
\eeq

\subsection{Equations for Carroll Hydrodynamics}
\label{Eqs_Carroll_Hydro}
Let us now compute the equations governing the dynamics of a Carroll fluid. As mentioned earlier, the approach we take in the present section is to start with the equations for a relativistic fluid and impose the $c\to 0$ limit to arrive at the equations for a Carroll fluid. Further, we do not assume the presence of any internal symmetries leading to additional conserved currents beyond the conservation of energy and momentum. A relativistic fluid is then governed by the conservation equation
\beq
\label{rel_hyd_1}
\nabla_\mu T^{\mu\nu} = 0,
\eeq
with $T^{\mu\nu}$ being the energy-momentum tensor for the fluid. The key underlying assumption of hydrodynamics is that the currents associated with conserved charges, such as $T^{\mu\nu}$, admit an expansion in the number of derivatives acting on the hydrodynamic degrees of freedom i.e.~the temperature $T$ and the fluid velocity $u^\mu$, with successive terms in the expansion involving more and more derivatives. These are termed ``constitutive relations'' for the fluid, and are necessary to form a closed set of dynamical equations that can be solved with the knowledge of the initial conditions. Since by assumption we are in a state near thermal equilibrium, one can further assert that successive terms in the derivative expansion become progressively smaller. Our focus in the present review will be on \emph{first order hydrodynamics}, where the constitutive relations for the currents are truncated at the first order in derivatives, neglecting second and higher order corrections. In particular, for a relativistic fluid, the constitutive relation for the energy-momentum tensor up to the first order in derivatives is given by
\beq
\label{rel_hyd_T}
T^{\mu\nu} = (\epsilon + P) \frac{u^\mu u^\nu}{c^2} + P g^{\mu\nu} - \eta \sigma^{\mu\nu} -\zeta \Delta^{\mu\nu} \Theta.
\eeq
Here $\epsilon, P, \eta$, and $\zeta$ respectively denote the energy density, pressure, shear viscosity and bulk viscosity of the relativistic fluid, all of which are state functions i.e.~functions of the temperature $T$. $\Delta_{\mu\nu} \equiv g_{\mu\nu}+\frac{u_\mu u_\nu}{c^2}$ is the projector orthogonal to the fluid velocity $u^\mu$. Further, $\Theta \equiv \nabla\cdot u$ denotes the fluid expansion, while $\sigma^{\mu\nu}$ is the symmetric-transverse-traceless shear tensor, given by
\beq
\label{rel_hyd_shear}
\sigma^{\mu\nu} \equiv \Delta^{\mu\alpha} \Delta^{\nu\beta} \nabla_{(\alpha} u_{\beta)} - \frac{\Theta}{d} \D^{\mu\nu}. 
\eeq

\medskip 

One can now impose the $c\to 0$ limit on the relativistic equation \eqref{rel_hyd_1} and keep the leading non-vanishing terms to arrive at the equations for a Carroll fluid. To do so, we make use of the PR parametrization for the metric and the fluid velocity, introduced in the previous section. Further, for the state dependent quantities we employ the following ansatz in the small-$c$ limit,
\beq
\begin{split}
&\epsilon \xrightarrow{c\to 0} \epsilon_{(0)} + c^2 \epsilon_{(2)} + \mathcal{O}(c^4)\, , \quad P \xrightarrow{c\to 0} p_{(0)} + c^2 p_{(2)} + \mathcal{O}(c^4)\, ,\\
&\eta \xrightarrow{c\to 0} \eta_{(0)} + c^2 \eta_{(2)} + \mathcal{O}(c^4)\, , \quad \zeta \xrightarrow{c\to 0} \zeta_{(0)} + c^2 \zeta_{(2)} + \mathcal{O}(c^4)\, .
\end{split}
\label{Carroll_Lims}
\eeq
In terms of these, and the geometric definitions introduced in section \ref{PR_para}, we arrive at the following equations for a Carroll fluid,
\begin{subequations}
\label{LOVF}
\begin{align}
    \hat{\partial}_{t}\epsilon_{(0)} &= - \theta\Big(\epsilon_{(0)} + p_{(0)} - \frac{\Xi}{d}\Big) + \xi^{ij}\Xi_{ij}, \label{LOVF1}\\
    \hat{\partial}_i p_{(0)} &=  - \varphi_i(\epsilon_{(0)} + p_{(0)}) - (\hat{\partial}_{t} + \theta)[(\epsilon_{(0)} + p_{(0)})\beta_i - \Xi_{ij}\beta^j] + (\hat{\nabla}_j + \varphi_j)\Xi^{j}_{\ i}. \label{LOVF2}
\end{align}
\end{subequations}
Here $\xi_{ij}$ is the ``Carrollian shear tensor,'' obtained via the $c\to 0$ limit from the relativistic shear tensor, $\sigma_{ij} \xrightarrow{c\to 0} \xi_{ij} + \mathcal{O}(c^2)$, with
\beq
\xi_{ij} = \hat{\gamma}_{ij} - \frac{\theta}{d}a_{ij}, \quad a^{ij}\xi_{ij} = 0.
\label{Carrollian-shear-tensor}
\eeq
Further, $\Xi_{ij}$ is the ``Carrollian viscous stress tensor,'' defined via
\begin{equation}
    \Xi_{ij} \equiv \eta_{(0)}\xi_{ij} + \zeta_{(0)}\theta a_{ij}, \quad \Xi \equiv a^{ij}\Xi_{ij} = \zeta_{(0)}\theta d.
    \label{Xi_Def}
\end{equation}

\medskip

Given an equation of state relating the pressure with the energy density, $p_{(0)} \equiv p_{(0)}(\epsilon_{(0)})$, as well as the shear and bulk viscosities $\eta_{(0)}, \zeta_{(0)}$, \eqref{LOVF} provides us with a set of dynamical equations to determine the spacetime evolution of the energy density and the field $\beta^i$ for the Carroll fluid from the knowledge of the initial conditions. Needless to say, if the viscous effects can be neglected, then the fluid becomes an \emph{ideal Carroll fluid}, with the hydrodynamic equations taking the much simpler form
\begin{subequations}
\label{LOPF}
\begin{align}
\hat{\partial}_{t} \epsilon_{(0)} &= - \theta(\epsilon_{(0)} + p_{(0)}), \label{LOPF1} \\
\hat{\partial}_i p_{(0)} &= - \varphi_i(\epsilon_{(0)} + p_{(0)}) - (\hat{\partial}_{t} + \theta)\big[(\epsilon_{(0)} + p_{(0)})\beta_i\big],\label{LOPF2}
\end{align}
\end{subequations}
which can be obtained from \eqref{LOVF} by setting $\eta_{(0)} = \zeta_{(0)} = 0$.

\medskip

{\ding{112}} \underline{\emph{Local Carroll boost invariance of the Carroll fluid equations}}\\
The manifold on which the Carroll fluid lives respects Carrollian symmetries locally in the tangent space. Of particular significance amongst these are local Carroll boosts. The hydrodynamic equations \eqref{LOVF} (or their ideal fluid analogues \eqref{LOPF}) are invariant under local Carroll boosts. This can be seen as follows. Under a local Carroll boost transformation with parameters $\lambda_i$ the degenerate metric $h_{\mu\nu}$, its kernel $k^\mu$, the dual form $\vartheta_\mu$ and the inverse metric $h^{\mu\nu}$ transform via \cite{Hartong:2015xda, Armas:2023dcz}\,\footnote{A more detailed discussion about the geometric aspects of Carroll manifolds appears in sec.~\ref{sec:Carroll-Goldstone}.}
\be
\delta_c h_{\mu\nu} = 0, \quad \delta_c k^\mu = 0, \quad \delta_c\vartheta_\mu = \lambda_\mu, \quad \delta_c h^{\mu\nu} = 2 \lambda_\rho h^{\rho(\mu} k^{\nu)}.
\label{LCTrans}
\ee
Here $\delta_c$ denotes variation under the local Carroll boost, while $\lambda_\mu = e_\mu^{~\,i} \lambda_i$, with $e_\mu^{~\,i}$ being the spatial vielbein. In the PR parametrization of the manifold \eqref{Carroll-structure} one has
\begin{equation}
    \label{eq:LO-PRgauge}
	k^{\mu} = (\Omega^{-1},\vec{0})^{\rm T}, \quad \vartheta_{\mu} = (-\Omega, b_i), \quad h_{\mu\nu} = \begin{pmatrix}
		0 & \vec{0} \\
		\vec{0} & a_{ij}
	\end{pmatrix}, \quad h^{\mu\nu} = \begin{pmatrix}
		\Omega^{-2}b^2 & \Omega^{-1}b^j \\
		\Omega^{-1}b^i & a^{ij}
	\end{pmatrix}.
\end{equation}
Using \eqref{LCTrans}, the various quantities in the PR parametrization transform vis-à-vis a local Carroll boost via
\begin{equation}
\label{LCPRTrans}
    \delta_c \Omega=0,\quad \delta_c a_{ij}=0,\quad \delta_c a^{ij}=0,\quad \delta_cb_i=\lambda_i.
\end{equation}
Further, using the fact that the Carroll fluid energy-momentum tensor is local Carroll boost invariant, it is straightforward to work out the transformation properties of various fluid variables, which turn out to be
\begin{equation}
\label{CFTrans}
    \delta_c \epsilon_{(0)}=0,\quad \delta_cp_{(0)}=0,\quad \delta_c \beta_i=-\delta_c b_i=-\lambda_i.
\end{equation}
Using the transformations in \eqref{LCPRTrans} and \eqref{CFTrans}, one can easily check that the hydrodynamic equations \eqref{LOVF} and \eqref{LOPF} for a Carroll fluid are indeed local Carroll boost invariant. The interested reader may refer to the Appendix A of \cite{Kolekar:2024cfg} for further details. 

\subsection{Subleading corrections}
\label{subleading_hydro}
Eqs.~\eqref{LOVF} for a viscous Carroll fluid (or their ideal fluid avatar \eqref{LOPF}) arose by keeping the leading terms in the strict $c\to 0$ limit of the relativistic hydrodynamic equation \eqref{rel_hyd_1}. However, necessitated by the physical situation at hand, it might be appropriate to keep subleading terms as well in a small-$c$ expansion around the strict Carroll limit $c\to 0$. This is known as the \emph{Carrollian regime} \cite{deBoer:2023fnj}. The hydrodynamic equations that follow from \eqref{rel_hyd_1} in the Carrollian regime by keeping the next-to-leading order (NLO) terms intact in a small-$c$ expansion about $c\to 0$, with the leading order given by \eqref{LOVF}, were derived and discussed extensively in \cite{Kolekar:2024cfg}. To wit, in the presence of viscous effects, the equations satisfied by the NLO quantities $(\epsilon_{(2)}, p_{(2)})$ in \eqref{Carroll_Lims} turn out to be
\begin{subequations}
\label{NLOVF}
\begin{align}
\hat{\partial}_{t}\epsilon_{(2)} &= - \theta\Big(\epsilon_{(2)} + p_{(2)}-\frac{\tilde{\Xi}}{d}\Big) + \xi^{ij}\tilde{\Xi}_{ij} - (\hat{\partial}_{t} + \theta)[\beta^2(\epsilon_{(0)} + p_{(0)})-\beta^i\beta^j\Xi_{ij}]\label{NLOVF1} \\
&- \hat{\nabla}_i[(\epsilon_{(0)} + p_{(0)})\beta^i-\Xi^i_{\ j}\beta^j] - (\epsilon_{(0)} + p_{(0)})\Big(2\vec{\beta}\cdot\vec{\varphi} + \hat{\gamma}_{ij}\beta^i\beta^j\Big) + \Xi_{ij}\Big(2\beta^i\varphi^j + \beta^i\hat{\gamma}^j_k\beta^k\Big), \nonumber\\
\hat{\partial}_i p_{(2)} &= - \varphi_i(\epsilon_{(2)} + p_{(2)}) - (\hat{\nabla}_j + \varphi_j)\Big[\beta^j\beta_i(\epsilon_{(0)} + p_{(0)})-\tilde{\Xi}^{j}_{\ i} -\frac{1}{2}\beta_i\Xi^j_k\beta^k -\frac{1}{2} \beta^j\Xi_{ik}\beta^k\Big] \nonumber \\
&\quad- (\hat{\partial}_{t} + \theta)\Big[\beta_i\beta^2(\epsilon_{(0)} + p_{(0)})+(\epsilon_{(2)} + p_{(2)})\beta_i-\tilde{\Xi}_{ij}\beta^j - \frac{1}{2}\beta_i\Xi_{jk}\beta^j\beta^k - \frac{\beta^2}{2}\Xi_{ik}\beta^k\Big] \nonumber \\
&\quad- (\epsilon_{(0)} + p_{(0)})(\varphi_i\beta^2 + \beta^j f_{ji}) + \Xi_{jk}(\varphi_i\beta^j\beta^k + \beta^j f^k_{\ \ i}). \label{NLOVF2}
\end{align} 
\end{subequations}
Here $\tilde{\Xi}_{ij}$ is the ${\cal O}(c^2)$ viscous stress tensor, defined via
\begin{equation}
    \tilde{\Xi}_{ij} = \eta_{(0)}\tilde{\xi}_{ij} + \eta_{(2)}\xi_{ij} + (\zeta_{(0)}\tilde{\theta} + \zeta_{(2)}\theta)a_{ij}, \quad \tilde{\Xi} \equiv a^{ij}\tilde{\Xi}_{ij} = d(\zeta_{(0)}\tilde{\theta} + \zeta_{(2)} \theta),
\end{equation}
with $\tilde{\theta}$ and $\tilde{\xi_{ij}}$ respectively being the $c^2$-terms in the small-$c$ expansion of the relativistic expansion and the shear tensor, i.e.~$\Theta \xrightarrow{c\to 0} \theta + c^2 \tilde{\theta} + \mathcal{O}(c^4)$ and $\sigma_{ij} \xrightarrow{c\to 0} \xi_{ij} + c^2 \tilde{\xi}_{ij} + \mathcal{O}(c^4)$. Explicitly,
\begin{equation}
\label{NLO-data}
\begin{split}
&\tilde{\theta} = \hat{\nabla}_i\beta^i + \vec{\beta}\cdot\vec{\varphi} + \big(\hat{\partial}_{t} + \theta\big)\frac{\beta^2}{2}\, , \\
&\tilde{\xi}_{ij} = \hat{\nabla}_{(i}\beta_{j)} + \beta_{(i}\hat{\nabla}_{t}\beta_{j)} + \beta_{(i}\varphi_{j)} + \frac{\beta^2}{2}\hat{\gamma}_{ij} - \frac{\tilde{\theta}}{d}a_{ij}, \quad a^{ij}\tilde{\xi}_{ij} = 0.
\end{split}
\end{equation}
Once again, the ideal (i.e.~non-viscous) form of the NLO equation \eqref{NLOVF} can be obtained by setting $\eta_{(0)}, \eta_{(2)}, \zeta_{(0)}, \zeta_{(2)} = 0$.

\newpage

\section{Carroll hydrodynamics II: The symmetry based approach}
\label{hydro_II}
The previous section lays down the formulation of Carroll hydrodynamics as the $c\to 0$ limit of relativistic hydrodynamics. Interestingly, there is an alternate approach to arrive at the equations of Carroll hydrodynamics, based solely on symmetries \cite{Armas:2023dcz}. The primary feature of the symmetry based approach is to take into account the dynamics of the Goldstone fields of spontaneously broken Carroll boost invariance in constructing the hydrodynamic description. Macroscopic thermal states generically break boost invariance spontaneously,  due to the presence of a preferred frame, the rest frame of the fluid. Following Goldstone's theorem, this leads to new massless degrees of freedom in the low energy spectrum whose dynamics should be incorporated into the hydrodynamic description. As argued in \cite{Armas:2023dcz}, the symmetry based approach leads to a more general class of Carroll fluids, encompassing the ones obtained from the $c\to 0$ limit of relativistic fluids, and will be the subject of our discussion in the present section.

\medskip

Before we discuss Carroll hydrodynamics from the perspective of spontaneously broken Carroll boost symmetry, let us try to understand why the analogous dynamics of Goldstone fields for spontaneously broken Lorentz boost invariance in a relativistic fluid is absent in the standard discussions of relativistic hydrodynamics. Consider the hydrodynamic regime of a Lorentz invariant field theory in thermal equilibrium on a curved background i.e. Lorentz boosts are realised as a symmetry in the tangent space. Equilibrium is characterized by the background admitting a timelike Killing vector $V^{\mu}$, $\pounds_V g_{\mu\nu} = 0$, under which the macroscopic hydrodynamic degrees of freedom, the temperature and fluid velocity, do not evolve, $\pounds_V T = 0$ and $\pounds_V u^\mu = 0$.\footnote{If the background were flat, then in suitable coordinates one can set $V^\mu = (1, \vec{0})$. The statements $\pounds_V T = \pounds_V u^\mu = 0$ then simply translate to the temperature and fluid velocity being time-independent, as the case should be for a system in thermal equilibrium.} One can identify the temperature and fluid velocity in terms of the metric and the timelike Killing vector as \cite{Jensen:2012jh}
\be
\label{def_rel_Tu}
T \equiv \frac{T_0}{\sqrt{-V^2}}\, , \quad u^\mu \equiv \frac{V^\mu}{\sqrt{-V^2}}\, ,
\ee
where $T_0$ is a constant setting the temperature scale, and $V^2 \equiv g_{\mu\nu} V^\mu V^\nu$. The definitions in \eqref{def_rel_Tu} automatically meet the requirement of vanishing Lie derivative with respect to $V^\mu$, as well as the normalization $u^\mu u_\mu = -1$. Now, as mentioned above, the presence of a preferred frame aligned with the thermal vector $u^\mu/T$ (i.e.~the rest frame) breaks the invariance under Lorentz boosts spontaneously. This is most straightforwardly realised as some vector operator in the underlying theory acquiring a non-vanishing timelike expectation value in the thermal state \cite{Nicolis:2015sra}, $\langle O_a(x)\rangle = \delta _a^0$, which acts as an order parameter for the broken symmetry, with $a, b, \ldots \in \{0,1,2,\ldots,d\}$ denoting the tangent space indices. The associated Goldstone fields of spontaneously broken Lorentz boost invariance appear in local boosts of the order parameter,
\be
O_a(x) = \Big(e^{i \vec{\phi}(x)\cdot \mbox{}\vec{\mathcal{B}}}\Big)_a^{~b} \langle O_b(x)\rangle \, ,
\ee
where $\phi^i(x)$ are the Goldstone fields and $\mathcal{B}_i$ are the generators of Lorentz boosts. From the above, it is clear that one can neatly package the three Goldstone fields in terms of a vector field $\varphi_a \equiv (e^{i \vec{\phi}(x)\cdot \vec{\mathcal{B}}})_a^{~0}$, with the expectation value $\langle \varphi_a\rangle = \delta_a^0$. The associated spacetime vector field $\varphi_\mu = e_\mu^{~\,a} \varphi_a$ then satisfies $\varphi^\mu \varphi_\mu = -1$, which can be thought of as the statement that $\varphi^\mu$ contains only three degrees of freedom, corresponding to the three Goldstones.  

\medskip

The equilibrium partition function (or equivalently, the generating functional \cite{Banerjee:2012iz, Jensen:2012jh}) in the presence of the boost Goldstones $\varphi_\mu$ can be written as
\be
W[g_{\mu\nu}, \varphi_\mu] = \int d^4x \sqrt{-g} \left[P(T,u\cdot\varphi,\varphi\cdot\varphi) + \lambda (\varphi\cdot\varphi + 1)\right],
\label{gen_func_Gold}
\ee
where $P$ is the fluid pressure, which is a function of the three possible zeroth-order scalars, and $\lambda(x)$ is a Lagrange multiplier that enforces the constraint $\varphi\cdot\varphi \equiv \varphi^\mu \varphi_\mu = -1$. The equation of motion for the Goldstone fields that follows from \eqref{gen_func_Gold}, ${\delta W}/{\delta \varphi_\mu} = 0$, is
\be
u^\mu \, \frac{\partial P}{\partial (u\cdot \varphi)}  + 2 \varphi^\mu \left(\lambda+\frac{\partial P}{\partial (\varphi\cdot \varphi)}\right) = 0\, .
\label{goldstone_eom}
\ee
Solving for the Lagrange multiplier $\lambda(x)$ from the above equation leads to
\be
\lambda(x) = - \frac{1}{2} \, \frac{u\cdot\varphi}{\varphi\cdot\varphi} \, \frac{\partial P}{\partial (u\cdot \varphi)} - \frac{\partial P}{\partial (\varphi\cdot \varphi)}\, .
\label{lm_sol}
\ee
Substituting this in the equation of motion for the Goldstone fields \eqref{goldstone_eom} along with the constraint $\varphi\cdot\varphi=-1$, leads to the condition $u^\mu = - \varphi^\mu \, u\cdot\varphi$, which has the solution $\varphi^\mu = - u^\mu$.\footnote{Refs.~\cite{Nicolis:2015sra, Armas:2023dcz} arrive at the same conclusion without recourse to the technique of Lagrange multipliers, but by using an appropriate projector on the equation of motion for the Goldstone fields to impose the constraint $\varphi\cdot\varphi=-1$.} Thus, for a relativistic fluid, the degrees of freedom in the Goldstone fields associated with the spontaneous breaking of Lorentz boosts are tied to the fluid four-velocity, and do not acquire an independent dynamics of their own. This is the underlying reason for the absence of boost Goldstones in relativistic hydrodynamics.\footnote{For relativistic superfluid hydrodynamics \cite{Bhattacharya:2011eea, Bhattacharya:2011tra, Bhattacharyya:2012xi, Hoult:2024cyx}, the Goldstone fields from the spontaneous breaking of Lorentz boosts can be expressed in terms of the Goldstone field associated with the spontaneous breakdown of the internal global $U(1)$ symmetry via inverse Higgs constraints \cite{Nicolis:2013lma, Nicolis:2015sra}, and thus are not independent.}

\medskip

The situation is entirely different, however, for Carroll hydrodynamics, where the Goldstone fields associated with the spontaneous breaking of Carroll boost invariance are not subservient to any other degree of freedom \cite{Armas:2023dcz}. Thus their dynamics needs to be accounted for in constructing the equilibrium partition function as well as in studying out-of-equilibrium phenomena, as we discuss in the following subsections. 

\subsection{The Carroll boost Goldstone}
\label{sec:Carroll-Goldstone}
Let us begin by introducing the requisite mathematical machinery that will be helpful for the subsequent discussion. As already mentioned, a (weak) Carroll structure consists of a degenerate metric $h_{\mu\nu}$ with signature $(0,+1,+1,\ldots,+1)$, also referred to as the ``spatial metric,'' and a nowhere vanishing vector field $k^\mu$ that serves as its kernel i.e.~$h_{\mu\nu}k^\nu = 0$. One further defines their inverses: the symmetric tensor $h^{\mu\nu}$, referred to as the ``co-metric,'' and the one form $\vartheta_{\mu}$, called the ``clock form,'' such that $\vartheta_{\mu}$ serves as the kernel of $h^{\mu\nu}$ i.e.~$\vartheta_\mu h^{\mu\nu} = 0$, and 
\be
k^\mu \vartheta_\mu = -1\, , \quad h^{\mu\rho} h_{\rho\nu} = \delta^{\mu}_\nu + k^\mu \vartheta_\nu\, .
\label{struct_rels}
\ee
One can express the degenerate metric in terms of the spatial vielbeins $e_\mu^{~\,i}$ as $h_{\mu\nu} = e_\mu^{~\,i} e_\nu^{~\,j} \delta_{ij}$, with $i,j,\ldots \in \{1,2,\ldots,d\}$ denoting the tangent space (spatial) indices. Similarly, the co-metric $h^{\mu\nu}$ can be expressed in terms of the inverse vielbeins $e^\mu_{~i}$ as $h^{\mu\nu} = e^\mu_{~i} e^\nu_{~j} \delta^{ij}$. The completeness relation in \eqref{struct_rels} in terms of the vielbein fields takes the form
\be
\label{struct_rels_2}
e^\mu_{~i} e_\nu^{\ i} = \delta^\mu_\nu + k^\mu \vartheta_\nu\, ,
\ee
while $e^\mu_{~i} e_\mu^{~\,j} = \delta^j_i$ as usual. 

\medskip

Under a local (i.e.~tangent space) Carroll boost transformation, the spatial vielbeins are invariant i.e.~$\delta_c e_\mu^{~\, i} = 0$, while the inverse vielbeins transform via $\delta_c e^\mu_{~ i} = k^\mu \lambda_i$, where $\lambda_i$ are the boost parameters \cite{Hartong:2015xda}. This implies $k^\mu \lambda_\mu = 0$, with $\lambda_\mu = e_\mu^{~\,i} \lambda_i$. Further, while the degenerate metric and its kernel that form the Carroll structure are invariant under local Carroll boosts, their inverses are not. One has
\be
\label{struct_rels_3}
\delta_c h_{\mu\nu} = 0, \quad \delta_c k^\mu = 0, \quad \delta_c\vartheta_\mu = \lambda_\mu, \quad \delta_c h^{\mu\nu} = 2 \lambda_\rho h^{\rho(\mu} k^{\nu)},
\ee
which were also quoted in \eqref{LCTrans}. 

\medskip

Let us now introduce the Goldstone fields associated with the spontaneous breaking of Carroll boost invariance, denoted by $\varrho^\mu$, such that under a local Carroll boost they transform via \cite{Armas:2023dcz}
\be
\delta_c \varrho^\mu = - h^{\mu\nu} \lambda_\nu .
\label{Gold_Boost}
\ee
This encapsulates the fact that only the spatial part of $\varrho^\mu$ is physical. This can be enforced by demanding the setup to be invariant under a timelike St\"{u}ckelberg symmetry under which the boost Goldstones transform as
\be
\delta_s \varrho^\mu = \xi k^\mu\, ,
\label{stuckelberg}
\ee
with $\xi$ being an arbitrary function. Now, using the boost Goldstones $\varrho^\mu$ and their transformation properties, one can construct alternate versions for the co-metric and the associated clock form via
\be
\tilde{h}^{\mu\nu} \equiv \tilde{e}^\mu_{~i} \tilde{e}^\nu_{~j} \delta^{ij}\, , \quad \tilde{\vartheta}_\mu \equiv \vartheta_\mu + h_{\mu\nu} \varrho^\nu\, , 
\label{new_inverses}
\ee
which are now local Carroll boost as well as St\"{u}ckelberg invariant. In writing \eqref{new_inverses} we have made use of the local Carroll boost and St\"{u}ckelberg invariant inverse vielbein fields $\tilde{e}^\mu_{~i} \equiv e^\mu_{~i}+k^\mu \varrho^\nu e_{\nu i}$. Taken together, the elements $(h_{\mu\nu}, \tilde{h}^{\mu\nu},k^\mu,\tilde{\vartheta}_\mu)$ satisfy
\be
h_{\mu\nu} k^\nu = 0\, , \quad \tilde{h}^{\mu\nu} \tilde{\vartheta}_\nu = 0\, , \quad k^\mu \tilde{\vartheta_\mu} = -1\, , \quad \tilde{h}^{\mu\rho} h_{\rho\nu} = \delta^\mu_\nu+k^\mu \tilde{\vartheta}_\nu\, ,
\label{eq:aristotle}
\ee
and effectively constitute a (weak) \emph{Aristotelian structure} \cite{Penrose:1968ar, deBoer:2017ing, deBoer:2020xlc, Figueroa-OFarrill:2020gpr, Bergshoeff:2022eog}, which is partly dynamical due to the presence of the boost Goldstones $\varrho^\mu$.\footnote{A weak Aristotelian structure can be thought of as comprising simultaneously a weak Carrollian and a weak Newton-Cartan structure \cite{Figueroa-OFarrill:2020gpr, Bergshoeff:2022eog}. In the construction above, the original spatial metric and the kernel $(h_{\mu\nu}, k^\mu)$ constitute a weak Carroll structure, while the modified (and independent) co-metric and clock-form  $(\tilde{h}^{\mu\nu}, \tilde{\vartheta}_\mu)$ constitute the weak Newton-Cartan part of the effective weak Aristotelian geometry.} 

\medskip

The crucial role played by the boost Goldstone fields in Carroll hydrodynamics can be appreciated now. For a Carroll fluid in thermal equilibrium, characterized by the presence of a spacetime Killing vector $V^\mu$ such that $\pounds_V h_{\mu\nu} = 0, \pounds_V k^\mu = 0$, one can attempt to define the temperature in the usual way it is done for non-Lorentzian systems, $T = T_0/{V^\mu \vartheta_\mu}$, with $T_0$ being a constant setting the temperature scale \cite{Jensen:2014ama, deBoer:2017ing, deBoer:2017abi, Armas:2019gnb, Jain:2020vgc, Armas:2020mpr, Armas:2023ouk}. However, this cannot be the correct definition, as it is manifestly non-invariant under local Carroll boost transformations, since $\delta_c(V^\mu \vartheta_\mu) = V^\mu \lambda_\mu \neq 0$. The correct definition is instead provided by making use of the modified clock form \eqref{new_inverses} defined in terms of the boost Goldstones, $T=T_0/{V^\mu \tilde{\vartheta}_\mu}$, which is invariant under local Carroll boosts, and thus provides an observer independent notion of temperature for the Carroll fluid in thermal equilibrium.

\subsection{The equilibrium partition function and currents}
\label{sec:eq_part_fuc}
Let us now move on to the construction of the equilibrium partition function for a Carroll fluid \cite{Armas:2023dcz}. As mentioned before, an equilibrium Carroll background $(h_{\mu\nu}, k^\mu)$ is characterized by the presence of a Killing vector $V^\mu$, satisfying $\pounds_V h_{\mu\nu} = 0, \pounds_V k^\mu = 0$. In addition, one also needs a set of fields $(\Lambda_\mu, \Xi)$, which parametrize the changes in the co-metric, the clock form and the Goldstone field when transported along $V^\mu$, via
\be
\pounds_V h^{\mu\nu} = - 2 \Lambda_\rho h^{\rho(\mu} k^{\nu)}\, , \qquad \pounds_V \vartheta_\mu = - \Lambda_\mu\, , \qquad \pounds_V \varrho^\mu = h^{\mu\nu}\Lambda_\nu-\Xi \, k^\mu\, .
\label{geom_transport}
\ee
Under an infinitesimal diffeomorphism generated by $\chi^\mu$, along with a local Carroll boost with parameters $\lambda_\mu$ and  a St\"{u}ckelberg transformation with parameter $\xi$, the fields $(\Lambda_\mu, \Xi)$ transform as
\be
\delta \Lambda_\mu = \pounds_\chi \Lambda_\mu - \pounds_V \lambda_\mu\, , \qquad \delta \Xi = \pounds_\chi \Xi - \pounds_V \xi\, .
\ee

\medskip

For constructing the partition function, one also needs to specify a derivative counting scheme. Following the usual conventions, the background geometric data $(h_{\mu\nu}, h^{\mu\nu}, k^\mu, \vartheta_\mu)$ is counted as zeroth-order in derivatives. Additionally, we count the Goldstone fields $\varrho^\mu$ as zeroth-order as well, implying that the local Carroll boost and St\"{u}ckelberg invariant versions of the co-metric and the clock form, $(\tilde{h}^{\mu\nu}, \tilde{\vartheta}_\mu)$, are zeroth-order too. 

\medskip

With the derivative counting scheme in place, the next step is to identify independent  zeroth-order invariant scalars on which the equilibrium pressure can depend. Apart from the temperature of the Carroll fluid $T = T_0/V^\mu \tilde{\vartheta}_\mu$ introduced earlier, we also have the modulus squared of the spatial fluid velocity, denoted by $\vec{\mathfrak{u}}^2 \equiv h_{\mu\nu} \mathfrak{u}^\mu \mathfrak{u}^\nu$, constructed out of the vector field $\mathfrak{u}^\mu = V^\mu/V^\nu \tilde{\vartheta}_\nu$, satisfying $\mathfrak{u}^\mu \tilde{\vartheta}_\mu = 1$.\footnote{The arrow notation in $\vec{\mathfrak{u}}^\mu$ is to highlight its spatial nature.} The vector field $\mathfrak{u}^\mu$ can in general be decomposed as $\mathfrak{u}^\mu \equiv - k^\mu + \vec{\mathfrak{u}}^\mu$, with $\vec{\mathfrak{u}}^\mu = \tilde{h}^\mu_{~\nu} \mathfrak{u}^\nu$, where we have introduced the shorthand notation $\tilde{h}^\mu_{~\nu} \equiv \tilde{h}^{\mu\rho} h_{\rho\nu}$. One can further define the spatial velocity with a lower index as $\vec{\mathfrak{u}}_\mu \equiv h_{\mu\nu} \mathfrak{u}^\nu = h_{\mu\nu} \vec{\mathfrak{u}}^\nu$. Note that $\vec{\mathfrak{u}}^\mu \neq {h}^{\mu\nu} \vec{\mathfrak{u}}_\nu$, but rather $\vec{\mathfrak{u}}^\mu = \tilde{h}^{\mu\nu} \vec{\mathfrak{u}}_\nu$. Also, one gets $\vec{\mathfrak{u}}^2 = \vec{\mathfrak{u}}^\mu \vec{\mathfrak{u}}_\mu$, ensuring notational consistency.

\medskip

In terms of the invariant equilibrium scalars $(T, \vec{\mathfrak{u}}^2)$, the partition function at the zeroth-order in derivatives takes the form
\be
W[h_{\mu\nu}, \vartheta_\mu, \varrho^\mu] = \int d^{d+1}x \, e\, \mathcal{P}(T, \vec{\mathfrak{u}}^2)\,,
\label{part_fuc_1}
\ee
where $e\equiv {\rm det}(\vartheta_\mu, e_{\mu}^{~i})$ is the integration measure, and $\mathcal{P}$ is the equilibrium pressure of the Carroll fluid. The variation of the partition function with respect to the sources has the generic form
\be
\delta W = \int d^{d+1}x \, e \, \left(\frac{1}{2} \mathcal{T}^{\mu\nu} \delta h_{\mu\nu} - E^\mu \delta\vartheta_\mu - G_\mu \delta\varrho^\mu\right).
\label{var_part_fuc}
\ee
Here $E^\mu$ is the energy current, $\mathcal{T}^{\mu\nu}$ is the stress-momentum tensor, and $G_\mu$ denotes the response of the system to the variation of the Goldstones. In particular, $G_\mu = 0$ is the equation of motion for the Goldstone fields. Invariance of the equilibrium partition function under local Carroll boost and St\"{u}ckelberg transformations yields, respectively,
\be
h_{\mu\nu} E^{\nu} = G_\mu\, , \qquad k^\mu G_\mu = 0\, .
\ee
Going onshell with respect to the Goldstone fields leads to $h_{\mu\nu} E^{\nu} = 0$. Thus, the Carroll boost Ward identity has become the equation of motion for the Goldstone fields. Further, one can define the equilibrium energy-momentum tensor comprising the energy current and the stress-momentum tensor via
\be
T^\mu_{~\,\nu} = - E^\mu \vartheta_\nu + \mathcal{T}^{\mu\rho} h_{\rho\nu}\, .
\label{Carroll_EMT}
\ee
Stated as is, it turns out that the energy-momentum tensor \eqref{Carroll_EMT} is neither Carroll boost nor St\"{u}ckelberg invariant. One has
\be
\delta_c T^{\mu}_{~\,\nu} = h^{\mu\rho} \lambda_{\rho} G_\nu\, , \qquad \delta_s T^\mu_{~\,\nu} = - \xi k^\mu G_{\nu}\, .
\ee
The Carroll boost and St\"{u}ckelberg invariance of the energy-momentum tensor thus require the Goldstone fields to be onshell i.e.~$G_\mu = 0$, which is also the statement of Carroll boost Ward identity, as mentioned above. 

\medskip

Now, for the zeroth-order equilibrium partition function \eqref{var_part_fuc}, the currents are given by
\be
\begin{split}
\mathcal{T}^{\mu\nu} &= \mathcal{P} h^{\mu\nu} + m \mathfrak{u}^\mu \mathfrak{u}^\nu-2(Ts+m\vec{\mathfrak{u}}^2) \mathfrak{u}^{(\mu} \varrho^{\nu)}\, ,\\
E^\mu &= \mathcal{P} k^\mu + (Ts + m\vec{\mathfrak{u}}^2) \mathfrak{u}^\mu\, ,\\
G_\mu &= (Ts+m\vec{\mathfrak{u}}^2) \vec{\mathfrak{u}}_\mu\, .
\end{split}
\label{carr_currents}
\ee
In the expressions above, the entropy density $s$ and the mass density $m$ are defined via the thermodynamic relation
\be
d\mathcal{P} = s \, dT + m \, d\vec{\mathfrak{u}}^2\, .
\ee
The equation of motion for the Goldstone fields $G_\mu = 0$ constrains the dynamics by demanding
\be
(Ts+m\vec{\mathfrak{u}}^2) \vec{\mathfrak{u}}_\mu = 0.
\label{G_eom}
\ee
By defining the energy density of the Carroll fluid as $\mathcal{E} \equiv E^\mu \tilde{\vartheta}_\mu$, the constraint \eqref{G_eom} allows for two classes of solutions, and thereby two classes of Carroll fluids:
\be
{\rm Class~I:}~\mathcal{E} + \mathcal{P} = Ts + m \vec{\mathfrak{u}}^2 = 0\, , \qquad {\rm Class~II:}~\vec{\mathfrak{u}}_\mu = 0\, .
\label{Carr_Classes}
\ee
The equilibrium energy-momentum tensors \eqref{Carroll_EMT} for the two classes of Carroll fluids take the form:
\be
{\rm Class~I:}~T^\mu_{~\,\nu} = \mathcal{P} \delta^{\mu}_\nu + m \mathfrak{u}^\mu \vec{\mathfrak{u}}_\nu\, , \qquad {\rm Class~II:}~T^\mu_{~\,\nu} = \mathcal{P} \delta^\mu_\nu - (\mathcal{E}+\mathcal{P})\mathfrak{u}^\mu\tilde{\vartheta}_\nu\, .
\label{EMT_Classes}
\ee
These are Carroll boost and St\"{u}ckelberg invariant, as the Goldstone fields are onshell.

\subsection{Getting the $c\to 0$  Carroll fluid from symmetries}
\label{sec:one_from_other}
In the previous subsection, we discussed how properly accounting for the spontaneous breaking of local Carroll boost invariance by including the dynamics of the Goldstone fields in constructing Carroll hydrodynamics leads to two broad classes of Carroll fluids \eqref{Carr_Classes}. Following \cite{Armas:2023dcz}, we will now argue that the class of Carroll fluids that satisfy $\vec{\mathfrak{u}}_\mu = 0$ is the one reached by taking the $c\to 0$ limit of relativistic hydrodynamics, discussed in sec.~\ref{carroll_hydro}.\footnote{We restrict the discussion here to ideal fluids. The interested reader may refer to \cite{Armas:2023dcz} for a detailed discussion including first-order dissipative effects as well as the structure and properties of the associated hydrodynamic modes. The analysis gets simplified by using the effective Aristotelian nature of the geometry \eqref{eq:aristotle}. Aristotelian fluids lack boost invariance and are thereby less constrained. See e.g.~\cite{deBoer:2017ing, deBoer:2020xlc, Armas:2020mpr, Ciambelli:2018xat, Petkou:2022bmz, Gouteraux:2023uff} for discussions on hydrodynamics without boost invariance in various physical settings.}

\medskip

Consider the energy-momentum tensor for an ideal relativistic fluid, given by
\beq
\label{rel_hyd_T_ideal}
T^{\mu}_{~\,\nu} = \frac{\epsilon + P}{c^2} u^\mu u_\nu + P \delta^{\mu}_{\nu} .
\eeq
Borrowing notation from sec.~\ref{carroll_hydro}, $(\epsilon, P)$ are respectively the energy density and pressure of the relativistic fluid, while $u^\mu$ is the fluid four-velocity, normalized such that $u^\mu u_\mu = -c^2$. To consider the $c\to 0$ limit, we introduce the decomposition of the background metric and its inverse in terms of \emph{pre-ultralocal} (PUL) variables via \cite{Hansen:2021fxi}
\be
g_{\mu\nu} = - c^2 T_\mu T_\nu + \Pi_{\mu\nu}\, , \qquad g^{\mu\nu} = -\frac{1}{c^2} K^\mu K^\nu + \Pi^{\mu\nu}\, .
\label{pul_defs}
\ee
The PUL variables satisfy the relations
\be
K^\mu T_\mu = -1\, , \quad \Pi_{\mu\nu} K^\nu = 0\, , \quad \Pi^{\mu\nu} T_\nu = 0\, , \quad \Pi^{\mu\rho} \Pi_{\rho\nu} = \delta^{\mu}_{\nu} + K^\mu T_\nu\,.
\label{pul_conds}
 \ee
In terms of the Lorentzian vielbein fields $E_\mu^{~\,a}$ and their inverses $E^\mu_{~a}$, the PUL decomposition corresponds to\footnote{Recall that the tangent space indices $a,b,\ldots \in \{0,1,2,\ldots d\}$ while $i,j,\ldots \in \{1,2,\ldots d\}$.}
\be
E_\mu^{~\,a} = (c T_\mu, E_\mu^{~\,i})\, , \quad E^\mu_{~a} = (-c^{-1}K^\mu, E^\mu_{~i})\,.
\label{viel_pul}
\ee
Under a local Lorentz boost denoted by ${\Lambda^0_{~i} \equiv c\,\Lambda_i}$ (equivalently, $\Lambda^i_{~0} \equiv c \, \Lambda^i)$, the Lorentzian vielbeins and their inverses transform via
\be
\delta_{\Lambda} E _\mu^{~\,a} = \Lambda^a_{~b} E_\mu^{~\,b}\, , \quad \delta_\Lambda E^\mu_{~a} = - \Lambda_{~a}^{b} E^\mu_{~\,b} \, ,
\label{viel_trans}
\ee
which induces the following Lorentz boost transformations on their components \eqref{viel_pul},
\be
\label{pul_boost}
\delta_\Lambda K^\mu = c^2 \Lambda^i E ^\mu_{~i}\, ,\quad  \delta_\Lambda T_\mu = \Lambda_i E_\mu^{~\,i}\, , \quad \delta_\Lambda  E_\mu^{~\,i} = c^2 \Lambda^i T_\mu \, , \quad \delta_\Lambda  E^\mu_{~i} = \Lambda_i K^\mu\, .
\ee
Also, the fluid four-velocity in terms of the PUL variables has the form
\be
u^\mu = - K^\mu - c^2 \ell^\mu \, , \quad u_\mu = - c^2 (T_\mu + \vec{\ell}_\mu)\,,
\label{velo_pul}
\ee
for some $\ell^\mu$, such that $T_\mu \ell^\mu = 0$. In \eqref{velo_pul} we have introduced the notation $\vec{\ell}_\mu \equiv \Pi_{\mu\nu} \ell^\nu$. Finally, the energy-momentum tensor \eqref{rel_hyd_T_ideal} expressed in PUL variables takes the form \cite{Armas:2023dcz}
\be
\label{rel_hyd_T_pul}
T^\mu_{~\,\nu} = (\epsilon+P)\left[K^\mu T_\nu + K^\mu \vec{\ell}_\nu + c^2 \ell^\mu T_\nu + c^2 \ell^\mu \vec{\ell}_\nu \right] + P \delta^\mu_\nu\, .
\ee

\medskip

Let us now consider the $c\to 0$ limit. The advantage of working with the PUL variables, beyond the obvious fact that one can use fully covariant notation, is that in the $c\to 0$ limit, the leading order terms in the expansion of the geometric data on the original pseudo-Riemannian manifold correspond to the geometric data on the Carroll manifold, i.e.
\begin{align}
K^\mu = k^\mu + \mathcal{O}(c^2) \, , \quad &T_\mu = \vartheta_\mu + \mathcal{O}(c^2) \, , \quad \Pi_{\mu\nu} = h_{\mu\nu} + \mathcal{O}(c^2) \, , \quad \Pi^{\mu\nu} = h^{\mu\nu} + \mathcal{O}(c^2) \, ,\nonumber\\
&E_\mu^{~\,i} = e_\mu^{~\,i} + \mathcal{O}(c^2)\, , \quad E^\mu_{~i} = e^\mu_{~i} + \mathcal{O}(c^2)\, .  \label{pul_carroll}
\end{align}
Further, as $u^\mu$ is local Lorentz boost invariant, from \eqref{pul_boost} we see that $\delta_\Lambda \ell^\mu = - \Lambda^i E^\mu_{~\,i}$. Therefore, when $c\to 0$, we get $\delta_c\ell^\mu = - h^{\mu\nu} \lambda_\nu + \mathcal{O}(c^2)$, where we have used $\Lambda_i = \lambda_i + \mathcal{O}(c^2)$, with $\lambda_i$ denoting the local Carroll boost parameters. Comparing with \eqref{Gold_Boost}, this implies that one can identify the leading term in the spatial part of $\ell^\mu$ with the spatial part of the Carroll boost Goldstone i.e. $\vec{\ell}_\mu \equiv \Pi_{\mu\nu} \ell^\nu = h_{\mu\nu} \varrho^\nu + \mathcal{O}(c^2)$. From \eqref{velo_pul}, this implies that $u_\mu = -c^2 (\vartheta_\mu + h_{\mu\nu}\varrho^\nu) + \mathcal{O}(c^4) = -c^2 \tilde{\vartheta}_\mu + \mathcal{O}(c^4)$. Using these results in \eqref{rel_hyd_T_ideal}, the leading order energy-momentum tensor in the limit $c\to 0$ becomes \cite{Armas:2023dcz}\footnote{Note that $k^\mu = -\mathfrak{u}^\mu$ when $\vec{\mathfrak{u}}_\mu = 0$.}
\be
T^\mu_{~\,\nu} = \mathcal{P} \delta^{\mu}_\nu - (\mathcal{E}+\mathcal{P}) \mathfrak{u}^\mu \tilde{\vartheta}_\nu\, ,
\label{carr_pul_emt}
\ee
where $\epsilon = \mathcal{E}+\mathcal{O}(c^2)$ and $P = \mathcal{P}+\mathcal{O}(c^2)$, with $\mathcal{E}, \mathcal{P}$ denoting the energy density and pressure for the Carroll fluid. The Carrollian energy-momentum tensor \eqref{carr_pul_emt}, obtained from the $c\to 0$ limit of the relativistic one \eqref{rel_hyd_T_ideal}, is identical to that for class II in \eqref{EMT_Classes}, corresponding to Carroll fluids with $\vec{\mathfrak{u}}_\mu = 0$. 

\medskip

As argued in \cite{Armas:2023dcz}, one can further show that the $c\to 0$ limit of the relativistic hydrodynamic equations $\nabla_\mu T^\mu_{~\,\nu} = 0$ expressed in PUL variables yields the Carroll hydrodynamic equations obtained in sec.~\ref{carroll_hydro}, after specializing to the PR gauge on the Carroll manifold, along with identifying the spatial part of the Carroll boost Goldstone $\vec{\varrho}_i$ with the PR variable $\beta_i$ via $\vec{\varrho}_i = - \beta_i$. 
This establishes that the class II of Carroll fluids constructed from symmetries, sec.~\eqref{sec:eq_part_fuc}, is identical to the one obtained by taking the $c\to 0$ limit of relativistic hydrodynamics.

\newpage

\section{Carroll hydrodynamics III: Applications to ultrarelativistic flows}
\label{heavy_ion}
The equations for the hydrodynamics of a Carroll fluid discussed in the previous sections portray an abstract appearance, and one may wonder whether they are of relevance to an actual physical setup. Interestingly, as was reported in \cite{Bagchi:2023ysc, Bagchi:2023rwd}, the Carroll fluid equations can be mapped to the equations that govern the dynamics of the quark-gluon plasma (QGP) produced in ultrarelativistic heavy-ion collisions, for specific choices of the geometric data $(\Omega, b_i, a_{ij})$ of the Carroll manifold.\footnote{In harmony with \cite{Bagchi:2023ysc, Bagchi:2023rwd}, we present the mapping between Carroll fluids and the Bjorken/Gubser flow models using the hydrodynamic equations expressed in the PR gauge (sec.~\ref{Eqs_Carroll_Hydro}).} The QGP models being referred to are the Bjorken flow \cite{Bjorken:1982qr} and Gubser flow \cite{Gubser:2010ze, Gubser:2010ui} models. As such, the dynamics of the hot expanding QGP produced in heavy-ion collisions is quite intricate. However, experiments at the Relativistic Heavy Ion Collider (RHIC) and the Large Hadron Collider (LHC) have shown that the QGP behaves like an almost perfect fluid, with a small shear viscosity to entropy density $\eta/s$ ratio \cite{Gale:2013da}. The phenomenological Bjorken and Gubser flow models provide hydrodynamic equations for the QGP assuming it to be  an ultrarelativistic fluid satisfying several simplifying symmetries. The common symmetry assumption present in both the models is that of \emph{boost-invariance} along the beam axis, or more precisely, the independence of the flow from the spacetime rapidity parameter, which is well justified in the central-rapidity region of heavy-ion collisions. Apart from boost invariance, the models make further simplifying assumptions to arrive at equations for the hydrodynamics of the QGP, and have been of immense importance in developing an intuitive understanding of the complex physical phenomena involved. 

\medskip

In the following subsections, we provide an overview of Bjorken and Gubser flow, and furnish the geometric data for a Carroll fluid appropriate for mapping to these phenomenological models. Note that we restrict the discussion below to $3+1$ dimensions.

\subsection{Carroll fluids and Bjorken flow}
\label{Bjorken}
One of the earliest hydrodynamic models attempting to describe the spacetime evolution of the hot and dense matter produced in ultrarelativistic heavy-ion collisions is Bjorken flow \cite{Bjorken:1982qr}. The model relies on several simplifying phenomenological assumptions. The pivotal assumption underlying Bjorken's model is the boost-invariance of the flow along the beam axis. The model further assumes translation and rotation invariance of the flow in the transverse plane. The spacetime evolution of the flow thus effectively becomes two-dimensional. To further elaborate the consequences of these symmetry assumptions, we need to choose a convenient coordinate system to work with. Without loss of generality, one can choose the $z$-axis to lie along the beam direction, while the $x, y$ axes form the transverse plane. We further assume that the collision between the two heavy ions occurs at the origin at time $t=0$. Demanding invariance of the flow under a particular spacetime transformation $x^\mu \to x^\mu +\xi^\mu(x)$ amounts to imposing the constraint $\pounds_\xi u^\mu = 0$, where $\pounds_\xi$ denotes the Lie derivative with respect to $\xi^\mu$, and $u^\mu(x)$ is the fluid four-velocity profile, normalized such that $u^\mu u_\mu  = -1$.\footnote{We work in natural units for Bjorken and Gubser flow.} For Bjorken flow, demanding the desired symmetries uniquely fixes the fluid four-velocity profile to be $u^\mu = (\gamma, 0,0,\gamma v)$, with $v=z/t$ and the Lorentz factor $\gamma = 1/\sqrt{1-v^2}$. It is beneficial to work with the Milne coordinates $(\tau, \rho)$ instead of the Cartesian $(t,z)$, with the proper time $\tau$ and rapidity $\rho$ given by (see fig.~\ref{fig:Milne})
\be
\label{def_tau_rho}
\tau = \sqrt{t^2 - z^2} \, , \quad \rho = \tanh^{-1} v = \tanh^{-1}\left(\frac{z}{t}\right) = \frac{1}{2} \log \left(\frac{t+z}{t-z}\right) .
\ee
The metric becomes
\be
ds^2 = - d\tau^2 + \tau^2 d\rho^2 + dx^2 + dy^2.
\label{Milne_Metric}
\ee
In terms of the Milne coordinates, the phenomenological assumptions of Bjorken flow lead to the fluid four-velocity being $u^\mu = (1, \vec{0})$ i.e.~the QGP appears to be at rest in these coordinates. In particular, there is no dependence on the rapidity $\rho$, which is the statement of boost-invariance along the beam axis in Milne coordinates.

\begin{figure}[t]
    \centering
    \includegraphics[scale=1.15]{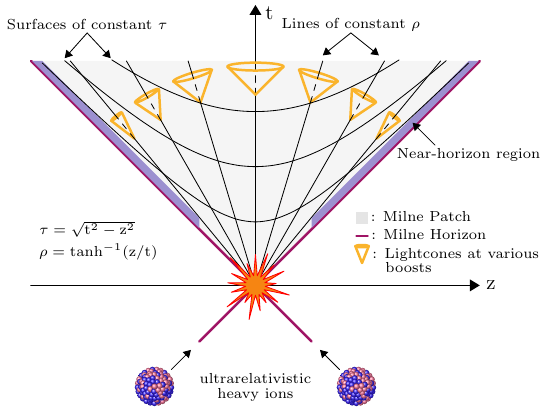}
    \caption{Depiction of the Milne patch covering part of the Minkowski spacetime.}
    \label{fig:Milne}
\end{figure}

With the velocity profile in hand, it is straightforward to work out the hydrodynamic equations for Bjorken flow. Working in Milne coordinates, and substituting $u^\mu = (1, \vec{0})$ in \eqref{rel_hyd_1} with the energy-momentum tensor given by \eqref{rel_hyd_T}, one finds that Bjorken flow is governed by the dynamical equation\footnote{This arises from the $\nu = \tau$ component of \eqref{rel_hyd_1}. The other equations following from \eqref{rel_hyd_1}, corresponding to $\nu = \rho, x$, and $y$, are respectively $\partial_\rho P = \partial_x P = \partial_y P = 0$ i.e.~they lead to the pressure being independent of $\rho, x$ and $y$. This is the manifestation of the phenomenological symmetries underlying Bjorken flow, implying that the pressure could only be a function of $\tau$. Since the equation of state relates the pressure to the energy density, this implies that the energy density too can only be a function of $\tau$.} 
\be
\frac{d\epsilon}{d\tau} = - \frac{\epsilon+P}{\tau} + \frac{1}{\tau^2} \left(\frac{2\eta}{3}+\zeta\right).
\label{Bjorken_Eq}
\ee
Given the equation of state $P = P (\epsilon)$, as well as the state dependence of the shear and bulk viscosities, $\eta = \eta(\epsilon)$ and $\zeta = \zeta(\epsilon)$, one can solve the first-order equation above to obtain the spacetime evolution of the energy density $\epsilon(\tau)$ for the QGP, with appropriate initial conditions.  

\medskip

{\ding{112}} \underline{\emph{Map from Carroll fluids to Bjorken flow}}

\medskip
Let us now move on to state the mapping between Carroll fluids and Bjorken flow. As was observed in \cite{Bagchi:2023ysc}, by choosing the geometric data on the Carroll manifold in PR parametrization \eqref{Carroll-structure} with the coordinate choice $x^\mu = (\tau, \rho,x,y)$ such that
\be
\Omega = 1, \quad b_i = - \beta_i, \quad a_{ij} dx^i dx^j = \tau^2 d\rho^2 + dx^2 +dy^2,
\label{BC_Data}
\ee
the Carroll fluid equations \eqref{LOVF} become
\be
\partial_\tau\epsilon_{(0)} = - \frac{\epsilon_{(0)}+p_{(0)}}{\tau} + \frac{1}{\tau^2} \left(\frac{2\eta_{(0)}}{3}+\zeta_{(0)}\right), \quad \partial_i p_{(0)} = 0.
\label{BC_Eqn}
\ee
The second equation above implies that the pressure for the Carroll fluid is independent of the rapidity $\rho$ as well as the transverse directions $x$ and $y$. From the equation of state, this immediately implies that the energy density too is independent of the rapidity as well as the transverse directions. Further, the first equation in \eqref{BC_Eqn} is nothing but the Bjorken flow equation \eqref{Bjorken_Eq}. Thus, both the dynamical equation for Bjorken flow as well as the phenomenological assumptions are captured by equations of Carroll hydrodynamics, simply by making an appropriate choice for the geometry of the underlying Carroll manifold. 

\medskip

It is also important to highlight here that the mapping \eqref{BC_Data} is invariant under local Carroll boosts, which can be checked easily using the transformations given in equations \eqref{LCPRTrans} and \eqref{CFTrans}.

\subsection{Carroll fluids and Gubser flow}
\label{Gubser}
Gubser flow \cite{Gubser:2010ze, Gubser:2010ui} provides a generalization of Bjorken flow, retaining the phenomenological assumptions of boost and rotation invariance along the beam axis, while trading off the over-restrictive translation invariance in the transverse plane with invariance under a combination of translations and special conformal transformations. The relaxed symmetry assumptions allow the hydrodynamic variables such as the energy density and  pressure to have a nontrivial radial profile as well, as opposed to Bjorken flow, where everything could only be a function of proper time due to the symmetries imposed. Gubser flow thus provides a more realistic model for the spacetime evolution of the QGP produced in heavy-ion collisions. The assumption of conformal invariance in Gubser flow can be justified based on the fact that at the ultrahigh energies involved in heavy-ion collisions, the masses of quarks can be neglected, making the dynamics effectively conformal. 

\medskip

In Cartesian coordinates, the symmetry generators for Gubser flow are (notation: $\xi \equiv \xi^\mu \partial_\mu$): 
\vspace{-5mm}
\begin{subequations}
\label{Gubser_Symms_1}
\begin{align}
&\circ\textrm{Boost-invariance along the beam axis:} \quad \xi_{\rm boost} = z\partial_t + t \partial_z\, , \label{Gubser_Boost}\\
&\circ\textrm{Rotation invariance about the beam axis:} \quad \xi_{\rm rot} = x\partial_y - y\partial_x\, ,  \label{Gubser_Rot}\\
&\circ\textrm{Symmetries in the transverse plane:}\left\{
\begin{aligned}
    \xi_1 = \partial_x + q^2(2xx^\mu\partial_\mu - x^\mu x_\mu \partial_x) \, ,\\
    \xi_2 = \partial_y + q^2(2yx^\mu\partial_\mu - x^\mu x_\mu \partial_y)\, ,\\
\end{aligned}\right.\label{Gubser_TSCT}
\end{align}
\end{subequations}
where we have once again chosen the beam direction to be the $z$-axis, while $(x, y)$ axes form the transverse plane. The generators $\xi_{\rm boost}$ and $\xi_{\rm rot}$ are also the symmetries of Bjorken flow, as discussed earlier in subsection \ref{Bjorken}. The new symmetry generators in Gubser flow are $(\xi_{1}, \xi_2)$, given in \eqref{Gubser_TSCT}, which are linear combinations of the translation generators in the transverse plane, $(\partial_{x}, \partial_y)$, and the special conformal generators $2b^\nu x_\nu x^\mu \partial_\mu - x^\mu x_\mu b^\nu \partial_\nu$, with $b^\nu = \delta^\nu_x/\delta^\nu_y$ for the two cases, respectively. The quantity $q$ in $(\xi_{1}, \xi_2)$ is a tunable parameter, with dimensions of inverse length. In particular, the limit $q\to 0$ reduces $(\xi_{1}, \xi_2)$ to translation generators in the transverse plane, and one may therefore expect to recover (conformal) Bjorken flow from Gubser flow in this limit, which is indeed the case. The transverse plane symmetry generators $(\xi_1, \xi_2, \xi_{\rm rot})$ satisfy the algebra
\be
[\xi_1, \xi_2] = -4q^2 \xi_{\rm rot}\, , \quad [\xi_1, \xi_{\rm rot}] = \xi_2\, , \quad [\xi_2, \xi_{\rm rot}] = - \xi_1\, ,
\ee
forming an $SO(3)_q$ subgroup of the full conformal group in four dimensions, $SO(4,2)$. Further, the $SO(3)_q$ generators commute with the generator $\xi_{\rm boost}$ of the $SO(1,1)$ subgroup of $SO(4,2)$.

\medskip

As was the case for Bjorken flow, it is easier to impose the desired symmetries \eqref{Gubser_Symms_1} on the fluid four-velocity $u^\mu$ using the Milne coordinates $(\tau, \rho, r, \phi)$, with the proper time $\tau$ and rapidity $\rho$ defined in \eqref{def_tau_rho}, while $r$ is the radial coordinate measuring distance from the beam axis in the transverse plane, $r\equiv\sqrt{x^2+y^2}, r\in[0,\infty)$, and $\phi \equiv \tan^{-1}\frac{y}{x}, \phi \in [0,2\pi]$ is the azimuthal angle. The metric in these coordinates takes the form
\be
ds^2 = - d\tau^2 + \tau^2 d\rho^2 + dr^2 + r^2 d\phi^2.
\label{Milne_Metric_2}
\ee
The symmetry generators now take the form $\xi_{\rm boost} = \partial_\rho$, $\xi_{\rm rot} = \partial_\phi$, and 
\begin{subequations}
\begin{align}
\xi_1 &= 2q^2 \tau r \cos\phi\,\partial_\tau + [1+q^2(\tau^2+r^2)] \cos\phi \, \partial_r - \frac{1+q^2(\tau^2 - r^2)}{r} \sin\phi \, \partial_\phi\, , \\
\xi_2 &= 2q^2 \tau r \sin\phi \, \partial_\tau + [1+q^2(\tau^2+r^2)] \sin\phi \, \partial_r + \frac{1+q^2(\tau^2 - r^2)}{r} \cos\phi \, \partial_\phi\, .
\end{align}
\end{subequations}

\medskip

By demanding the boost and rotation invariance of the fluid four-velocity, $\pounds_{\xi_{\rm boost}} u^\mu = 0$ and $\pounds_{\xi_{\rm rot}} u^\mu = 0$, along with $\mathbb{Z}_2$ symmetry under $\rho \leftrightarrow -\rho$ (equivalent to $z\leftrightarrow-z$ in Cartesian coordinates), the Gubser flow four-velocity profile gets fixed to
\be
\label{Gubser_velo_1}
u^\mu = (\cosh \kappa(\tau, r), 0, \sinh\kappa(\tau,r), 0).
\ee
Now, it turns out that there does not exist any choice for the function $\kappa(\tau,r)$ such that the four-velocity in \eqref{Gubser_velo_1} also becomes invariant under the generators $\xi_1, \xi_2$. In fact, given that $\xi_1, \xi_2$ generate conformal isometries for the background \eqref{Milne_Metric_2}, 
\be
\pounds_{\xi_a} g_{\mu\nu} = \frac{1}{2} \left(\nabla_\alpha \xi^\alpha_a\right) g_{\mu\nu}\, , 
\label{x1x2Milne}
\ee
with $a=\{1,2\}$, one only demands the fluid four-velocity to be invariant up to a conformal factor under the action of $\xi_1, \xi_2$. This leads to a unique solution for the function $\kappa(\tau, r)$, given by
\be
\label{def_kappa}
\kappa(\tau, r) = \tanh^{-1} \left[\frac{2q^2\tau r}{1+q^2(\tau^2+r^2)}\right],
\ee
such that
\be
\pounds_{\xi_1} u^\mu = -\frac{1}{4}\left(\nabla_\nu \xi^\nu_1\right) u^\mu\, , \quad \pounds_{\xi_2} u^\mu = -\frac{1}{4}\left(\nabla_\nu \xi^\nu_2\right) u^\mu .
\ee
To summarize, the final form of the four-velocity profile for Gubser flow, fixed purely by using symmetries, is given by \eqref{Gubser_velo_1}, with $\kappa(\tau, r)$ defined in \eqref{def_kappa}.

With the four-velocity profile fixed, it is now straightforward to compute the hydrodynamic equations for Gubser flow. From \eqref{rel_hyd_1}, using the energy-momentum tensor \eqref{rel_hyd_T}, along with the conformal equation of state $P = \epsilon/3$ and the fact the the bulk viscosity vanishes for a conformal fluid, one gets the following equations for Gubser flow,
\begin{subequations}
\label{Gubser_Flow_1}
\begin{align}
\partial_\tau \epsilon &= \frac{4\epsilon}{3} \left(\frac{\cosh 2\kappa - 2}{\tau} - \frac{\sinh 2\kappa}{r}\right) + \frac{2\eta_o \epsilon^{3/4}}{3~{\rm sech}^3\kappa} \left(\frac{1}{\tau}-\frac{\tanh \kappa}{r}\right)^2,\\
\partial_r \epsilon &= \frac{4\epsilon}{3} \left(\frac{\cosh 2\kappa - 1}{r} - \frac{\sinh 2\kappa}{\tau}\right) - \frac{2\eta_o \epsilon^{3/4}\sinh \kappa}{3~{\rm sech}^2\kappa} \left(\frac{1}{\tau}-\frac{\tanh \kappa}{r}\right)^2,
\end{align}
\end{subequations}
which are respectively the $\nu = \tau, r$ components of \eqref{rel_hyd_1}. The $\nu = \rho, \phi$ components of \eqref{rel_hyd_1} lead to the independence of the energy density from the rapidity $\rho$ and the azimuthal coordinate $\phi$, implying $\epsilon = \epsilon(\tau,r)$, which was built into the system by the choice of symmetries imposed. Note that in writing \eqref{Gubser_Flow_1}, we have used the scaling $\eta = \eta_o \epsilon^{3/4}$ for the shear viscosity of a conformal fluid, with $\eta_o$ being a dimensionless parameter. 

\medskip

It turns out that one can recast the Gubser flow equations \eqref{Gubser_Flow_1} into a simpler form by working on the ${\rm dS}_3\times\mathbb{R}$ background \cite{Gubser:2010ui}, instead of the Milne patch. To achieve this, one has to first perform a Weyl rescaling of the metric in \eqref{Milne_Metric_2} via $ds^2 \to ds^2/\tau^2$, followed by a change of coordinates from $(\tau, r) \to (\varsigma,\psi)$ such that
\be
\sinh\varsigma = -\frac{1-q^2(\tau^2-r^2)}{2q\tau}\, , \quad \tan \psi = \frac{2qr}{1+q^2(\tau^2 - r^2)}\, .
\label{ctrf}
\ee
This leads to the metric
\be
ds^2 = -d\varsigma^2 + \cosh^2\varsigma (d\psi^2+\sin^2\psi \, d\phi^2) + d\rho^2.
\label{ds3R_metric}
\ee
This is the metric of ${\rm dS}_3\times\mathbb{R}$ spacetime, with the coordinates $\varsigma \in \mathbb{R}, \psi \in [0,\pi], \phi \in [0,2\pi]$ covering the global three-dimensional de Sitter submanifold, while the rapidity $\rho \in \mathbb{R}$ as before. Note that the conformal symmetries $SO(3)_q$ of the Milne metric \eqref{Milne_Metric_2}, generated by $(\xi_1, \xi_2, \xi_{\rm rot})$, have now become exact isometries of the ${\rm dS}_3$ submanifold, as is evidenced by the presence of the two-sphere parametrized by $(\psi, \phi)$ in \eqref{ds3R_metric}. With $\xi_{\rm rot} = \partial_\phi$ remaining unchanged, the isometry generators $\xi_1, \xi_2$ now take the form
\be
\xi_1 = 2q\cos\phi \, \partial_\psi - 2q\cot\psi \sin\phi\, \partial_\phi\, , \quad \xi_2 = 2q\sin\phi \, \partial_\psi + 2q\cot\psi \cos\phi\, \partial_\phi\, .
\ee
Apart from the $SO(3)_q$ isometries, the $SO(1,1)$ boosts are trivially an isometry of \eqref{ds3R_metric}, generated by $\xi_{\rm boost} = \partial_\rho$. Thus, the advantage offered by working on the ${\rm dS}_3\times\mathbb{R}$ background  \eqref{ds3R_metric} is that the phenomenological symmetries of Gubser flow have now become exact isometries of the background, as opposed to working in the Milne patch \eqref{Milne_Metric_2}, where $(\xi_1, \xi_2)$ generated conformal symmetries, \eqref{x1x2Milne}. Now, demanding the fluid four-velocity profile to be invariant under the $SO(3)_q\times SO(1,1)\times \mathbb{Z}_2$ symmetries of the ${\rm dS_3}\times \mathbb{R}$ background leads to the fluid being at rest i.e.~$u^\mu=(1,0,0,0)$. Using this velocity profile in the hydrodynamic equations \eqref{rel_hyd_1}, the $\nu = \psi, \phi, \rho$ components simply lead to the independence of the energy density from these coordinates, simply as consequences of the symmetries imposed, implying $\epsilon = \epsilon(\varsigma)$. The only nontrivial dynamical equation follows from the $\nu = \varsigma$ component of \eqref{rel_hyd_1}, given by
\be
\frac{d\epsilon}{d\varsigma} = - \frac{8\epsilon}{3}\tanh\varsigma + \frac{2}{3}\eta_o \epsilon^{3/4} \tanh^2\varsigma\, .
\label{Gubser_flow_dS}
\ee
Compared to the formulation on the Milne background, equation \eqref{Gubser_Flow_1}, the formulation of Gubser flow on the ${\rm dS}_3\times\mathbb{R}$ background is thus much simpler. One can always invert the coordinate transformation in  ~\eqref{ctrf} and undo the Weyl rescaling to go back to Milne from the ${\rm dS}_3\times\mathbb{R}$ formulation.

\medskip

{\ding{112}} \underline{\emph{Map from Carroll fluids to Gubser flow}}

\medskip

The Carroll hydrodynamic equations \eqref{LOVF} are capable of capturing both the phenomenological assumptions as well as the dynamical equations corresponding to Gubser flow, purely from a geometric standpoint, as was also the case for Bjorken flow. To see this, let us first specialize to a conformal Carroll fluid, by setting $p_{(0)} = \epsilon_{(0)}/3$ and $\zeta_{(0)} = 0$ in \eqref{LOVF}. This yields the following equations for a conformal Carroll fluid in the PR parametrization,\footnote{With $\zeta_{(0)} = 0$, from \eqref{Xi_Def} one has $\Xi_{ij} = \eta_{(0)} \xi_{ij}$ and $\Xi = 0$.}
\begin{subequations}
\label{LOVCF}
\begin{align}
    \hat{\partial}_{t}\epsilon_{(0)} &= - \frac{4}{3} \theta\epsilon_{(0)} + \eta_{(0)}\xi^{ij}\xi_{ij}, \\
    \hat{\partial}_i \epsilon_{(0)} &=  - 4\varphi_i\epsilon_{(0)}  - (\hat{\partial}_{t} + \theta)[4\epsilon_{(0)}\beta_i - 3\eta_{(0)}\xi_{ij}\beta^j] +3 (\hat{\nabla}_j + \varphi_j)(\eta_{(0)}\xi^{j}_{\ i}). 
\end{align}
\end{subequations}
Next, let us consider the mapping of conformal Carroll hydrodynamics to Gubser flow described in Milne coordinates. With the coordinates $x^\mu$ on the Carroll manifold \eqref{Carroll-structure} adapted to the Milne coordinates, $x^\mu = (\tau, \rho, r, \phi)$, followed by choosing the geometric data for the manifold such that 
\be
\begin{split}
\label{Gubser_Data_1}
&\Omega = \cosh \kappa\, , \quad b_r = -\beta_r + \sinh\kappa \, , \quad b_\rho = - \beta_\rho\, , \quad b_\phi = -\beta_\phi\, ,\\
&a_{ij}dx^i dx^j = \tau^2 d\rho^2 + [1+2q^2(\tau^2+r^2) + q^4 (\tau^2 - r^2)^2](dr^2 + r^2 d\phi^2)\, ,
\end{split}
\ee
the conformal Carroll fluid equations \eqref{LOVCF} become
\be
\label{Gubser_Carroll_1}
\begin{split}
\partial_\tau \epsilon_{(0)} &= \frac{4\epsilon_{(0)}}{3} \left(\frac{\cosh 2\kappa - 2}{\tau} - \frac{\sinh 2\kappa}{r}\right) + \frac{2\eta_o \epsilon_{(0)}^{3/4}}{3~{\rm sech}^3\kappa} \left(\frac{1}{\tau}-\frac{\tanh \kappa}{r}\right)^2,\\
\partial_r \epsilon_{(0)} &= \frac{4\epsilon_{(0)}}{3} \left(\frac{\cosh 2\kappa - 1}{r} - \frac{\sinh 2\kappa}{\tau}\right) - \frac{2\eta_o \epsilon_{(0)}^{3/4}\sinh \kappa}{3~{\rm sech}^2\kappa} \left(\frac{1}{\tau}-\frac{\tanh \kappa}{r}\right)^2,\\
\partial_{\rho} \epsilon_{(0)} & = 0\, , \quad \partial_{\phi} \epsilon_{(0)} = 0\, ,
\end{split}
\ee
where we have used $\eta_{(0)} = \eta_o \epsilon_{(0)}^{3/4}$ which follows from the $c\to 0$ limit of the corresponding relativistic relation. The first two equations for the conformal Carroll fluid in \eqref{Gubser_Carroll_1} are indeed the dynamical equations for Gubser flow in Milne coordinates, \eqref{Gubser_Flow_1}, while the last two equations in \eqref{Gubser_Carroll_1} are nothing but the statements of independence of the energy density from the rapidity and the azimuthal coordinate i.e.~rotations about the beam axis. Thus, the choice of the geometric data \eqref{Gubser_Data_1} for a conformal Carroll fluid maps it to Gubser flow, establishing the equivalence between the two.

\medskip

The equivalence is of course not limited to Milne coordinates, and can also be seen to work for Gubser flow on the ${\rm dS}_3\times\mathbb{R}$ background. Adapting the coordinate chart on the Carroll manifold \eqref{Carroll-structure} such that $x^\mu = (\varsigma, \rho, \psi, \phi)$, and choosing the geometric data on the manifold such that
\be
\label{Gubser_Data_2}
\Omega = 1\, , \quad b_i = -\beta_i \, , \quad a_{ij}dx^i dx^j = \cosh^2\varsigma\, (d\psi^2+\sin^2\psi \, d\phi^2) + d\rho^2\, ,
\ee 
the conformal Carroll fluid equations \eqref{LOVCF} become
\be
\label{Gubser_Carroll_2}
\partial_\varsigma\epsilon_{(0)} = - \frac{8\epsilon_{(0)}}{3}\tanh\varsigma + \frac{2}{3}\eta_o \epsilon_{(0)}^{3/4} \tanh^2\varsigma\, , \quad \partial_\rho \epsilon_{(0)} = 0\, , \quad \partial_\psi \epsilon_{(0)} = 0\, , \quad \partial_\phi \epsilon_{(0)} = 0\, .
\ee
The first equation above is clearly the dynamical equation for Gubser flow on the ${\rm dS}_3\times\mathbb{R}$ background, \eqref{Gubser_flow_dS}, while the remaining ascertain the independence of energy density from $\rho, \psi$ and $\phi$, which are nothing but the $SO(3)_q\times SO(1,1)\times \mathbb{Z}_2$ phenomenological symmetries of Gubser flow, arising purely out of geometric considerations on the Carroll side. This acts as as a testament to the robustness of the mapping between Carroll hydrodynamics and Gubser flow, and more generally to ultra-relativistic boost-invariant models of heavy-ion collisions.

\medskip

Finally, we should note that the mapping \eqref{Gubser_Data_1} between Carroll hydrodynamics and Gubser flow on the Milne background, as well as the mapping \eqref{Gubser_Data_2} between Carroll hydrodynamics and Gubser flow on the ${\rm dS}_3\times \mathbb{R}$ background, are invariant under tangent space local Carroll boost transformations on the associated Carroll manifold. This can be checked straightforwardly by using the local Carroll boost transformation properties of the various objects involved, which were derived in equations \eqref{LCPRTrans} and \eqref{CFTrans}.

\subsection{Pointers to literature}
\label{hydr_pointers}
Carroll hydrodynamics is a subject in its infancy and remains largely unexplored. Following are some directions which we have not addressed in the present review.
\begin{itemize}
\item \emph{Carroll hydrodynamics and black holes}
\begin{itemize}
    \item[$\star$] The black hole membrane paradigm \cite{PhysRevD.18.3598, PhysRevD.33.915} models the dynamics of the event horizon of a black hole as a fluid. It was argued in \cite{Donnay:2019jiz} that the associated hydrodynamic equations are that of a Carroll fluid, with further work appearing in \cite{Freidel:2022vjq, Redondo-Yuste:2022czg}. This is natural when viewed from the perspective of the fact that Carroll symmetries generically arise on null hypersurfaces, which black hole horizons indeed are. See some further discussion on this in sec.~\ref{sec:stretched}.
    \item[$\star$] Another approach for constructing Carroll hydrodynamics from a symmetry based perspective appears in \cite{Freidel:2022bai}. It does not incorporate the dynamics of the Goldstone fields associated with the spontaneous breakdown of Carroll boost invariance, but rather introduces and makes use of a new set of symmetries, the near-Carrollian diffeomorphisms, that are analogues of spacetime diffeomorphisms in the vicinity of the event horizon of a black hole. 
\end{itemize}
\item \emph{One-dimensional Carroll fluids:} Refs.~\cite{Athanasiou:2024lzr, Athanasiou:2024ykt, Petropoulos:2024jie} investigate Carroll fluids in one spatial dimension, which happen to have a dual correspondence with one dimensional Galilean fluids, owing to the isomorphism between Carrollian and Newton-Cartan manifolds in 2D. Aspects such as the well-posedness of the hydrodynamic equations and criteria for the boundedness of their solutions have been studied.
\item \emph{Carrollian fluid/gravity correspondence:} One of the most concrete realisations of the holographic AdS/CFT duality is the fluid/gravity correspondence \cite{Bhattacharyya:2007vjd}, which relates the hydrodynamic regime of $d$-dimensional holographic CFTs in (conformally) flat spacetimes with gravitational dynamics in $d+1$-dimensional asymptotically AdS spacetimes. Refs.~\cite{Ciambelli:2018wre, Ciambelli:2020eba, Ciambelli:2020ftk} discuss the limiting behaviour of the fluid/gravity correspondence when the AdS bulk cosmological constant is tuned to zero, which is holographically tantamount to taking $c\to 0$ in the boundary CFT, to construct a Carrollian fluid/gravity correspondence between the hydrodynamic regime of a holographic Carrollian CFT and its gravitationally dual asymptotically flat spacetime.
\end{itemize}

\newpage

\section{Carroll and Condensed Matter}
\label{carroll_cond_matt}
As per the last section, we have been discussing tools emerging from Carrollian dynamics to study some real life phenomena. In recent years, more and more disparate physical situations have been found to be connected to Carrollian physics. As mentioned previously, it has been very clear that as long as a characteristic (or effective) velocity of the system goes to zero, we can think of an emergent Carrollian symmetry. For Carroll hydrodynamics in the last section, a similar notion of effective velocity has been presented. In the case of condensed matter physics, this characteristic velocity is often given by the \textit{Fermi Velocity} of the system, especially if we are thinking of strongly coupled electrons. 
\medskip

But in general, the appearance of Carroll symmetries in condensed matter systems is further widespread. Due to the inherent ultra-local nature of fields it brings about, the underlying symmetries can be mapped to that of other theories, including that of \textit{Fractons}: a (quasi)particle which is unable to move in response to applied forces if set in isolation. Fractons can however move when we form dipoles, and this procedure effortlessly embeds into the representation theory of Carrollian dynamics. Also, as mentioned in earlier sections, there are Carroll particles which actually \textit{can move}, and they may move following a generalized Hall law, partially breaking the Carroll boost symmetry. 
\medskip

Perhaps the most interesting application of Carrollian symmetries in condensed matter physics in recent times emerged as its connection with \textit{flat dispersion relations} in quantum systems. The infinite dimensional supertranslation symmetry, when applied to Hamiltonian systems has a very particular consequence, namely the Hamiltonian densities of such systems at two different spatial points (Poisson) commute \cite{Bagchi:2022eui}. As discussed in earlier sections, this is a direct sign of emergence of Carroll symmetries. This is return guarantees, in the momentum space, that energy is independent of momentum, giving rise to flat dispersion relations. These situations are very important ones, especially for many-body systems in a condensed matter physics context, and can happen as the group velocity of a excitation localised in real space vanishes everywhere on a lattice due to destructive interference. 
\medskip

Fuelled by actual material applications, the study of flat band systems has started an exciting flurry of works to study characteristics of such structures \footnote{See \cite{2024Nanop..13.3925D} for a recent review on the applications and for a more exhaustive list of references.}. This includes a vast
spectrum of systems and phenomena, ranging from Moiré patterns in multi-layer graphene \cite{Moire1, Cao2018,PhysRevLett.122.106405}, superconductivity \cite{Volovik2018,Aoki2020}, fractional Quantum Hall Effect \cite{PARAMESWARAN2013816}, non-Hermitian quantum
systems \cite{PhysRevB.96.064305} etc. Even more excitingly, flat dispersions appear in the dynamics of shallow water waves when written as an effective gauge theory \cite{Tong:2022gpg}, which prompts one to expand the scope of Carroll dynamics beyond the many-body paradigm.
\medskip

The seminal realisation that bilayer graphene systems and flat bands at magic angles were directly linked with Carroll physics, came in \cite{Bagchi:2022eui}, adding to the rich literature in this direction. Furthermore, due to the intricate relationship between flat dispersions and Carroll symmetries, this opens up the possibility of actually designing explicitly supertranslation invariant lattice systems, that in turn may find real-world applications. This is what was precisely spearheaded in \cite{Ara:2024fbr}, where the key ingredient of this construction, i.e. the identification of so called \textit{Compact Localised States} (CLSs) \cite{PhysRevLett.118.166803,2014EL....10530001F}, was elucidated. It was clearly shown that CLS states appear naturally from the choice of a supertranslation invariant basis which demonstrates ultra-local correlation functions.  

\medskip
Some of these structures discussed above will be the main topic of discussion in this section, while we will only be touching upon some others, although many details will surely be omitted. Most of what we will discuss is still at a very nascent stage, and excitement is only slowly building up around these ideas. A detailed pointer to other related literature is provided at the end of the section, and interested readers are requested to consult the mentioned works and references therein.

\subsection{Fractons, restricted mobility and Carroll dynamics}
Quite recently, a novel class of particles, long conjectured \cite{Chamon:2004lew, Haah:2011drr, Vijay:2015mka, Vijay:2016phm}, has been discovered that features not only charge conservation but also conservation of dipole moment, resulting in severe restrictions on the mobility of the particles. They form a very new phase of matter, fuelled by truly exotic symmetries.  
The demand for explicit dipole symmetry conservation puts constraint on the form of the action of such associated fields and leads to non-Gaussian theories. The non-Gaussian term immediately breaks the infinite multipole symmetry of the free field theory down to only the dipole symmetry. One can get rid of this non-Gaussianity at the expense of linearly realised dipole symmetry or by removing the presence of spatial derivatives in the Lagrangian \cite{Bidussi:2021nmp, Marsot:2022imf}. The latter case gives us a Carrollian theory since usual particle symmetries are enhanced.  
 
\medskip

To start with, consider a complex scalar field $\Phi$, which describes a set of conserved particles. The demand of particle number conservation is basically saying that the action of this field will be invariant under a global $U(1)$ transformation $\Phi \to e^{i\a}\Phi$, for constant $\a$. The charge density operator is given by, $\rho = \Phi^\dagger\Phi$ and because of $U(1)$ invariance total charge $\int d^dx\, \rho$ is constant. Now let's further demand that this theory conserves dipole moment i.e. $\int d^dx\, (\rho \vec{x}) = \text{constant}$. This implies an additional invariant transformation $\Phi \to e^{i\vec{\beta}\cdot \vec{x}}\Phi$ for some constant vector $\vec{\beta}$. So what happens is that instead of a change by a global constant, the phase of the scalar field can now changes by a linear function in position. So we can write a general transformation $\Phi \to e^{i\a(x)}\Phi$, where $\a(x)$ is now at most a linear function in $x$.
\medskip

Now one asks, what are the possible terms in the field Lagrangian we can write down which respect this symmetry transformation? It is obvious that any number of spatial derivatives acting only on single $\Phi$ will not transform covariantly. Rather we must have the form $\mathcal{X}_{ij} = \Phi\d_i\d_j\Phi - \d_i\Phi\d_j\Phi$, quartic in spatial derivatives, which transforms covariantly under this transformation. This leads us to the following invariant Lagrangian at lowest order 
\begin{equation}\label{fracLag}
    \mathcal{L}= |\d_t\Phi|^2 - \lambda \mathcal{X}_{ij} \mathcal{X}_{ij}^*- \tilde{\lambda}\mathcal{X}_{ii} \mathcal{X}_{jj}^*
\end{equation}
Here $\lambda$ and $\tilde{\lambda}$ are arbitrary couplings for the non-gaussian terms. One can also add a mass term of the form $m^2|\Phi|^2$. This describes a field theory for fractons \cite{Pretko:2018jbi}. Such fracton field theories evidently have a characteristic non-Gaussian term that reflects the fact that fractons necessarily interact with each other even in the absence of any mediating gauge field \footnote{Once we allow for our theory to have a non-linearly realised dipole symmetry, it is indeed possible to have both Gaussian terms and added spatial derivatives of fields. For an example involving a symmetry breaking scenario, see \cite{Bidussi:2021nmp}.}. Even though a fracton cannot move on its own, it can achieve restricted mobility by pushing off other fractons in the system through the exchange of virtual dipoles as shown in Fig.~\ref{fracton}.

\medskip

To gain a better understanding of the algebraic structure for the action \eqref{fracLag}, we look at the dipole algebra spanned by the following generators: electric charge $Q^{(0)}$, dipole charge $Q_i^{(1)}$, fracton energy $H_f$, linear momentum $P_i$, angular momentum $J_{ij}$. They satisfy the following non-vanishing commutation relations \cite{Gromov:2018nbv}
\begin{eqnarray}
    && [J_{ij},J_{kl}] = 4\delta_{[i[k}J_{j]l]}\,, \quad [J_{ij},P_k] = 2\delta_{k[i}P_{j]} \\
    && [J_{ij},Q_k^{(1)}] = 2\delta_{k[i}Q^{(1)}_{j]}\,,\quad [P_i,Q_j^{(1)}] = \delta_{ij}Q^{(0)} \,.
\end{eqnarray}
The explicit forms of these generators can be found by computing the Noether charges and assuming canonical commutation relations related to the symmetries of the Lagrangian \cite{Bidussi:2021nmp}. 
\medskip

\begin{figure}[t]
\centering
\includegraphics[scale=1]{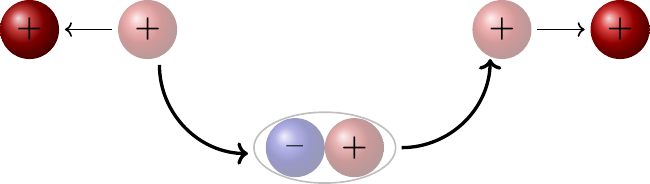}
\caption{Interaction of fractons via virtual dipole exchange.}
\label{fracton}
\end{figure}

Notice the similarity of this with the Carroll algebra \eqref{carral} when one replaces $Q^{(0)} \rw H$ (time translation generator), $Q_i^{(1)} \rw C_i(\text{Carroll boost generator})$ and neglect the additional generator $H_f$ which basically commutes with all the other generators. Therefore, it can be said, that the symmetries of the fracton field theories with conserved $U(1)$ and dipole charge are equivalent to Carroll symmetries up to an additional central element, leading to Carroll/fracton correspondence \cite{Figueroa-OFarrill:2023vbj, Figueroa-OFarrill:2023qty}. This puts the associated mobility restrictions in a physical context. 
\medskip

Exactly like with the Hamiltonian in the Carroll case, the fractonic charge $Q^{(0)}$ divides the fracton representations on two classes. When the charge vanishes the theory correspond to (immobile) monopoles, but when the charge is non-zero the associated theory is of  neutral fractons, in particular dipoles which are mobile. Fracton monopoles do not move and are stuck to a spatial point, which mirrors the sense of lightcones closing in the Carroll case, albeit in this case it happens due to manifest dipole conservation. The second class of representations, i.e. moving dipoles, can often be compared to magnetic-type Carroll particles that one gets from Poincaré tachyonic representations.

\subsection{Flattening of bands and supertranslations}

We now move to flat bands and the role of Carrollian symmetries in this context. 

Before we go into technical arguments and calculations regarding origins of flat-band physics, it is better to give a quick and simple physical argument of why one should expect Carroll invariance, or supertranslations in general, to appear in the context of flattening of bands in electronic systems.
Recall that flat bands basically emerge in a limit where the Dirac cone in momentum space opens up completely, and energy in this case is not dependent on canonical momentum. This complete opening up of the Dirac cone in momentum space, by virtue of Fourier expansions, turns out to be equivalent to actual position space lightcones closing down. Although note that, in the case of condensed matter systems, these local Minkowski frames are to be constructed using Fermi velocity as the characteristic one.

\medskip

This setup then clearly mimics the physics of the Carrollian limit and hence whenever one encounters flat bands in the systems we will be discussing in this section, one should always encounter Carrollian structures. This also makes sense from the point of view that flat bands emanate from localisation of excitations in a lattice system, which we will discuss in more detail. From a Carroll perspective, this localization in space is a manifest consequence of the closing up of the effective lightcone as the characteristic velocity goes to zero. Although this might sound very heuristic and hand-waiving, this simple argument actually makes the emergent flat band picture very lucid. See Fig.~\ref{collapse} for an illustration of this process. 
\medskip

\begin{figure}[h]
\begin{subfigure}{0.5\textwidth}
\includegraphics[width=0.8\linewidth, height=5.5cm]{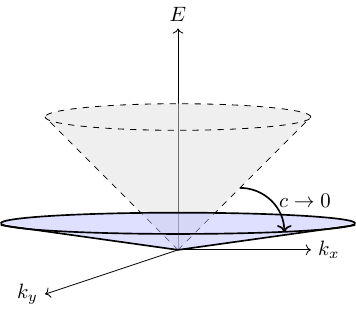} 
\end{subfigure}
\begin{subfigure}{0.5\textwidth}
\includegraphics[width=0.8\linewidth, height=5.5cm]{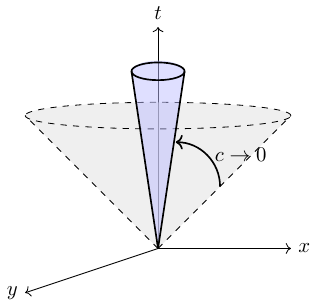}
\end{subfigure}
\caption{\textbf{Left:} Flattening of the dispersion relation in momentum space.~~\textbf{Right:} Shrinking of the space-time light cone as $c\to 0$.}
\label{collapse}
\end{figure}

Let us now move into some more technical reasoning for the same band flattening. For this, we should start with the Dirac-Schwinger condition, as discussed around \eqref{dsc}. Recall that, in relativistic classical (quantum) field theories, this condition translates into Poisson (or commutator) brackets for the Hamiltonian density computed in two different spatial points: 

\begin{equation}
\left\{\mathcal{H}(x), \mathcal{H}\left(x^{\prime}\right)\right\}_{\mathrm{PB}}
=2 \cP^k(x) \partial_{k}\delta\left(x-x^{\prime}\right)
+\left(\partial_k \cP^k(x')\right) \delta\left(x-x^{\prime}\right),
\end{equation}

where $\cP_k$ are the momentum density of the theory and canonical conjugate relations are assumed. Now as discussed in preceding sections, the hallmark of Carrollian dynamics to emerge is the vanishing of this Poisson bracket between Hamiltonian densities. So a rule of thumb to discern Carroll invariant systems would be to look for the sacrosanct condition:
\be\label{HPB}
\{\mathcal{H}(x),\mathcal{H}(x')\}_{\rm PB} = 0.
\ee
This inherently Carrollian condition has important physical significance along the lines we discussed earlier. As discussed in \cite{Bagchi:2022eui}, the vanishing of Dirac-Schwinger condition implies the emergence of supertranslations. Quantum mechanically, this also implies an emergent degeneracy in the eigenmanifold, that goes hand-in-hand with flat band physics. 
\medskip

To enunciate more, let us consider a generic free theory Hamiltonian, expressed in terms of infinite number of Heisenberg picture oscillators:
\begin{eqnarray}\label{first_master}
H = \sum_n E_n a^{\dagger}_n a_n ,
\end{eqnarray}
we will also consider $n$ to be a continuous variable so we can work with a uncountably infinite set, and convert the above sum to an integral when it is of opportune. All other charges in the system are also supposed to be represented by the mode number $n$.
\medskip

Let us start with the case of a fermionic theory, since that is more important for, say, physics of spin-chains. The Carroll version of fermions have already been discussed in detail in the section \eqref{spinsec}. For us the fermions are canonical to start with $i.e.$ they have the usual equal-time anti-commuting oscillators 
\be{}
\{ a^{\dagger}_n, a_m\} = \delta_{mn}
\ee 
 the Kronecker delta here works for only discrete mode numbers, and is to be replaced by a continuous Dirac delta when we take the relevant continuum limit. Now let us work with a self-adjoint operator $\mathfrak{H}$, deemed to be the single particle Hamiltonian with suitable boundary conditions implied on the eigenfunctions. By simple QFT rules we should have  the eigenvalues of $\mathfrak{H}$ as $E_n$ that appears in  free theory Hamiltonian \eqref{first_master}. Note that here $u_n(\vec{r})$ are orthonormal single particle eigenfunctions: 
\be{}
\mathfrak{H} u_n(\vec{r}) = E_n u_n(\vec{r}).
\ee 
Hence the Hamiltonian can be written in this basis, in terms of second quantised eigenfunctions:
\begin{eqnarray} \label{repack}
&& H = \sum_{m,n} \int d^d\vec{r}\,~ a^{\dagger}_m(t) u^{\star}_m (\vec{r}) ~\mathfrak{H} \left(a_n(t) u_n (\vec{r}) \right) = \int d^d\vec{r} ~\, \Psi^{\dagger} (\vec{r},t )\mathfrak{H} \Psi (\vec{r},t ) \\
&& \mbox{where } \Psi (\vec{r},t ) := \sum_n a_n(t) u_n(\vec{r}). \nonumber
\end{eqnarray}
The definition of the field operator $\Psi$ above, in terms of the oscillators directly imply the following non-zero anti-commutator with a Dirac delta on the RHS:
\begin{eqnarray}
\{ \Psi (\vec{r},t) , \Psi^{\dagger}(\vec{r}',t) \} = \delta^d (\vec{r} - \vec{r}'),
\end{eqnarray}
which is the usual canonical relation one would expect for spinor fields.
At the classical level, the same action can be written as a phase space one :
\begin{eqnarray}
S = \int dt\, d^d \vec{r}~ \Psi^{\star} \left( i\partial_t - \mathfrak{H} \right) \Psi .
\end{eqnarray}
In 4D, i.e. for $d=3$ Weyl fermions, the dispersion for the Hamiltonian is $E_{\vec{k}} = c |\vec{k}|$, while for massive Dirac fermions we have the usual relativistic dispersion, $E_{\vec{k}} = c \left( \vec{k}^2+ m^2 c^2 \right)^{1/2}$. Note that $\mathfrak{H}$ is either a 2 and 4 dimensional matrix valued differential operator respectively for these cases. This is very natural as the non-relativistic or the large $c$ limit of the later gives us: $E_{\vec{k}} \approx \frac{\vec{k}^2}{2m}$, which describes the Schr{\"o}dinger field theory and the eigenoperator described by $\mathfrak{H} = -\frac{1}{2m} \nabla^2$, on single component fermions.

\medskip

To come  back to our case of interest, let us focus on a single component continuum field $\Psi(\vec{r},t)$ to describe the real space degrees of freedom. We now impose the \textit{supertranslation} transformations on it, which depends on an arbitrary real function $f$ on $\mathbb{R}^d$:
\begin{eqnarray}\label{suptra}
\delta_{f} \Psi(\vec{r},t) = f(\vec{r}) \partial_t \Psi,
\end{eqnarray}
 As described earlier, $f=1$ is the time translation and $f= \vec{b} \cdot \vec{r}$, for any $d$-vector $\vec{b}$, is associated to the Carrollian boost. Since $f$ can be arbitrary, we have an infinite dimensional symmetry arising here. Note that this don't have an inhomogeneous term because this is a single component case of fermion. Now recall that on the space of Heisenberg picture oscillators, time translations imply following temporal evolution: 
\be{}
a_m(t) = e^{-iE_m \,t} a_m (0),
\ee 
hence this supertranslation transformation acts as:
\begin{eqnarray} \label{a_tran}
\delta_f a_m (t) = -i  \sum_{n} E_n \int d^d \vec{r}\, f(\vec{r}) \, u_n (\vec{r})  u^{\star}_m (\vec{r}) a_n(t) 
\end{eqnarray}
and similarly on $a^{\dagger}_m$. A little algebra tells one the transformation of the fundamental anti-commutator:
\begin{eqnarray}
\delta_f \{a_m, a^{\dagger}_n \} = - i \int d^d \vec{r} \,f(\vec{r}) u_n (\vec{r})  u^{\star}_m (\vec{r}) (E_n -E_m)
\end{eqnarray}
Note that the above expression vanishes identically for $f =1$ due to orthonormality of the eigenfunctions $u_n$. This is inherently expected, since by definition the system is time translation invariant.  A sufficient condition for the variation of the anti-commutator to vanish for an arbitrary function $f$ is  $$E_n = E_m \quad \forall \, m,n,$$ 
 i.e., the energy dispersion for the system in question must be trivial. In the regime of condensed matter systems, this trivialization of the energy dispersion has very particular significance. One says in this case the  Hamiltonian has to be a \textit{flat-band} one. The invariance of the brackets $\{a_m, a_n\}$ and $\{a^{\dagger}_m, a^{\dagger}_n\}$ does not add anything do this dynamics. One can further check that with the flat-dispersion Hamiltonian $H = E \sum_{n} a^{\dagger}_n a_n$, the system is also explicitly invariant under the supertranslation transformations:
\begin{eqnarray}
\delta_f H  = 0&=& i \sum_{n,m} \int d^d \vec{r} \, f(\vec{r})\,E_n \left(  u^{\star}_n (\vec{r})  u_m (\vec{r}) a^{\dagger}_n a_m - u_n (\vec{r})  u^{\star}_m (\vec{r})a^{\dagger}_m a_n \right) 
\end{eqnarray}
The canonical structure and the Hamiltonian for the flat-band system remaining invariant under these supertranslations is of utmost significance here, as this implies one can design systems with flat-bands given  infinite number of Carrollian supertranslation symmetries are emergent. Thus, our differential operator with inherent supertranslation invariance simply gives rise to flat-band spectra.

\medskip

Before delving into particular examples, we try to understand connection between flat-bands and Carroll symmetry starting from Lorentz covariant systems having trivial open boundary conditions. In our discussions in the preceding sections, we have encountered many examples of intrinsically Carrollian systems. Now the question is whether it is natural to identify a Carroll theory as non-dispersive one? To answer, we take the Hamiltonian of a free massive Lorentz covariant theory, whose Hamiltonian in terms of Fourier basis oscillator modes become:
\begin{eqnarray} \label{master}
H = \int d^d \vec{k} ~\sqrt{c^2 \vec{k}^2 + m^2 c^4} \,~ a^{\dagger}_{\vec{k}} a_{\vec{k}} .
\end{eqnarray}
We have just one field component and all other indices on the oscillators have been suppressed. The dispersion relation $E_{\vec{k}} = \sqrt{c^2 \vec{k}^2 + m^2 c^4}$ bears the fact that the theory is manifestly Lorentz covariant. Written in real space, the integrand in the Hamiltonian will surely involve spatial derivatives in fields, which will make sure Hamiltonian densities do not commute, and the Dirac-Schwinger condition kicks in.
\medskip

But, as discussed earlier, a direct consequence of Carroll invariance is commuting Hamiltonian densities, as per \cite{Henneaux:2021yzg}. Therefore it follows that theories which have no spatial derivative in real space Hamiltonian, i.e. no $\vec{k}$ dependence in the Fourier version, hence giving rise to  non-dispersive Hamiltonians in terms of Fourier modes. These, as one can deduce, should automatically have Carrollian symmetry. These are the usual Electric Carroll theories we have seen earlier \eqref{CarrKG}. The attentive reader will now point out that we have also seen that there are Carrollian theories having explicit spatial derivatives, like in the case of Magnetic Carroll theories (see Sec. \ref{carrscalar}). But still the Hamiltonian densities commute and there is an emergent Carrollian symmetry for those theories. However, there is no manifest momentum independence of Energy as in the case discussed above.

\medskip

One straightforward way to construct such an ``Electric" theory is starting with relativistic dispersion $E_{\vec{k}}= \sqrt{c^2\vec{k}^2 + m^2 c^4}$ and promptly take $c \rightarrow 0$ with $m c^2 = E_0$ remaining finite. $E_0$ defines the UV cut-off at which the theory is defined. Hence $E_k = E_0$ is momentum independent, and it consequently becomes dispersion-less (or flat-band). For the special case of $m=0$, the dispersionless Hamiltonian is identically zero.
\medskip

With this, it should be clear that physics of dispersionless Hamiltonians directly ties in with supertranslation symmetry, and subsequently to Carroll symmetry. One must however remember that in many-body systems, which are of main interest in condensed matter systems, appearance of flat bands may be more intricate than simple transformations at the level of oscillators, which we have been doing so far for continuum systems. We will later come back to examples of such systems. 

\subsection{Flat bands: Examples}
\subsubsection{Warm up: Creutz Ladder}
Our warm-up exercise is that of the $1+1$ dimensional lattice model of fermions known as the Creutz ladder \cite{RevModPhys.73.119}. There are multiple such ladder model of fermions used by condensed matter theorists to model many-body phenomenon. Consider a specific  ladder consisting of a couple of parallel one dimensional chains of arbitrary length, each having fermions sitting at lattice sites with lattice parameter $A$, see the Figure for clarification.  Each electron on the chain can experience nearest neighbour hopping with energy $t$ and further there are inter-chain hoppings, for both nearest ($\gamma$) as well as next-to-nearest neighbour ($\gamma'$) interactions (see Fig.~\ref{fig:ladd}).
\medskip

\begin{figure}
 \centering
    \includegraphics[scale=1.15]{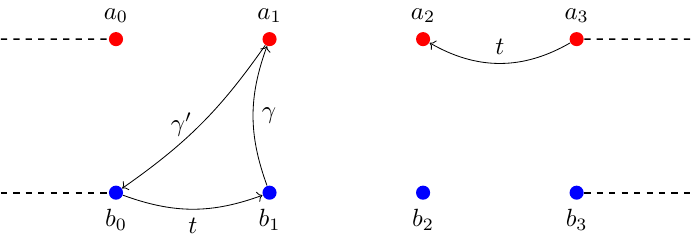}
  \caption{The Creutz ladder with two chains of fermions, given by red ($a$) and blue ($b$) site local states as in \eqref{1Dham}. The hopping terms are shown by arrows.}
    \label{fig:ladd}
\end{figure}

The Hamiltonian in real space oscillator basis is:
\begin{eqnarray} \label{1Dham}
H =\sum_{p} t (a^{\dagger}_{p+1} a_p + b^{\dagger}_{p+1} b_p  ) + \gamma a^{\dagger}_{p} b_p  + \gamma' ( a^{\dagger}_{p+1} b_p + a^{\dagger}_{p} b_{p+1} ) + \mathrm{h.c.}
\end{eqnarray}
For periodic boundary conditions, one may go to the Fourier basis defined by 
\be{}
c_k = \sum_{p} a_p e^{ikAp}, d_k = \sum_{p} b_p e^{ikAp} \quad \text{for} \quad k\in (-\pi/A , \pi/A).
\ee 
As a consequence, the Fourier space Hamiltonian becomes:
\begin{eqnarray} \label{kernel_creutz}
H =\sum_k
\begin{pmatrix}
c_k^{\dagger} & d_k^{\dagger}
\end{pmatrix}   \begin{pmatrix}
2t\, \cos(kA) & \gamma + 2 \gamma' \cos(kA) \\
\gamma + 2 \gamma' \cos(kA) & 2t\, \cos(kA) 
\end{pmatrix} \begin{pmatrix}
c_k \\
d_k
\end{pmatrix}.
\end{eqnarray}
Diagonalizing, we get two branches or bands:
\be{}
E^{\pm} (k) = 2 (t \pm \gamma') \cos(k) \pm \gamma.
\ee 
We readily observe that for $t= \gamma'$, $E^-$ becomes independent of $k$ and hence this solution leads to a flat-band. Hence for $t= \gamma'$, in the diagonal basis, we can write
\eqref{1Dham} $H = H_{\mathrm{flat}} + H'$, with
\be{}
H_{\mathrm{flat}} = - \gamma \sum_k \tilde{c}^{\dagger}_k \tilde{c}_k ~~~\mbox{ and } ~~~~ H' = \sum_k  \left(4 t \cos(k) + \gamma \right)\tilde{d}^{\dagger}_k \tilde{d}_k
\ee
Now with the choice of $\gamma > 2t$, the bands don't cross and the flat-band Hamiltonian contains the lowest energy modes.
Since the modes of $H_{\mathrm{flat}}$ contains the lowest lying modes, the effective dynamics is driven by it. In the continuum limit, the lattice constant $A \rightarrow 0$, the Brillouin zone ($k \in [-\pi/A,  \pi/A]$) covers the whole of real line and $\tilde{c}_k \rightarrow \tilde{c}(k)$. Defining the Fourier transform of these modes as the single component fermionic field $\psi(x)$, we have the Hamiltonian describing the low energy dynamics as:
\begin{eqnarray}
H_{\mathrm{flat}} = - \gamma \int dx\, \psi^{\dagger} \psi,
\end{eqnarray}
with the hopping strength $\gamma$ playing the role of the mass term. This above Hamiltonian is very clearly the Hamiltonian for lower Carroll fermions \refb{lowerH}. So our idea of mapping Carroll symmetries to flat-bands works in this case as well. 

\subsubsection{Magic angles and Moiré patterns}
Our next example would be a more involved one, that of bi-layer graphene twisted at particular angles called magic angles. The starting point is the effective low energy description of electrons in monolayer graphene, which is both scale (massless) and `Lorentz invariant' with respect to a characteristic velocity, with the Hamiltonian describing the free dynamics of a two component spinor $\psi$:
\begin{eqnarray} \label{mlham}
H= i v_F \int d^2 \mathbf{r} \, \Psi^{\dagger} (\mathbf{r},t) \,   \pmb{\sigma}\cdot  \pmb{\nabla}\, \Psi(\mathbf{r},t),
\end{eqnarray}
giving rise to linear dispersion $E( \mathbf{k}) = v_F |\mathbf{k}|$, with $v_F$ being the Fermi velocity. As mentioned earlier, in such a non-Lorentzian system this effective velocity plays the role of speed of light, albeit much smaller than actual $c$. Then the `effective Minkowski' frame attached to the system experiences a contraction as the $v_F\to 0$. This is the main idea behind the setup of this section.
\medskip

Bilayer graphene is an intriguing physical system which is exactly what the name suggests, two layers of graphene sheets on top of each other, but having a twist angle $\theta$ between them (see Fig.~\ref{moi}). Electron transport in this system is different that the monolayer cousin as one needs to take care of the interlayer couplings, and effect if the twist parameter directly changes the Hamiltonian. The low energy continuum model capturing these features and  is the so-called `chiral' one presented in the seminal paper \cite{PhysRevLett.122.106405}:
\begin{eqnarray}
\label{blg_ham}
\tilde{H} = v_F \int d^2 \mathbf{r} \, \Phi^{\dagger} (\mathbf{r},t) D \Phi  (\mathbf{r},t).
\end{eqnarray}

\begin{figure}[!t]
\centering
\includegraphics[width=.6\linewidth]{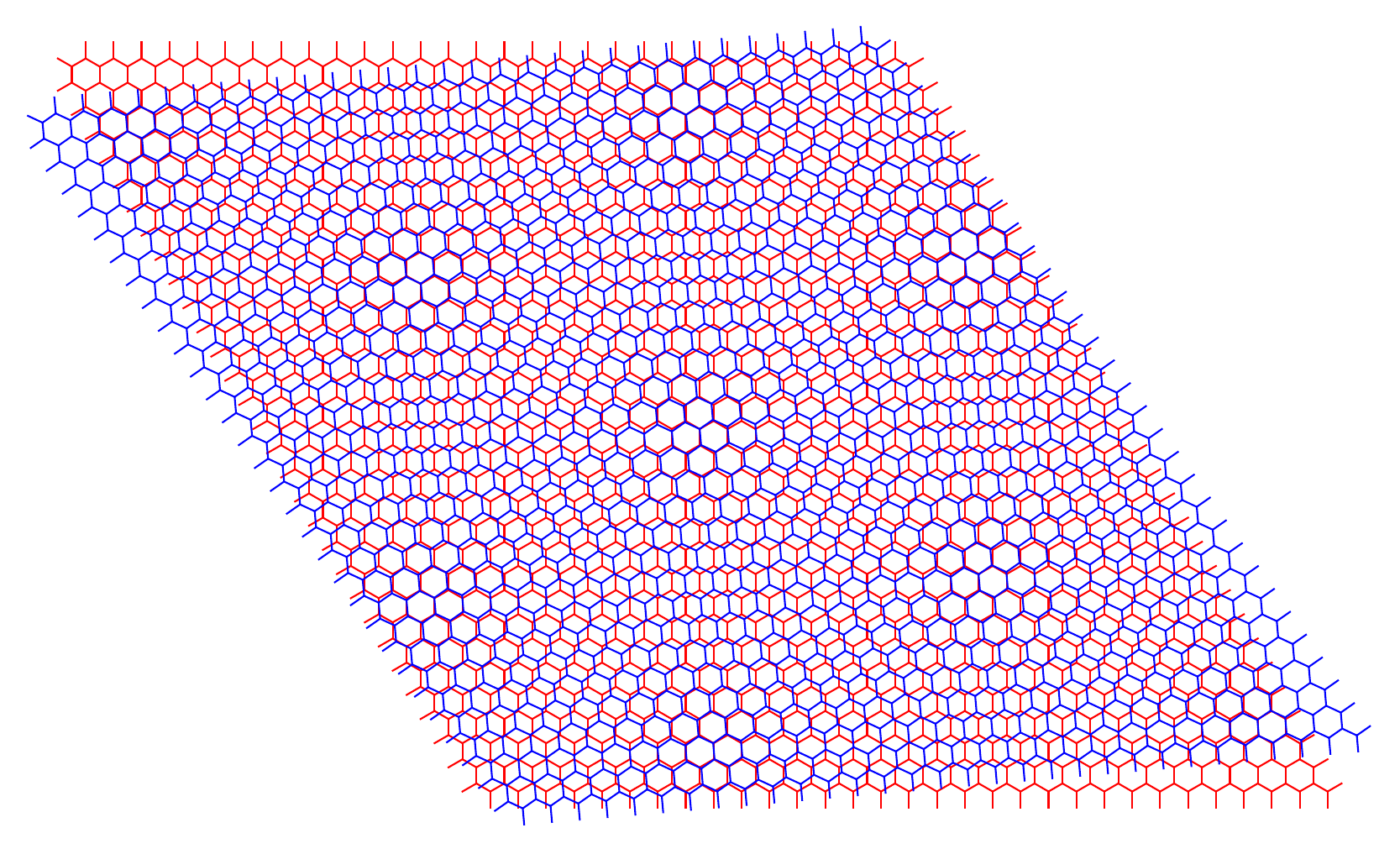}
\caption{An illustration for Moiré patterns for a section of bilayer hexagonal lattices (like Graphene). The twist angle here is $5^{\circ}$.}
\label{moi}
\end{figure}

Here, $\Phi^{\dagger} = \begin{pmatrix} \psi_1^{\dagger} & \psi_2^{\dagger} & \chi_1^{\dagger} &  \chi_2^{\dagger} \end{pmatrix} $ is a 4-component spinor, containing 2 components for each $(\psi, \chi)$ graphene layer and $D$ is the matrix valued differential operator that defines the action of spatial derivatives: 
\begin{eqnarray}
\label{Dop}
&& D =\begin{pmatrix} 0  &  \mathcal{D}^{\ast}( -\mathbf{r}) \\
  \mathcal{D}( \mathbf{r}) & 0 \end{pmatrix}, ~~  \mbox{ where } ~~ \mathcal{D}( \mathbf{r})  = \begin{pmatrix}
 -2i \bar{\partial} & \alpha \mathfrak{U}(\mathbf{r}) \\
 \alpha \mathfrak{U}(-\mathbf{r}) & -2i \bar{\partial}
 \end{pmatrix}\\
 \end{eqnarray}
 Here $\mathfrak{U}(\mathbf{r})$ is the crystal potential, given by the vectors defining the $\theta$ twist direction
 \begin{eqnarray}    
&& \mathfrak{U}(\mathbf{r}) = \sum_{j=0}^{2} e^{i\left( \frac{2\pi j}{3} - \mathbf{q}_j \cdot \mathbf{r}\right)} \mbox{ for }
  \mathbf{q}_0 = \frac{8\pi}{3a} \sin(\theta/2) (0,-1), \mathbf{q}_{1,2} =  \frac{8\pi}{3a} \sin(\theta/2) (\pm \sqrt{3}/2, 1/2) \nonumber
\end{eqnarray}
and $a$ is the lattice constant. Note that any generic crystal potential breaks the symmetries for monolayer graphene systems as it is explicitly twist dependent. Any two of the $\mathbf{q}$ appearing in the equation above serve as the basis vectors in the `mini'-Brillouin zone (mBZ), i.e. the twisted sector. The dimensionless parameter $\alpha$ controls the coupling between the layers through the interlayer hopping strength, and also gives rise to the mBZ. 
\medskip

At the very particular values of the $\theta$, the story becomes much more interesting. In fact, we can compute the fermi velocity $v_F(\alpha)$ of electrons at the Dirac points for this setup, and this is always given in terms of the zero modes of the perturbed Hamiltonian. An involved computation, involving the momentum space Green's function, leads to $v_F ( \alpha = 1/\sqrt3) =0$, which makes sure the Dirac cones are flattened out.  This happens for the angle $\theta \approx 1.05^{ \circ } $ \cite{PhysRevLett.122.106405}, which is famously known as the \textit{`magic angle'} \cite{PhysRevLett.122.106405}. This makes sure we have pristine flat bands present for all the zero modes, and the Hamiltonian actually vanishes. In principle, this is where Carrollian dynamics emerges in this theory.

\subsubsection{Compact Localized States}
Our last construction will be arguably the most important one with respect to Carroll symmetry and flat bands, since this provides us with an organizing principle to design actual quantum lattice systems with flat-band spectra. The key player in this game, as mentioned in the introduction, is the idea of a Compact Localized State (CLS), which is the manifestation of the exact destructive interference between sub-lattice Bloch wavefunctions. Such CLS units make sure that local excitations do not propagate beyond a few local states, and this confinement leads to ultra-locality. This much should be already enough for the reader to get excited about the connection of such states with Carroll symmetry, where such ultra-locality is the mantra we chant. This is what was done in \cite{Ara:2024fbr}. 
\medskip

Before we begin, it is important to state that despite a large volume of work done concerning CLSs over the last few years \footnote{See \cite{2018AdPhX...373052L} for a comprehensive review.}, a local symmetry based principle that defines the existence of CLSs have been absent from the literature. Of course, various analytical approaches have been tried (see for example \cite{Dias2015,Xu2015,PhysRevB.97.035161}), but a robust Lie algebraic technique has been missing so far. And there comes our old friend supertranslations into the picture, whose magic makes such a feat possible. 
\medskip

{\ding{112}} \underline{\em{Carroll fermions on a chain}}

\medskip

Carroll fermions in (1+1)D were previously studied in detail in \cite{Banerjee:2022ocj, Hao:2022xhq} for continuum spinors, and the representation theory was worked out as well. On a one spatial dimensional chain, the spinor action becomes much simpler, with nearest neighbour (NN) hopping: 
\begin{align} \label{ham1}
H= \sum \mathcal{H}_j; \mbox{ where } \mathcal{H}_j = \psi_{j+1}^\dagger q \psi_j + \mbox{h.c.}
\end{align}
where $\mathcal{H}_j$ are the discrete version of Hamiltonian density and $q$ is a 2$\times$2 matrix, analogous to a Dirac gamma matrix. For a generic matrix $q$ this is a standard free fermion theory with non-zero correlation length, whose effective continuum theory describing fluctuations around the ground state has spatial derivatives. We of course want one where there are no spatial derivatives.
\medskip

Provided canonical anti-commutation structure for the fermions: $\{\psi_{i, \alpha}, \psi_{j,\beta}^{\dagger}\} = \delta_{ij} \delta^{\alpha \beta}$ etc. ($\alpha, \beta =1,2$) one arrives following Heisenberg equation of motion,both at first and second order:
\begin{align} \label{heisen}
    \dot{\psi}_j =  -i  \left (q \psi_{j-1} + q^{\dagger} \psi_{j+1}\right), ~~~ 
    \ddot{\psi}_j = -\{q,q^{\dagger}\} \psi_j.
\end{align}
The single derivative equation simply says that the time evolution of the fermion wavefunction should spread to more and more lattice points, which is completely expected. 
Now let us demand something special for $q$. Notice that if we have a \textit{nilpotent} $q$, there always exists a $\kappa \in \mathbb{R}$, such that the anti-commutator in the double derivative equation $\{q,q^{\dagger}\} = \kappa^2 \bm{1} $\footnote{A generic nilpotent representation could be \begin{align} \label{q_param}
    q = \tau \begin{pmatrix}
    1 & \alpha\\
    -1/\alpha & -1\end{pmatrix}.
\end{align}}. This makes the stationary states associated to the fermion completely localised in space, or in other words, ultra-local! Not only that, going ahead, we can explicitly see that just by this nilpotent choice, the discrete Hamiltonian densities at two different lattice points commute: 
\begin{align}
    [\mathcal{H}_i, \mathcal{H}_j ] = 0,
\end{align}
which is reminiscent of a discrete version of the Carrollian Dirac-Schwinger condition.
Hence, for any given lattice function $f$ defined at a site, the operator:
\begin{align} \label{sucharge}
    Q_f = \sum_{j} f_j \mathcal{H}_j
\end{align}
is a conserved charge. Comparing with \eqref{suptra} we can call this the discrete version of supertranslation generator, which comes weighted by these arbitrary functions at every site, and are self commuting 
\begin{equation}
    [Q_f,Q_g] = 0
\end{equation}
Now we would like to find a very particular basis $\xi$ where supertranslations act as a local gauge transformation, i.e. $\delta_f \xi_j = f_j \dot{\xi}_j$ \footnote{Which is analogous to how continuum supertranslations act.}, and again owing to the nilpotency of $q$, we can easily choose: 
\begin{equation}
    \xi _j = \frac{1}{\kappa}\left(q\psi_j +q^{\dagger} \psi_{j+1}\right).
\end{equation}
Here the normalization factor $\kappa = \sqrt{\{q,q^{\dagger}\}}$ is used for canonical transformations. Now these fermions $\xi$ turn out to be the single-particle CLS states. To see this, one has to transform the original Hamiltonian \eqref{ham1} in terms of CLS basis representation, which reads: 
\begin{align} \label{ham2}
    H = \sum_j \xi^{\dagger}_j \left( q+ q^{\dagger}\right) \xi_j.
\end{align}
One can easily argue that all Hamiltonians of the above form gives rise to a set of two non-dispersive or flat bands. And just like that, we have identified the chronology CLS $\to$ flat band!

\medskip 

Thus a very simple recipe given by nilpotent matrices can give rise to flat dispersion bands in lattice systems. But this is not really a coincidence. In fact the $q$ matrix can be though of a nilpotent representation of the Carrollian Clifford algebra in two dimensions \cite{Banerjee:2022ocj}. Moreover, one could generalise the construction to higher dimensions, larger number of flat bands and longer hopping ranges by tinkering with such nilpotent matrices. 
\medskip

{\ding{112}} \underline{\em{Exact model and interactions}}
\medskip

\begin{figure}
    \centering
\includegraphics[scale=1]{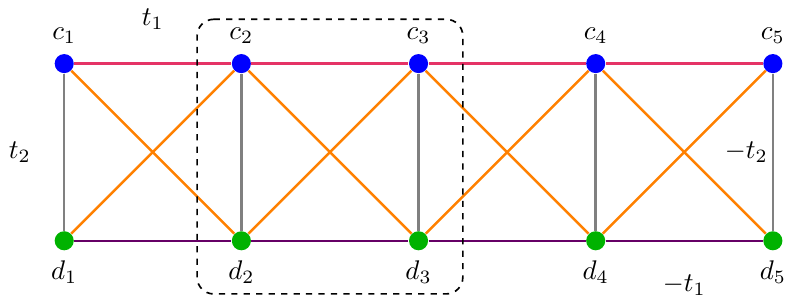}
    \caption{Representation of a Compact Localised State in a ladder system with two kinds of fermions (blue and green). This is the same system represented by \eqref{H2}. The dotted loop shows a compact state as in \eqref{canalbet} made out of four site local states.}
    \label{fig:enter-label}
\end{figure}

 In \cite{Ara:2024fbr}, an exact supertranslation invariant lattice model was chosen for the purposes of this discussion. This stems from the parent lattice with appropriate hopping terms:
\begin{align}
    \label{H2}
H = \sum_{j}\left( t_1 \left(c^{\dagger}_{j+1} c_j -d^{\dagger}_{j+1} d_j \right) +t_2 \left(c^{\dagger}_{j+1} d_j -d^{\dagger}_{j+1} c_j \right) + \mbox{h.c.}\right).
\end{align}
Note that this is similar to the ladder model considered in \eqref{1Dham} but here we will not have only one flat band. 
The dispersion bands given by this Hamiltonian can be found by going into the Fourier basis and diagonalizing the Kernel:
\begin{equation}
E_{\pm} = \pm \sqrt{2}\sqrt{t^2_1 + t^2_2 + (t^2_1 - t^2_2) \cos({2ka})},
\end{equation}
 Note here, the bandwidth of the bands vanishes at $t_2 \rightarrow t_1$, the specific limit we are interested in, and the one where the pristine flat bands occur. One can show in the continuum fermion theory, this is also the limit where spatial derivatives vanish. 
\medskip

Now using the choice of CLS basis augmented by the supertranslations, we can simply choose the CLS modes as two-component fermions out of the site-local modes $\psi^{\dagger} = ( c^{\dagger} \, \, d^{\dagger})$,
\begin{eqnarray}\label{canalbet}
    \alpha_j = \frac{1}{2}\left(c_j +d_j -c_{j+1} +d_{j+1} \right), ~~ \beta_j = \frac{1}{2}\left(c_j +d_j +c_{j+1} -d_{j+1} .\right)
\end{eqnarray}
So written in terms of CLS, and at the exactly tuned flat band parameters, we get
\begin{eqnarray} \label{h_CLS}
    H = 2 \tau \sum_j \left(\alpha^{\dagger}_j \alpha_j - \beta^{\dagger}_j \beta_j\right).
\end{eqnarray}
Which is nice and diagonal, and more importantly, directly gets you a hint about the ground state of the Hamiltonian, which is either populated entirely by $\beta$ oscillators or $\alpha$ oscillators. 
The ultra-locality of the CLS in the $\beta$ filled ground state then implies: 

\begin{equation} 
\label{free_corr_cls} \langle \beta^{\dagger}_i (t) \beta_j (0) \rangle =   e^{-2i\tau\,t}\delta_{i,j}, \langle \alpha^{\dagger}_i (t) \beta_j (0) \rangle = 0= \langle \alpha^{\dagger}_i(t) \alpha_j (0) \rangle. \end{equation}

Now, this Kronecker delta in this lattice case is nothing but the Dirac delta in the continuum correlations, something which we have encountered in  \eqref{f}, albeit in $1+1$ D.
\medskip

Note that, we can always take this free theory to an arbitrarily supertranslated frame, namely the CLS modes change like:
\begin{equation}
\alpha_m \rightarrow e^{i Q_f} \alpha_m e^{-iQ_f} = \alpha^f_m = e^{-2i\tau f_m} \alpha_m, \, \text{similarly,}~~\beta_m \rightarrow \beta^f_m = e^{2i\tau f_m} \beta_m,
\end{equation}
It is important to stress that under supertranslations the ultra-local (CLS like) modes would only receive position-dependent phases, which are nothing but the lattice functions, making these just local gauge transformations. The nice feature of this construction is that irrespective of the lattice function $f_m$ used, the oscillators will change in a way that correlation functions always remain invariant and ultra-local. 
\medskip

 Due to the construction having such robust structure, it is imperative to add interactions to these systems, since in the strict condensed matter sense, flat bands are defined only for interacting systems. We of course want these interaction terms to be supertranslation invariant as well, which means site-local 4-fermion interactions like $c^\dagger c d^\dagger d$ will not work, and imposing symmetry constraints, we see that the only such choice involving four spinless fermions is:
$$ \sim \sum_j \alpha^{\dagger}_j \alpha_j \beta^{\dagger}_j \beta_j,$$
 Now we can impose periodic boundary conditions on the lattice, and write the interacting Hamiltonian in the form:
\begin{equation}\label{intmodel}
    H_{int} =
\sum_j \left(V n^{\alpha}_j n^{\beta}_j +2\tau (n^{\alpha}_j-n^{\beta}_j) \right).
\end{equation}
Note that $n^{\alpha}_j = \alpha^{\dagger}_j \alpha_j$ etc. are number operators, and $V$ is the strength of interaction added to free theory. We now have two parameters $(\tau, V)$ to play with in the theory. This means we could have non-trivial regimes for the dynamics. However, just from the Hamiltonian, one could be sure that the system is exactly solvable in all parameter space, augmented by the continuing presence of infinite number of supertranslation charges. The above, is one of the very few fully interacting Carroll invariant quantum systems.
\medskip

This is a non-trivial interacting theory which leads to various interesting quantum phases depending on values of the free parameters, and the filling constraint for the fermions. As seen in the free theory, if we tune $\tau$ from a positive to a negative value, one can see that the ground state switches from the one having all $\beta$ oscillators excited to the one having all the $\alpha$ ones excited. In fact, depending on the parameter values of $V,\tau$ one can have a much more interesting phase structure (see Fig.~\ref{phased}). In the half filling case, the gap between the ground state and the first excited one in this case vanishes along $\tau <0, V = 4\tau$ line and also along $\tau >0, V = -4\tau$, on top of the expected $\tau =0, V>0$ line. In the continuum theory, these are the phase boundaries where pure conformal Carrollian theory takes over, which can be seen from the correlation functions. For $\tau< 0, V <4 \tau$, however, one reaches the phase where the lowest energy level is highly degenerate. This is termed as the `exotic' phase as the density of states are just decoupled delta functions here, however the supertranslation invariance remains robust\footnote{For more pertinent related questions about this very robustness of the Carroll symmetry under (Carroll-)irrelevant deformations, see \cite{Ara:2024fbr} and \cite{upcome}.}. 
\medskip

\begin{table}
\begin{center}
\begin{tabular}{|c|c|c|l|l|}
\hline
\textbf{Phase}  & \textbf{Parameter space} & \textbf{Ground State}   \\ \hline
All $\alpha$ & $\tau <0$, $V>4\tau$ & $|n_j^\alpha =1,n_j^\beta = 0, j=1,...,N\rangle $ \\ \hline
All $\beta$ & $\tau >0$, $V>-4\tau$ & $|n_j^\alpha =0,n_j^\beta = 1, j=1,...,N\rangle $   \\ \hline
Exotic & $V<0, -V/4 > \tau > V/4$ &  $|n_j^\alpha =n_j^\beta , j=1,...,N\rangle$;
~~$\sum_{j=1}^N n_j^\alpha =\frac{N}{2}=\sum_{j=1}^N n_j^\beta$\\ \hline
\end{tabular}
\end{center}
\caption{Comparative characteristics of the three phases associated to half-filling case. The ground states are indicated in each of these cases. }\label{tab1}
\end{table}

\begin{figure}[ht]
\centering
\includegraphics[scale=1.15]{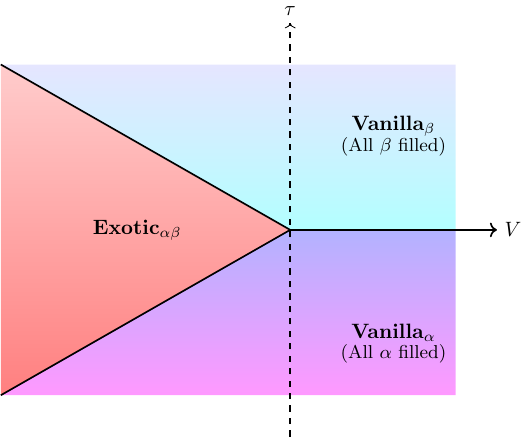}
\caption{The phases associated to the interacting model \eqref{intmodel} plotted in the $(V-\tau)$ plane. In this figure, adapted from \cite{Ara:2024fbr}, three distinct quantum phases as in Table \ref{tab1} have been shown. }
\label{phased}
\end{figure}

\subsection{Phase separation and Luttinger Liquids}
In this section, we discuss an intriguing application of Carroll dynamics to critical condensed matter systems. Like other connections in this part of our discussion, this particular one is equally novel and surprising. Since Carroll dynamics represents a situation where the characteristic particle velocities go to zero, but scaling symmetries persist, there have been vague murmurs of a connection with phase transitions. Of course, as we saw in the last section, quantum phase transitions (QPTs) do happen in infinite dimensional Carroll invariant systems, but the bigger question is much more subtle i.e.: \textit{Can critical systems in condensed matter physics, generally devoid of relativistic structures, be described by larger classes of theories than usual CFTs?}
\medskip

In \cite{Biswas:2025dte} a first step was taken towards answering this question, by showing how ultralocal field dynamics may appear in the parameter space of Tomonaga-Luttinger liquids \cite{GiamarchiB, FradkinB, Voit1995, NagaosaQFT, Senechal2004} (LL or TLL). LL is a very well-known model in the many-body physics context \cite{10.10631.1704046} and represents the low energy dynamics of a plethora of 1D and quasi-1D (fermionic and bosonic) systems\footnote{See \cite{gogolin2004bosonization} for various examples and insights into LL systems.}. In the continuum (bosonized) description LL can be effectively written down as a massless scalar field theory, more precisely a $c=1$ CFT. The Hamiltonian description in this limit is very straightforward: 
\begin{eqnarray}\label{masterH}
	\mathcal{H}  = \frac{1}{2 \pi} \int dx \Big[ uK  (\pi \Pi(x))^2 +  \frac{u}{K}  (\partial_x \phi(x))^2\Big],
\end{eqnarray}
with $u$ and $K$ being the effective velocity (renormalised Fermi velocity) and Luttinger parameter. The system is written in terms of dual bosonic fields $\phi(x)$ and $\theta(x)$ such that the conjugate momentum to $\phi(x)$ is $\Pi(x) = \frac{1}{\pi}\partial_x\theta(x)$, making it equivalent to a simple scalar with different scalings on the derivatives.
\medskip

The parameter space associated to this theory is mostly well understood since CFT techniques can be used to read off observables, but intriguingly they are known to fail at a very particular regime, known as the \textit{Phase Separation} (PS). In terms of the Hamiltonian written above, the PS region is a special point achieved by taking $u \to 0$, such that $uK \to 1$ but $u/K \to 0$ (equivalently $u \to 0, \, K \to \infty$) and the Hamiltonian is just $\sim \Pi(x)^2$. In this region, due to the apparent singularity, CFT techniques do not work \cite{PS1,PS2,PS3,PS4} and the physics on two sides of this separation line differ drastically. The claim in \cite{Biswas:2025dte} is that the new paradigm needed to explain the physics of PS is none other than the Carroll one.
\medskip

To understand more about this, we need an exact model where one can probe this region, and an observable to probe it with. A good candidate model is the so-called $t-V$ model \cite{GiamarchiB} which deals with a lattice of spinless fermions of the same kind, given by oscillators ($c^{\dagger}_j,c_j$), with the lattice Hamiltonian:
\begin{align}\label{H_tV}
\mathcal{H}_{t-V} &= -t \sum_i ( c^{\dagger}_{i} c_{i+1} + c^{\dagger}_{i+1} c_{i} ) - V \sum_i n_i n_{i+1},
\end{align}	
where $n_j = c^{\dagger}_j c_j$ is the fermion number operator (with  $\{c_i,c_j^{\dagger}\} = \delta_{ij}$) and $V$ is (nearest neighbour) interaction strength. Putting $t=1$, and taking the long wavelength limit, one can bosonize the theory to connect with the scalar field Hamiltonian in \eqref{masterH} \cite{Senechal2004}. This mapping reveals that the LL regime of exact solvability is only valid upto $|V|<2$, after which the PS region sets in. In this regime, we have to do hardcore numerics to understand the dynamics.
\medskip

As far as the observable is concerned, we focus on correlation functions, namely the \textit{density-density correlator}. The effective theory of LL actually describes the dynamics of leading order fluctuations of fermionic theories, and the densities are basically bilinears of these fluctuations. One can write the bosonic version of the fluctuation as:
\begin{eqnarray}
	\rho(x)= \bigg(\rho_0 - \partial_x\phi(x)\bigg)\sum\limits_{p\in \mathbb{Z}}e^{2\pi ip(\rho_0x - \phi(x))}.
\end{eqnarray}
This can be understood better by thinking from the continuum model, as $\rho(x)-\rho_ 0 =\delta\rho(x)\approx \partial_x\phi$ is basically the primary operator in the bosonic CFT in \eqref{masterH}. In the LL region correlators of these operators $\langle \rho(x) \rho(x')\rangle $ will then have the usual $\sim x^{-2}$ scaling. If we further claim that beyond $V\geq 2$ a Carrollian field theory takes over, then these correlations would change into the ultra-local ones (see \eqref{f} for example.). It is often instructive to calculate these correlators in the momentum space, for both LL and beyond. In the simplest form, it turns out analytically we can expect:
\begin{equation}
\langle \rho(q) \rho(-q)\rangle_{\{V < 2\}} \sim -\frac{q}{2K}\text{sign}(q),~~~~\langle \rho(q) \rho(-q)\rangle_{PS} \sim -\frac{1}{\sqrt{2\pi}}q^2.
\end{equation}
So, such a density-density correlator $S(q)$, sometimes called the \textit{structure factor}\cite{PS2} in the literature, would change from a linear dependence in momenta to a quadratic dependence as we move from CFT (LL) region to Carrollian field theory (PS) region. 
\medskip

\begin{figure}[t]
\centering
\includegraphics[scale=0.9]{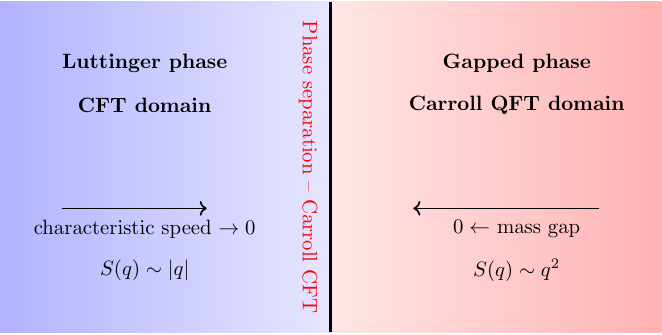}
\caption{An illustration of emergent Carroll symmetry at phase separation region. The PS can be thought of as the regime where renormalised Fermi velocity becomes zero, and CCFT structures emerge from CFT. On the other side of the PS, the structure factor keeps scaling quadratically with momentum. }
\label{14}
\end{figure}

To confirm this transition structure on the lattice model \eqref{H_tV} itself, we can define a discrete version of the structure factor in momentum space:
\begin{equation}
	\begin{aligned}\label{Sq}
		S (q) &&= \frac{1}{L} \sum_{i,j} \big[ \langle n_i n_j \rangle -\langle n_i \rangle \langle n_j \rangle \big] e^{iq(i-j)}.
	\end{aligned}	
\end{equation}
where the volume factor is just the system size $L$. In \cite{Biswas:2025dte}, some of the present authors extracted this correlation numerically, using a Density Matrix Renormalisation Group (DMRG) approach. Although the details are beyond the scope of this review, one can check that the structure factor in the $t-V$ lattice model exactly gives us what we had expected from the exponent of momenta: it seamlessly and sharply moves from $1$ to $2$ as we probe across LL to PS regime. One can, however, have a naive physical understanding of this effect, since at $V\geq 2$ the attractive interaction potential starts to overcome the delocalisation due to particle hopping, resulting in a bunched-up or `pinned' phase. The emergent degeneracy of the vacuum results from a clear particle-hole separation of the excitations, where the energy configurations remain the same given all possible such separations, or domain walls present. Note that deep into the $V> 2$ region this phase survives as well, and one could perhaps attribute this to the presence of a gapped Carroll phase where the spatial delta function in the correlation remains unchanged\footnote{A simple explanation of this can be found in the massive Carroll scalar field theory in 1+1 D \cite{Banerjee:2023jpi}, where the time ordered correlation reads $ \langle \mathcal{T} (\phi(t,x) \phi(t',x'))\rangle = \frac{1}{2m}e^{-im|t-t'|}\delta(x-x').$ Hence the momentum space density-density correlation for this theory is predicted to still have the $\sim q^2$ scaling.}. See Fig.~\ref{14} for an illustration. 
\medskip

We must note here that this transition is not an one-off example. A large of class spin models have LL theory as the long wavelength limit, and it is likely that all of them admit a Carrollian description across the PS region. It is, however, hard to prove rigorously that a PS region separates relativistic and Carrollian theories on both sides. But still, this investigation is an impressive start, and owing to the exponential progress in simulating spin models, one might actually be able to capture signatures of Carroll physics in PS regions. 
 
\subsection{Shallow water gauge theory}

In this section, we will look at the surprising emergence of Carroll structures in \textit{shallow water waves}. Shallow water approximation applies to fluid systems for which the height is much smaller compared to its horizontal span. Our oceans and the atmosphere fall into this class, and they admit a linearized description when the shallow approximation is used. It has recently been shown \cite{Tong:2022gpg} that the non-linear shallow water equations map to that of a $2+1$ D Abelian gauge theory, where the electric and magnetic field corresponds to conserved height and conserved vorticity of the fluid. When linearized, the solutions split into two main classes, the \textit{geostrophic flat band} and \textit{Poincaré waves}. One can then construct effective actions for these two cases. This is our starting point and we will map these effective actions to the \textit{Electric} and \textit{Magnetic} versions of Carrollian electrodynamics \cite{Duval:2014uoa}, which has been discussed priorly. 
\medskip

We start with fluids in $D=2+1$ dimensions. Following \cite{Tong:2022gpg}, the non-linear shallow water system is described by two dynamical fields: $h(x^i,t)$ i.e. the height of the fluid and $\vec{u}(x^i,t)$, the horizontal fluid velocity, with $i=1,2$. The EOMs are
    \begin{eqnarray}
        \frac{Dh}{Dt}=-h\nabla.\Vec{u}, \quad \frac{Du^i}{Dt}=f\epsilon^{ij}u^j-g\frac{\partial h}{\partial x^i}, \,  i=1,2. \label{eq:1}
    \end{eqnarray}
The material derivative, $\frac{D}{Dt}=\frac{\partial }{\partial t}+\Vec{u}.\nabla$ measures the change in height in relation to the flow,
with $g$ being the gravitational constant and $f$ the Coriolis parameter. The theory admits two conserved charges given by zero modes of the currents
\begin{eqnarray}\label{concharge}
    J^0=h, \, \Vec{J}=h\Vec{u}; \,\quad  \Tilde{J}^0=\zeta+f, \, {\Vec{\Tilde{J}}}=(\zeta+f)\Vec{u}.
\end{eqnarray} 
Here $\zeta=\epsilon^{ij}\partial_i u_j$ is the vorticity of the shallow water. 
\medskip

In \cite{Tong:2022gpg}, a $u(1)\times u(1)$ gauge theory for shallow water was formulated. Writing the  2+1 dimensional gauge fields as $A_{\mu}$ with $\mu=0,1,2$ , the electric field $E_i$ and the magnetic field $B$ are $E_i=\partial_{[t} A_{i]} $ and $B=\partial_{[1} A_{2]}$
and same goes for the other set of fields $(\tilde{E}_i,\tilde{B})$. Without going into the details of the non-linear theory, we start by linearising \eqref{eq:1}, by assuming that variations of $h,u^i$ are small. This yields  $h(x^j,t)=H+\eta(x^j,t)$ and $u^i(x^j,t)=0+u^i(x^j,t)$, where $H$ is a constant height and $\eta \ll H$ \cite{Sheikh-Jabbari:2023eba}. Retaining terms up to linear order in the variations, the linearised shallow water equations read: 
\begin{eqnarray} \label{6}
    \frac{\partial \eta}{\partial t}+H\nabla.\Vec{u}=0, \quad \frac{\partial u^i}{\partial t}=f\epsilon^{ij}u^j-g\frac{\partial \eta}{\partial x^i}. \label{eq:6b}
\end{eqnarray}
Next, we put in the ansatz for these fields, which read: $u_i=\hat{u}_ie^{i(\omega t-k.x)}$ and $\eta=\hat{\eta}e^{i(\omega t-k.x)}$. This
leads to solutions with two distinct dispersion relations having: $\omega=\pm \sqrt{v^2k^2+f^2}$ and $\omega=0$.
The first type, known as Poincar\'{e} waves, has dispersion relation akin to a relativistic theory with an effective mass $f$ and characteristic speed $v^2=gH$. The second class has $\omega=0$, and are called flat-bands\footnote{These are in a sense similar to the flat bands we have been discussing in previous section, since the energy does not depend on momentum modes.}. These are additional, time independent equilibrium solutions beyond the $h=$ constant ones. They have a non-trivial spatial profile with gravitational force balanced by a corresponding velocity profile $fu=-\partial h/\partial y$, which in turn generates a Coriolis force. This particular equilibrium is known as \textit{geostrophic balance}.
\medskip

Now consider the linearisation of the associated gauge theory, i.e., we put in $A_\mu=\hat{A}_{\mu}+\delta A_{\mu}$  with $\hat{A}_0=0$ and $\partial_1 \hat{A}_2-\partial_2 \hat{A}_1=H.$
Fluid variables can now be mapped to gauge fields using $B=\eta$ and $E_i=H\epsilon_{ij}u^{j}.$
Our \eqref{eq:6b} in terms of fluid variables read:
\begin{equation}
    \Dot{E_i}=\epsilon_{ij}\left(fE_j-gH\partial_j B\right).
    \label{eq:10}
\end{equation}
Using this, one can write two different classes of linearised effective gauge theory actions corresponding to both flat bands and the Poincaré-like dispersion relation.

\medskip

{\ding{112}} \underline{\em{Effective actions}}
\medskip

Geostrophic flat bands are non-trivial time-independent solutions of (\ref{eq:10}), given by $A_0=-\frac{v^2}{f}B$. 
In \cite{Tong:2022gpg} an effective action for the flat band was proposed: 
\begin{equation}
    S_{geos}=\int d^3x \frac{1}{2H}\left(E_i-\frac{v^2}{f}\partial_iB\right)^2.
    \label{eq:13}
\end{equation}
EOMs require that $E_i-(v^2/f)\partial_iB$ is constant, reproducing the geostrophic balance condition.
To understand this, consider the fluid velocity to be time-independent in \eqref{6}, which leads to $\epsilon^{ij}\partial_t E_j = 0$. Which means either both $E^i$'s vary with time in the same way or $E^i$'s are explicitly time-independent. The latter i.e. $\partial_t E_i = 0$, is a peculiar theory that does not have symmetries of the relativistic Maxwell electric field.
Note that RHS of \eqref{eq:6b} in this case reflects the EOM for the flat band.
\medskip

For Poincaré waves, a gauge $A_0=0$ is chosen. The linearized action in this case is
\begin{equation}\label{CSPoincaré}
    S=\int d^3x \frac{1}{2H}\left(\Dot{A_i}^2-v^2B^2+f\epsilon^{ij}A_i\Dot{A_j}\right).
\end{equation}
 Note that if we reinstate $A_0$, the action above becomes the familiar relativistic Maxwell-Chern-Simons action, albeit defined with the characteristic velocity $v$. Here we assume an effective Poincaré invariant metric in 2+1 D, $g_{\mu\nu} = \text{diag}(-v^2,1,1)$, given the characteristic speed $v^2 = gH$. 
Consequently, EOM associated to \eqref{CSPoincaré} has an interesting form:
\begin{equation}\label{eqnPoinc}
	\partial_t\left(\dot{A_i} - f\epsilon^{ij}A_j\right) = -v^2 \epsilon^{ij}\partial_j B, \quad
\end{equation}
This looks like a 2+1 dimensional version of Faraday's law (sans the $v^2$), if one identifies $\left(\dot{A_i} - f\epsilon^{ij}A_j\right)$ as a re-imagined Electric field. Clearly that is not possible in the relativistic paradigm owing to symmetry considerations. 
\medskip

We now focus on the other half of the story, i.e. Carrollian electrodynamics in $D=2+1$ dimension following \cite{deBoer:2021jej}. Using the machinery explained before (see section \eqref{gaugeexp}), we find the leading order Lagrangian:
\begin{equation}
    \mathcal{L}_0=\frac{1}{2}\left(F_{0i}^{(0)}\right)^2 =\frac{1}{2}\left(E_i^{(0)}\right)^2\qquad i=1,2\label{ced1}\\
\end{equation}
with $F_{0i}=E_i^{(0)}=\partial_tA_i^{(0)}-\partial_iA_t^{(0)}$. This is nothing but the electric Carroll Electrodynamics Lagrangian and this is naturally Carroll invariant. The next to leading order Lagrangian is 
\begin{equation}
    \mathcal{L}_1=F_{0i}^{(0)}F_{0i}^{(1)}-\frac{1}{4}\left(F_{ij}^{(0)}\right)^2 =\frac{1}{2}\left(E_i^{(0)}\hat{E}_i^{(1)}-\left(B^{(0)}\right)^2\right) \label{ced2}
\end{equation}
where $F_{ij}^{(0)}F^{ij(0)}=F_{12}^{(0)}F^{12(0)}+F_{21}^{(0)}F^{21(0)}=2\left(B^{(0)}\right)^2$ 
and $F_{0i}^{(1)}=E_i^{(1)}=\partial_tA_i^{(1)}-\partial_iA_t^{(1)}=\frac{1}{2}\hat{E}_i^{(1)}$. This is a pure Lagrange multiplier field.
It should be noted here that though $\mathcal{L}_0$ is Carroll invariant, the subleading magnetic Lagrangian $\mathcal{L}_1$ is not. In order to restore Carroll symmetry at the subleading order, we introduce a constraint that implements the leading order equations of motion. 

\medskip

{\ding{112}} \underline{\em{Mapping from Carroll to Water waves}}
\medskip

We can now see how to map between the actions for flat bands and Poincaré waves map to leading and subleading order Carroll electrodynamics. The actions are already mentioned in \eqref{ced1} and \eqref{ced2}. In the table below (Table.~\ref{actioncomp}) we show the explicit mapping between the two sets of actions in terms of gauge fields.
\medskip
\begin{table}[]
\centering
\renewcommand{\arraystretch}{3.5}
\resizebox{\textwidth}{!}{
\begin{tabular}{|c|c|c|}
\hline
 Theory & Gauge field action & Carroll action \\[10pt]
\hline
   \textbf{Flat Band} & $ S_{geos}=\int d^3x \frac{1}{2H}\left(E_i-\frac{v^2}{f}\partial_iB\right)^2$ & $S^{(0)}=\int d^3x\,\frac{1}{2}\left(\partial_t A_i^{(0)}-\partial_i A_t^{(0)}\right)^2$ \\[5pt]
\hline
Mapping:& \multicolumn{2}{|c|}{$A_t^{(0)} = \frac{1}{\sqrt{H}}\left(A_0 + \frac{v^2}{f}B\right),~~~
 A_i^{(0)} = \frac{1}{\sqrt{H}}A_i$ $\implies E_i^{(0)} = \frac{1}{\sqrt{H}} \left({E_i}-\frac{v^2}{f}\partial_i B\right).$} \\[5pt]
\hline
  \textbf{Poincaré Waves} & $S_{poin}=\int d^3x \frac{1}{2H}\left(\dot{A_i}^2-v^2B^2+f\epsilon^{ij}A_i\dot{A_j}\right)$  & $S^{(1)}=\frac{1}{2}\int d^3x \left(c^2\hat{E}_{i}^{(1)} \partial_t A_i^{(0)} -c^2(B^{(0)})^2\right).$  \\[5pt]
\hline
Mapping:&  \multicolumn{2}{|c|}{$A_i^{(0)}= \frac{A_i}{\sqrt{H}}  \,\text{ , }\, B^{(0)}[\{A_i^{(0)}\}]= \pm B[\{\frac{A_i}{\sqrt{H}}\}],~~c\equiv v$, ~~~${c^2}\hat{E}_{i}^{(1)} \equiv \partial_t A_i^{(0)}-f\epsilon^{ij}A_j^{(0)}$ }  \\[7pt]
\hline
\end{tabular}}
\caption{Mapping between fields for linearised gauge theory associated to shallow water waves to electric and magnetic Carroll electrodynamics in 2+1 D. Note that $[\{.\}]$ denotes a functional of the gauge field. }
\label{actioncomp}
\end{table}

Let us now briefly discuss the consequences of this mapping. The mapping of the flat band action and the electric Carroll electrodynamics is absolutely straightforward. One can further show that this is a gauge invariant identification by doing $A_t^{(0)} \to A_t^{(0)} + \partial_t \lambda_1\,,\quad A_i^{(0)} \to A_i^{(0)} + \partial_i \lambda_2$ provided $\lambda_1 - \lambda_2 = \text{spacetime constant}$. In fact, using the identification, which leads to the flat band condition, we find that $T^i_t=0$ from the fluid action. This is a tell-tale signature of Carroll boost invariance.
\medskip

The mapping for Poincaré waves is certainly more involved, albeit extremely illuminating since the placeholder for speed of light ($c$) has to be identified with the characteristic velocity ($v$). The mapping between fields as in Table \eqref{actioncomp} leads to the equivalence of the actions (and EOM) for sure, but we have to remember that subleading Carroll magnetic theory is Poincaré invariant at first glance, but only
after imposing the leading-order EOM as a constraint on the system, Carroll symmetries emerge. This, in our case, translates to setting $E_i^{(0)}=0$. Once this constraint is imposed, the action transforms as $\delta \mathcal{L}_1=b^kx_k\partial_t \mathcal{L}_1$, providing invariance under Carroll symmetry. It can also be checked $T^i_{\, t}$ component of the stress tensor vanishes once the leading order EOM are imposed (see \cite{Bagchi:2024ikw} for more). This, as we emphasised before, is a signature of Carroll boost invariance.

\subsection{Pointers to literature}
\begin{itemize}
\item{\em{More on Fractons}}
\begin{itemize}
 
 \item[$\star$] Coupling of Fracton gauge fields to curved spacetimes were discussed first in \cite{Jain:2021ibh}, and related formalism was developed. In \cite{Hartong:2024hvs} the authors studied dipole Chern–Simons theory with and without a cosmological constant in $2+1$ D, and used the underlying Aristotelian geometry to map it to a Fracton gauge theory. Higher dimensional generalisations of the same construction were also discussed. More discussions can be found in \cite{Afxonidis:2025wce}, where the authors generalised the notion of using dipole-symmetric gauge fields and coupling them to curved manifolds.

\item[$\star$] In \cite{Pena-Benitez:2023aat} the authors found out that the dipole-conserving Fracton algebra can be found as an Aristotelian (and also pseudo-Carrollian) contraction of the Poincaré algebra, but in one higher dimension. Using this method one can dimensionally reduce a higher dimensional relativistic gauge theory, with a symmetry broken phase, to get a Fracton electrodynamics theory. 

\item[$\star$] A duality between between fracton gauge fields and non-Lorentzian particles (both Carroll and Galilean) have been discussed in a recent paper \cite{Ahmadi-Jahmani:2025iqc}. This basically makes clarifies the connections between static Fractons and electric Carroll particles, and also between Fracton dipoles and electric/magnetic Galilean particles. 

\end{itemize}

     \item{\textit{Subsystem symmetries:}} An interesting connection was made in \cite{Kasikci:2023tvs} between Carrollian physics and spacetime subsystem symmetries. Subsystem symmetries are intrinsic ones that remain conserved on substructures of the original internal space associated to the system. The authors in the aforementioned work focused on the scalar field theory model of \cite{Baig:2023yaz}, which was itself proposed based on fractonic lattice models. The subsystem symmetries manifest in the spacetime was shown to be a direct consequence of Carroll boost invariance. 

    \item{\textit{Ultra-local fermions:}} Highly localised fermions in a lattice context were also discussed in \cite{Ara:2023pnn}, where the authors considered a lattice fermion model with  extremely large value of interaction strength, where one can neglect hopping terms, and get ultra-local modes. This regime of the dynamics also enjoys infinite number of conserved symmetries, and the ground state is highly degenerate as well. 

    \item{\textit{Hall physics:}} We have seen time and again that Carroll dynamics describes a limit of relativistic physics where the speed of light $c \to 0$. In this regime, time evolution effectively "freezes", and particles become non-propagating in the conventional sense. Despite this, nontrivial dynamics can still arise, especially in the presence of background fields—this is where the Hall effect comes into play. This topic was explored recently in \cite{Marsot:2022imf}, where it has been shown that a Carroll particle in background electromagnetic field can move following generalised Hall law. 
\end{itemize}
\newpage

\part{Concluding Remarks}

\bigskip

\bigskip

\bigskip

\section*{Outline of Part IV}
Part IV of our trilogy is about what may have been if we had infinite time and energy. It outlines what we would not cover, but don't want to leave completely unsaid. These are applications of Carroll to various other places including to black hole horizons, cosmology, string theory and subjects such as Carroll gravity. Having looked into the kaleidoscope in Fig \ref{fig:Avatar}, it is hard to unsee it. So we take a glimpse at it again, albeit briefly. 

\newpage

\section{What we have not discussed}\label{penult}
We are at the penultimate chapter of our review and there are numerous things we have not discussed. To keep this review to a reasonable length, we will only be briefly touching up on some other applications of Carrollian and Conformal Carrollian symmetries. We also provide some references to relevant work in these very short summaries below. 

\subsection{Black hole horizons and Carrollian hydrodynamics}
\label{sec:stretched}
We encountered the possibility of Carrollian structures emerging on the event horizon of black holes \cite{Penna:2018gfx, Donnay:2019jiz} at the very beginning of the review, but we have not come back to this point. We do so briefly here. Just as a reminder, Carrollian symmetries are expected to appear on generic null surfaces and black hole event horizons are null surfaces. Below we would see a more intricate relation between black hole horizons and Carrollian hydrodynamics. 

\medskip

In section \ref{carroll_hydro}, we saw how Carroll Hydrodynamics arise in physically relevant situations such as QGP. We now turn to the dynamics of black hole horizons. It turns out one can better understand the membrane paradigm \cite{PhysRevD.18.3598, PhysRevD.33.915} equations of the ``stretched-horizon'' through Carroll hydrodynamics \cite{Penna:2018gfx, Donnay:2019jiz, Redondo-Yuste:2022czg}. The main idea of \cite{Donnay:2019jiz} is to interpret the near-horizon limit of the stretched horizon as a $c \to 0$ ultra-relativistic limit with the radial coordinate of the horizon $\rho$ being identified as $\rho = c^2$. To make contact with what we discussed all the way in Sec. \ref{ssec:intro2}, this radial coordinate is the effective speed of light. Suppose if one considers a black hole geometry spanned by coordinates $(v,\rho,x^A)$, then the near horizon geometry can be encapsulated by a Gaussian null coordinate system \cite{Moncrief1983} (see also Appendix A of \cite{Bhattacharyya:2016xfs}),
\begin{equation}
\label{eq:ngmetric}
    ds^2 = -2 \kappa \, \rho dv^2 + 2 d\rho \, dv + 2 \theta_A \, \rho dv\, dx^A + (S_{AB} + \lambda_{AB} \rho)\, dx^A\,dx^B + \mathcal{O}(\rho^2) \, ,
\end{equation}
where $\kappa$ is the surface gravity, $\theta_A$ is proportional to the twist (thus it is non-zero only for rotating spacetimes), and $S_{AB}$ is the intrinsic spatial metric on the horizon $\rho=0$. The extrinsic geometry is captured by $\kappa$, $\theta_A$, the horizon expansion $\Theta$ and the shear tensor $\sigma_{AB}$ given by
\begin{equation}
    \Theta = \partial_v \ln \sqrt{S} \, , ~~~~~~ \sigma_{AB} = \dfrac{1}{2}\partial_v S_{AB} - \dfrac{\Theta}{D-2} \, S_{AB} \, , 
\end{equation}
where $S$ is the determinant of $S_{AB}$. 

\medskip

One can impose vacuum Einstein's equations on the metric \eqref{eq:ngmetric} and project the equations onto the horizon to obtain the null Raychaudhuri equation and Damour equation given respectively by
\begin{eqnarray}\label{eq:membraneeqs}
    & \partial_v \Theta - \kappa \, \Theta + \dfrac{\Theta^2}{D-2} + \sigma_{AB} \sigma^{AB} = 0 \, , \\
    & \left( \partial_v + \Theta \right) \theta_A + 2\nabla_A \left(\kappa + \dfrac{D-3}{D-2}\,\Theta \right) - 2 \nabla_B \sigma^B_{~\,A} = 0 \, .
\end{eqnarray}
We have seen how the PR parametrization \eqref{PRmetric} naturally encodes the Carroll diffeomorphisms. Comparing \eqref{eq:ngmetric} to \eqref{PRmetric}, we see that one can make the following identifications
\begin{equation}\label{eq:bhcarid}
    c^2 = \rho \, , ~~~~~ \Omega = \sqrt{2 \kappa} \, , ~~~~~ b_A  = \dfrac{\theta_A}{\sqrt{2\kappa}} \, .
\end{equation}
From the identification \eqref{eq:bhcarid}, one can construct quantities that are covariant under the Carroll diffeomorphisms \eqref{Carroll_diffeos}. 

\medskip

The ``stretched membrane'' is a hypersurface $\Sigma_{\rho}$ near $\rho=0$ with unit normal given by
\begin{equation}
    n = \dfrac{d\rho}{\sqrt{2 \kappa \rho}} \, .
\end{equation}
One can now define the Energy-momentum tensor of the membrane via
\begin{equation}
    T_{\mu \nu} = \dfrac{1}{8 \pi G} \left( K \, h_{\mu\nu} - K_{\mu\nu} \right) \, ,
\end{equation}
where $K^{\mu}_{\,~\nu} = h_{\nu}^{\,~\alpha} \mathcal{D}_{\alpha} n^{\mu}$ ($\mathcal{D}_{\alpha}$ is the covariant derivative compatible with the full metric) is the extrinsic curvature of $\Sigma_{\rho}$, $K$ is trace $K= K^{\mu}_{~\,\mu}$ and $h_{\mu\nu} = g_{\mu\nu} - n_{\mu} n_{\nu}$. Einstein's equations imply the stress tensor is conserved:
\begin{equation}\label{eq:conseqstress}
    \bar{\nabla}_j T^{ji} = 0 \, ,
\end{equation}
where $i = (v,x^A)$ and $\bar{\nabla}_i$ is the covariant derivative compatible with the induced metric of \eqref{eq:ngmetric} at $\rho =0$. \cite{Donnay:2019jiz} argues that \eqref{eq:conseqstress} in the $\rho \to 0$ limit can be interpreted as a conservation equation for the ``Carrollian momenta'' (built out of geometric quantities with identification \eqref{eq:bhcarid}) defined on the horizon. By construction these Carollian momenta are covariant under the Carroll diffeos \eqref{Carroll_diffeos}. For a hypersurface $\Sigma_{\rho}$ at finite $\rho$, \eqref{eq:conseqstress} describes the dynamics of a relativistic fluid. Thus, the identification \eqref{eq:bhcarid} implies that the near-horizon limit implements a $c \to 0$ Carrollian limit on the relativistic fluid (see \S\ref{Eqs_Carroll_Hydro}). Thus, the dynamics of the horizon is mapped to the dynamics of a Carrollian fluid that lives on the horizon.

\subsection{Carroll gravity}

In section \ref{sec:carfieldtheories}, we have seen how a $c \to 0$ expansion of Lorentzian field theories results in theories that realise Carroll symmetries. An interesting question would be how to apply this Carroll expansion to gravitational theories. This is the opposite limit of the Post-Newtonian limit that is typically studied in relevant astrophysical scenarios \cite{Blanchet:2002av,Levi:2018nxp}. Rather than being a mathematical curiosity, recent work \cite{Hansen:2021fxi,Ecker:2023uwm} has suggested that this limiting case of gravity dubbed Carroll gravity has ``black hole solutions'' which in particular demonstrate a version of Hawking effect \cite{Aggarwal:2024yxy}. It was even suggested that this limit might be useful to describe the near-singularity dynamics of GR or equivalently the so called BKL limit \cite{Hansen:2021fxi,Oling:2024vmq}. The Carroll expansion in the metric formulation of GR is implemented through a ``Pre-Ultra local'' (PUL) parametrization \cite{Hansen:2021fxi} (note the similarities to the PR parametrization \eqref{PRmetric}) where the metric is split as
\begin{equation}
    g_{\mu\nu} = - c^2 \mathcal{T}_{\mu} \mathcal{T}_{\nu} + \Pi_{\mu\nu} \, , ~~~~~ g^{\mu\nu} = - \dfrac{1}{c^2} \mathcal{V}^{\mu} \mathcal{V}^{\nu} + \Pi^{\mu\nu} \, ,
\end{equation}
where $\mathcal{T}_{\mu}$ and $\mathcal{V}^{\mu}$ are `time-like' and $\Pi_{\mu\nu}$ is `space-like' and they satisfy
\begin{equation}
    \mathcal{T}_{\mu} \mathcal{V}^{\mu} = -1 \, , ~~~ \mathcal{T}_{\mu} \Pi^{\mu\nu} = 0 \, , ~~~ \Pi_{\mu\nu} \mathcal{V}^{\nu} = 0 \, , ~~~~ \delta^{\mu}_{~\,\nu} = - \mathcal{V}^{\mu} \mathcal{T}_{\nu} + \Pi^{\mu\rho} \Pi_{\rho \nu} \, .
\end{equation}
The idea now is to expand these quantities ($\mathcal{T}_{\mu}, \mathcal{V}^{\mu}, \Pi^{\mu\nu}$) in powers of $c^2$ assuming that these quantities are analytic in $c^2$ \footnote{This is merely a choice. For an implementation of the expansion in odd powers of $c$, see \cite{Ergen:2020yop}.}. As a result, the expansion for Schwarzschild and Kruskal-Szekeres is different because of non-analyticity in $c^2$. Implementation of the expansion for the Einstein-Hilbert action results in
\begin{equation}\label{eq:ehactionexp}
    \begin{split}
        S &= \dfrac{c^3}{16 \pi G} \int d^{d+1}x \, \sqrt{-g} \, R \\
        &= \dfrac{c^2}{16 \pi G} \int d^{d+1}x \, E \, \left[ (\mathcal{K}^{\mu\nu} \mathcal{K}_{\mu\nu} - \mathcal{K}^2) + c^2 \Pi^{\mu\nu} \overset{\text{\tiny ($\mathcal{C}$) }}{\mathcal{R}}_{\mu\nu} + \dfrac{c^4}{4} \Pi^{\mu\nu} \Pi^{\rho\sigma} (d \mathcal{T})_{\mu\rho} (d\mathcal{T})_{\nu\sigma} \right] \, ,
    \end{split}
\end{equation}
where $\mathcal{K}_{\mu\nu}$ is the extrinsic curvature associated with $\Pi_{\mu\nu}$ ($\mathcal{K}$ is it's trace), $\overset{\text{\tiny ($\mathcal{C}$)}}{\mathcal{R}}_{\mu\nu}$ denotes the Ricci tensor associated with the affine connection at $\mathcal{O}(c^0)$. The expansion \eqref{eq:ehactionexp} organizes as
\begin{equation}
    S = c^2 S_{\text{LO}} + c^4 S_{\text{NLO}} + \mathcal{O}(c^6) \, .
\end{equation}
The leading order LO theory is called the `electric' limit \cite{Henneaux:1979vn,Henneaux:2021yzg} and the next leading order NLO theory contains information about the `magnetic' limit \cite{Henneaux:2021yzg}. \eqref{eq:ehactionexp} can be considered as a Lagrangian approach to studying small $c$ expansion. Both `electric' and `magnetic' limits have been approached from the Hamiltonian perspective in \cite{Henneaux:2021yzg} and reviewed in Sec. \ref{sec:carfieldtheories}. One can also approach the expansion in the first order formulation \cite{Bergshoeff:2017btm}. Finally, similar to how one can obtain Riemannian geometry from gauging the Poincar\'e algebra, one can obtain an ultra-relativistic limit of the Riemannian geometry by gauging the Carroll algebra \cite{Hartong:2015xda,Bekaert:2015xua,Niedermaier:2020jdy}. This is the geometry of a null hypersurface embedded in a Lorentzian geometry in one higher dimension. In this context, there has been work on understanding asymptotic structure of the electric and magnetic Carrollian limits of Einstein gravity as well as the coupled Einstein-Yang-Mills system.\cite{Fuentealba:2022gdx}.

\medskip

One could work out the solutions of the LO and NLO theories. LO theory equations of motion are constraints and evolution equations akin to a $3+1$ split Hamiltonian formulation of GR. The solutions of the NLO theory are those with non-zero mass \cite{Hansen:2021fxi}. For instance, the Carroll limit of the Schwarzschild metric can be shown to be the solution of the NLO theory. One can incorporate the cosmological constant by considering an appropriate scaling with $c$. The $c \to 0$ limit effectively means there is no light cone structure and thus there is no notion of event horizon. Carroll extremal surfaces (suitable analogues of Lorentzian extremal surfaces) were argued to replace this Lorentzian notion of horizon \cite{Ecker:2023uwm}, and thus a Carroll black hole is
\begin{equation}
    \text{Carroll black hole} ~ = ~ \text{Carroll extremal surface}~ + ~ \text{Carroll thermal properties} .
\end{equation}
The thermal properties are inferred from 2D Carroll-Dilaton gravity in the so called PSM formalism. Related literature on Carroll gravity include \cite{Dautcourt:1997hb,Matulich:2019cdo,Ravera:2019ize,Gomis:2019nih,Ciambelli:2019lap,Grumiller:2020elf,Gomis:2020wxp,Aviles:2022xyx,Perez:2021abf,Concha:2021jnn,Guerrieri:2021cdz,Lovrekovic:2021dvi,Figueroa-OFarrill:2022mcy,Campoleoni:2022ebj,Sengupta:2022rbd,Miskovic:2023zfz,Musaeus:2023oyp,Pekar:2024ukc,March:2024zck,Tadros:2024bev,Vigneron:2025vgn,Blitz:2025spc,Kolar:2025ebv,PhysRevD.111.124007,Aviles:2025ygw}.  Carroll (covariant) geodesic action is another topic of interest, since it gives one the scope to  probe (curved) Carrollian spacetimes with test particles. Such intrinsic Carroll version of geodesics have been discussed recently in \cite{Ciambelli:2023tzb}, and non-trivial orbits of such particles in the Carroll–Schwarzschild black hole background \cite{Hansen:2021fxi,Perez:2021abf} has been considered. In this context, there has been work on understanding asymptotic structure of the electric and magnetic Carrollian limits of Einstein gravity as well as the coupled Einstein-Yang-Mills system \cite{Fuentealba:2022gdx}. 
%The connection between AdS$_3$-Carroll gravity and FSC have been realised in \cite{Aviles:2025ygw}.

\newpage

\subsection{Carroll in cosmology}
We now briefly review the role of Carroll in cosmology following \cite{deBoer:2021jej}. Cosmological dynamics in the Carroll limit leads to a drastically constrained causal structure and has deep implications for dark energy and inflation. We start with the standard flat Friedmann–Lemaître–Robertson–Walker (FLRW) metric:
\begin{eqnarray}
    ds^2 = -c^2 dt^2 + a(t)^2 dx^idx^j\delta_{ij}
\end{eqnarray}
where $a(t)$ is the cosmological scale factor. Assuming a perfect fluid energy–momentum tensor with energy density $\rho$ and pressure $p$, the Einstein field equations reduce to two key dynamical Friedmann's equations:   
\begin{eqnarray}\label{Friedmann}
    \left( \frac{\dot{a}}{a} \right)^2 = \frac{8\pi G}{3c^2} \rho \,,\quad \frac{\ddot{a}}{a} = -\frac{4\pi G}{3c^2} \left( \rho + 3p \right)\,.
\end{eqnarray}
Here $a(t)$ is fixed by the equation of the state of the form $p = w\rho$, where $w$ is a parameter which differs upon the type of matter present in the universe (such as for non-relativistic matter $w=0$, and $w=1/3$ for ultra-relativistic ones). Solving \eqref{Friedmann}, it can be checked that for any value of $w$ other than $-1$, we will have a spatially flat universe, whereas for $w=-1$, we will have an exponentially expanding universe. 

\medskip

To explain phenomena such as inflation, dark energy, and early universe dynamics, it is often preferred a cosmological model based on a single scalar field. The scalar field behaves like a perfect fluid with\footnote{In cosmology, we usually assume a homogeneous and isotropic universe. Under this assumption, the scalar field depends only on time, not on space. So spatial gradients vanishes in this context. However, spatial derivatives do matter in contexts such as structure formation, CMB studies etc.} 
\begin{eqnarray}
    \rho = \frac{1}{2c^2}\dot{\phi}^2 + V(\phi)\,,\quad p = \frac{1}{2c^2}\dot{\phi}^2 - V(\phi)\,.
\end{eqnarray}
Then we will have 
\begin{eqnarray}
    w = \frac{\frac{1}{2c^2}\dot{\phi}^2 - V(\phi)}{\frac{1}{2c^2}\dot{\phi}^2 + V(\phi)} = \frac{\frac{c^2}{2}\pi_{\phi}^2 - V(\phi)}{\frac{c^2}{2}\pi_{\phi}^2 + V(\phi)} \approx -1 + c^2\frac{\pi_{\phi}^2}{V(\phi)} + \mathcal{O}(c^4)\,.
\end{eqnarray}
Here $\pi_{\phi}$ is the conjugate momentum of the field $\phi$ and we have performed a small c-expansion, where it has been assumed that the potential $V(\phi)$ term is c-independent. As can be seen, in the leading order, we will have $w=-1$, which corresponds to a perfect fluid. Using Friedmann's equations, for any $w$ one can get the following identity
\begin{equation}
    \rho + p = \frac{3c^2}{8\pi G}(1+w)H^2(t)
\end{equation}
where $H(t)$ is the Hubble function, defined as $H(t) = \dot{a}/a$. It follows that in the Carroll limit, $\rho + p \to 0$, implying a necessary condition for dark energy. However, it is not sufficient as individually both $\rho$ and $p$ can vanish in the Carroll limit. Therefore to model dark energy one must have non-zero energy density, with $w=-1$. Further, one can also consider the scenario with scalar field inflation where $w$ is time-dependent. It can be shown that under Carroll limit, it will tend towards $w=-1$. For details, please look at \cite{deBoer:2021jej}.  

\medskip

The key observation is that when we want to explore the universe through an expansion in the speed of light $c$, we conceptualize spacetime as being composed of small Hubble cells, each with a characteristic size given by the Hubble radius $cH^{-1}$. As we move away from the Carroll point, these Hubble cells become localized patches, in which observers reside.  In the small-$c$ regime, the physics becomes ultra-local and lightcones effectively collapse — a defining feature of Carrollian geometry. As $c$ increases, the Hubble radius expands, allowing more degrees of freedom to enter each Hubble patch, thereby gradually restoring the relativistic structure of spacetime.

\subsection{String theory and Carroll}
String theory remains one of the most promising avenues to a framework of quantum gravity. One of the central reasons for the success of string theory is the presence of 2D conformal invariance on the worldsheet. The Polyakov action
\begin{align}
    S_P = \frac{1}{2\pi \a'}\int d^2\xi \, \sqrt{-\gamma} \, \gamma^{ab} \, \partial_a X^\mu \partial_b X^\nu \, \eta_{\mu\nu}
\end{align}
describes the motion of a string in a flat $D$-dimensional background with the metric $\eta_{\mu\nu}$. In the above, $\a'$ is the string length, which can be related to the tension of the string as $T=1/2\pi\a'$, and $\gamma^{ab}$ is the worldsheet metric. One can look at a covariant quantisation of the system, where$\gamma^{ab}$ is fixed to be flat and in this ``conformal'' gauge the the worldsheet still has residual gauge symmetries which turn out to be two copies of the Virasoro algebra. In this Old Covariant Quantisation (OCQ) formulation, the string theory is quantised as a constrained system where the Virasoro symmetry is imposed by the vanishing of the conformal stress tensor on the physical states of the string Hilbert space:
\begin{align}
    \<phy'| T(\omega)|phy\> = 0 = \<phy'| {\bar{T}}(\bar{\omega})|phy\>. 
\end{align}
In the above, $T(\omega) = \sum \L_n e^{in\omega}, \quad {\bar{T}}(\bar{\omega}) = \sum \bL_n e^{in \bar{\omega}}$ are the holomorphic and anti-holomorphic stress tensors of the 2D worldsheet CFT, now parametrized on the cylinder $(\w, \bar{\omega}) = \t \pm \sigma$. The power of the 2D  CFT living on the string worldsheet allows us to calculate many things of interest in string theory and plays a pivotal role in understanding the structure of quantum gravity through the lens of string theory. 

\medskip

Carrollian structures can appear in string theory on both the worldsheet theory as well as in target space. Below we discuss both aspects.  

\medskip

{\ding{112}} \underline{\emph{Tensionless strings}}

\medskip

The very high energy regime of string theory has been one shrouded in mystery and intrigue. The string partition function diverges beyond a certain limiting temperature known as the Hagedorn temperature \cite{Atick:1988si}. It is expected that here string theory transitions into a different phase, called the Hagedorn phase beyond this temperature, where degrees of freedom would be very different from perturbative string theory. In this very high energy limit, string scattering amplitudes are also known to simplify \cite{Gross:1987kza, Gross:1987ar} and there are recursion relations among them \cite{Gross:1988ue} which indicate the presence of a very rich symmetry structure. It has been speculated that higher spin structures emerge in string theory in this regime and connections to higher spin gravity \cite{Vasiliev:2004qz} have been explored \cite{Chang:2012kt}. 

\medskip

This very high energy limit is diametrically opposite to the point-particle limit of string theory where the string shrinks to a point reducing string theory to its supergravity avatar. If we work in the non-interacting string, there is only one tuneable parameter in the game, the string length $\a'$. $\a'\to0$ is the point particle limit mentioned above. The very high energy limit is thus the limit where the string length goes to infinity $\a'\to\infty$ or equivalently, the tensionless limit $T\to 0$. 

\medskip

When one considers point particles, the massless limit means the particles travel at the speed of light, on null geodesics. The worldlines of these particles are null. Tension for a string is mass per unit area. The tensionless limit of the string is thus analogous to the massless limit of the point particle. The worldsheet of the string becomes a 2D  null surface in keeping with this observation and Carrollian structures naturally appear on the worldsheet. 

\medskip

The work on null strings was initiated by Schild in the 1970s \cite{Schild:1976vq}. The major advancement in terms of understanding the theory in terms of an action came in \cite{Isberg:1993av} where from a Hamiltonian perspective the action was written down as
\begin{align}
    S_{ILST} = \int d^2\xi \, V^a V^b \partial_a X^\mu \partial_b X^\nu \, \eta_{\mu\nu}
\end{align}
In the above, the subscript ILST represents the authors \cite{Isberg:1993av}. We see that from the Polyakov action there has been the following change in the $T\to0$ limit:
\begin{align}
    T\sqrt{-\gamma} \gamma^{ab} \to V^a V^b
\end{align}
where $V^a$ are vector densities and the RHS naturally has a vanishing determinant reflecting the null nature of the 2D  surface. The action has worldsheet diffeomorphism invariance and hence we need to pick a gauge. In the analogue of the tensile conformal gauge, we choose $V^a=(1,0)$. It is then straightforward to see that one is left with 
\begin{align}
    S_{ILST}^{gauge-fixed} = \int d^2\xi \, \partial_\tau X^\mu  \partial_\tau X_\mu .  
\end{align}
which is precisely the action of a free massless electric Carroll scalar theory. One can readily show that in this gauge, there exists a residual gauge symmetry now on the tensionless worldsheet which is given by the 2D  Conformal Carroll or the BMS$_3$ algebra. Although the algebra itself was discovered in \cite{Isberg:1993av}, the identification with BMS and the ultra-relativistic limit of the Virasoro algebra was first made about 20 years later in \cite{Bagchi:2013bga}. Carrollian conformal structures in this context have been thoroughly explored in a number of works following \cite{Bagchi:2013bga}. An incomplete list of references include \cite{Bagchi:2015nca, Bagchi:2016yyf, Casali:2016atr, Casali:2017zkz, Bagchi:2017cte, Bagchi:2019cay, Bagchi:2020fpr, Bagchi:2020ats}. A detailed review of tensionless strings and associated Carrollian structures would appear elsewhere \cite{Bagchi:2026wcu}. 

\medskip

Connections to other limiting `corners' of string theory to tensionless strings have also been discussed in the literature, starting with \cite{PhysRevLett.132.161603}, which chalked out a duality web that connects various non-Lorentzian sectors. Null strings have been used recently to count microstates of BTZ black holes in a novel way \cite{Bagchi:2022iqb}, reproducing the correct leading and (more surprisingly) subleading Bekenstein-Hawking entropy. 
Another rather intriguing recent development is the hint of a connection between tensionless strings and holography in asymptotically flat spacetimes briefly alluded to in \cite{Kervyn:2025wsb} following earlier work in \cite{Stieberger:2018edy}. This is a tantalizing hint that Carrollian structures on the worldsheet of the tensionless can be related to Carrollian holography, also indicating how to formulate holography in AFS in a top-down manner. Some related questions are currently under investigation. 

\medskip

{\ding{112}} \underline{\emph{Carroll string and String Carroll}}
\medskip

\underline{\emph{Carroll string:}} In our discussion above, Carrollian structures appeared as residual gauge symmetry on the worldsheet of tensionless strings. In parallel avenues of progress, string propagating in Carrollian (more generally non-Lorentzian) target spaces have been discussed in the literature, in many guises, for last decade or so, first arising in the pursuit of extending Carrollian particle actions (see Sec. \eqref{section2}). We will briefly talk about strings in the Carrollian target spaces (or `Carroll strings') in what follows, and for related progress in Non-Relativistic string theories, the reader is pointed to the review \cite{Oling:2022fft}.
\medskip

The study of Carroll strings was initiated in \cite{Cardona:2016ytk} where Carroll symmetries were dynamically introduced in canonical string action by taking limits akin to the particle Carroll case. In the simplest setting, one starts with the canonical phase space action for a relativistic string with conformal constraints: 
\begin{equation}
S=\int d^2\sigma\ \Big[\Pi\cdot\dot{X}-N\mathcal{H}-N_\phi \mathcal{J}]
\end{equation}
 $\Pi_\mu=\frac{\delta \mathcal{L}}{\delta \dot{X}^\mu}$ being the canonical momentum, and dot/prime are usual worldsheet time/space derivatives. Further $\mathcal{H}=\Pi^2+\mathbb{T}^2X^{\prime2}$ and $\mathcal{J}=\Pi\cdot X^\prime$ are the Hamiltonian and momentum constraints involving the tension $\mathbb{T}$. $N$ and $N_\phi$ are the Lagrange multipliers (lapse and shift, respectively).  The Carrollian limit on this action is taken by considering the rescaling of only the \textit{logitudinal} directions $a = (0,1)$ by some dimensionelss parameter $\omega$:
\be
X^\mu \to \frac{X^a}{\omega},~~~\Pi^a \to \omega \Pi^a
\ee
Now one can put the scaled canonical variables back into the action and take the limit $\omega \to \infty$. Finiteness of the action would also require that we scale the Lagrange multipliers by $N \to N/\omega^2,~~N_\phi \to N_\phi$. Note this is subtly different from a Carroll particle limit, and is often called the `stringy' Carroll limit. In this limit, it was shown \cite{Cardona:2016ytk} that canonical constraints generate an algebra similar to the BMS algebra, with Hamiltonians commuting at different spatial points. Similarly, Carroll p-brane actions, from dynamics perspective, were also constructed. In all cases, the claim was that extended Carroll objects can only have trivial dynamics.
\medskip

Various versions of such Carrollian strings/branes have been discussed since. Note that, in most cases, the interest was around theories with relativistic worldvolumes propagating in Carrollian target spaces. This includes canonical analyses of different classes of extended Carroll objects and their symmetries \cite{Kluson:2017fam, Roychowdhury:2019aoi, Kluson:2022jxh}, extension to extended worldsheet symmetry algebras and `magnetic' versions of Carroll strings \cite{Harksen:2024bnh} etc.~Furthermore, it has been shown that starting with different relativistic parent string theories, one reaches inequivalent Carroll strings in the limit \cite{PhysRevD.110.066008}. However, insight into the physics of such strings have been only limited due to usual triviality of their dynamics. 

\medskip

\bigskip

\underline{\emph{String Carroll expansions:}} In dealing with field theories on Carroll manifolds, we employed a Carroll expansion in the powers of the speed of light $c$. This has been dubbed the ``particle''-Carroll expansion in literature, to distinguish it from more intricate expansions, where one considers not one, but several Carrollian or null directions in the underlying non-Lorentzian manifold. One may think that this is a very artificial situation, but very much like the fact that initial explorations of Carroll were considered purely mathematical and now we see that Carroll structures are ubiquitous in physical applications, this line of thought turns out to be very misleading. We will elucidate with an example. 

Consider a generic non-extremal black hole. It is well known that the near-horizon limit of this non-extremal black hole admits a 2D  Rindler spacetime. Consider the simple case of the Schwarzschild black hole:
\begin{align}
    ds^2 = -\left(1-\frac{2M}{r}\right) dt^2 + \left(1-\frac{2M}{r}\right)^{-1} dr^2 + r^2 d\Omega^2.
\end{align}
The near horizon limit is achieved by zooming near the event horizon $r=2M$. To do this we use:
\begin{align}
     r-2M = \frac{\e}{2M}\rho^2. 
\end{align}
The near horizon expansion thus can be written as 
\begin{align}
    ds^2 = 4M^2 d\Omega^2 + \e \left[-\rho^2 d\t^2 + d\rho^2 + 2\rho^2 d\Omega^2 \right] + \mathscr{O}(\e^2)
\end{align}
where we have scaled $t=2M\t$. Hence to first order in $\e$, the spacetime is 2D Rindler times a 2-sphere. This is thus a Carroll manifold with two null directions \footnote{This may be a little hard to follow with the sphere factor sitting in the $\mathscr{O}(\e)$ term. For details, please see \cite{Bagchi:2024rje}.}. The near horizon expansion thus takes the form of a ``string''-Carroll expansion \cite{Bagchi:2023cfp, Bagchi:2024rje} where there is a ``speed of light'' associated with not one, but two spacetime directions, giving a vanishingly small Rindler or an exponentially large sphere in the strict limit. String theory near black holes thus assumes a Carroll flavour and this has been discussed in detail in \cite{Bagchi:2023cfp, Bagchi:2024rje}. 

\newpage

\subsection{Other extensions}
We have discussed the vanilla Carroll and Conformal Carroll symmetries in our review. But Carroll comes in various flavours. Here is a pointer to some of the other varieties. 

\begin{itemize}
    \item {\em $N$-conformal Carroll}: We had encountered the possibility of more generic conformal Carroll structures when we discussed conformal isometries of Carroll manifolds in Sec.~\ref{section2}. We refer the reader back to the equation \eqref{con-car}. For $N\equiv - \frac{\lambda_1}{\lambda_2} \neq 2$, the conformal isometries close to form what is known as the $N$-conformal Carroll algebras which are not isomorphic to BMS. These general symmetries have been found on black hole horizons \cite{Donnay:2016ejv, Donnay:2015abr, Grumiller:2019fmp} and more recently in the context of possible duals to plane wave spacetimes \cite{Despontin:2025dog} for the case of $N\to\infty$. It is important to mention here that this limit of $N\to\infty$ leads to an isomorphism with what is called a {\em Warped Conformal Field Theory} \cite{Hofman:2011zj, Detournay:2012pc}. 

  \item {\em Supersymmetric Conformal Carroll}: Superconformal versions of Carroll symmetry have been investigated in the literature. Perhaps the first such algebra was written down in the context of null superstrings \cite{Lindstrom:1990qb, Lindstrom:1990ar} but connections to Carroll or BMS were not made till much later. Depending on how one scales the relativistic SUSY algebra, there are several different versions of Carroll Conformal SUSY that one could end up in. In $D=2$, there are two distinct possibilities which we called the Homogeneous and Inhomogeneous algebras. We provide some details in Appendix B. These have been realised as asymptotic symmetry algebras of 3D supergravity in AFS \cite{Barnich:2014cwa, Lodato:2016alv} as well as the worldsheet symmetries of different null superstrings \cite{Lindstrom:1990qb, Bagchi:2016yyf, Bagchi:2017cte}. Superconformal Carroll algebras in higher dimensions were constructed in \cite{Bagchi:2022owq} and followed up in \cite{Lipstein:2025jfj, Zheng:2025cuw}. 

  \item {\em Higher spin and Carroll}: Carrollian higher spin theories were first introduced in the context of asymptotic symmetries of 3D higher spin gravity in AFS in \cite{Afshar:2013vka} followed closely by \cite{Gonzalez:2013oaa}. We discuss the $W_3$ equivalent of the Carroll Conformal algebra (rather wonderfully named the BMW algebra!) in Appendix B. Some more work in Carroll higher spin theories include \cite{Campoleoni:2021blr}. 
    
\end{itemize}

\newpage

\section{Conclusions: One symmetry to rule them all!}

\begin{figure}[t]
\centering
\includegraphics[width=.25\linewidth]{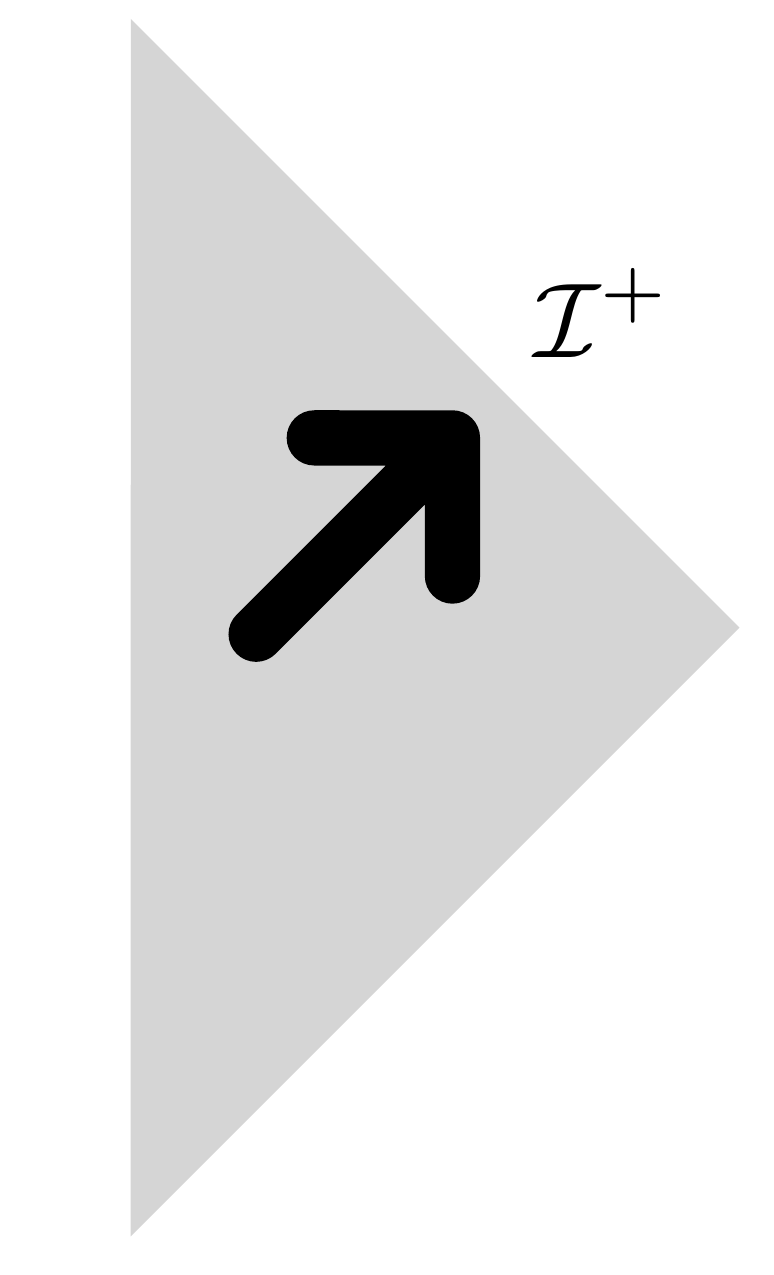}
\caption{The future is Carroll!}
\label{future}
\end{figure}

{\it{One ring to rule them all, \\ One ring to find them, \\ One ring to bring them all, \\ And in the Darkness, bind them.}}

\smallskip

Our endeavour in this review has not been as sinister as Sauron's, but the repeated occurrence of Carrollian symmetries in various, apparently unrelated branches of theoretical physics, indeed calls for a need to bind them (perhaps not in darkness).  We have previously remarked that relativistic conformal symmetry has been a unifying language in diverse physical situations like the study of critical phenomena in statistical mechanics, the understanding of quantum field theories in terms of flows from fixed points governed by conformal symmetry, to the understanding of holography via the AdS/CFT correspondence and the study of quantum gravity in the context of string theory, where conformal symmetry shows up on the worldsheet. As we have seen, relativistic CFTs become particularly powerful in $D=2$ where the infinite dimensional Virasoro algebra comes into play and one can rely entirely on symmetries to constrain quantities of interest, without even resorting to the underlying Lagrangian descriptions. 

\medskip

The central message of the present review is that there is a new ring of power in the game now. And this one, Carrollian symmetry and its conformal cousin, may be even more powerful than the original relativistic conformal symmetry, since these symmetries are infinite dimensional in {\it all spacetime dimensions}, not just $D=2$. We have discussed applications of Carroll and Conformal Carroll symmetries in the context of holography for asymptotically flat spacetimes in Part 2 and to hydrodynamics and condensed matter physics in Part 3. In Sec.~\ref{penult}, we briefly discuss some of the other applications which we did not cover in the main body of this review. Carroll symmetry is indeed the One symmetry to rule them all!

\medskip

The directions of future work are too many to list in this review. We would just like to end by saying that to a very good approximation, the future looks Carrollian (see Fig.~\ref{future} above).

\medskip

\bigskip

\subsection*{Acknowledgments}
The Carroll adventure has been (and continues to be) a long winding road and there are numerous people we are indebted to. We thank Sudipta Dutta for initial collaboration on this review. 

\medskip

AB wishes to thank his friends and collaborators on this wonderful joyride. A special shout-out to Daniel Grumiller, Stephane Detournay and Joan Simon for a remarkable collaboration that is ongoing for the last dozen years and more. Thanks to the people with whom the papers that form the backbone of this review have been written: Shamik Banerjee, Rudranil Basu, Shankhadeep Chakrabortty, Sudipta Dutta, Reza Fareghbal, Kedar Kolekar, Max Riegler, Zodinmawia and to other collaborators Ritankar Chatterjee, Mirah Gary, Emil Have, Jelle Hartong, Arthur Lipstein, Taniya Mandal, Mangesh Mandlik, Aditya Mehra, Wout Merbis, Hisayoshi Muraki, Poulami Nandi, Pulastya Parekh, Jan Rosseel, Amartya Saha, Pushkar Soni.

\smallskip

A special mention to Shahin Sheikh-Jabbari without whose invitation and constant encouragement this review would never have been written. 

\smallskip

AB is partially supported by a Swarnajayanti Fellowship from the Science and Engineering Research Board
(SERB) under grant SB/SJF/2019-20/08 and also by an ANRF grant CRG/2022/006165. 

\smallskip

ABan is supported in part by an OPERA grant
and a seed grant NFSG/PIL/2023/P3816 from BITS Pilani, and further an early career research grant ANRF/ECRG/2024/002604/PMS from ANRF India. He also acknowledges financial support from the Asia Pacific Center for Theoretical Physics (APCTP) via an Associate Fellowship.

\smallskip

PD is supported by an NSERC Discovery grant.

\smallskip

SM is supported by a Fellowship for Academic and Research Excellence (FARE) at IIT Kanpur.

\newpage

\section*{APPENDICES}

\appendix
\section[A Celestial crash-course: Infrared triangle and celestial holography]{A Celestial crash-course: Infrared triangle and celestial holography\symfootnote{This appendix was written in collaboration with Sudipta Dutta.}}
\label{ap:celreview}

In this appendix, we quickly summarize the main elements of the Celestial approach to flat space holography. We will be brief (and thus not exhaustive with referencing) and redirect the reader to many excellent reviews which are available in the literature \cite{Strominger:2017zoo,Pasterski:2021rjz,Raclariu:2021zjz,Pasterski:2021raf}. The basic premise is that there exists a correspondence between bulk physics in $(3+1)D$ asymptotically flat spacetimes and the 2D Celestial sphere at null infinity. The approach gained prominence when a triangular set of relations were uncovered between asymptotic symmetries, soft theorems, and the memory effect \cite{He:2014laa,Strominger:2014pwa,Lysov:2014csa,Pasterski:2015tva,Pasterski:2015zua}. For instance, soft theorems turned out to be a consequence of the asymptotic symmetries of flat spacetimes \cite{He:2014laa,Strominger:2014pwa}. The memory effect is a DC effect resulting in a permanent shift of the relative position of the detectors when you have gravitational/electromagnetic radiation passing through them \cite{Braginsky:1987kwo,PhysRevLett.67.1486,Bieri:2013hqa}. The memory effect could be understood a transition between spontaneously broken vacua of the infinite dimensional asymptotic symmetry group \cite{Strominger:2017zoo}. The soft theorems are related to the memory effect through a Fourier transform since the Fourier transform of a pole in frequency space (coming from the soft factor) is a temporal step function \cite{Strominger:2014pwa}. Thus, deep in the infrared (IR), there is a notion of universality to the physics in asymptotically flat spacetimes which is solely captured by symmetries. For example, owing to the relation between soft theorems and memory effect, the IR physics of black hole collisions and the scattering of fundamental particles is essentially identical. One of the main observables of flat spacetimes is the scattering amplitude or the $S$-matrix. In \eqref{eq:mellinsmatrix}, we have seen how the action of asymptotic symmetries is more transparent in a different basis. Since in this basis, the conformal properties are more manifest, we will briefly review how the Ward identities of the 2D CFT encode the soft theorems in the bulk \cite{He:2014laa}. Before that, we will recap some constraints imposed by the asymptotic symmetries on some of the IR observables \cite{Strominger:2017zoo}.

\subsection{Infrared structures of asymptotically flat spacetimes}
\label{apssec:softsmatrix}

Soft theorems are universal statements regarding the low energy limit of scattering amplitudes involving massless particles. To be precise, in the limit when the energy of one of the massless particles goes to zero (goes soft), the scattering amplitude admits a universal factorization that is independent of the details of the other particles which are involved. Such statements were first discovered in QED by Low in 1958 \cite{PhysRev.96.1428,PhysRev.110.974} and they were later generalised to gravity by Weinberg \cite{PhysRev.140.B516}. Although the QED subleading soft theorem was known in 60s \cite{PhysRev.110.974}, its existence in gravity was discovered only in 2014 \cite{Cachazo:2014fwa}. We illustrate these statements and how they can be understood to arise from constraints of asymptotic symmetries below.

\medskip
  
Consider a scattering amplitude of $(n+1)$ particles involving one photon $\mathcal{M}_{n+1}$. Using Feynman diagrammatics, one can show that this amplitude admits the following factorization in the limit when the energy of the photon tends to zero \cite{PhysRev.110.974}:
\begin{align}
	\mathcal{M}_{n+1}(p_i,k)=\left[\frac{1}{\omega}\sum_{i=1}^{n}\frac{\epsilon_{\mu}e_ip_i^\mu}{p_i.\hat{k}}+\sum_{i=1}^{n}\frac{e_i\epsilon_{\mu}\hat{J}^{\mu\nu}k_{\nu}}{p_i.\hat{k}}+\mathcal{O}(\omega)\right]\mathcal{M}_n(p_i) \, .
\end{align}
In the above expression, $p_i$ and $k$, respectively, denote the momentum of hard and soft particles. $e_i$ denotes the charge of the particles. $\omega$  and $\epsilon_{\mu}$ are the energy and polarisation vector of the photon. $\hat{J}^{\mu\nu}$ is the angular momentum. 
Analogously, for gravitons, the expansion of scattering amplitudes is given by \cite{PhysRev.140.B516}
\begin{align} \label{gravisoft}
	\mathcal{M}_{n+1}(p_i,k)=\left[\frac{1}{\omega}\sum_{i=1}^{n}\frac{\epsilon_{\mu\nu}p_i^\mu p_i^\nu}{p_i.\hat{k}}+\sum_{i=1}^{n}\frac{\epsilon_{\mu\nu}p_i^\mu\hat{J}^{\nu\lambda}k_{\lambda}}{p_i.\hat{k}}+\mathcal{O}(\omega)\right]\mathcal{M}_n(p_i) \, .
\end{align}
We will now recap the notion of asymptotic symmetries as it was historically discovered by BMS \cite{Bondi:1962px,Sachs:1962wk,Sachs:1962zza} and explicitly show how the charges lead to a factorization of $S$-matrix as in \eqref{gravisoft}.

 \medskip

 We will discuss asymptotic symmetries in the context of Bondi gauge for asymptotically flat spacetimes with coordinates $(u,r,x_A)$ ($x^A$ denotes the coordinates on the sphere):
 \begin{align} \label{Bondi}
 	ds^2=-Udu^2-2e^{2\beta}dudr+g_{AB}\left(dx^A+\frac{1}{2}U^Adu\right)\left(dx^B+\frac{1}{2}U^Bdu\right) \, .
 \end{align}
 The gauge fixing conditions are $g_{rr}=g_{rA}=0$ and the condition on $g_{AB}$: 
\begin{equation}
	\partial_r\text{det}\left(\frac{g_{AB}}{r^2}\right)=0 \, .
\end{equation}
These four conditions would completely fix four diffeomorphisms. We now impose the notion of asymptotic flatness through
\begin{equation}
	\lim_{r\to \infty} ds^2 \to -du^2+2dudr+2 r^2 \gamma_{z\bar{z}} \, dz \,d\bar{z} \, .
\end{equation}
Here $\gamma_{z\bar{z}}$ is the metric on two sphere. However, there is no one-size-fits-all approach to determing the fall off conditions. As long as the conditions allow all the physically relevant solutions and eliminate the unphysical ones, they can be considered as appropriate boundary conditions. The original choice made by BMS \cite{Bondi:1962px,Sachs:1962wk,Sachs:1962zza} in the large $r$ falloff conditions are
\begin{equation}\label{falloff}
    \begin{split}
        g_{uu}=-1+\mathcal{O}\left(\frac{1}{r}\right), \qquad  g_{ur}=-1+\mathcal{O}\left(\frac{1}{r}\right), \qquad g_{uz}=\mathcal{O}(1)  \\ 
	g_{zz}=\mathcal{O}(r), \qquad g_{z\bar{z}}=r^2\gamma_{z\bar{z}}+\mathcal{O}(1), \qquad g_{rr}=g_{rz}=0 \, .
    \end{split}
\end{equation}
These specific fall off conditions constrains \eqref{Bondi} to be
\begin{align} \label{afs}
	ds^2=&-du^2-2du\,dr+2r^2\gamma_{z\bar{z}}\,dz\,d\bar{z} \\  \nonumber
	&+\frac{2m_B}{r}du^2+rC_{zz}\,dz^2+rC_{\bar{z}\bar{z}}\,d\bar{z}^2+D^zC_{zz}\,du\,dz+D^{\bar{z}}C_{\bar{z}\bar{z}}\,du\,d\bar{z}  \\   \nonumber
	&+\frac{1}{r}\left(\frac{4}{3}(N_z+u\partial_zm_B)-\frac{1}{4}\partial_z(C_{zz}C^{zz})\right)\,du\,dz+c.c.+...
\end{align}
\eqref{afs} is just a geometric constraint on asymptotically flat spacetimes because we have not imposed Einstein's equations. $m_B,N_z$ and $C_{zz}$ are arbitrary functions of $u,z,\bar{z}$ as of now. $m_B$ and $N_z$ are called Bondi mass aspect and angular momentum aspect respectively. $C_{zz}$ is the asymptotic shear tensor that characterizes the gravitational waves. These functions comprise the asymptotic data. The set of diffeomorphisms that preserves the fall off conditions in \eqref{falloff}, are the asymptotic symmetries \footnote{One should also mod out the set by trivial diffeomorphisms.} of 4 dimensional asymptotically flat spacetimes. The asymptotic Killing vector fields take the form \cite{Barnich:2009se,Barnich:2010ojg}
\begin{equation}\label{eq:asymkilling}
    \begin{split}
        \xi &= \left(f(z,\z) + \dfrac{u}{2} D_z Y^z \right) \partial_u + \left[ \left( 1 + \dfrac{u}{2r} \right) Y^z - \dfrac{1}{r} D^z f \right] \partial_z \\ &\qquad - \left[ \dfrac{u}{2r} D^\z D_z Y^z + \dfrac{1}{r} D^\z f \right] \partial_\z
         + \left[ D^z D_z f - \dfrac{1}{2}(u+r) D_z Y^z  \right] \partial_r + c.c. \, .
    \end{split}
\end{equation}
%The Killing vectors when taken all the way to the null boundary take the following form
%\begin{align}
	%\xi=f(z,\bar{z})\partial_u+ \big( Y^z\partial_z+\frac{u}{2}D_zY^z\partial_u\big)+\big(Y^{\bar{z}}\partial_{\bar{z}}+\frac{u}{2}D_{\bar{z}}Y^{\bar{z}}\partial_u\big) \, .
%\end{align}
 Here, $f(z,\bar{z})$ are arbitrary functions on 2-sphere and parametrise the supertranslations. $Y^z$ and $Y^{\bar{z}}$ are holomorphic and anti-holomorphic functions respectively. They are parametric functions associated to the superrotations. $D_z$ is the covariant derivative with respect to sphere metric $\gamma_{z\bar{z}}$.  Here, we have only discussed the case of $\mathscr{I}^{+}$, a similar story would follow for $\mathscr{I}^{-}$ as well.

 \medskip

 We now study the consequences of the Ward identities associated with these symmetries. We make a simplifying assumption by restricting the Killing vector of \eqref{eq:asymkilling} to be bounded in an orthonormal frame as you approach $\mathscr{I}^+$. This allows us to just separate out supertranslations. We will show below how they are responsible for the leading soft graviton theorem. The argument for superrotations encoding the sub-leading soft theorem is entirely analogous \cite{Kapec:2014opa,Kapec:2016jld}. The action of the supertranslations on the data of $\mathscr{I^{\pm}}$ is given by
 \begin{equation}
     \begin{split}
         \mathcal{L}_f  m_B&=f\partial_um_B+\frac{1}{4}[N^{zz}D^2_zf+2D_zN^{zz}D_zf+c.c.] \, ,  \\
 	\mathcal{L}_f C_{zz}&=f\partial_uC_{zz}-2D^2_zf \, .
     \end{split}
 \end{equation}
 $N_{zz}=\partial_u C_{zz}$, is the Bondi News tensor. Notice that if we start off with a vacuum solution with $m_B=C_{zz}=0$, then the supertranslations result in a spacetime with non-zero $C_{zz}$ (the asymptotic shear tensor gets shifted by $D^2_zf$). Now, a simple diffeomorphism should not change the curvature, and this demand leads to
\begin{equation}
	C_{zz}=-2D^2_z C(z,\bar{z}) \, , \qquad \mathcal{L}_f C(z,\bar{z})=f(z,\bar{z}) \, .
\end{equation}
$C(z,\bar{z})$ is the Goldstone mode (as it parametrizes classically inequivalent vacua) associated with broken supertranslation symmetry. As the operator $D^2_z$ kills the four global translations, these symmetries remain unbroken.

\medskip

We are now ready to define the scattering problem in GR, but there is a problem because the maximal Cauchy development of the data at $\mathscr{I}^-$ using Einstein's equation is not enough. This is because the development would determine the data at $\mathscr{I}^+$ up to a BMS$^+$ \footnote{BMS$^+$ and BMS$^-$ denotes the asymptotic symmetry group at $\mathscr{I}^{\pm}$ respectively.} frame. In \cite{Strominger:2013jfa}, Strominger proposed a matching condition that fixes the BMS$^+$ frame in terms of BMS$^-$. These conditions are given by 
\begin{equation} \label{matching conditions}
	C(z,\bar{z})|_{\mathscr{I}^+_-}=	C(z,\bar{z})|_{\mathscr{I}^-_+}\, , \qquad m_B(z,\bar{z})|_{\mathscr{I}^+_-}=m_B(z,\bar{z})|_{\mathscr{I}^-_+} \, .
\end{equation}
These are known as the antipodal matching conditions. Since the gravitational scattering is now well defined, the symmetries of gravitational scattering are only the ones that respect this antipodal matching. These conditions thus reduce the symmetries from $\text{BMS}^+ \times \text{BMS}^-$ to a diagonal subgroup that respects
\begin{equation}
		f(z,\bar{z})|_{\mathscr{I}^+_-}=	f(z,\bar{z})|_{\mathscr{I}^-_+} \, .
\end{equation}
 This matching condition implies an infinite number of conserved charges at celestial sphere on $\mathscr{I}^+_-$ and $\mathscr{I}^-_+$.  
 \begin{align}
 	&Q^+=Q^- \, ,   \qquad  \text{where} \\ \nonumber
 	&Q^+=\frac{1}{4\pi G}\int_{\mathscr{I}^+_-}d^2z \, \gamma_{z\bar{z}}\,f(z,\z) \,m_B(z,\z) \, , \quad Q^-=\frac{1}{4\pi G}\int_{\mathscr{I}^-_+}d^2z \,\gamma_{z\bar{z}} \, f(z,\z) \, m_B(z,\z) \, .
 \end{align}
In a quantum theory, these charges should commute with the $S$-matrix elements of flat spacetimes \cite{Kapec:2014opa} as the associated symmetries are respected by gravitational scattering: 
\begin{equation}
	Q^+\mathcal{S}-\mathcal{S}Q^-=0  \implies  \langle in|Q^+\mathcal{S}-\mathcal{S}Q^-|out\rangle=0 \, .
\end{equation}
 The above Ward identity can be shown to be equivalent to the leading soft graviton theorem \cite{He:2014laa} in \eqref{gravisoft}, i.e.
 \begin{equation} \label{leading}
 \lim_{\omega \to 0}	\omega \mathcal{M}_{n+1}(k,p_i)=\sum_{i=1}^{n}\frac{\epsilon_{\mu\nu}p_i^\mu p_i^\nu}{p_i.\hat{k}} \mathcal{M}_n(p_i) \, .
 \end{equation}
 Similarly conserved charges associated with superrotations follow from the anti-podal matching condition involving $N_z$. These charges are given by 
 \begin{equation}
 	Q^\pm=\frac{1}{8\pi G}\int_{\mathscr{I}^{\pm}}d^2z \, (Y^zN_z+Y^{\bar{z}}N_{\bar{z}}) \, .
 \end{equation}
The subleading soft graviton theorem follows from the associated Ward identity \cite{Kapec:2014opa,Kapec:2016jld}.
 
 \subsection{Hint of 2D CFT from the infrared } \label{Celestial CFT}
 
 In this subsection, we shall discuss how the soft theorems of scattering amplitudes play a role in constraining the holographic dual of flat spacetimes. From the action of the Lorentz group on the Celestial sphere, we have motivated the Mellin basis \eqref{eq:mellinsmatrix} which diagonalizes the boost generators: 
 \begin{align} \label{scattering}
 	\<\mathcal{O}_{1}(z_1,\bar{z}_1)\mathcal{O}_2(z_2,\bar{z}_2)...\mathcal{O}_n(z_n,\bar{z}_n)\>\sim \int_{0}^{\infty} \left( \prod_{i=1}^n d \omega_i \, \omega_{i}^{\Delta_i-1} \right)\mathcal{S}_n(\omega_i,z_i,\bar{z}_i,\sigma_i) \, .
 \end{align}
This is the principal holographic map in the Celestial approach \cite{Pasterski:2016qvg,Pasterski:2017kqt}. 
\medskip

 Let us now try to understand how the bulk soft theorems are realised in the Mellin space. In Mellin space, the energy is traded by the conformal weights $\Delta$ \eqref{eq:mellinweights}. It turns out that via the celestial map \eqref{scattering}, the poles in the soft expansion, are mapped to the poles in the conformal dimension $\Delta$ of the associated primary operators. For instance, one can extract the leading and subleading soft contribution $G_0(z,\bar{z})$ and $G_1(z,\bar{z})$  from a graviton conformal primary operator $G_{\Delta}(z,\bar{z})$ as \footnote{We choose to only work with the positive helicity graviton.}\cite{Donnay:2018neh}
 \begin{equation}\label{eq:celsoftprimaries}
 	G_0(z,\bar{z})=Res_{\Delta=1}G_{\Delta,\sigma=2}(z,\bar{z})\, , \qquad G_1(z,\bar{z})=Res_{\Delta=0}G_{\Delta,\sigma=2}(z,\bar{z}) \, .
 \end{equation}
  Using these leading and subleading graviton conformal primaries, one can define the following operators of the dual CFT \cite{Strominger:2013jfa,He:2014laa,Donnay:2018neh,Barnich:2013axa}
  \begin{align} \label{softop}
  	P(z,\bar{z})=\partial_{\bar{z}}G_0(z,\bar{z})\, , \qquad \bar{T}(\bar{z})=\int d^2w \frac{1}{(z-w)^4}G_1(w,\bar{w}) \, .
  \end{align}
 The integral transformation on $G_1(z,\z)$ is called the Shadow transformation \cite{Pasterski:2017kqt}. One peculiarity of this transformation is that it maps a primary operator of holomorphic weight $(h,\bar{h})$ to another one of $(1-h,1-\bar{h})$. Notice that the subleading soft graviton $G_1(z,\bar{z})$ has weights $(1,-1)$. Thus, after the shadow transformation it gives rise to another primary of weight $(0,2)$. The weights of the operator $P(z,\bar{z})$ are $(\frac{3}{2},\frac{1}{2})$.

\medskip
 
The leading and subleading soft theorems in \eqref{gravisoft} would correspond to the Ward identities of these 2D CFT operators defined in \eqref{softop}. If we plug in the leading soft graviton theorem \eqref{leading} in the holographic map  \eqref{scattering}, the insertion of $P(z,\bar{z})$ in the correlation functions leads to the following Ward identity \cite{Donnay:2018neh}
 \begin{equation} \label{celestialkac}
 	\langle P(z,\bar{z})\prod_{i=1}^{n} \Phi_{h_i,\bar{h}_i }(w_i,\bar{w}_i) \rangle =\sum_{i=1}^{n}\frac{1}{z-w_i} \,	\langle \prod_{i=1}^{n} \Phi_{h_i+\frac{1}{2},\bar{h}_i +\frac{1}{2}}(w_i,\bar{w}_i) \rangle \, .
 \end{equation}
 Subsequently, from the subleading soft theorem,  we shall have 
 \begin{equation} \label{Shadow}
 	\langle \bar{T}(\bar{z})\prod_{i=1}^{n} \Phi_{h_i,\bar{h}_i }(w_i,\bar{w}_i) \rangle= \sum_{i=1}^{n}\Big(\frac{\bar{h}_i}{(\bar{z}-\bar{w}_i)^2}+\frac{1}{\bar{z}-\bar{w}_i}\partial_{\bar{w}_i}\Big)\langle \prod_{i=1}^{n} \Phi_{h_i,\bar{h}_i }(w_i,\bar{w}_i) \rangle \, .
 \end{equation}
Thus, the subleading soft theorem implies the existence of a Virasoro stress tensor in Celestial CFT \cite{Kapec:2016jld}. This stress tensor defined in \eqref{softop} is known as the Shadow stress tensor. Similarly, the leading soft graviton theorem demands that the dual Celestial CFT must contain a current of weight $(\frac{3}{2},\frac{1}{2})$. However, it should be noted that the associated Ward identity \eqref{celestialkac} is unusual from the point of view of a 2D CFT as it shifts the conformal dimension of other hard particle insertions.

\newpage
\section{Extensions of Carroll: Supersymmetry and Higher Spin}

\subsection{Superconformal Carroll}
In $D=2$, one can look to supersymmetrize the conformal Carroll algebra. This can be done either in intrinsic way i.e. starting from the 2D CCA and grading the algebra or via limiting approach; starting with the superconformal algebra and taking appropriate scaling limit over the supercharges. We will follow the later one in this appendix. Our starting point would be 2D $\mathcal{N}=(1,1)$ superconformal algebra
\begin{align}
    &[\mathcal{L}_n, \mathcal{L}_m] = (n - m)\mathcal{L}_{n+m} + \frac{c}{12}(n^3 - n)\delta_{n+m, 0}\,, \nonumber\\
    &[\mathcal{L}_n, \mathcal{Q}_r] = \left(\frac{n}{2} - r\right) \mathcal{Q}_{n+r}\,, \\
    &\{\mathcal{Q}_r, \mathcal{Q}_s\} = 2\mathcal{L}_{r+s} + \frac{c}{3}\left(r^2 - \frac{1}{4}\right)\delta_{r+s, 0}\nonumber\,.
\end{align}
and similarly for the antiholomorphic part generated by the $\bar{\mathcal{L}}_n$ and $\bar{Q}_r$. Here $\mathcal{Q}_r,\bar{\mathcal{Q}}_r$ are the supercharges in the relativistic algebra. We have already seen the contraction of the bosonic part in section \ref{sec: Tale of two contraction}. Here we will discuss the fermionic sector. There exists two types of scaling: 
\subsection*{Homogeneous scaling}
As can be guessed from the nomenclature, both the supercharges are scaled in a similar way
\begin{eqnarray}
    Q_r^+ = \sqrt{\e}\mathcal{Q}_r\,,\quad Q_r^- = \sqrt{\e}\bar{\mathcal{Q}}_{-r}\,.
\end{eqnarray}
The resulting algebra including the bosonic part is given by, 
\begin{align}\label{homsuperbms}
&[L_n, L_m] = (n-m) L_{n+m} + \frac{c_L}{12} \, (n^3 -n) \delta_{n+m,0}, \nonumber\\
& [L_n, M_m] = (n-m) M_{n+m} + \frac{c_M}{12} \, (n^3 -n) \delta_{n+m,0}, \\
& [L_n, Q^\a_r] = \Big(\frac{n}{2} - r\Big) Q^\a_{n+r}, \quad \{Q^\a_r, Q^\b_s \} = \delta^{\a\b} \left[M_{r+s} + \frac{c_M}{6} \Big(r^2 - \frac{1}{4}\Big)  \delta_{r+s,0} \right]. \nonumber
\end{align}
All the other commutators vanish. One thing to note is that this algebra does not include a super-Virasoro subalgebra. 

\subsection*{Inhomogeneous scaling}
There is another way to scale the supercharges, done in the following way
\begin{eqnarray}
    G_r = \mathcal{Q}_r - i\bar{\mathcal{Q}}_{-r}\,,\quad H_r = \e\left(\mathcal{Q}_r + i\bar{\mathcal{Q}}_{-r}\right)\,.
\end{eqnarray}
This leads to the following algebra
\begin{align}\label{inhomsuperbms}
& [L_n, L_m] = (n-m) L_{n+m} + \frac{c_L}{12} (n^3 -n) \delta_{n+m,0}, \nonumber\\
& [L_n, M_m] = (n-m) M_{n+m} + \frac{c_M}{12} (n^3 -n) \delta_{n+m,0}, \\
& [L_n, G_r] = \Big(\frac{n}{2} -r\Big) G_{n+r}, \ [L_n, H_r] = \Big(\frac{n}{2} -r\Big) H_{n+r}, \ [M_n, G_r] = \Big(\frac{n}{2} -r\Big) H_{n+r}, \nonumber\\
& \{ G_r, G_s \} = 2 L_{r+s} + \frac{c_L}{3} \Big(r^2 - \frac{1}{4}\Big)   \delta_{r+s,0}, \ \{ G_r, H_s \} = 2 M_{r+s} + \frac{c_M}{3} \Big(r^2 - \frac{1}{4}\Big)   \delta_{r+s,0}.\nonumber
\end{align}
Notice that, in contrast to the algebra \eqref{homsuperbms}, this contains a super-Virasoro subalgebra. 

Both algebras \eqref{homsuperbms} and \eqref{inhomsuperbms} have appeared as asymptotic symmetries of 3D supergravity in AFS \cite{Barnich:2014cwa, Lodato:2016alv, Bagchi:2018ryy} and as worldsheet symmetries of tensionless superstrings \cite{Lindstrom:1990qb, Bagchi:2016yyf, Bagchi:2017cte, Bagchi:2018wsn}. 

\subsection{Higher spin}
The conformal Carroll algebra naturally extends to higher spin versions. This was first explored in terms of asymptotic symmetries of higher spin theories in 3D AFS \cite{Afshar:2013vka}. We present below the spin-three extension of the 2D CCA. The starting point in \cite{Afshar:2013vka} was to consider a $sl(3)_k \oplus sl(3)_{-k}$ Chern-Simons theory, $k$ being the Chern-Simons level. A canonical analysis led to the asymptotic symmetry algebra of flat space spin-3 gravity. This algebra, also called the BMW algebra is given by:
\begin{align}
[L_m, L_n] &= (m - n) L_{m+n} + \frac{c_L}{12} (m^3 - m)\, \delta_{m+n,0}  \\
[L_m, M_n] &= (m - n) M_{m+n} + \frac{c_M}{12} (m^3 - m)\, \delta_{m+n,0}  \\
[L_m, U_n] &= (2m - n) U_{m+n}  \\
[L_m, V_n] &= (2m - n) V_{m+n}  \\
[M_m, U_n] &= (2m - n) V_{m+n} \\
[U_m, U_n] &= (m - n)(2m^2 + 2n^2 - mn - 8) L_{m+n} 
+ \frac{192}{c_M} (m - n) \Lambda_{m+n} \\
&\quad - \frac{96(c_L + \frac{44}{5})}{ c_M^2} (m - n) \Theta_{m+n}
+ \frac{c_L}{12} m (m^2 - 1)(m^2 - 4)\, \delta_{m+n,0}  \\
[U_m, V_n] &= (m - n)(2m^2 + 2n^2 - mn - 8) M_{m+n}
+ \frac{96}{c_M} (m - n) \Theta_{m+n}  \\
&\quad + \frac{c_M}{12} m (m^2 - 1)(m^2 - 4)\, \delta_{m+n,0} \,.
\end{align}
Here, $L_n, M_n$ are the bms$_3$ generators, which generate superrotations and supertranslations respectively. $U_n, V_n$ are the spin-3 version of superrotations and supertranslations respectively. In the above algebra $\Lambda_m = \sum_p :L_p M_{m-p}: - \frac{3}{10}(m+2)(m+3) M_m\quad \text{and}\quad 
\Theta_m = \sum_p M_p M_{m-p}$ are the composite operators. The central charges can be found to be  
\begin{eqnarray}
    c_L =0\,,\quad c_M = 12k\,.
\end{eqnarray}
This algebra also arises from an Inönü–Wigner contraction of two copies of the $W_3$ algebra, similar to how BMS$_3$ contracts from Virasoro $\times$ Virasoro. In \cite{Fuentealba:2015wza}, the asymptotic structure of 3D hypergravity was studied, where it was shown that, in the absence of cosmological constant, gravity coupled to a spin $5/2$ field leads to an asymptotic symmetry algebra given by a hypersymmetric non-linear extension of BMS$_3$.

\newpage

\newpage
\bibliographystyle{JHEP}
\bibliography{review-refs}

\end{document}